\documentclass[journal]{IEEEtran}

\usepackage{amsmath}
\usepackage{amsfonts}
\usepackage{amssymb}
\usepackage{verbatim}
\usepackage{multirow}
\usepackage{graphicx}
\usepackage{subfigure}
\usepackage{algorithm}
\usepackage{algorithmic}
\usepackage{color}
\usepackage{cite}

\usepackage{pstricks}
\usepackage{pst-node}
\usepackage{booktabs}
\usepackage{threeparttable}
\usepackage{rotating}
\usepackage{url}
\usepackage{epsfig}
\usepackage{epstopdf}

\begin{document}

\title{Pansharpening via Detail Injection Based Convolutional Neural Networks}

\author{
        Lin He,~\IEEEmembership{Member,~IEEE,}
        Yizhou Rao,~\IEEEmembership{}
        Jun Li,~\IEEEmembership{Senior Member,~IEEE}
        and Antonio Plaza,~\IEEEmembership{Fellow,~IEEE}
        Jiawei Zhu,~\IEEEmembership{}

\thanks{Lin He, Yizhou Rao and Jiawei Zhu are with the School of Automation Science and Engineering, South China University of Technology, Guangzhou, 510640, China.
Jun Li is with the Guangdong Provincial Key Laboratory of Urbanization and Geo-simulation, School of Geography and Planning, Sun Yat-sen University, Guangzhou, 510275, China.
Antonio Plaza is with the Hyperspectral Computing Laboratory, Department of Technology of Computers and Communications, Escuela Polit\'{e}cnica, University of Extremadura, C\'{a}ceres, E-10071, Spain.
}

}

\markboth{Journal of \LaTeX\ Class Files,~Vol.~xx, No.~x, xx~20xx}%
{Shell \MakeLowercase{\textit{et al.}}: Bare Demo of IEEEtran.cls for Journals}

\maketitle

\begin{abstract}
Pansharpening aims to fuse a multispectral (MS) image with an associated panchromatic (PAN) image, producing a composite image with the spectral resolution of the former and the spatial resolution of the latter. Traditional pansharpening methods can be ascribed to a unified detail injection context, which views the injected MS details as the integration of PAN details and band-wise injection gains. In this work, we design a detail injection based CNN (DiCNN) framework for pansharpening, with the MS details being directly formulated in end-to-end manners, where the first detail injection based CNN (DiCNN1) mines MS details through the PAN image and the MS image, and the second one (DiCNN2) utilizes only the PAN image. The main advantage of the proposed DiCNNs is that they provide explicit physical interpretations and can achieve fast convergence while achieving high pansharpening quality. Furthermore, the effectiveness of the proposed approaches is also analyzed from a relatively theoretical point of view. Our methods are evaluated via experiments on real-world MS image datasets, achieving excellent performance when compared to other state-of-the-art methods.
\end{abstract}


\IEEEpeerreviewmaketitle

\section{Introduction}

Due to the physical characteristics of multispectral (MS) image sensors, they generally acquire MS images with limited spatial resolution. However, high spatial resolution MS images are required in many applications, such as classification, target detection, scene interpretation and spectral unmixing \cite{Survey:Vivone2015,Survey:Alparone2007}. Therefore, pansharpening has been an active area of research, drawing significant attention in remotely sensed image processing. The pansharpening task aims at fusing a low spatial resolution MS image and a registered wide-band panchromatic (PAN) image, utilizing the detail information contained in the PAN image to sharpen the MS image, hence yielding a high spatial resolution MS image \cite{Survey:Vivone2015}. The task can be seen as a special reconstruction based on different types of data with different characteristics. For simplicity, low spatial resolution MS images and high spatial resolution MS images are hereinafter called LRMS images and HRMS images, respectively. A HRMS image pansharpened from the LRMS image is called pansharpened HRMS image hereinafter. Ideally, as a full resolution image, the pansharpened HRMS image should have the same spectral resolution as the original LRMS image and the same spatial resolution as the corresponding PAN image.

Over past decades, a wide variety of pansharpening methods have been proposed in the literature \cite{Survey:Vivone2015,Survey:Loncan2015,Survey:Alparone2007,Aiazzi2002Context}. Among such existing methods, component substitution (CS) and multi-resolution analysis (MRA) are two widely representative categories \cite{Survey:Vivone2015,Survey:Loncan2015,Survey:Alparone2007}. CS approaches usually replace certain components of the MS image with those from the PAN image in a given domain, which include principal component analysis (PCA) based pansharpening \cite{PCA:Jr1989,Shettigara1992A,Shah2008An},Brovey transform (BT) based pansharpening \cite{Brovey:Gillespie1987,Brovey:Tu2001} and Gram-Schimidt (GS) transform based pansharpening \cite{GS:Laben2000,GS:Aiazzi2007}, etc. In contrast, MRA methods exploit the spatial detail information through the multiresolution decomposition of the images, which generally involves detail extraction and detail integration in multiple scales. Examples are pansharpenings based on decimated wavelet transform (DWT)\cite{DWT:Khan2008}, undecimated wavelet transform (UDWT) \cite{Nason1995The}, \textit{a tr\'ous} wavelet transform (ATWT)\cite{ATWT:Ranchin2000,AWLP:Otazu2005,WDT:Vivone2014} and Laplacian pyramid (LP)\cite{MTF:Aiazzi2003,MTF:Aiazzi2006,MTF:Lee2010}. The aforementioned methods differ mainly in how spatial details are extracted from the PAN image and how they are injected into the pre-interpolated LRMS image. One major challenge for CS/MRA approaches is to preserve spatial details resolved from the PAN image as much as possible, while avoiding spectral distortion. This refers to the spectral deviation from an ideal spectrum, especially when PAN and MS images are acquired in spectral ranges that overlap only partially \cite{CNN-Pansharp:Masi2016,Survey:Vivone2015}. Unfortunately, existing CS/MRA methods are often prone to significant spectral distortion \cite{Survey:Loncan2015}, even under some improvement of fusion strategies, such as histogram matching \cite{Histogram:Thomas2008}, weighted detail injection \cite{WDT:Vivone2014}, or some hybrid intermediate processes \cite{GFPCA:Liao2015}. This is probably due to the fact that the details are not very effectively learned and injected, although CS/MRA approaches indeed aim to utilize the detail information.

Recently, convolutional neural networks (CNNs) start prevailing in image enhancement tasks such as super-resolution \cite{SuperRes-CNN:Dong2016,Dong2014Learning} and pansharpening \cite{CNN-Pansharp:Masi2016,Wei2017Boosting}. Super-resolution is, to some degree, a pansharpening-related task, as both super-resolution and pansharpening aim to enhance image resolution. There are however differences among them, since the former is usually a single input single output (SISO) process while the latter is a multiple input single output (MISO) case. Dong \textit{et al.} proposed a super-resolution CNN (SRCNN) which is a three-layer CNN to learn the mapping from the input low-resolution image to the output high-resolution image \cite{SuperRes-CNN:Dong2016}. Kim \textit{et al.} designed a deep CNN structure for super-resolution, where the residual component is learned \cite{SuperRes-CNN:Kim2016}. Whether or not details are injected from a PAN image to its associated LRMS image represents the major difference between pansharpening and super-resolution tasks. Considering such, Masi \textit{et al.} presented a pansharpening CNN (PNN) following the basic thread of SRCNN \cite{CNN-Pansharp:Masi2016}, where the pre-interpolated LRMS image is stacked with the PAN image at the input layer, and then a CNN process is used to learn the relationship between the input and the pansharpened HRMS image. Although PNN exhibits good performance on real remotely sensed data, difficulties arise from the long-time training iterations and the problem that it misses the domain specific pansharpening structure and roughly treats pansharpening as a black-box learning procedure. Afterwards,
Wei \textit{et al.} designed a CNN method for pansharpening \cite{Wei2017Boosting}. The method comprises the process of residual learning and the subsequent dimension reduction, which is faced with the problems that the learned residual has no explicit physical interpretation for pansharpening and there is an additional computation load related to dimension reduction. They also introduce strategies like multiscale kernels into the CNN based pansharpening\cite{Yuan2018A}.

In this paper, we develop a new technique aimed to address the limitations of existing works. Our main contributions are twofold. On the one hand, we build a new general detail injection pansharpening framework, called DiPAN, which aims at clear interpretability and intuitive motivation. On the other hand, in the context of our newlyde developed DiPAN framework, we develop detail injection based CNNs (DiCNNs) for MS detail learning. Contributions of our work can be summarized as follows:

\begin{enumerate}
\item The first method, called DiCNN1, adopts a framework in which the pathway of stacked convolutional layers only learns the MS details from the combination of the pre-interpolated LRMS image and the PAN image in an end-to-end manner, resulting in good initialization. The method, following the basic idea in our previous conference paper \cite{rao2017residual}, has clear interpretability in the detail injection context, and can greatly reduce the uncertainty of learning, thus achieving high computational efficiency and pansharpening quality. Detailed description, discussion and experimental results are provided in this work. Furthermore, we present a relatively theoretical analysis and proof of the effectiveness of DiCNN1. To the best of our knowledge, the effectiveness of a parsharpening CNN has not been previously explored from such a relatively theoretical point of view.

          \item The second method, called DiCNN2, works under the assumption that ideal MS detail is only relevant to the PAN image, and directly uses the PAN image as the input of the convolutional layer pathway. DiCNN2 exhibits some benefits with regards to DiCNN1. First, DiCNN2 can be used to realize transfer learning when there occur bad bands in test MS images. Second, DiCNN2 can achieve even higher computational efficiency than DiCNN1, since its input is a one-dimensional PAN image only, yielding less amount of CNN free parameters than DiCNN1 in the first convolutional layer.

              \item A relatively general detail injection formulation is summarized, which is able to accommodate CS/MRA pansharpening methods, as well as the proposed DiCNNs. The formulation can be used as a domain-specific structure to guide the design of new pansharpening methods.

        \end{enumerate}

The remainder of the paper is organized as follows. Section \ref{sec:DetailInjection} introduces the detail injection framework. Section \ref{sec:SuperPanMethods} summarizes major existing CNN-based super-resolution and pansharpening methods. Section \ref{sec:Methods} proposes our detail injection based CNN pansharpening methods and presents the corresponding complexity analysis. Section \ref{sec:ExperiResults} evaluates the proposed methods via experiments with real MS data sets. Section \ref{sec:Conclusions} concludes the paper with some remarks and hints at plausible future research lines.

\section{Detail Injection Framework} \label{sec:DetailInjection}

Let ${\bf P}\in\mathbb{R}^{H \times W}$ denote an observed PAN image with size $H \times W$; let ${\bf \widetilde{M}}\in\mathbb{R}^{H \times W\times N_b}$ be a pre-interpolated LRMS, which has been interpolated spatially to the scale of the PAN image (with $N_b$ being the number of bands); and let ${\bf\widehat{M}}$ be the pansharpened HRMS image.

\begin{figure*}[t]\scriptsize
\centering
\includegraphics[width=5in]{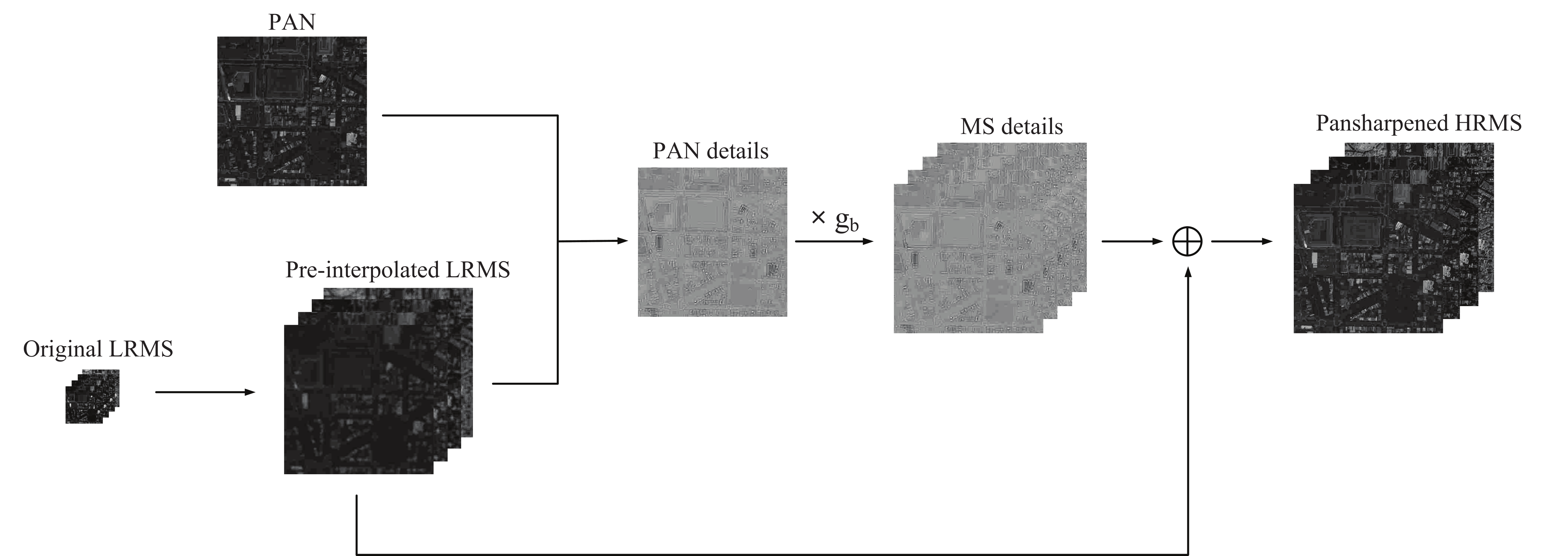}
\caption{Schematic diagram of the DiPAN framework.}
\label{fig:DetailInjection}
\end{figure*}

Traditionally, CS/MRA methods are viewed as two major groups of pansharpening methods \cite{Survey:Vivone2015}. CS category can be generally formulated as:
\begin{equation}
   \label{eq:CS1} 
   {\widehat {\mathbf {M}}}_b={ \widetilde {\mathbf{M}}}_b+ g_b\cdot(\mathbf{P}-\mathbf {I}_c),\:\:\:\:\:\:\:\:\: b=1,\cdots,N_b,
\end{equation}
where ${\widehat {\mathbf {M}}}_b$ and ${ \widetilde {\mathbf{M}}}_b$ are the $b$th bands of ${\bf \widehat{M}}$ and ${\widetilde {\mathbf{M}}}$, respectively, $g_b$ represents the injection gain associated with ${\bf \widetilde{M}}_b$, $N_b$ is the number of MS bands, and $\mathbf{I}_c$ is the intensity component of the MS image which is often a weighted sum $\mathbf{I}_c=\sum\limits_{b=1}^{N_b}\omega_b\widetilde {\mathbf{M}}_b$. To show the substitution process in CS methods, (\ref{eq:CS1}) can be reformulated as:
\begin{equation}
\begin{split}
   \label{eq:CS2} 
   {\widehat {\mathbf {M}}_b}
   =&{ \widetilde {\mathbf{M}}_b-\mathbf{I}_c}+ g_b\cdot(\mathbf{P}-\mathbf{I}_c)+\mathbf{I}_c\\
   =&(\widetilde {\mathbf{M}}_b-\mathbf{I}_c)+ g_b\cdot(\mathbf{P}-\frac{g_b-1}{g_b}\mathbf{I}_c),
   \end{split}
\end{equation}
which suggests that, in a CS method, the component $\mathbf{I}_c$ is substituted with the component $g_b\cdot(\mathbf{P}-\frac{g_b-1}{g_b}\mathbf{I}_c)$.
On the other hand, the general formulation of MRA methods is of the form \cite{Survey:Vivone2015}:
\begin{equation}
   \label{eq:MRA} 
   {\widehat {\mathbf {M}}_b}={ \widetilde {\mathbf{M}}_b}+ g_b\cdot(\mathbf{P}-\mathbf {P}_c),\:\:\:\:\:\:\:\:\: b=1,\cdots,N_b,
\end{equation}
where $\mathbf {P}_c$ denotes the low-frequency component of the PAN image, which is usually obtained in a MRA way. According to the representations in (\ref{eq:CS1}) and (\ref{eq:MRA}), both CS and MRA methods are normally based on two sequential phases: i) the extraction of MS details from the PAN image, which usually comprises intermediate processes of yielding PAN details and obtaining band injection gains, and ii) the injection of the MS details into the LRMS image to produce HRMS image. Therefore, such two categories of pansharpening methods can be represented in an unified detail injection framework, namely DiPAN, as follows:
\begin{equation}
   \label{eq:detail_injection} 
   \begin{split}
   {\widehat {\mathbf {M}}}_b &= { \widetilde {\mathbf{M}}}_b  + g_b\cdot\mathbf{d}\\
                              &= { \widetilde {\mathbf{M}}}_b  + {\mathbf D}_b,
   \end{split}
\end{equation}
where ${\mathbf d}$ represents the PAN details which are usually calculated by involving both the PAN image and the MS image with a certain criterion, ${\mathbf D}_b=g_b\cdot{\mathbf d}$ denotes the MS details which should complement the pre-interpolated LRMS image ${\bf \widetilde{M}}$, while $g_b$ stands for the associated injection gain responsible for transferring the PAN details to the MS details. A schematic diagram of DiPAN is given in Fig. \ref{fig:DetailInjection}, where it is indicated that the full-resolution pansharpened HRMS image ${\bf \widehat{M}}$ can be decomposed into the MS details and the LRMS approximation.

As DiPAN formulation in (\ref{eq:detail_injection}) has clear physical interpretability for the pansharpening process, it can be used as a pansharpening domain-specific structure to guide the design of new pansharpening methods.

\section{Super-resolution and Pansharpening using CNN strategy \label{sec:SuperPanMethods}}

Recently, CNNs were successfully applied in image super-resolution and pansharpening. CNNs are usually treated as the descendants of traditional artificial neural networks \cite{CNN:Fukushima1988,CNN:Lecun1989,CNN:Lecun1998}, in which assumptions such as limited receptive field (processing input only in a neuron's local neighborhood) and the spatial invariant weight (so-called weight sharing) are normally jointly employed.

The response of a convolutional layer in a CNN can be given by:
\begin{equation}
   \label{eq:convolution layer} 
    \mathbf{Y}_{l} =  \varphi (\mathbf{W}_{l}*\mathbf{X}_{l} + \mathbf{B}_{l}),
\end{equation}
where $*$ denotes the convolution operation, $\mathbf{X}_l$ and $\mathbf{Y}_l$ are the input and output of the $l$th layer, respectively, $\mathbf{W}_l$ and $\mathbf{B}_l$ are the weight and bias metrics, respectively, and $\varphi(\cdot)$ represents the activation function.
Due to the ability to mitigate gradient vanishing and its computational simplicity, the rectified linear unit (ReLU) \cite{Nair2010Rectified} is commonly used in CNNs, whose input-output relation is $\mathbf{Y}_l=\max(0,\mathbf{X}_l)$ \cite{CNN-Hyperspectral:Hu2015,SuperRes-CNN:Dong2016,CNN-Hyperspectral:Chen2016,CNN-Multiscale:Jia2017}.

Both image super-resolution and pansharpening are tasks to recover high-resolution images from the observed low-resolution data, with the major disparity being that one is a SISO process and the other one is MISO. In image super-resolution, usually the low spatial resolution image (as a single input) is processed to output a high spatial resolution image, while pansharpening utilizes the MS image with low spatial resolution and the PAN image with low spectral resolution as two separate data sources to recover the full resolution HRMS image. The two kinds of image resolution enhancements above are used as mathematical tools to minimize the loss function of expected square error:
\begin{equation}
\label{eq:losssr}
\ell(\pmb{\theta}) =  E\|\widehat{{\mathbf{H}}}(\mathbf{X};\pmb{\theta}) - \mathbf{Y}\|^2_F,
\end{equation}
where $\widehat{\mathbf{H}}$ is the predicted high-resolution image following a parametric structure, $\mathbf{Y}$ is the ideal high-resolution image, $\pmb{\theta}$ denotes the parameters used to infer the predicted image, and $\mathbf{X}$ is the low-resolution input, which means a low spatial resolution image for image super-resolution task that represents both the low spectral resolution PAN image and the associated LRMS image for pansharpening task.

Dong \emph{et al.} designed a three-layer CNN for image super-resolution able to directly learn the mapping between the low-resolution image and the high-resolution image, which is called super-resolution convolutional neural network (SRCNN) \cite{SuperRes-CNN:Dong2016}. Therein patch extraction and representation are used to improve computational efficiency and feature locality in the training phase. The objective is to minimize the following patch-wise mean square error:
\begin{equation}
\label{eq:losssrcnn}
\begin{split}
\ell(\pmb{\theta}) &= E \|\widehat{\mathbf{H}}(\mathbf{X};\pmb{\theta}) - \mathbf{Y}\|^2_F\\
&= \frac{1}{N_p} \sum_{i=1}^{N_p} \|\widehat{\mathbf{H}}^{(i)}(\mathbf{X}^{(i)};\pmb{\theta}) - \mathbf{Y}^{(i)}\|^2_F,
\end{split}
\end{equation}
where $i$ is the index of patches, $N_p$ denotes the number of total patches, $\pmb{\theta}$ represents the free CNN parameters to be optimized under the CNN context, $\mathbf{X}^{(i)}$ refers to the $i$th patch in the low-resolution image, and $\widehat{\mathbf{H}}^{(i)}$ stands for the $i$th patch in the predicted high-resolution image. As this CNN's counterpart for pansharpening, Masi  \emph{et al.} introduced a pansharpening CNN (PNN)\cite{CNN-Pansharp:Masi2016}, which stacks the pre-interpolated LRMS image and the PAN image together and then uses CNN to mine the mapping between this concatenation and real HRMS image.

The loss function to be minimized is:
\begin{equation}
\label{eq:losspnn}
\begin{split}
\ell(\pmb{\theta})  &= E\|{\widehat{\mathbf {M}}}({\mathbf{G}};\pmb{\theta}) - {\mathbf{Y}}\|^2_F\\
&= \frac{1}{N_p} \sum_{i=1}^{N_p} \|{\widehat {\mathbf {M}}}^{(i)}({\mathbf{G}^{(i)}};\pmb{\theta}) - {\mathbf{Y}^{(i)}}\|^2_F,
\end{split}
\end{equation}
where ${\bf{G} = ({\bf \widetilde{M}},\bf P})$ in the size $H \times W \times (N_b+1)$ denotes the concatenation of the pre-interpolated LRMS image $\bf \widetilde{M}$ and the PAN image $\bf P$ along the band dimension. Here, the target $\mathbf{Y}$ stands for the ideal HRMS for the pansharpening case. Considering that MS images are in 3D data arrangement, $\widehat{\mathbf {M}}$ and $\mathbf{Y}$ are originally 3-way or third-order tensors \cite{kolda2009tensor}. To accommodate matrix representation, $\widehat{\mathbf {M}}$ and $\mathbf{Y}$ in (\ref{eq:losspnn}) are unfolded as matrices, for example along the first mode and being denoted as $\widehat{\mathbf {M}}_{(1)}$ and $\mathbf{Y}_{(1)}$ \cite{kolda2009tensor}. But, for simplicity, $\widehat{\mathbf {M}}$ and $\mathbf{Y}$ in (\ref{eq:losspnn}) represent their unfolding, matrices $\widehat{\mathbf {M}}_{(1)}$ and $\mathbf{Y}_{(1)}$, respectively. If not stated otherwise, the remaining part of the paper follows the same expression routine when involving 3-way tensor representation.

The deep residual network (ResNet) has reached excellent performance in image classification \cite{Resnet:He2015}. Its success largely stems from attaching an identity skip connection to fit a residual mapping. Kim \emph{et al.} extended ResNet and proposed a deep network for super-resolution, which intends to learn the residual supplementary to the input low-resolution image instead of the predicted high-resolution image itself \cite{SuperRes-CNN:Kim2016}. The loss function is defined as shown below:
\begin{equation}
\label{eq:lossvdsr}
\begin{split}
\ell(\pmb{\theta})  &= E \|\widehat{\mathbf{R}}(\mathbf{X};\pmb{\theta}) + \mathbf{X} - \mathbf{Y}\|^2_F\\
&= \frac{1}{N_p} \sum_{i=1}^{N_p} \|\widehat{\mathbf{R}}^{(i)}(\mathbf{X}^{(i)};\pmb{\theta}) + \mathbf{X}^{(i)} - \mathbf{Y}^{(i)}\|^2_F,
\end{split}
\end{equation}
where $\mathbf{R}$ represents the residual need to learn. Later, Wei \emph{et al.} used a similar strategy for pansharpening, termed deep residual pansharpening neural network (DRPNN)\cite{Wei2017Boosting}. In the DRPNN, the concatenation of the pre-interpolated LRMS image and the PAN image pass through both stacked layers and a shortcut connection to yield the residual and then an additional convolutional layer is performed for dimensionality reduction. The connected objective is to minimize the follwoing loss:
\begin{equation}
\label{eq:DRPNN}
\begin{split}
\ell(\pmb{\theta})  &= \|{\omega}(\widehat{\mathbf{R}}(\mathbf{G};\pmb{\theta}) + \mathbf{G}) - {\mathbf{Y}}\|^2_F\\
&= \frac{1}{N_p} \sum_{i=1}^{N_p} \|{\omega}(\widehat{\mathbf{R}}^{(i)}(\mathbf{G}^{(i)};\pmb{\theta}) + \mathbf{G}^{(i)}) - {\mathbf{Y}^{(i)}}\|^2_F,
\end{split}
\end{equation}
where ${\omega}(\cdot)$ denotes a convolution operation for dimensional matching.

In comparison with the CS/MRA approaches, CNNs provide a new possibility to perform learning for pansharpening, where the details are driven from the context. However, in comparison with DiPAN, the main limitation of the aforementioned CNN-based pansharpening approaches is the lack of physical interpretability, and the fact that they do not use an appropriate domain-specific structure. The weaknesses are, specifically,
\begin{itemize}
\item PNN treats pansharpening merely as a black-box learning procedure, without considering the domain-specific structure useful to pansharpening, which results in a long-time training process and limited learning ability.
\item DRPNN involves the structure of residual and the subsequent dimension reduction, which is faced with the problem that the processed residual has no explicit physical interpretation in pansharpening context, and there is additional computational load for dimension reduction.
\end{itemize}

\section{Proposed Methods}\label{sec:Methods}

Based on the DiPAN framework in Section \ref{sec:DetailInjection}, we develop two detail injection based CNN (DiCNNs) for pansharpening. The advantages of the proposed DiCNNs are as follows.
\begin{itemize}
\item We take into consideration the detail structure used in traditional CS/MRA based pansharpening and then directly learn MS details, without separating the PAN details and the connected gains, which allows us to circumvent the intermediate process to learn such two kinds of information individually, thus reducing the model uncertainty.

\item Compared to existing CNN pansharpening methods, our newly proposed methods have clear and meaningful interpretation in the context of detail injection, which can also achieve excellent learning performance.

\end{itemize}

\subsection{DiCNN1}
Following DiPAN, our pansharpening method focuses on reconstructing the MS details in a CNN network manner. To achieve this goal, we build a feedforward neural network where a shortcut connection skips three stacked convolution layers and the output of the shortcut is added to the output of stacked layers to yield the predicted HRMS [as shown in Fig. \ref{fig:NetStructure}(a)]. This network employs the concatenation of the pre-interpolated LRMS and the PAN images as the input. However, only the pre-interpolated LRMS is propagated through the shortcut connection. In this way, the stacked layers utilize the interaction of the pre-interpolated LRMS and PAN images to yield only the MS details  that can further supplement the LRMS image to produce the pansharpened HRMS image. Specifically, our objective is to minimize the following loss function:

\begin{figure}[ht]
\centering
\subfigure[]{
\label{fig:subfig:b}
\includegraphics[width=8cm,trim=0   0   0   0, clip]{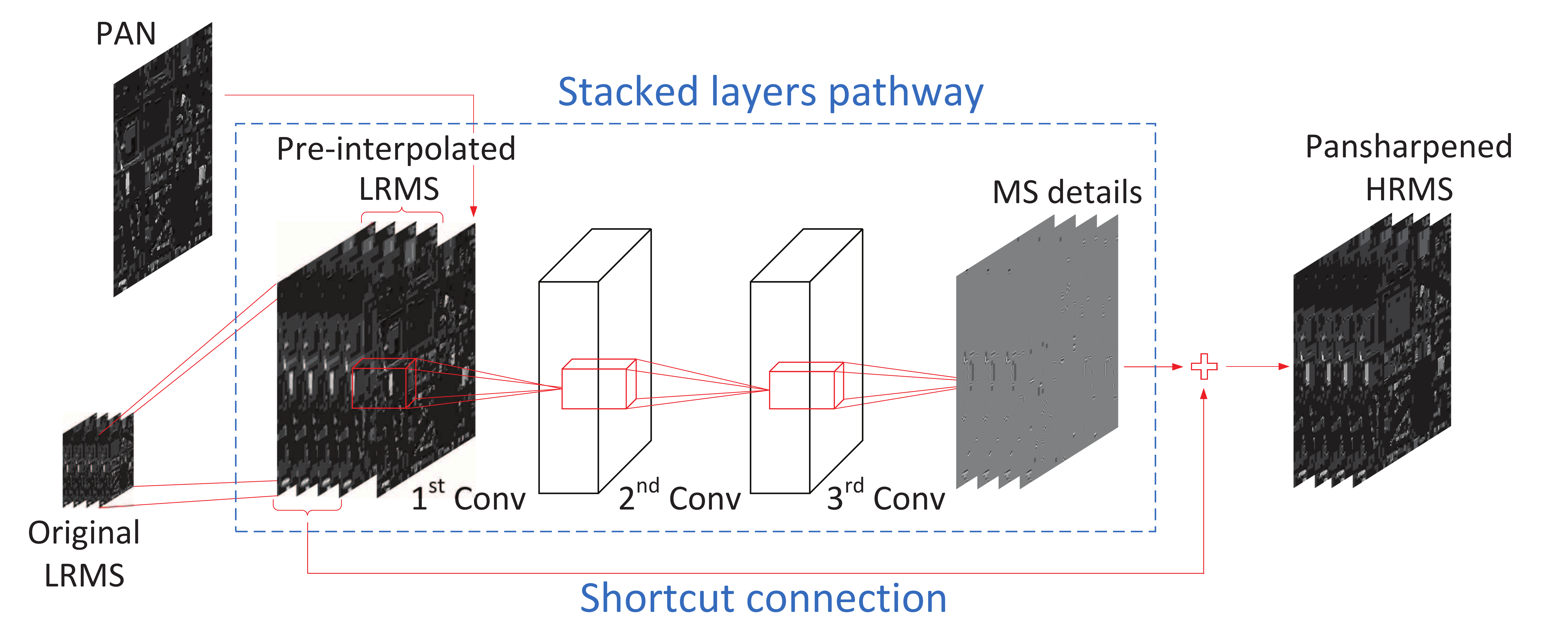}}
\\[-1pt]

\subfigure[]{
\label{fig:subfig:d}
\includegraphics[width=8cm,trim=0   0   0   0, clip]{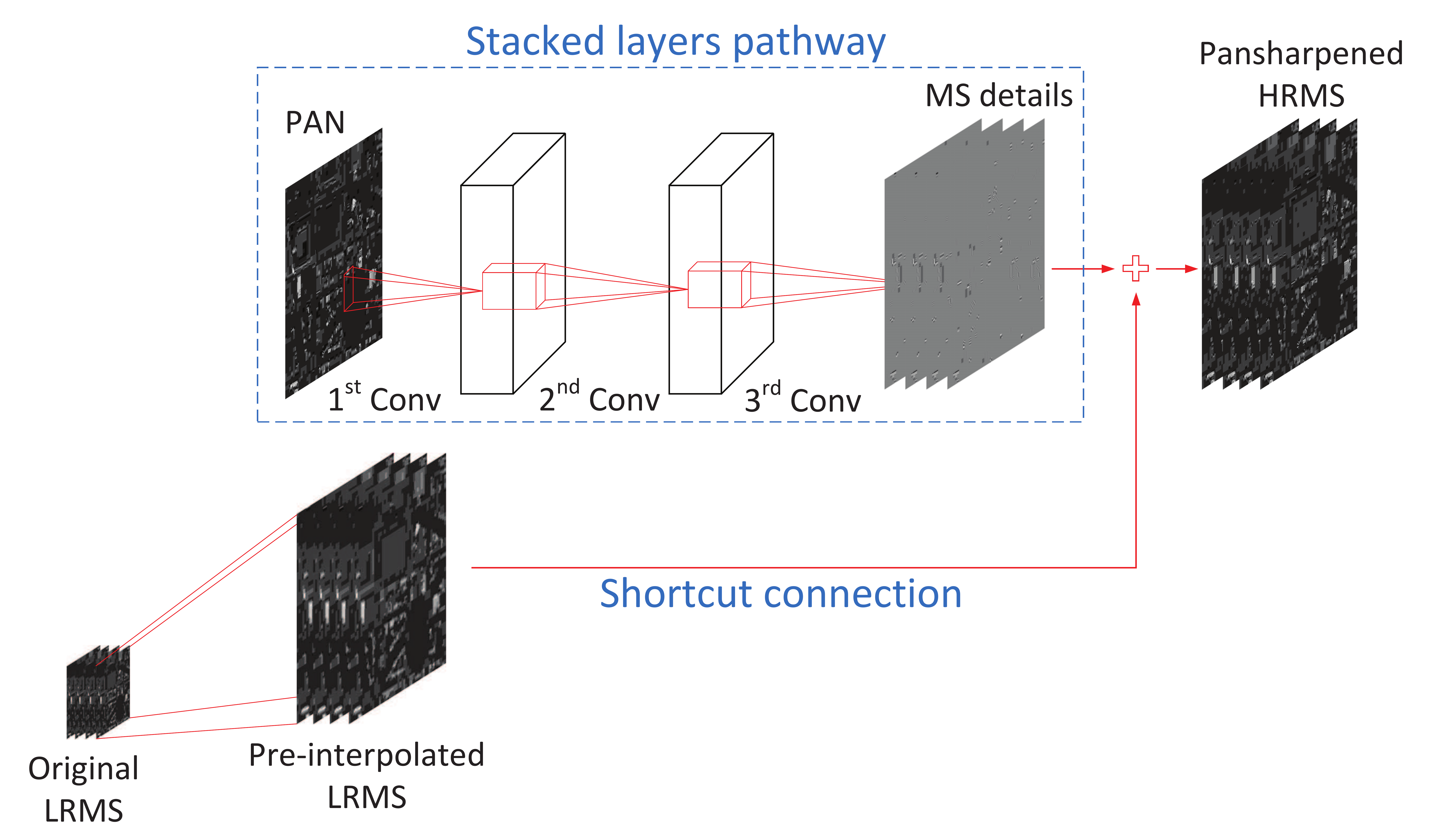}}
\\[-1pt]

\caption{Architectures of (a) DiCNN1 and (b) DiCNN2.}
\label{fig:NetStructure}
\end{figure}

\begin{equation}
\label{eq:DICNN1}
\begin{split}
\ell(\pmb{\theta})
=& \|\widehat{\mathbf{D}}(\mathbf{G};\pmb{\theta}) + \widetilde {\mathbf{M}} - {\mathbf{Y}}\|^2_F\\
=&\frac{1}{N_p} \sum_{i=1}^{N_p} \|\widehat{\mathbf{D}}^{(i)}(\mathbf{G}^{(i)};\pmb{\theta}) + \widetilde {\mathbf{M}}^{(i)} - {\mathbf{Y}^{(i)}}\|^2_F,
\end{split}
\end{equation}
where $\widehat{\mathbf{D}}$ represents the MS details reconstructed with the input ${\mathbf{G}}$, the concatenation of the LRMS image and the PAN image, and the parameter $\pmb{\theta}$.

\begin{figure*}[ht]\scriptsize
\centering
  \begin{tabular}{cccc}
   \includegraphics[height=2.5in]{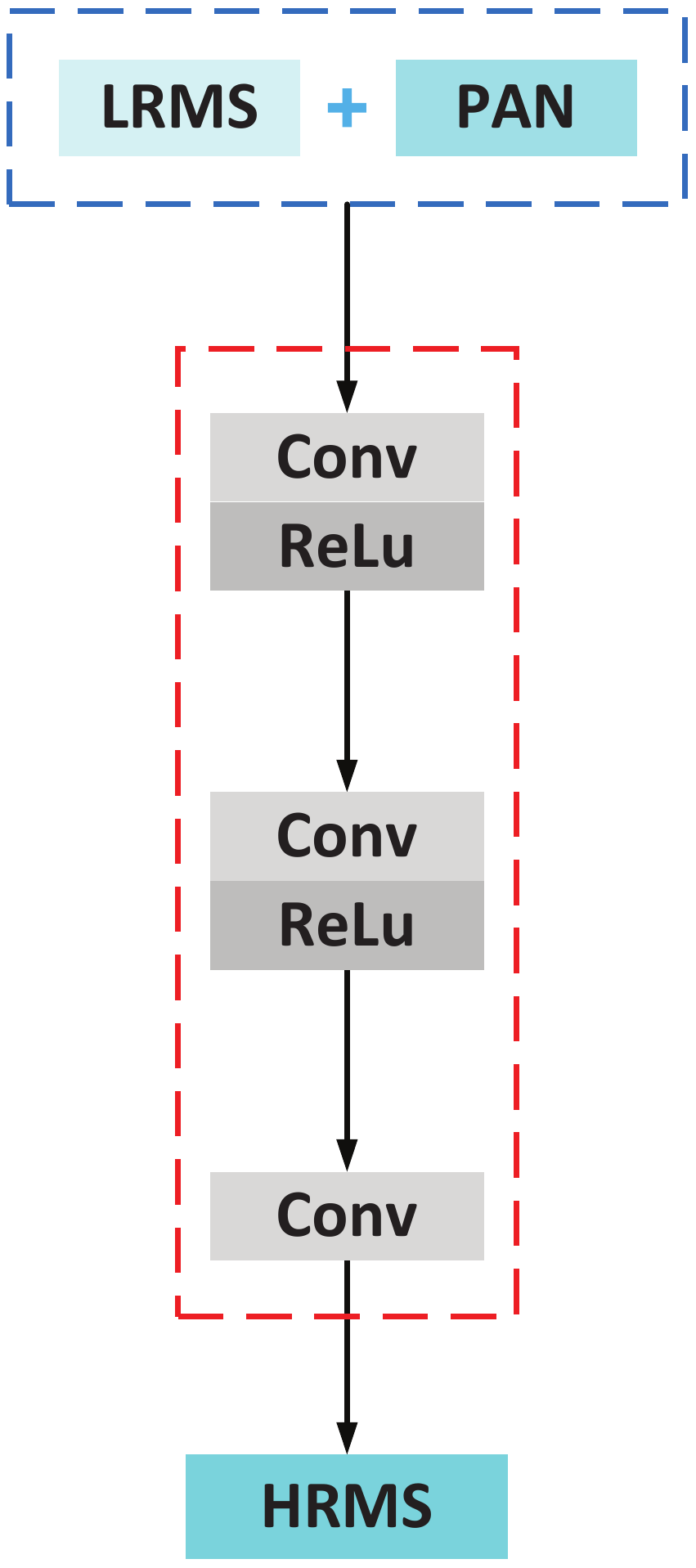} &
   \includegraphics[height=2.5in]{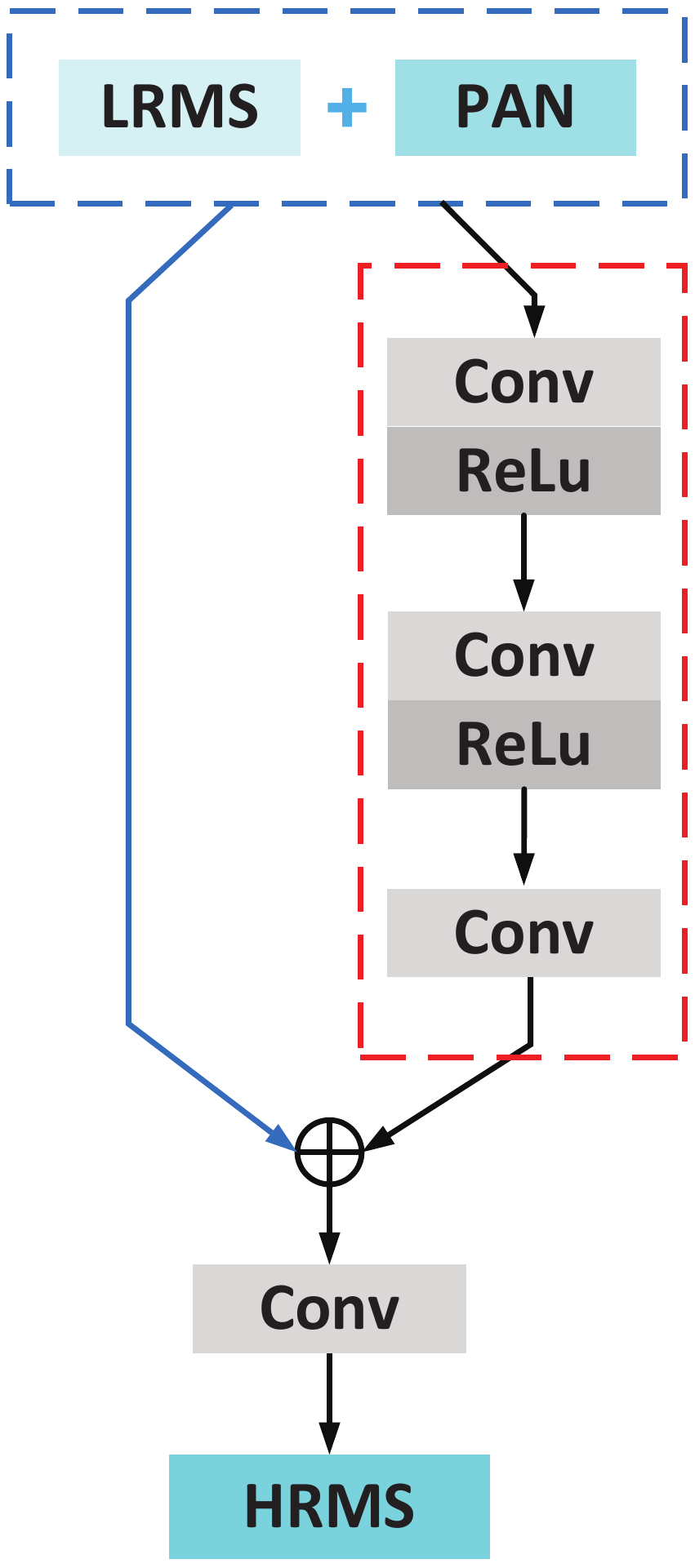} &
  \includegraphics[height=2.5in]{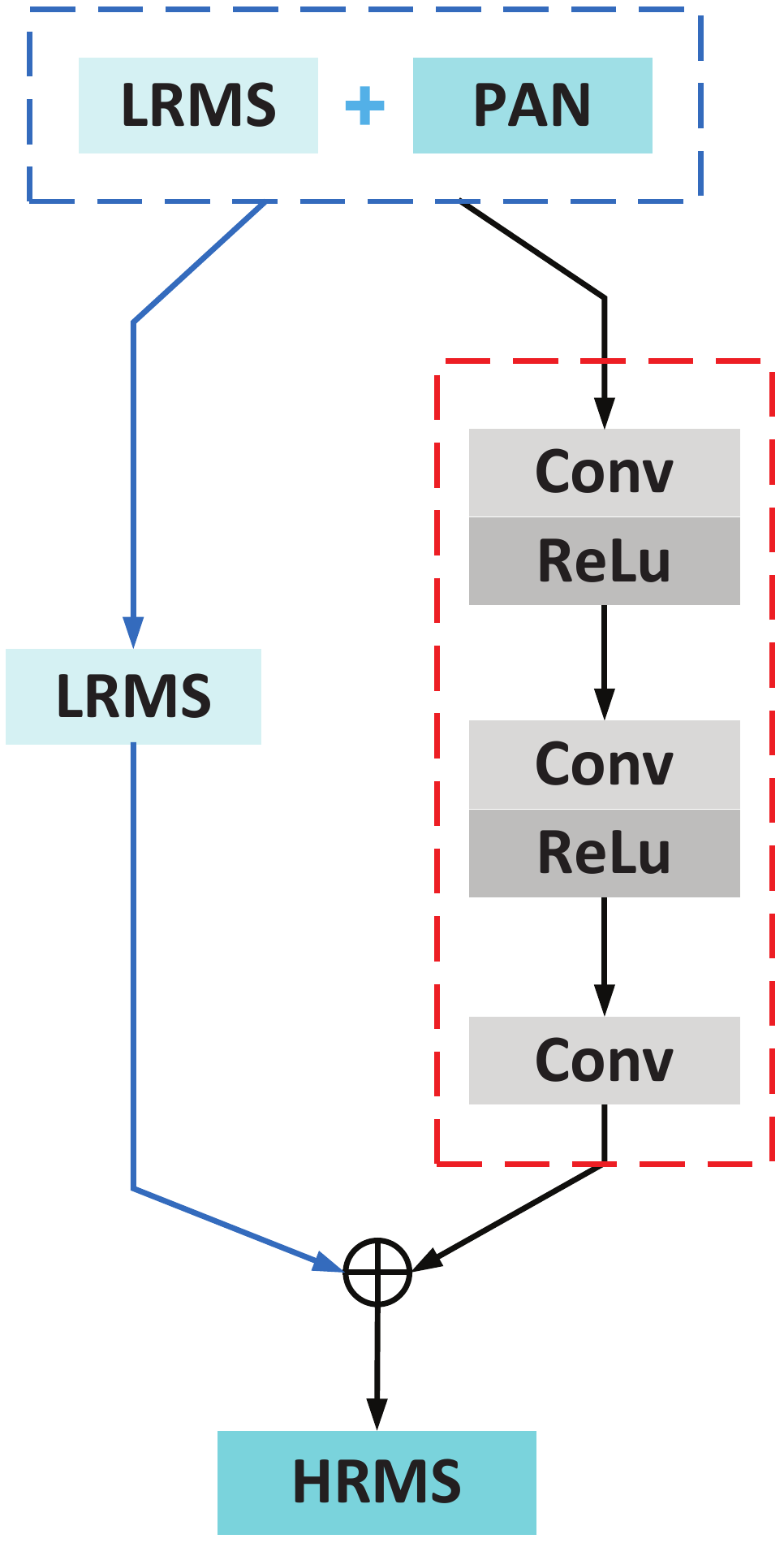}  &
  \includegraphics[height=2.5in]{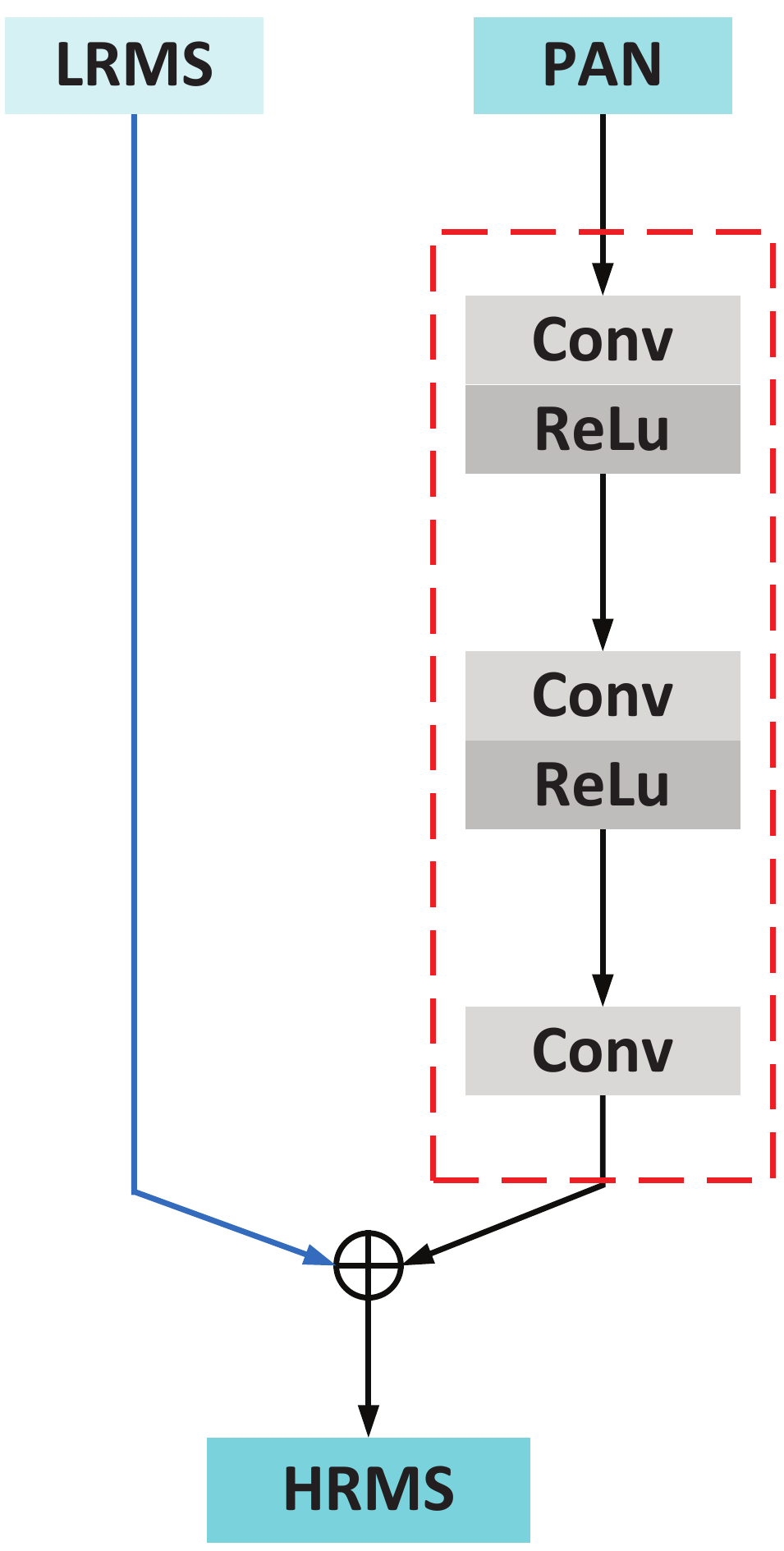}\\
\\
 (a) PNN	&  (b) DRPNN & (c) DiCNN1 & (d) DiCNN2\\
  \end{tabular}
\caption{Structure comparison of (a) PNN, (b) DRPNN, (c) DiCNN1, and (d) DiCNN2, where the red dashed-line box marks the convolutional layers pathway and $\bigoplus$ means the pixel-wise addition.}
\label{fig:StructComp}
\end{figure*}

Practically, pansharpening is ill-posed which means that many solutions exist for a given low-resolution input. This is mathematically connected to an underdetermined inverse problem, of which the solution is not unique. In theory, such a problem can be relieved by constraining the solution space with appropriate prior information, which influences the overall performance of pansharpening. Fig. \ref{fig:StructComp} depicts the basic structure of several CNN-based methods, with Fig. \ref{fig:StructComp}(a) and Fig. \ref{fig:StructComp}(b) representing the PNN and DRPNN (mentioned previously) and Fig. \ref{fig:StructComp}(c) representing our DiCNN1. As we can observe, the PNN directly learns the mapping between its input (the pre-interpolated LRMS image plus the PAN image) and the reconstructed HRMS image, without involving any prior knowledge on structure, regarding pansharpening just as a black-box learning problem. In the DRPNN, a residual structure is introduced into pansharpening [as shown in Fig. \ref{fig:StructComp}(b)], motivated by the residual learning process for image super-resolution in \cite{SuperRes-CNN:Kim2016}. However, this residual structure brings some inherent weaknesses when used for pansharpening. First, DRPNN uses the concatenation of the pre-interpolated LRMS image and the PAN image as its input. This input goes through the stacked layers and the shortcut connection simultaneously, which forces the output of stacked layers pathway to be of the same dimensionality as the input of the input concatenation, i.e. one dimension more than that of the pansharpened HRMS image, thus yielding a residual learning result that has no explicit physical interpretation in the pansharpening context. Second, this dimensionality mismatch has to rely on an extra convolutional layer to cope with such mismatch, which apparently aggravates the computational burden.

Different from PNN and DRPNN, our DiCNN1 takes into consideration the special detail structure based on the detail injection framework. It uses the concatenation of the pre-interpolated LRMS image and the PAN image as the input of the stacked layers, whereas the shortcut connection inputs only the pre-interpolated LRMS image. This strategy makes the output of stacked layers pathway be the MS details that can directly supplement the pre-interpolated LRMS image to produce the HRMS image, which guarantees that this CNN is able directly learn the MS details. This implies that DiCNN1 does introduce a domain-specific structure with meaningful interpretation, meanwhile excluding the additional computational burden. On the other hand, compared to detail injection based CS and MRA methods, DiCNN1 learns only the MS details \emph{per se}, avoiding to separately process the PAN details and the associated gains and hence reducing the model uncertainty.

\subsection{DiCNN2}

 When a pansharpening CNN model has been trained, the test MS images may be changed, for example, there arise bad bands. In this situation, can a pansharpening CNN model be transferred to pansharpen those different kinds of test images?

 As mentioned in previous sections, pansharpening utilizes the details mainly existing in the PAN image to supplement the LRMS image, so as to achieve the HRMS image. These details can be viewed as the result from a filtering process where certain low-frequency components are filtered out \cite{Wei2015Bayesian},
which is a common rule for pansharpening on various sorts of images. Under this rule, it therefore makes sense that, for a given CNN, different sets of network parameters suitable for pansharpening different kinds of images have certain inherent connections. As a result, it is possible to use a pre-trained CNN model on a kind of images for pansharpening other kinds of images. This is actually a transfer learning \cite{Yosinski2014How}. By close inspection of Fig. \ref{fig:NetStructure}(a), we can see that both the PAN image and the LRMS image are fed into the convolution layers pathway, which indicates that the LRMS image will significantly affect detail extraction when the type of the MS image varies and thus reduce the robustness of the model learning in the stack layers pathway. To address this issue, we have developed another pansharpening CNN, named DiCNN2 [as shown in Fig. \ref{fig:NetStructure}(b)]. In DiCNN2, only the PAN image is connected to the convolution layers pathway, which removes the influence of the LRMS image on detail extraction. Though this may also reduce the specificity of details for a certain kind of MS images, the shortcut connection still inputs the pre-interpolated LRMS image to force the convolution layers pathway to learn only the information about the MS details. The objective to minimize for DiCNN2 is:
\begin{equation}
\label{eq:DICNN2}
\begin{split}
\ell(\pmb{\theta})    &= \|\widehat{\mathbf{D}}({\mathbf{P}};\pmb{\theta}) + \widetilde {\mathbf{M}} - {\mathbf{Y}}\|^2_F\\
&= \frac{1}{N_p} \sum_{i=1}^{N_p} \|\widehat{\mathbf{D}}^{(i)}(\mathbf{P}^{(i)};\pmb{\theta}) + \widetilde {\mathbf{M}} - {\mathbf{Y}^{(i)}}\|^2_F.
\end{split}
\end{equation}
In real applications, once a CNN is trained, the network parameters in the convolution layers pathway are fixed, except for those on the last layer. When a new kind of images are input, only this last layer needs to be fine-tuned.

It is noteworthy that DiCNN2 is also a kind of detail injection based CNN. In addition to performing pre-training transfer, DiCNN2 can be seen as an alternative to DiCNN1 for usual pansharpening task where the data for training and prediction come from the same sensors. Fig. \ref{fig:StructComp}(d) depicts the simplified structure of such a pansharpening CNN, which suggests that DiCNN2 can provide similar benefits as DiCNN1, such as meaningful detail injection interpretation, high computational efficiency, and model simplification. Especially, DiCNN2 uses the PAN image as the input of the stacked convolutional layers, in contrast with the concatenation of the PAN image and multi-band LRMS image, thus leading to even higher computational efficaciency than DiCNN1.

\subsection{Analysis of Effectiveness}
The formulations of CNN models are usually non-convex optimization problems with many local minima \cite{choro2015loss,bottou2010large,haykin2009neural}. To solve such optimizations,  the iterative gradient descent method is often used, which involves some factors, where the initialization and the gradient are usually critical for the solution.

Intuitively, better iteration initializations are beneficial to attain better gradient descent solution. Let us investigate such an initialization issue in more detail. For the four pansharpening CNNs illustrated in Fig \ref{fig:StructComp}, the output of the stacked convolutional layers pathway is as follows:
\begin{equation}
\label{eq:finalout}
\begin{split}
\mathbf{Z}_3=\mathbf{W}_3*\varphi(\mathbf{W}_2*\varphi(\mathbf{W}_1*\mathbf{X}+\mathbf{B}_1)+\mathbf{B}_2)+\mathbf{B}_3,
\end{split}
\end{equation}
where $*$ denotes convolution, $\varphi(\cdot)$ represents the ReLU activation function, and $\mathbf{Z}_l=\mathbf{W}_l*\varphi(\mathbf{W}_{l-1}*\varphi(\mathbf{Z}_{l-1})$ denotes the output of the $l$th convolution layer. $\mathbf{Z}_l$ is in 3D data arrangement and thus a 3-way tensor, the concept that has been previously mentioned in the description of (\ref{eq:losspnn}). Note that $\mathbf{Z}_3$ has specific meanings for different pansharpening CNNs, where it represents the MS details $\widehat{\mathbf{D}}$ for our DiCNN1 and DiCNN2, the residuals $\widehat{\mathbf{R}}$ for DRPNN, and the pansharpened HRMS image $\widehat{\mathbf{M}}$ for PNN.

In this work, the initialization of CNN parameters $\mathbf{W}_l$ and $\mathbf{B}_l$ are assumed to follow an i.i.d. zero-mean random distribution and be independent of the neuron output of the $l-1$th layer $\mathbf{A}_{l-1}=\varphi(\mathbf{W}_{l-1}*\mathbf{A}_{l-2}+\mathbf{B}_{l-1})$. And, obviously, the CNN input $\mathbf{X}$ can be used as $\mathbf{A}_0$. For later use, we present a property about $\mathbf{Z}_3$ and its proof below.
\begin{normalsize}
\begin{equation}
\label{eq:finalexpectation2}
\begin{split}
&E\{\{\mathbf{Z}_3\}_{(1)}\mathbf{Y}\}\\
=&E\{\{\{\mathbf{W}_3*\varphi(\mathbf{Z}_{2})\}_{(1)}+\{\mathbf{B}_3\}_{(1)}\}\mathbf{Y}\}\\
=&E\{\{\mathbf{W}_3*\varphi(\mathbf{Z}_{2})\}_{(1)}\mathbf{Y}\}+E\{\{\mathbf{B}_3\}_{(1)}\mathbf{Y}\}\\
=&E\{\{\sum\limits_{m}\sum\limits_{n}\sum\limits_{l}\mathbf{W}_3(m, n, l)\\
&\varphi(\mathbf{Z}_{2}(m-x, n-y, l-b))\}_{(1)}\mathbf{Y}\}+E\{\{\mathbf{B}_3\}_{(1)}\}E(\mathbf{Y})\\
=&E\{\{\sum\limits_{m}\sum\limits_{n}\sum\limits_{l}\mathbf{W}_3(m, n, l)\\
&\{\varphi(\mathbf{Z}_{2}(m-x, n-y, l-b))\}_{(1)}\}\mathbf{Y}\}+\mathbf{0}\cdot E(\mathbf{Y})\\
=&E\{\sum\limits_{m}\sum\limits_{n}\sum\limits_{l}\mathbf{W}_3(m, n, l) \\
&\{\varphi(\mathbf{Z}_{2}(m-x, n-y, l-b))\}_{(1)}\mathbf{Y}\}\\
=&\sum\limits_{m}\sum\limits_{n}\sum\limits_{l}\{0\cdot
E\{\{\varphi(\mathbf{Z}_{2}(m-x, n-y, l-b))\}_{(1)}\mathbf{Y}\}\}\\
=&\mathbf{0},
\end{split}
\end{equation}
\end{normalsize}
where $\mathbf{Y}$ is a matrix not necessarily independent of $\mathbf{Z}_3$ and
$\{\cdot\}_{(1)}$ means the unfolding of a 3-way tensor along its first mode, and the equations
\begin{equation}
\label{eq:tensorconv}
\begin{split}
&\{\mathbf{W}_3*\varphi(\mathbf{Z}_{2})\}_{(1)}\\
=&\{\sum\limits_{m}\sum\limits_{n}\sum\limits_{l}\mathbf{W}_3(m, n, l)
\varphi(\mathbf{Z}_{2}(m-x, n-y, l-b))\}_{(1)}\\
=&\sum\limits_{m}\sum\limits_{n}\sum\limits_{l}\mathbf{W}_3(m, n, l)
\{\varphi(\mathbf{Z}_{2}(m-x, n-y, l-b))\}_{(1)}\\
\end{split}
\end{equation}
are utilized.

We will justify that our DiCNNs can achieve better initialization. First, consider DiCNN1. Its loss function $E(\|\widehat{\mathbf{D}} + \widetilde {\mathbf{M}} - {\mathbf{Y}}\|^2_F)$ can be rewritten as

\begin{equation}
\label{eq:CostInitDICNN1}
\begin{split}
&E(\|\widehat{\mathbf{D}} + \widetilde{\mathbf{M}} - {\mathbf{Y}}\|^2_F)\\
=&E\{\mathrm{Trace}\{(\widehat{\mathbf{D}}+\widetilde{\mathbf{M}}-\mathbf{Y})
(\widehat{\mathbf{D}}+\widetilde{\mathbf{M}}-\mathbf{Y})^T\}\}\\
=&E\{\mathrm{Trace}(\widehat{\mathbf{D}}\widehat{\mathbf{D}}^T)+\mathrm{Trace}(\widehat{\mathbf{D}}\widetilde{\mathbf{M}}^T)-\mathrm{Trace}(\widehat{\mathbf{D}}\mathbf{Y}^T)\\
&+\mathrm{Trace}(\widetilde{\mathbf{M}}\widehat{\mathbf{D}}^T)+\mathrm{Trace}(\widetilde{\mathbf{M}}\widetilde{\mathbf{M}}^T)-\mathrm{Trace}(\widetilde{\mathbf{M}}\mathbf{Y}^T)\\
&-\mathrm{Trace}(\mathbf{Y}\widehat{\mathbf{D}}^T)-\mathrm{Trace}(\mathbf{Y}\widetilde{\mathbf{M}}^T)+\mathrm{Trace}(\mathbf{Y}\mathbf{Y}^T)\}\\
=&E\{\mathrm{Trace}(\widehat{\mathbf{D}}\widehat{\mathbf{D}}^T)+2\mathrm{Trace}(\widehat{\mathbf{D}}\widetilde{\mathbf{M}}^T)-2\mathrm{Trace}(\widehat{\mathbf{D}}\mathbf{Y}^T)\\
&+\mathrm{Trace}(\widetilde{\mathbf{M}}\widetilde{\mathbf{M}}^T)-2\mathrm{Trace}(\widetilde{\mathbf{M}}\mathbf{Y}^T)+\mathrm{Trace}(\mathbf{Y}\mathbf{Y}^T)\}\\
=&\mathrm{Trace}\{E(\widehat{\mathbf{D}}\widehat{\mathbf{D}}^T)\}+2\mathrm{Trace}\{E(\widehat{\mathbf{D}}\widetilde{\mathbf{M}}^T)\}\\
&-2\mathrm{Trace}\{E(\widehat{\mathbf{D}}\mathbf{Y}^T)\}+\mathrm{Trace}\{E(\widetilde{\mathbf{M}}\widetilde{\mathbf{M}}^T)\}\\
&-2\mathrm{Trace}\{E(\widetilde{\mathbf{M}}\mathbf{Y}^T)\}+\mathrm{Trace}\{E(\mathbf{Y}\mathbf{Y}^T)\}\\
=&\mathrm{Trace}\{E(\widehat{\mathbf{D}}\widehat{\mathbf{D}}^T)\}+\mathrm{Trace}\{E(\widetilde {\mathbf{M}}\widetilde {\mathbf{M}}^T)\}\\
&-2\mathrm{Trace}\{E(\widetilde {\mathbf{M}}\mathbf{Y}^T)\}+\mathrm{Trace}\{E(\mathbf{Y}\mathbf{Y}^T)\},
\end{split}
\end{equation}
where the equations
\begin{equation}
\mathrm{Trace}\{E({\widehat{\mathbf{D}}}\widetilde{\mathbf{M}}^T)\}=\textbf{0}
\end{equation}
\begin{equation}
\mathrm{Trace}\{E({\widehat{\mathbf{D}}}\mathbf{Y}^T)\}=\textbf{0}
\end{equation}
are utilized, which can obtained through (\ref{eq:finalexpectation2}).

Consider PNN, its loss function $E(\|\widehat {\mathbf {M}} - {\mathbf{Y}}\|^2_F)$ can be transformed as
\begin{equation} \small
\label{eq:CostInitPNN}
\begin{split}
&E(\|\widehat{\mathbf {M}} - {\mathbf{Y}}\|^2_F)\\
=&E\{\mathrm{Trace}\{(\widehat{\mathbf{M}}-\mathbf{Y})(\widehat{\mathbf{M}}-\mathbf{Y})^T\}\}\\
=&E\{\mathrm{Trace}(\widehat{\mathbf{M}}\widehat{\mathbf{M}}^T-\widehat{\mathbf{M}}\mathbf{Y}^T-
\mathbf{Y}\widehat{\mathbf{M}}^T+\mathbf{Y}\mathbf{Y}^T)\}\\
=&E\{\mathrm{Trace}(\widehat{\mathbf{M}}\widehat{\mathbf{M}}^T)-2\mathrm{Trace}(\widehat{\mathbf{M}}\mathbf{Y}^T)
+\mathrm{Trace}(\mathbf{Y}\mathbf{Y}^T)\}\\
=&\mathrm{Trace}\{E(\widehat {\mathbf {M}}\widehat {\mathbf {M}}^T)\}-2\mathrm{Trace}\{E(\widehat {\mathbf {M}}\mathbf{Y}^T)\}
+\mathrm{Trace}\{E(\mathbf{Y}\mathbf{Y}^T)\}\\
=&\mathrm{Trace}\{E(\widehat {\mathbf {M}}\widehat {\mathbf {M}}^T)\}+\mathrm{Trace}\{E(\mathbf{Y}\mathbf{Y}^T)\},
\end{split}
\end{equation}
where the equation
\begin{equation}
\mathrm{Trace}\{E({\widehat{\mathbf{M}}}\mathbf{Y}^T)\}=\textbf{0}
\end{equation}
is involved, which can also be obtained via (\ref{eq:finalexpectation2}).

To compare the initialization of loss function of DiCNN1 shown in (\ref{eq:CostInitDICNN1}) with that of PNN shown in (\ref{eq:CostInitPNN}), we have
\begin{equation}
\label{eq:E(compare)}
\begin{split}
&E(\|\widehat{\mathbf{D}} + \widetilde {\mathbf{M}} - {\mathbf{Y}}\|^2_F) - E(\|\widehat {\mathbf {M}} - {\mathbf{Y}}\|^2_F)\\
=&\mathrm{Trace}\{E(\widehat{\mathbf{D}}\widehat{\mathbf{D}}^T)\}+\mathrm{Trace}\{E(\widetilde {\mathbf{M}}\widetilde {\mathbf{M}}^T)\}\\
&-2\mathrm{Trace}\{E(\widetilde {\mathbf{M}}\mathbf{Y}^T)\}+\mathrm{Trace}\{E(\mathbf{Y}\mathbf{Y}^T)\}\\
&-\mathrm{Trace}\{E(\widehat {\mathbf {M}}\widehat {\mathbf {M}}^T)\}-\mathrm{Trace}\{E(\mathbf{Y}\mathbf{Y}^T)\}\\
=&\mathrm{Trace}\{E(\widehat{\mathbf{D}}\widehat{\mathbf{D}}^T)\}+\mathrm{Trace}\{E(\widetilde {\mathbf{M}}\widetilde {\mathbf{M}}^T)\}\\
&-2\mathrm{Trace}\{E(\widetilde {\mathbf{M}}\mathbf{Y}^T)\}-\mathrm{Trace}\{E(\widehat {\mathbf {M}}\widehat {\mathbf {M}}^T)\}\\
=&\mathrm{Trace}\{E(\widetilde {\mathbf{M}}\widetilde {\mathbf{M}}^T)\}-2\mathrm{Trace}\{E(\widetilde {\mathbf{M}}\mathbf{Y}^T)\}\\
=&\mathrm{Trace}\{E(\widetilde {\mathbf{M}}\widetilde {\mathbf{M}}^T)\}-2\mathrm{Trace}\{E\{\widetilde {\mathbf{M}}(\mathbf{Y}^T+\widetilde {\mathbf{M}}^T-\widetilde {\mathbf{M}}^T)\}\}\\
=&-\mathrm{Trace}\{E(\widetilde {\mathbf{M}}\widetilde {\mathbf{M}}^T)\}-2\mathrm{Trace}\{E\{\widetilde {\mathbf{M}}(\mathbf{Y}^T-\widetilde {\mathbf{M}}^T)\}\}\\
=&2\mathrm{Trace}\{E\{\widetilde {\mathbf{M}}(\widetilde{\mathbf{M}}^T-\mathbf{Y}^T)\}\}-\mathrm{Trace}\{E(\widetilde {\mathbf{M}}\widetilde {\mathbf{M}}^T)\}\\
<& 0,
\end{split}
\end{equation}
where the equation
\begin{equation}
\mathrm{Trace}\{E(\widehat{\mathbf{D}}\widehat{\mathbf{D}}^T)\}=\mathrm{Trace}\{E(\widehat {\mathbf {M}}\widehat {\mathbf {M}}^T)\}
\end{equation}
is utilized during the derivation from step 2 to step 3, which is reasonable because, in the initial phases of these two CNNS, their convolutional layers pathways have similar structure, similar inputs and similarly distributed network parameters.
Moreover,
the diagonal entries of $\widetilde {\mathbf{M}}\widetilde {\mathbf{M}}^T$ are always greater than or equal to zero. But in the scenario of a real image, it is impossible that all of the diagonal entries are equal to zero. Accordingly, we have
\begin{equation}
\label{eq:LRMS-FNorm}
\mathrm{Trace}\{E(\widetilde {\mathbf{M}}\widetilde {\mathbf{M}}^T)\}>0.
\end{equation}
Taking a close inspection of the term $2\mathrm{Trace}\{E\{\widetilde {\mathbf{M}}(\widetilde{\mathbf{M}}^T-\mathbf{Y}^T)\}\}$ in the last equality of (\ref{eq:E(compare)}), we find $(\widetilde{\mathbf{M}}^T-\mathbf{Y}^T)$ is exactly the ideal MS details whose energy should account for small portion that of the HRMS image and thus we have
\begin{equation}
\mathrm{Trace}\{E(\widetilde {\mathbf{M}}\widetilde {\mathbf{M}}^T)\}>2|\mathrm{Trace}\{E\{\widetilde {\mathbf{M}}(\widetilde{\mathbf{M}}^T-\mathbf{Y}^T)\}\}|.
\label{eq:TraceComparison}
\end{equation}
Table \ref{table:TraceComparison} illustrates the values of two traces in  (\ref{eq:TraceComparison})
\[T_1=\mathrm{Trace}\{E(\widetilde {\mathbf{M}}\widetilde {\mathbf{M}}^T)\}\]
and
\[T_2=2|\mathrm{Trace}\{E\{\widetilde {\mathbf{M}}(\widetilde{\mathbf{M}}^T-\mathbf{Y}^T)\}\}|\]
computed on three datasets, which do verify the inequality in (\ref{eq:TraceComparison}).
\begin{table}[H]
\caption{Trace values}
\centering
\begin{tabular}{c|cc}
\hline
{}& $T_1$  & $T_2$ \\
\hline
IKONOS& 203.8785 &2.9\\
\hline
Quickbird& 108.138 &1.1619\\
\hline
Worldview-2 &607.1628 &20.2275\\
\hline
\end{tabular}
\label{table:TraceComparison}
\end{table}
Utilizing both (\ref{eq:LRMS-FNorm}) and (\ref{eq:TraceComparison}), we have the inequality in the last step of (\ref{eq:E(compare)}).

According to (\ref{eq:E(compare)}), we obtain
\begin{equation}
\label{eq:E(conclude)}
\begin{split}
E(\|\widehat{\mathbf{D}} + \widetilde {\mathbf{M}} - {\mathbf{Y}}\|^2_F) < E(\|\widehat {\mathbf {M}} - {\mathbf{Y}}\|^2_F).
\end{split}
\end{equation}

\begin{figure*}[htb]\scriptsize
\centering
\begin{tabular}{ccc}
\includegraphics[width=0.26\paperwidth]{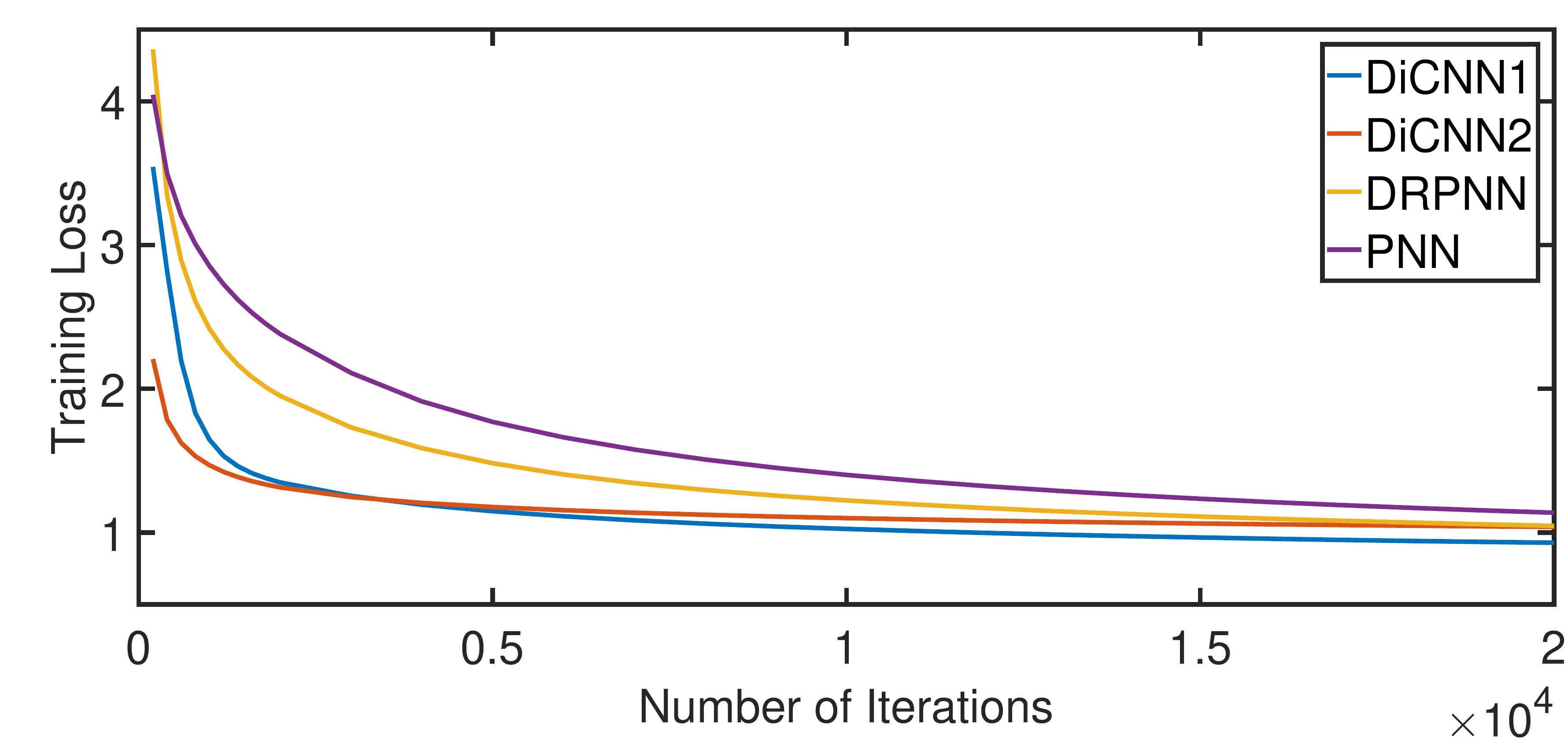}&
\includegraphics[width=0.26\paperwidth]{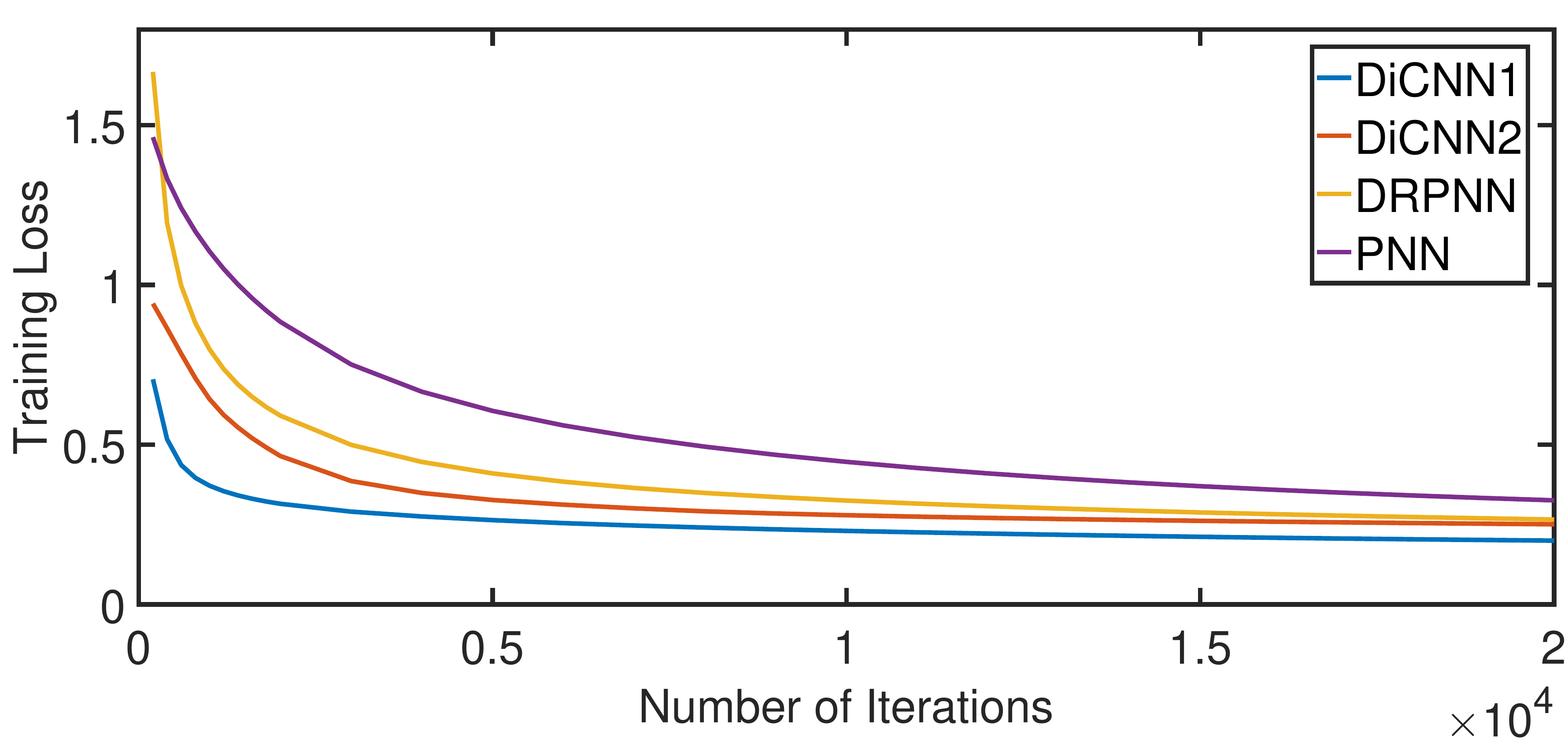}&
\includegraphics[width=0.26\paperwidth]{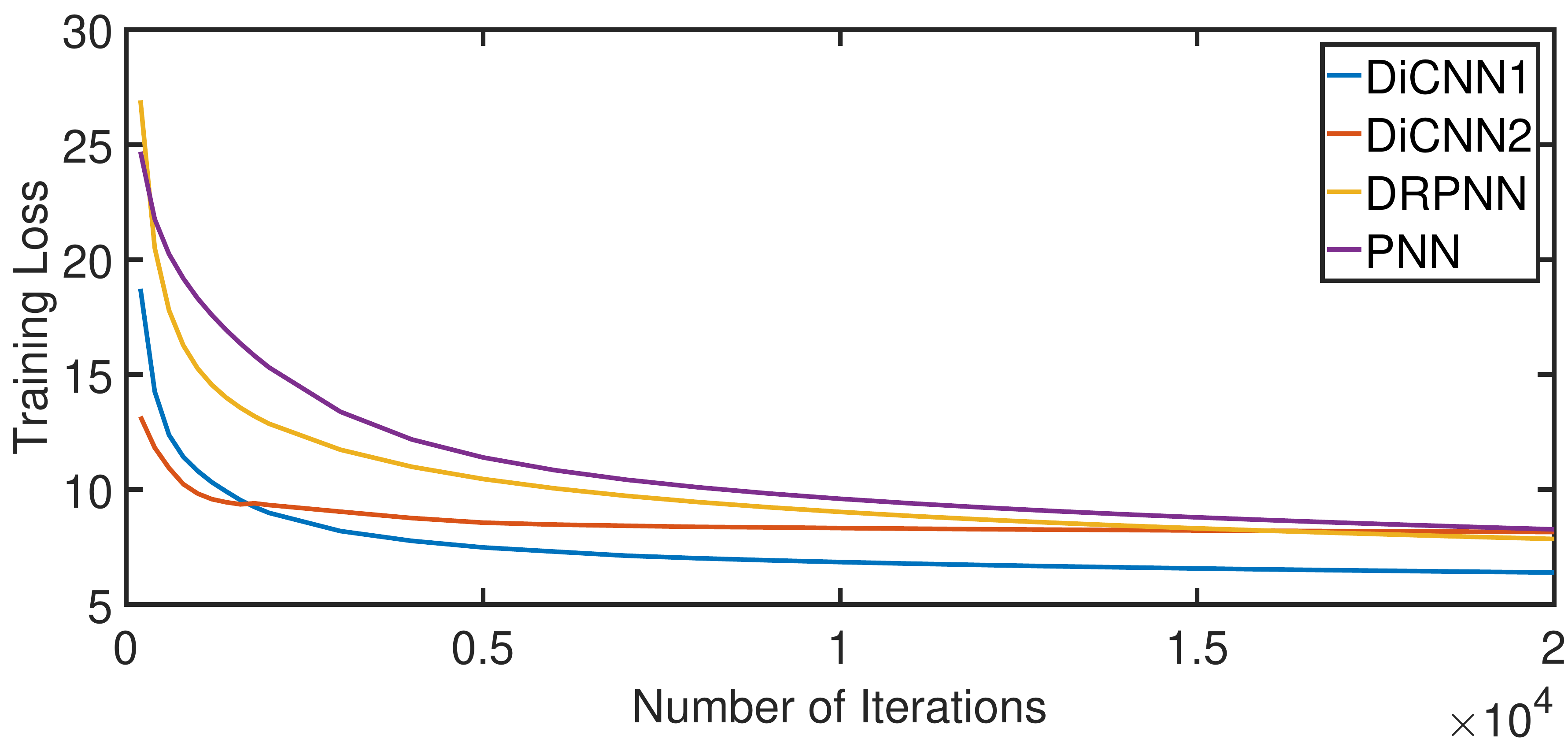}\\
(a) IKONOS image & (b) Quickbird image & (c) Worldview-2 image\\
\end{tabular}
\caption{Training losses of DiCNN1, DiCNN2, PNN and DRPNN}
\label{fig:loss}
\end{figure*}

As suggested in (\ref{eq:E(conclude)}), the initial loss of DiCNN1 is less than that of PNN. DiCNN1 is characteristic of better initialization than PNN. We can extend such an analysis to other pansharpening CNN methods. Fig. \ref{fig:loss} illustrates the losses over the iterations of gradient descent. As observed, our DiCNN1 and DiCNN2 always achieve less initial losses than PNN and DRPNN, i.e., DiCNN1 and DiCNN2 exhibit better initializations. Intuitively, better initialization is more beneficial to an iterative optimization.

Next, we examine the impact of gradient on the optimizations of four pansharpening CNNs. The parameters of a pansharpening CNN discussed before are updated with gradient descent essentially as
$\mathbf{W}^{t+1}_l=\mathbf{W}^{t}_l - \alpha \cdot \frac{\partial \ell(\pmb{\theta})}{\partial \mathbf{W}^{t}_l}$,
where $\alpha$ represents the learning rate
and the gradients can be represented as
\begin{equation}
\label{eq:chain rule}
\begin{split}
\frac{\partial \ell(\pmb{\theta})}{\partial \mathbf{W}_l} = \frac{\partial \ell(\pmb{\theta})}{\partial \mathbf{Z}_l} \frac{\partial \mathbf{Z}_l}{\partial \mathbf{W}_l},
\end{split}
\end{equation}
 where $\frac{\partial \mathbf{Z}_l}{\partial \mathbf{W}_l}=\mathbf{A}_{l-1}$. Recall that $\mathbf{A}_{l-1}$ represents neuron output after activation in the previous layer. Then, the sensitivity of the $l$th layer is $\pmb{\delta}_l = \frac{\partial \ell(\pmb{\theta})}{\partial \mathbf{Z}_l}$.
With the help of chain rule in calculus, we can obtain
\begin{equation}
\label{eq:sensitivity2}
    \begin{split}
\pmb{\delta}_l
    &= \frac{\partial \ell(\pmb{\theta})}{\partial \mathbf{Z}_{l+1}} \frac{\partial \mathbf{Z}_{l+1}}{\partial \mathbf{Z}_l}\\
    &= \pmb{\delta}_{l+1} \frac{\partial \mathbf{Z}_{l+1}}{\partial \mathbf{Z}_l}\\
    &=\mathbf{W}_{l+1} * \pmb{\delta}_{l+1}  \otimes \varphi'(\mathbf{Z}_l),
    \end{split}
\end{equation}
where $\varphi'(\cdot)$ represents the derivative of the activation function and $\otimes$ denotes element-wise multiplication. Let $L$ be the total number of convolutional layers.
According to (\ref{eq:sensitivity2}),  if the sensitivity of the final  $L$th layer is known, the sensitivities of all the other layers can be attained. The gradient in  (\ref{eq:chain rule}) can therefore be reformulated as
\begin{equation}
\label{eq:GeneralGradient}
\begin{split}
&\frac{\partial \ell(\pmb{\theta})}{\partial \mathbf{W}_l}\\
  =& \pmb{\delta}_l * \mathbf{A}_{l-1}\\
  =& \mathbf{W}_{l+1} * \pmb{\delta}_{l+1} \otimes \varphi'(\mathbf{Z}_l) * \mathbf{A}_{l-1}\\
  =& \mathbf{W}_{l+1}* \cdots ({\mathbf{W}_{L}} * {\pmb{\delta}_{L}} \otimes \varphi'(\mathbf{Z}_{L-1}) ) \cdots \otimes \varphi'(\mathbf{Z}_l) * \mathbf{A}_{l-1}.
\end{split}
\end{equation}

The sensitivities of the final convolution layers of DiCNN1 and DiCNN2 can be calculated with
\begin{equation}
\label{eq:sensitivitydicnn1}
\begin{split}
\pmb{\delta}_{L}=\frac{\partial \ell(\pmb{\theta})}{\partial \mathbf{Z}_{L}}=\mathbf{Z}_{L} + \mathbf{X} - \mathbf{Y}.
\end{split}
\end{equation}
Meanwhile, due to the same modality of loss function between PNN and DRPNN, their final layer sensitivities can be represented as:
\begin{equation}
\label{eq:sensitivitypnn}
\begin{split}
\pmb{\delta}_{L}=\frac{\partial \ell(\pmb{\theta})}{\partial \mathbf{Z}_{L}}=\mathbf{Z}_{L}  - \mathbf{Y}.
\end{split}
\end{equation}
Through (\ref{eq:GeneralGradient}), (\ref{eq:sensitivitydicnn1}) and (\ref{eq:sensitivitypnn}), we can obtain the gradients of the four pansharpening CNNs, as shown in Table \ref{table:gradientcomparison}.

\begin{table*}[htb]
\caption{Gradients involved in four pansharpening CNNs}
\centering
\begin{tabular}{c|ccc}
\hline
{}&Gradient&$\mathbf{A}_{L-1}$&$\mathbf{A}_0$\\
\hline
DiCNN1&  $\mathbf{W}_{l+1}* \cdots (\mathbf{W}_{L} * (\mathbf{Z}_{L} + \mathbf{X} - \mathbf{Y}) \otimes \varphi'(\mathbf{Z}_{L-1}) ) \cdots \otimes \varphi'(\mathbf{Z}_l) * \mathbf{A}_{l-1}$  &$\varphi(\mathbf{Z}_{L-1})$&$\mathbf{G}$\\
\hline
DiCNN2& $\mathbf{W}_{l+1}* \cdots (\mathbf{W}_{L} * (\mathbf{Z}_{L} + \mathbf{X} - \mathbf{Y}) \otimes \varphi'(\mathbf{Z}_{L-1}) ) \cdots \otimes \varphi'(\mathbf{Z}_l) * \mathbf{A}_{l-1}$ &$\varphi(\mathbf{Z}_{L-1})$&$\mathbf{PAN}$\\
\hline
PNN & $\mathbf{W}_{l+1}* \cdots (\mathbf{W}_{L} * (\mathbf{Z}_{L}  - \mathbf{Y}) \otimes \varphi'(\mathbf{Z}_{L-1}) ) \cdots \otimes \varphi'(\mathbf{Z}_l) * \mathbf{A}_{l-1}$ &$\varphi(\mathbf{Z}_{L-1})$&$\mathbf{G}$\\
\hline
DRPNN & $\mathbf{W}_{l+1}* \cdots (\mathbf{W}_{L} * (\mathbf{Z}_{L}  - \mathbf{Y})  ) \cdots \otimes \varphi'(\mathbf{Z}_l) * \mathbf{A}_{l-1}$ &$\mathbf{Z}_{L-1}+\mathbf{G}$&$\mathbf{G}$\\
\hline
\end{tabular}
\label{table:gradientcomparison}
\end{table*}

As observed in Table \ref{table:gradientcomparison}, it is difficult to quantitatively assess the influence of those gradients on the optimization processes of the four pansharpening CNNs. Here, we resort to empirical analysis. Fig. \ref{fig:loss} illustrates the training losses of the four CNN methods on three datasets. It is observable that the initial losses of DiCNN1 and DiCNN2 are less than those of PNN and DRPNN, corresponding to the theoretical analysis presented earlier in subsection \ref{sec:Methods}.C, which means that DiCNN1 and DiCNN2 achieve better initializations. PNN not only has worse initialization, but also its iteration process (involving its gradient) does not change the inferior tendency of its loss. During the iterative process, PNN always yields loss higher than DiCNN1 and DiCNN2. That is, the impact of gradient-based iteration process is not strong enough to compensate for the loss of inappropriate initialization. DRPNN has the worst initialization. Although its gradient-involved iterative process makes its loss drop fast, its loss is still always higher than that of DiCNN1 during the iteration.

\section{Experimental results}\label{sec:ExperiResults}
This section evaluates the performance of our pansharpening methods, where three real-world remotely sensed image datasets are considered. These datasets were acquired with WorldView-2, IKONOS and Quickbird sensors. During the evaluation, we conduct reduced-resolution and full-resolution experiments, as well as transfer learning experiments.

In the case of reduced-resolution assessments, we set experiments using Wald's protocol \cite{Wald:Wald1997}. The MS image and the PAN image were degraded to lower resolution by Gaussian filter with a factor of 4\cite{Alparone2006MTF}, and then the degraded MS image was pre-interpolated to the same spatial size as the degraded PAN image using a polynomial kernel (EXP)\cite{EXP:Aiazzi2002}.
The criteria used for the assessment include x-band extension of universal image quality index (Qx) \cite{Alparone2004A}, spatial correlation coefficient (SCC) \cite{J1998A}, spectral angle mapper(SAM) \cite{Yuhas1992Descrimination}, and \emph{Erreur Relative Globale Adimensionnelle de Synth\`{e}se} (ERGAS) \cite{Wald2002Data}. These indexes are widely used to measure the qualities of pansharpened images, with the original MS image as the ground-truth.

For fair comparison, we apply consistent parameter setting to different CNN-based pansharpening methods.
Particularly, the number of convolutional layers in convolution pathway for all pansharpening CNNs used in comparison are set to three, so that we can compare the achievement of only using the basic network structure while avoiding the influence of the deepness of hidden layers. The number of training iterations is set to  $3.0\times10^5$.

The learning properties of the compared CNN-based methods are summarized in Table \ref{table:comparison}, which show only our DiCNN1 and DiCNN2 are built in pansharpening detail injection context. CNN based pansharpening methods were trained using a GPU (Nvidia GTX 1060 3GB with CUDA 8.1 and CUDNN V5) through Caffe \cite{Caffe:Jia2014} in Ubuntu 14.04 operating system, and tested on MATLAB R2016b via CPU mode (laptop with Intel I7 and 8GB RAM ) through the deep learning framework Matconvnet \cite{MatconvNet:Vedaldi2015} in Windows 10 operating system.

\begin{table}[h]
\small
\caption{COMPARISON AMONG CNN-BASED METHODS}
\centering
\begin{tabular}{c|cccc}
\hline
{}&PNN&DRPNN&DiCNN1&DiCNN2\\
\hline
Detail Learning& No &No&\textbf{Yes} &\textbf{Yes}\\
\hline
Residual Learning& No &\textbf{Yes} &No &No \\
\hline
Transfer Learning & No &No&No &\textbf{Yes}\\
\hline
\end{tabular}
\label{table:comparison}
\end{table}

In addition to DiCNN1, DiCNN2, PNN and DRPNN, several representative CS/MRA methods, including Gram Schmidt adaptive (GSA) \cite{GS:Aiazzi2007}, partial replacement adaptive component substitution (PRACS) \cite{Choi2010A}, \textit{a tr\'ous} wavelet transform (ATWT) \cite{WDT:Vivone2014}, Band-Dependent Spatial-Detail (BDSD) \cite{Garzelli2007Optimal} and Generalized Laplacian Pyramid with Context-Based Decision (GLP-CBD) \cite{Alparone2007Comparison} are also run for comparison.

\subsection{Experiment 1: WorldView-2 Washington Dataset}
\begin{table}[htp]
\small
\caption{Quality indexes of different pansharpening methods under reduced-resolution quality assessment on a $256\times256$ subscene of Worldview-2 data sets.}
\centering
\begin{tabular}{c|ccccc}
\hline
{}&Q8&SAM& ERGAS &SCC&Time(s)\\
\hline
Refrence&1 &0 &0 &1&{}\\
\hline
EXP&0.6726 &7.9558 &8.0358 &0.5127 &{}\\
\hline
\hline
GSA&0.9151 &7.5830 &4.3501 &0.8973 &0.85\\
\hline
PRACS&0.8682 &7.7322 &5.2648 &0.8650 &1.43\\
\hline
ATWT&0.8974 &7.2241 &4.7585 &0.8926 &\textbf{0.84}\\
\hline
BDSD&0.9178 &8.1158 &4.5293 &0.8993 &1.14\\
\hline
GLP-CBD&0.9148 &7.5004 &4.3438 &0.8981 &\textbf{0.84}\\
\hline
\hline
PNN &0.9243 &7.6205 &4.2924 &0.8966&2.20\\
\hline
DRPNN &0.9325 &7.2175 &3.9664 &0.9149 &1.17\\
\hline
DiCNN1 &\textbf{0.9492} &\textbf{6.2771} &\textbf{3.6487} &0.9281 &1.13\\
\hline
DiCNN2 &0.9448 &7.2012 &3.7063 &\textbf{0.9299}&0.98\\
\hline
\end{tabular}
\label{table:qualitywv}
\end{table}

\begin{figure*}[t]\scriptsize
\centering
  \begin{tabular}{ccccc}
\includegraphics[width=0.14\paperwidth]{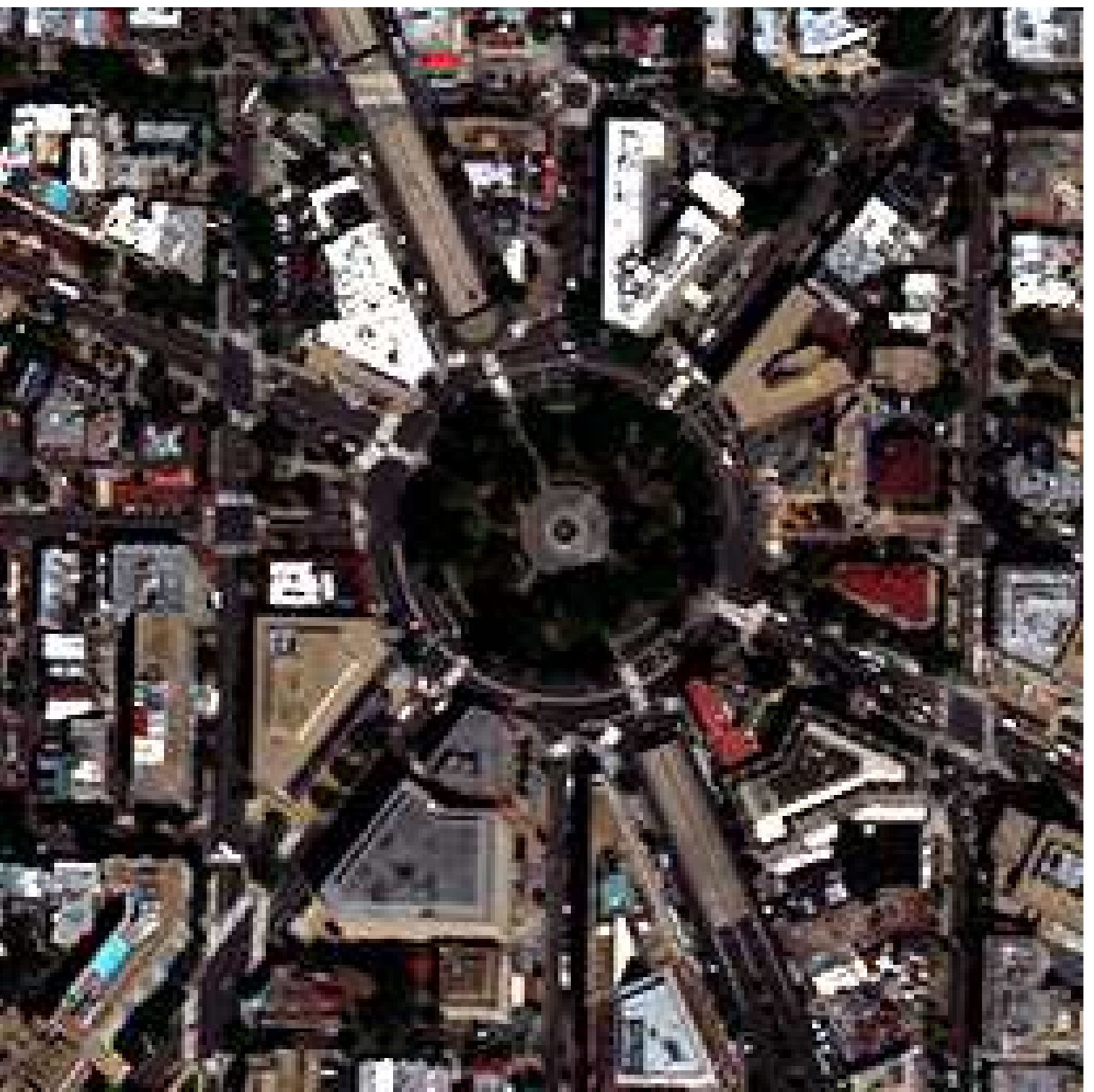} &
\includegraphics[width=0.14\paperwidth]{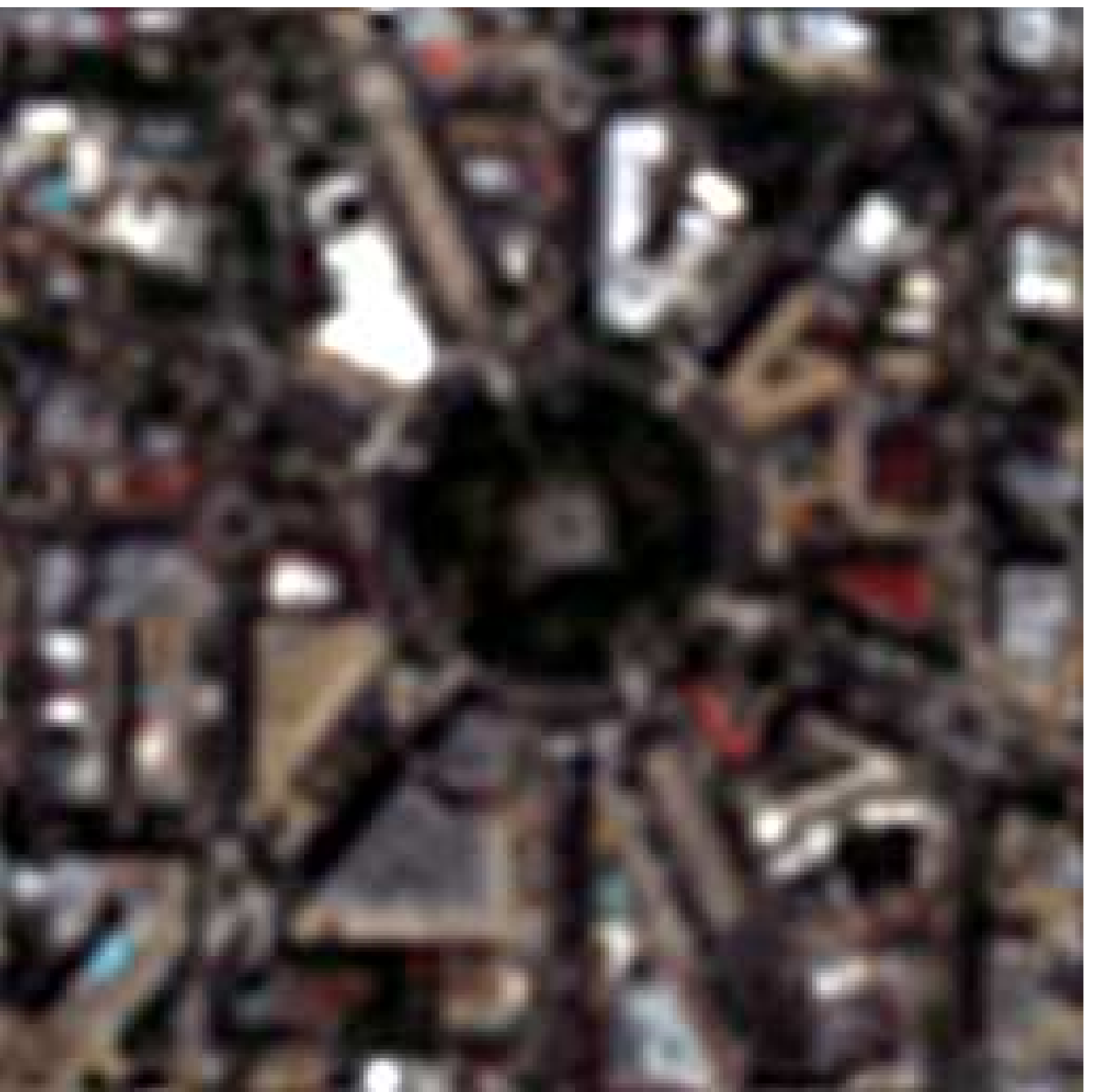} &
\includegraphics[width=0.14\paperwidth]{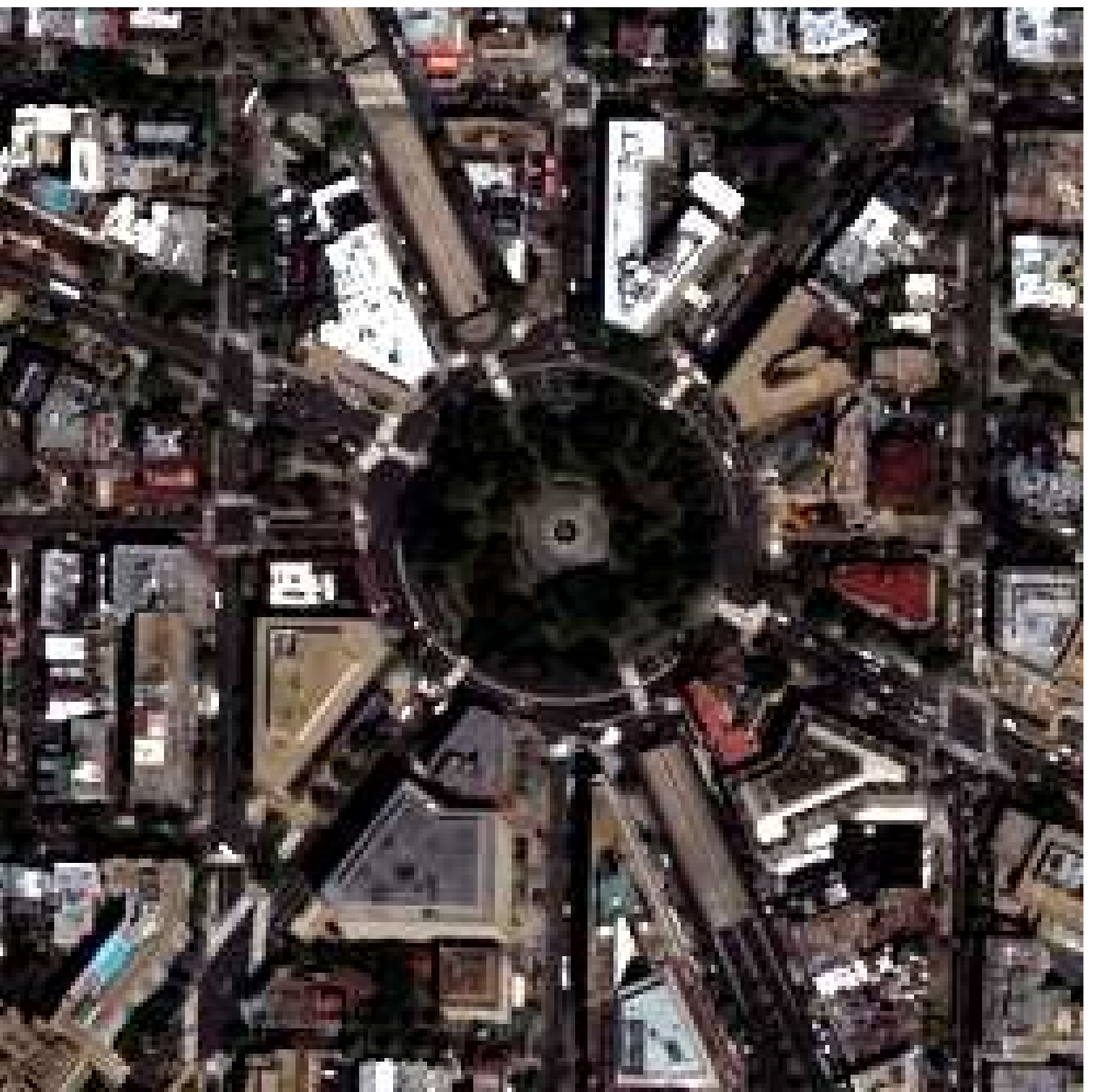} &
\includegraphics[width=0.14\paperwidth]{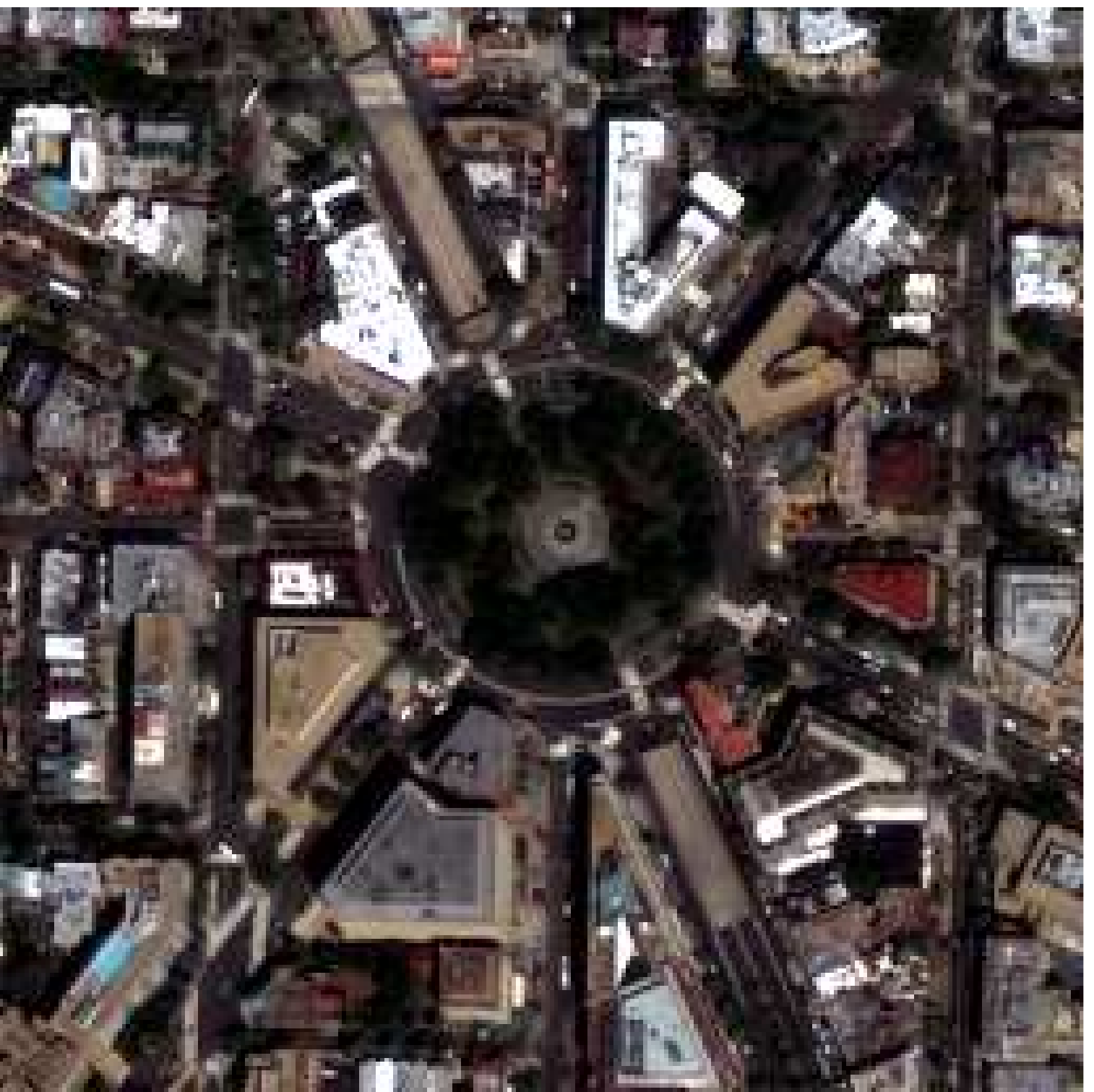} \\
(a) & (b) & (c) &(d) \\
\includegraphics[width=0.14\paperwidth]{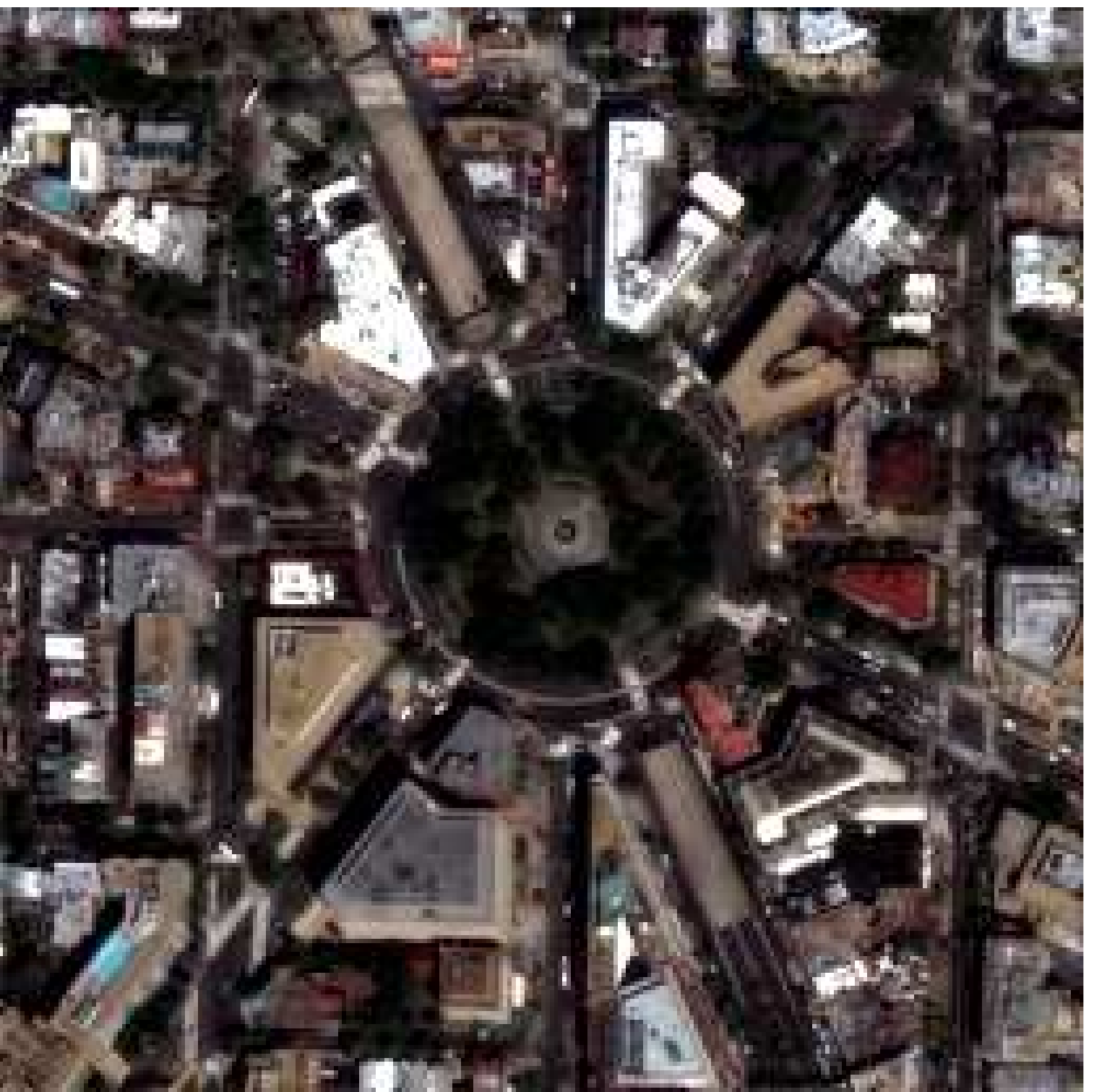} &
\includegraphics[width=0.14\paperwidth]{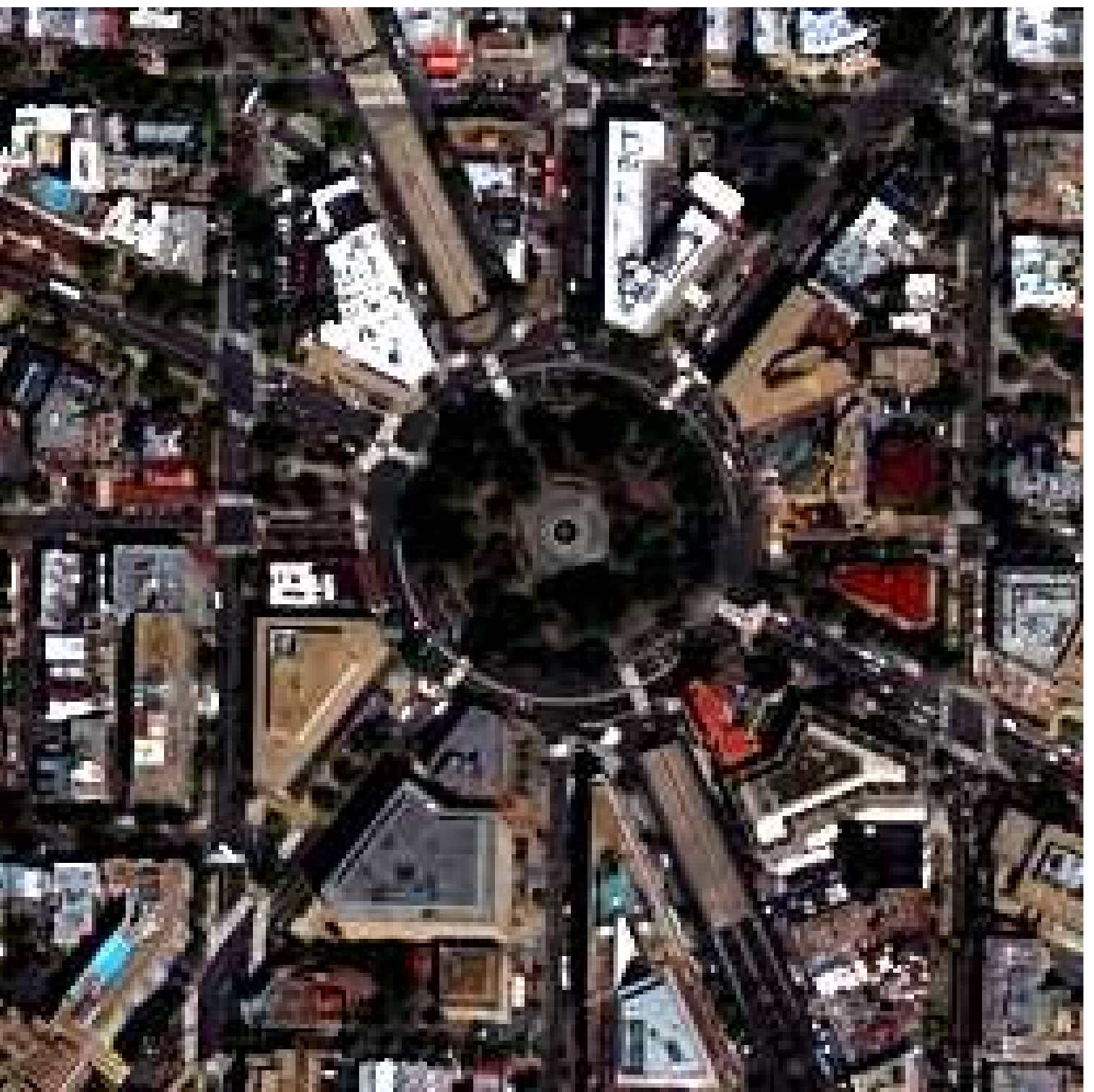} &
\includegraphics[width=0.14\paperwidth]{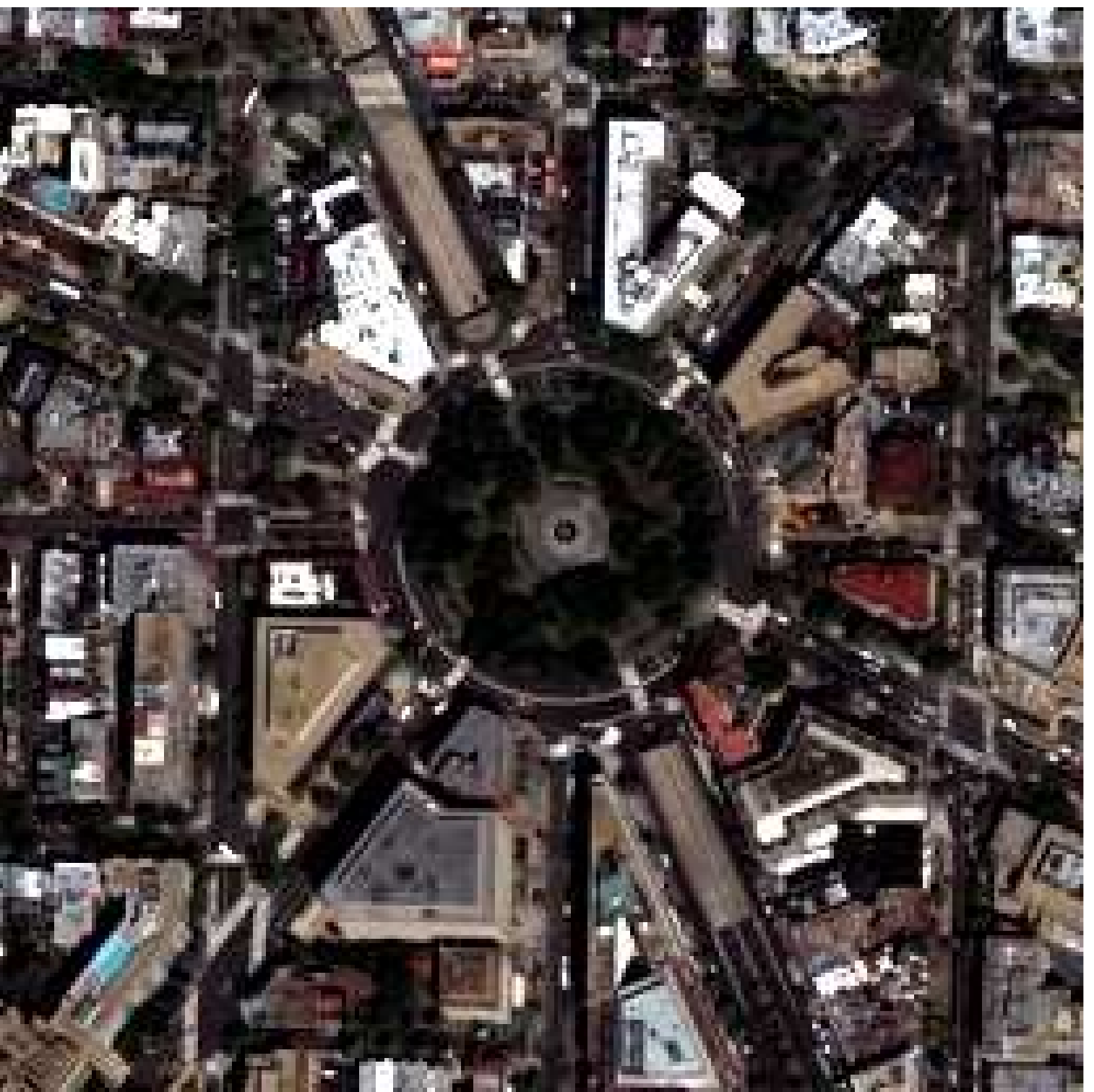} &
\includegraphics[width=0.14\paperwidth]{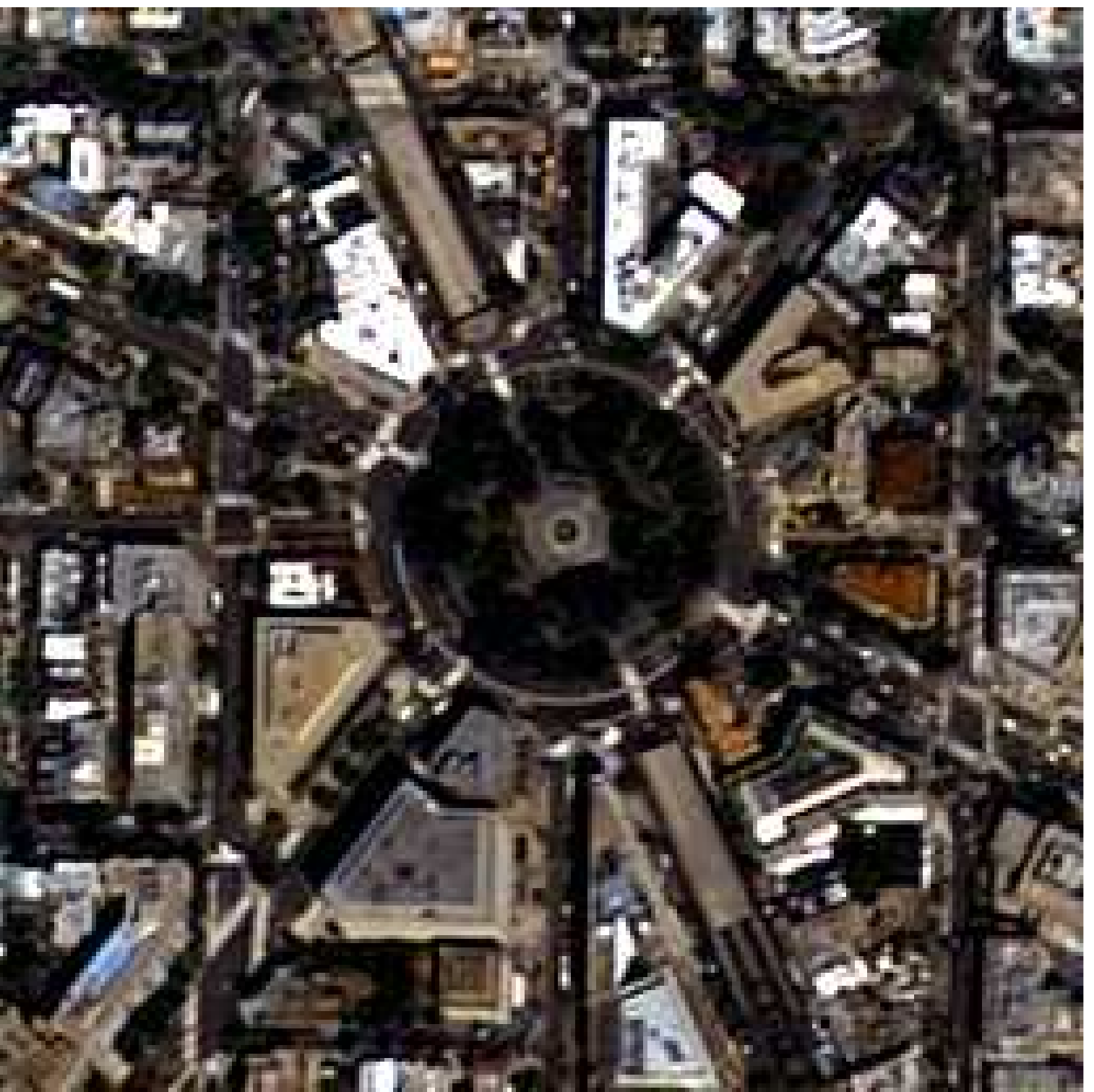} \\
(e) & (f) & (g) &(h)  \\
\includegraphics[width=0.14\paperwidth]{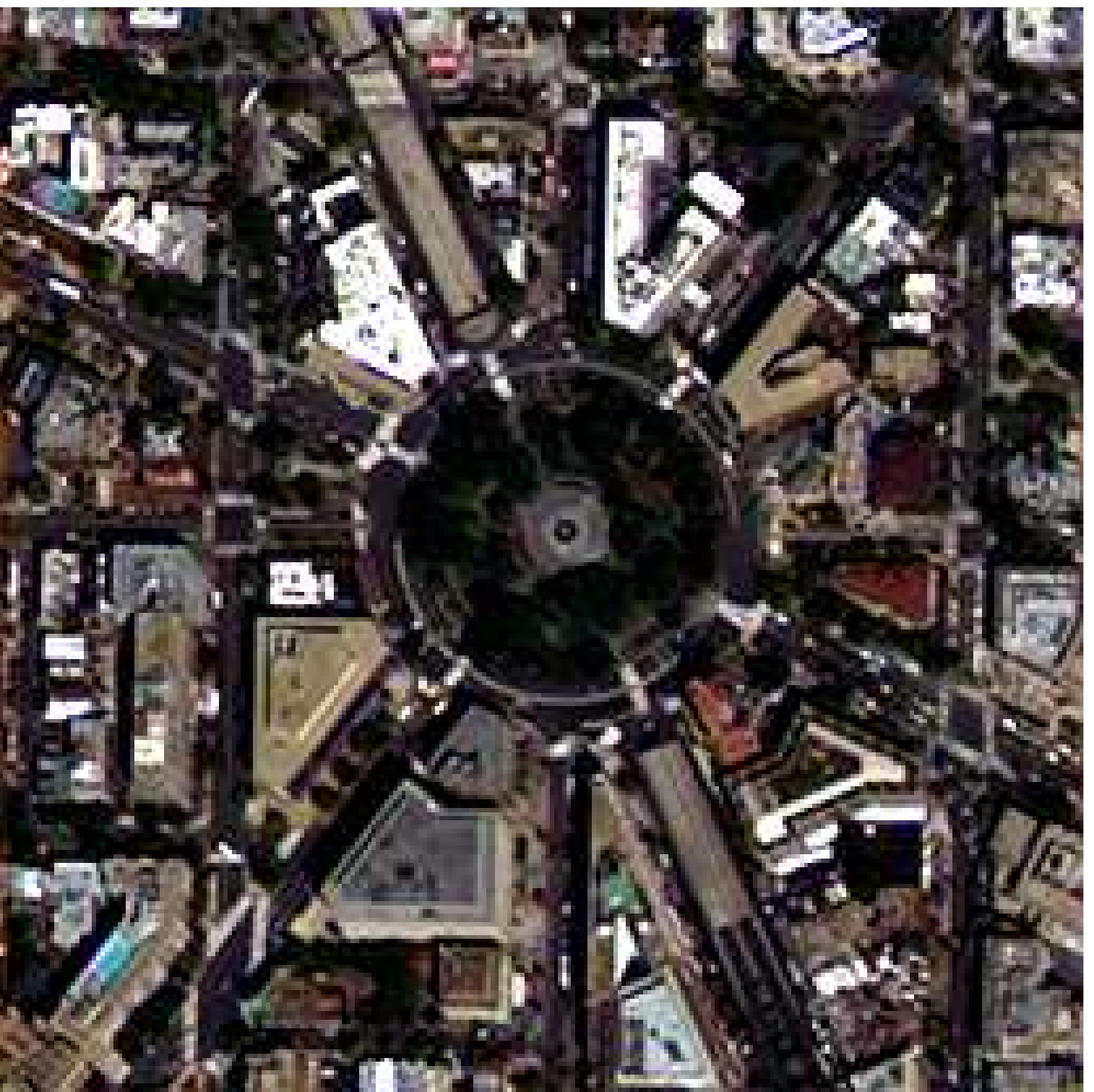} &
\includegraphics[width=0.14\paperwidth]{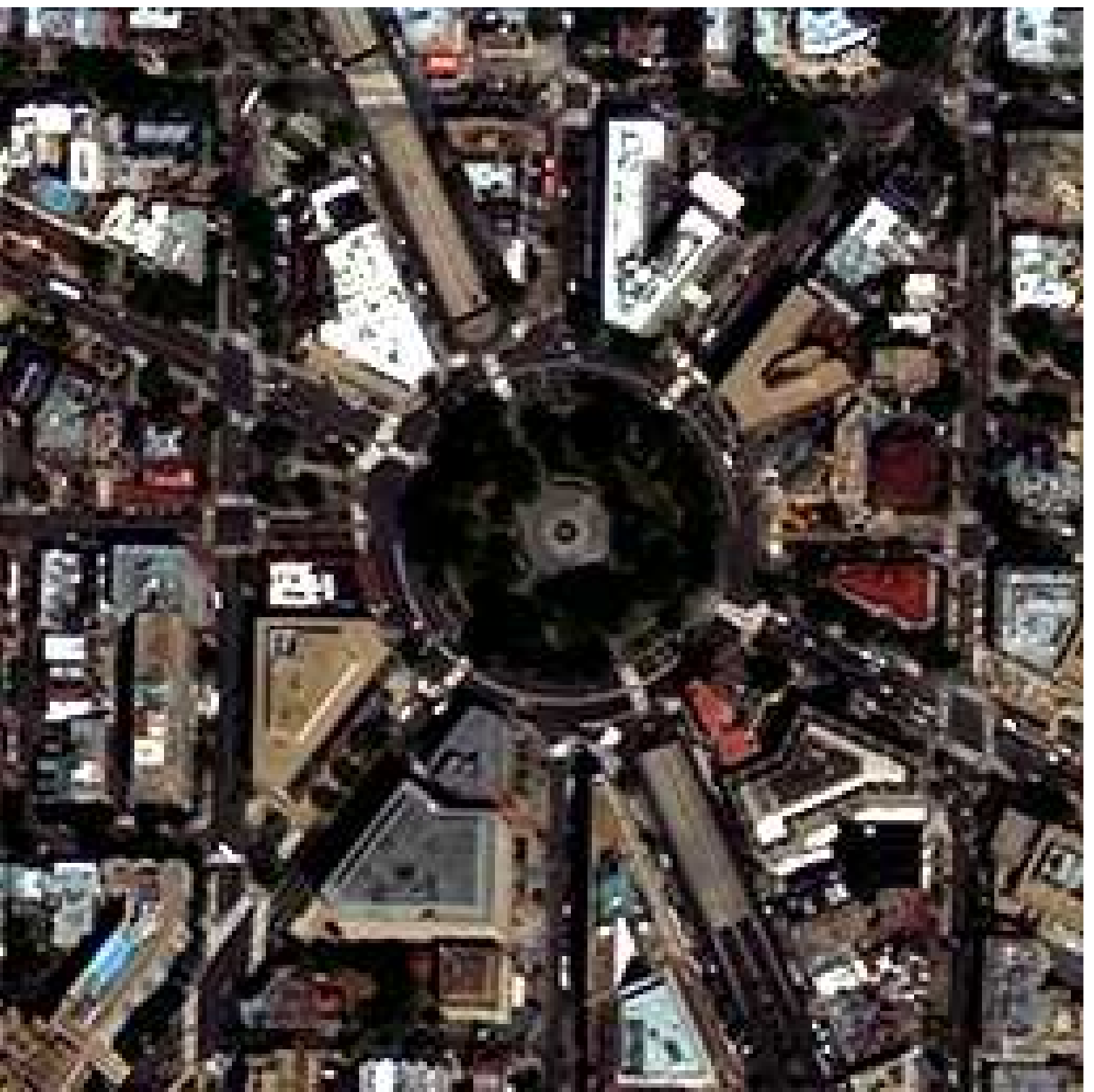} &
\includegraphics[width=0.14\paperwidth]{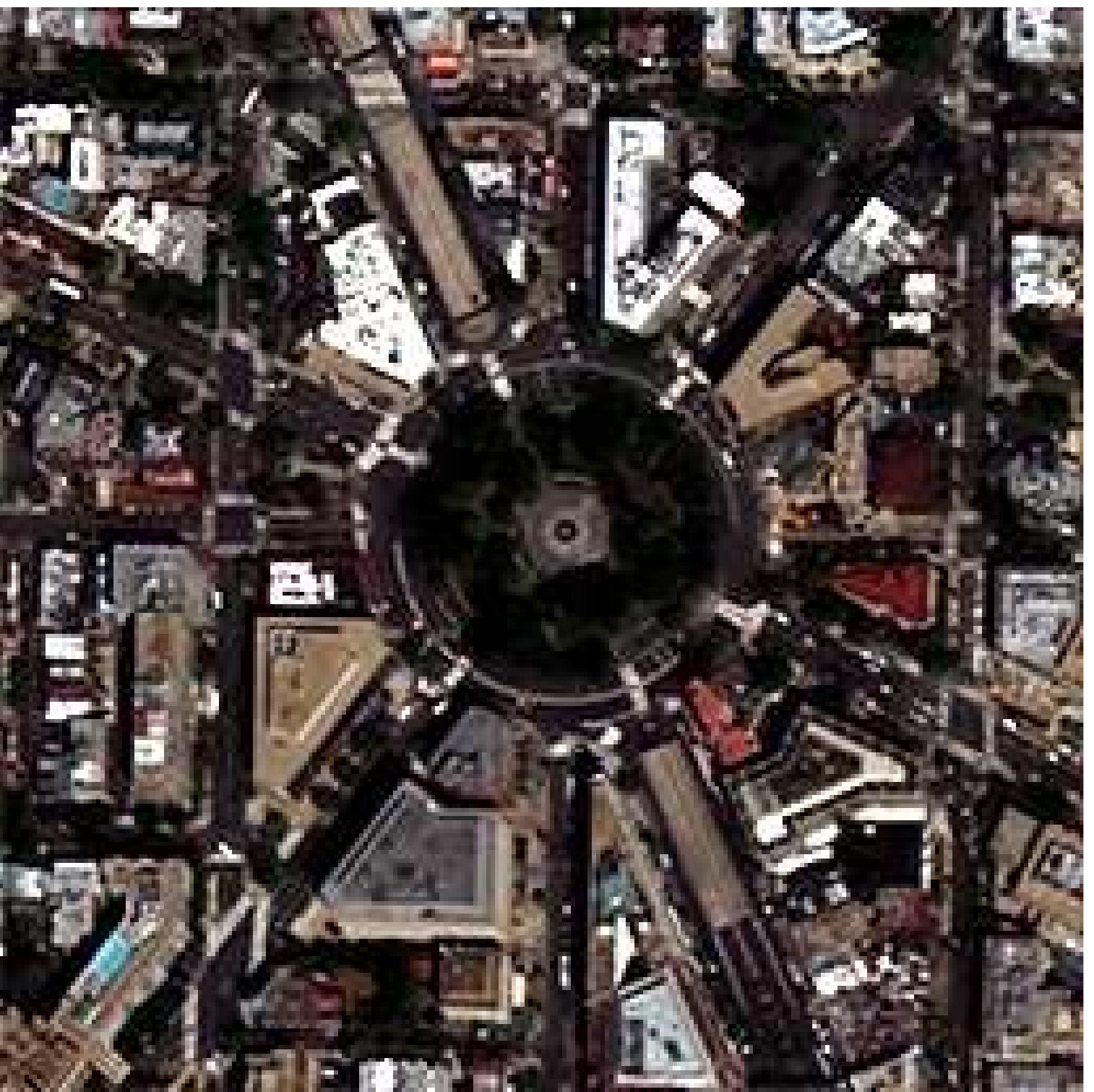} \\
(i) &(j) & (k)  \\
\\
\end{tabular}
\caption{Pansharpening results for Worldview-2 dataset (composited with red, green, blue bands). (a) Ground-truth; (b)EXP; (c)GSA; (d)PRACS; (e)ATWT; (f)BDSD; (g)GLP-CBD; (h)PNN; (i)DRPNN; (j)DiCNN1; (k)DiCNN2.}
\label{figure:reducewv2}
\end{figure*}

\begin{figure*}[t]\scriptsize
\centering
\begin{tabular}{ccccc}
\includegraphics[width=0.14\paperwidth]{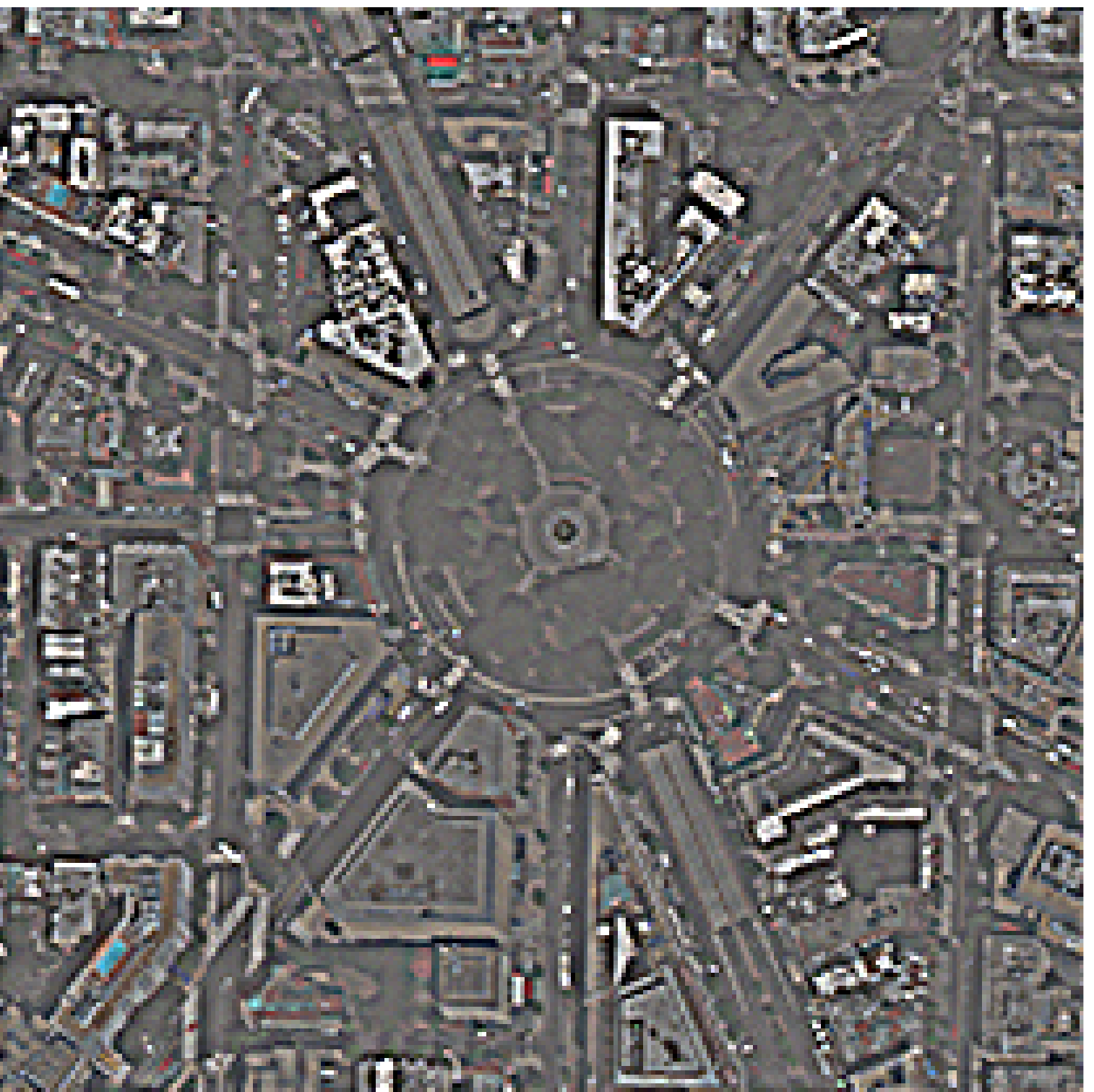} &
\includegraphics[width=0.14\paperwidth]{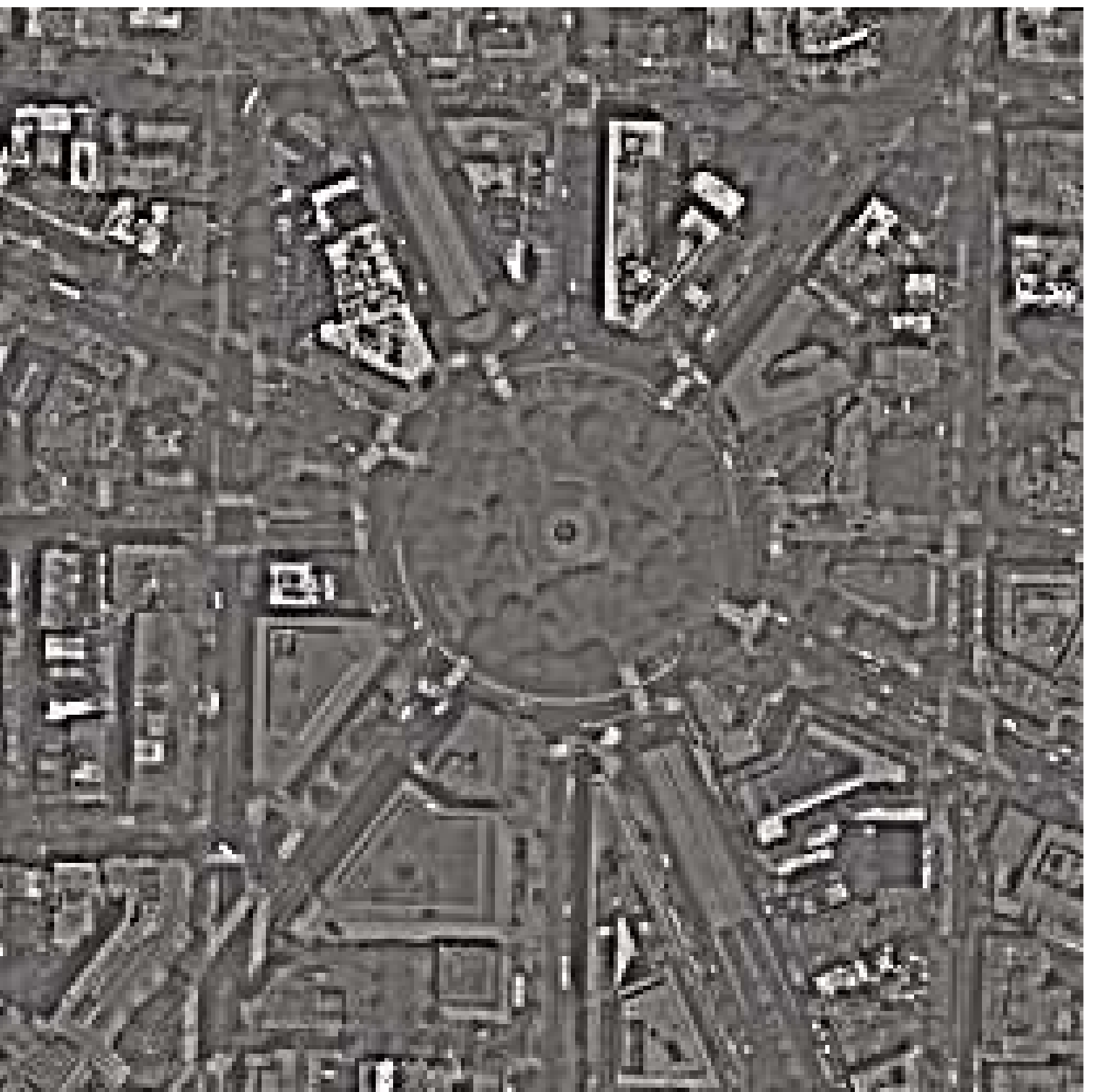} &
\includegraphics[width=0.14\paperwidth]{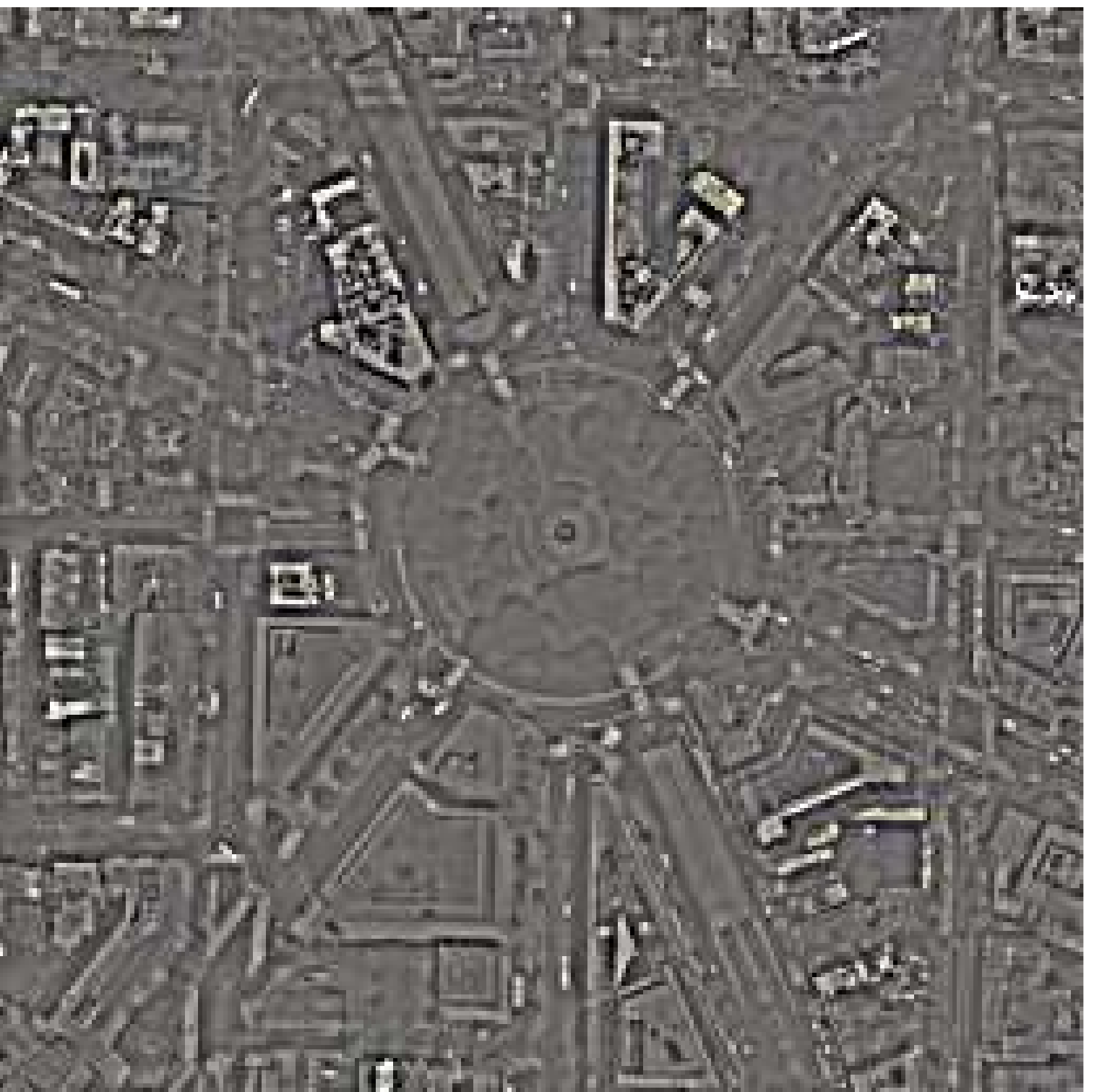} &
\includegraphics[width=0.14\paperwidth]{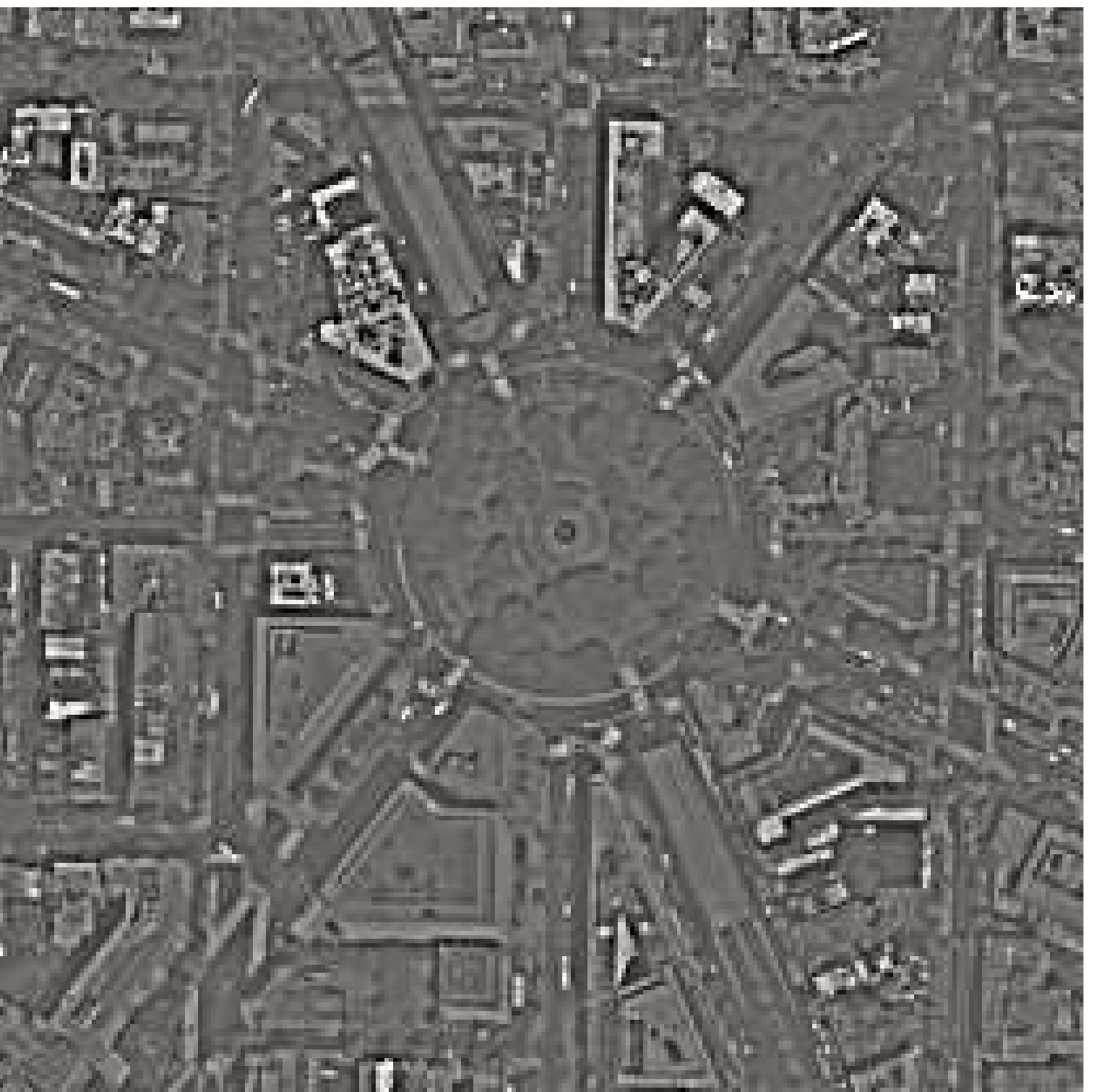} &
\includegraphics[width=0.14\paperwidth]{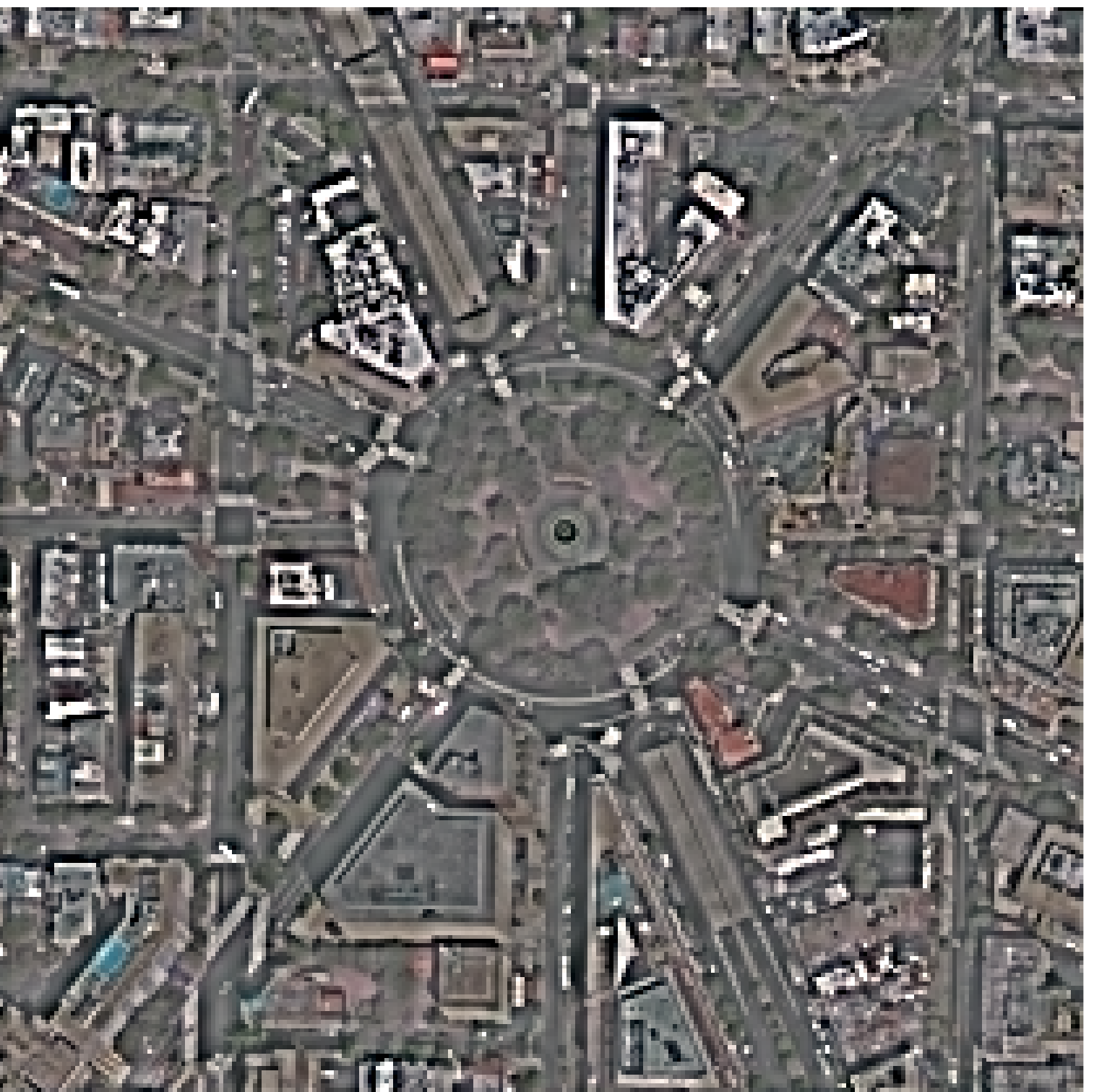} \\
(a) & (b) & (c) &(d) &(e)  \\
\includegraphics[width=0.14\paperwidth]{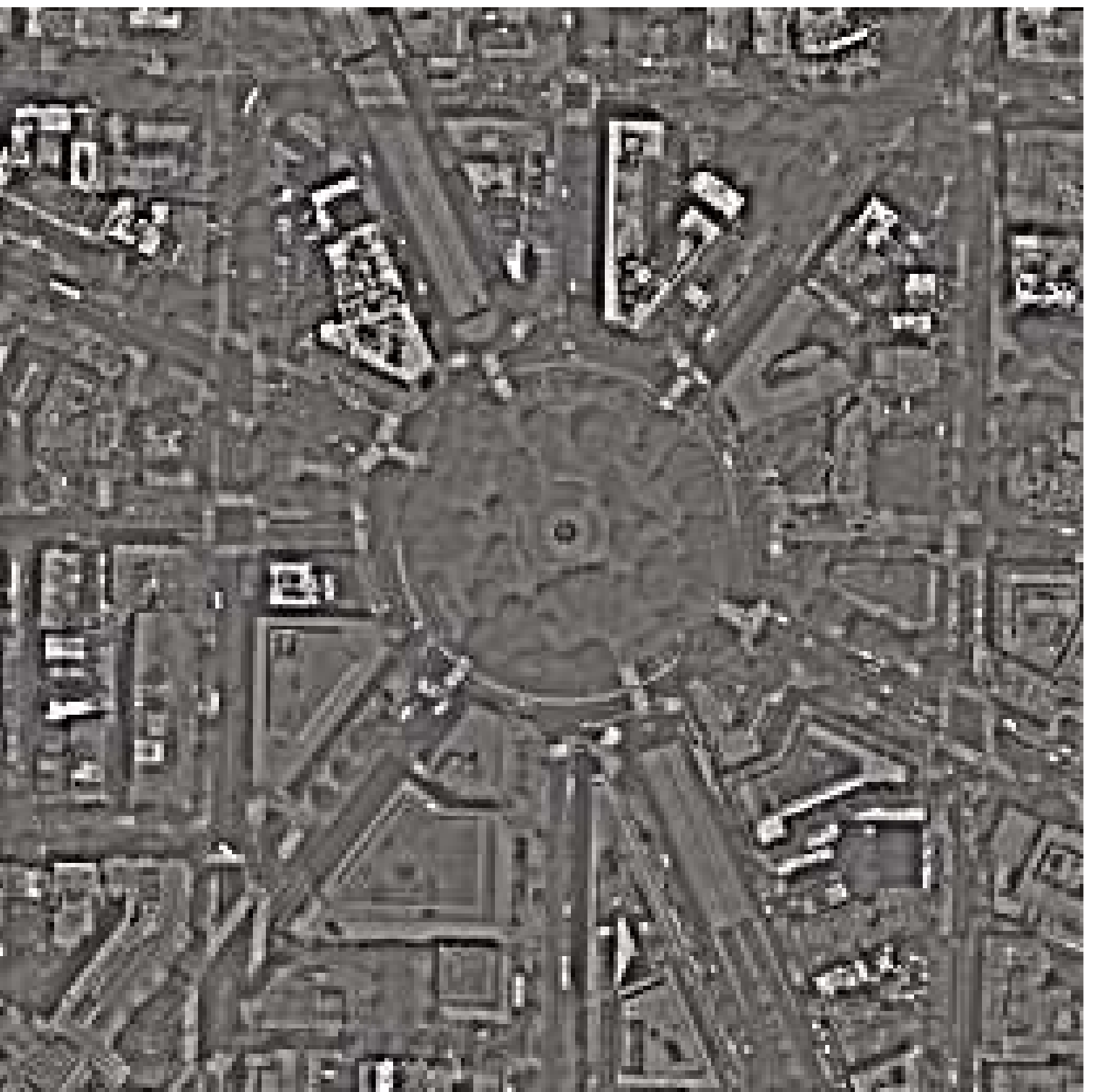} &
\includegraphics[width=0.14\paperwidth]{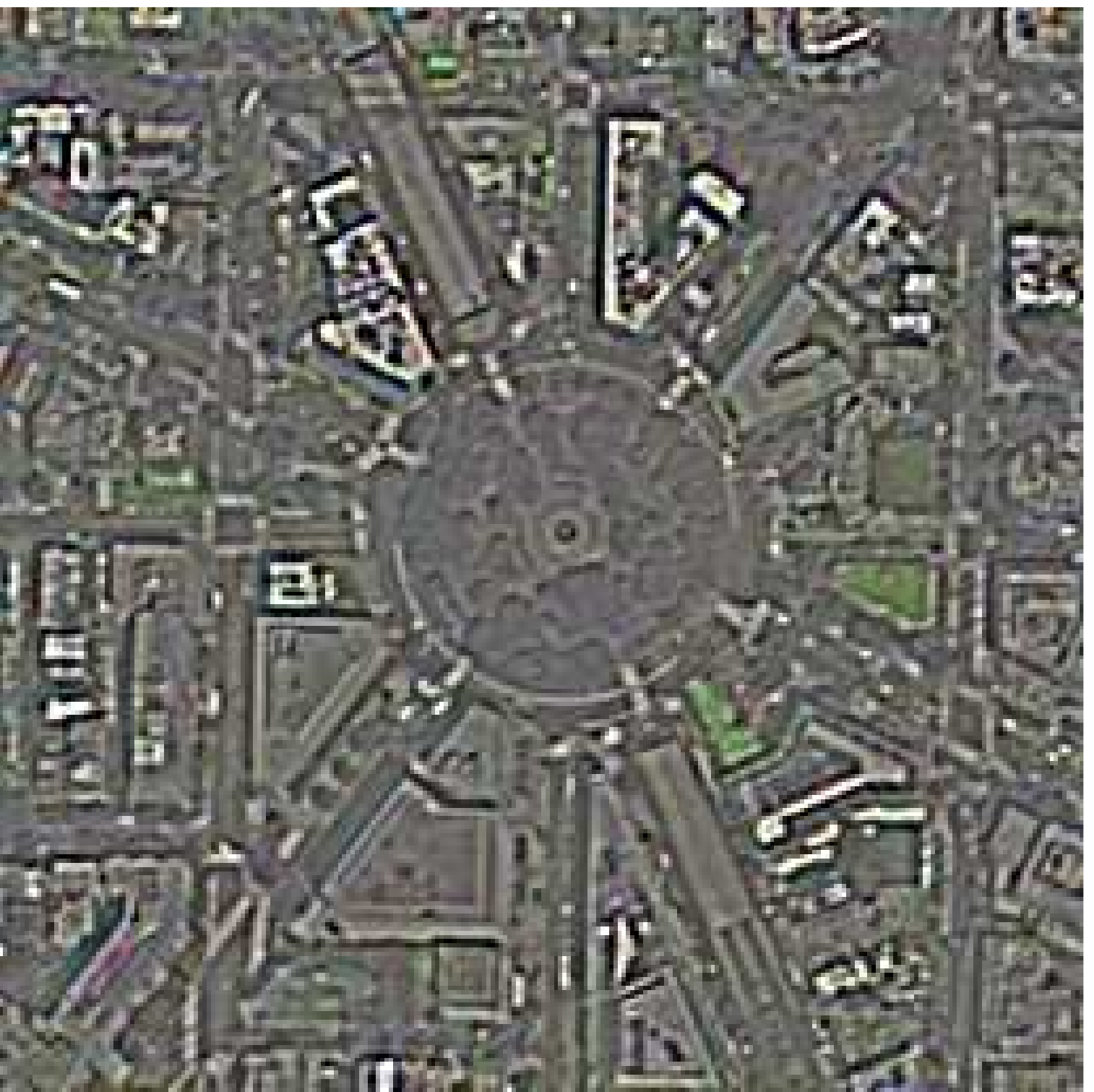} &
\includegraphics[width=0.14\paperwidth]{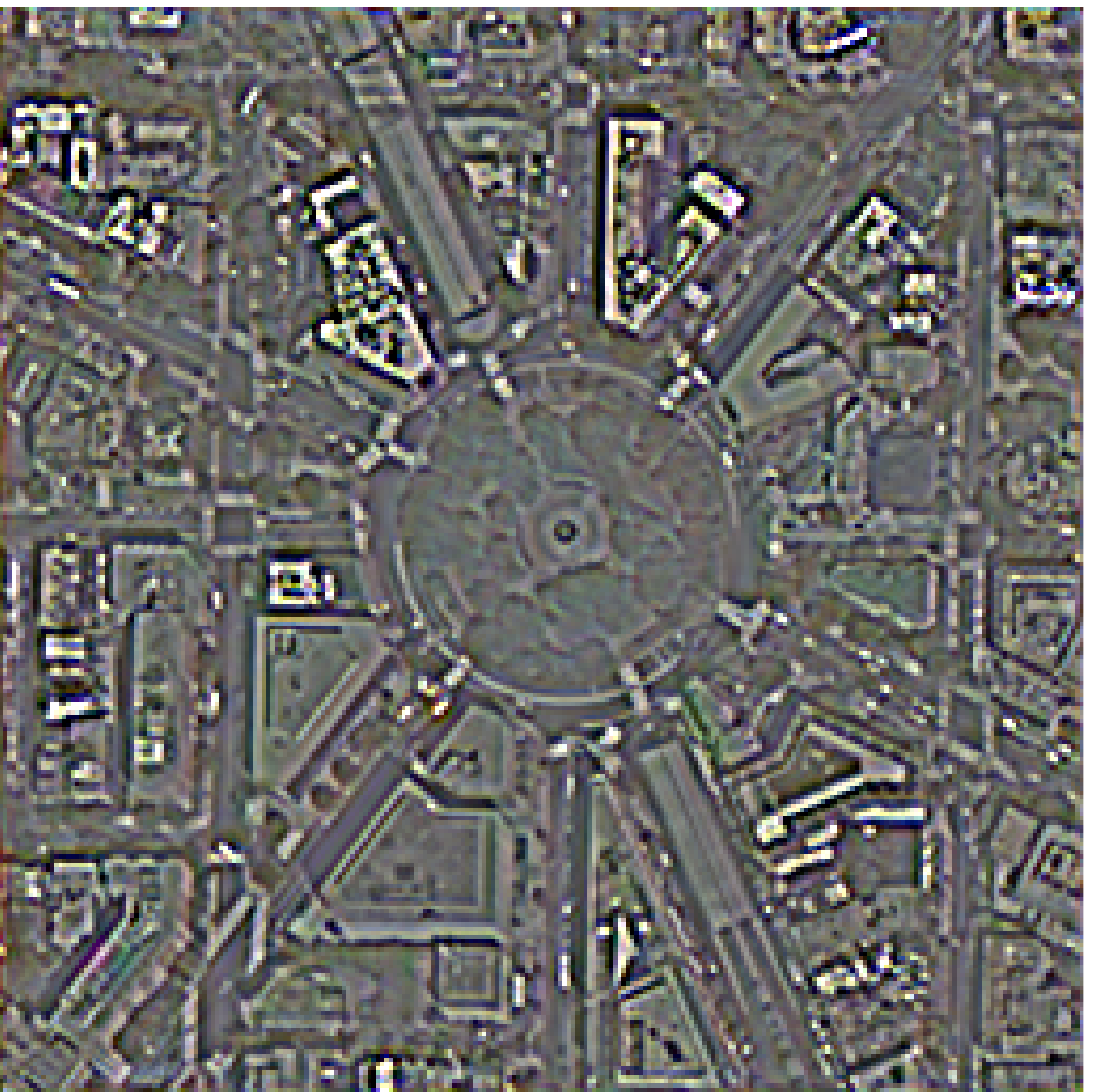} &
\includegraphics[width=0.14\paperwidth]{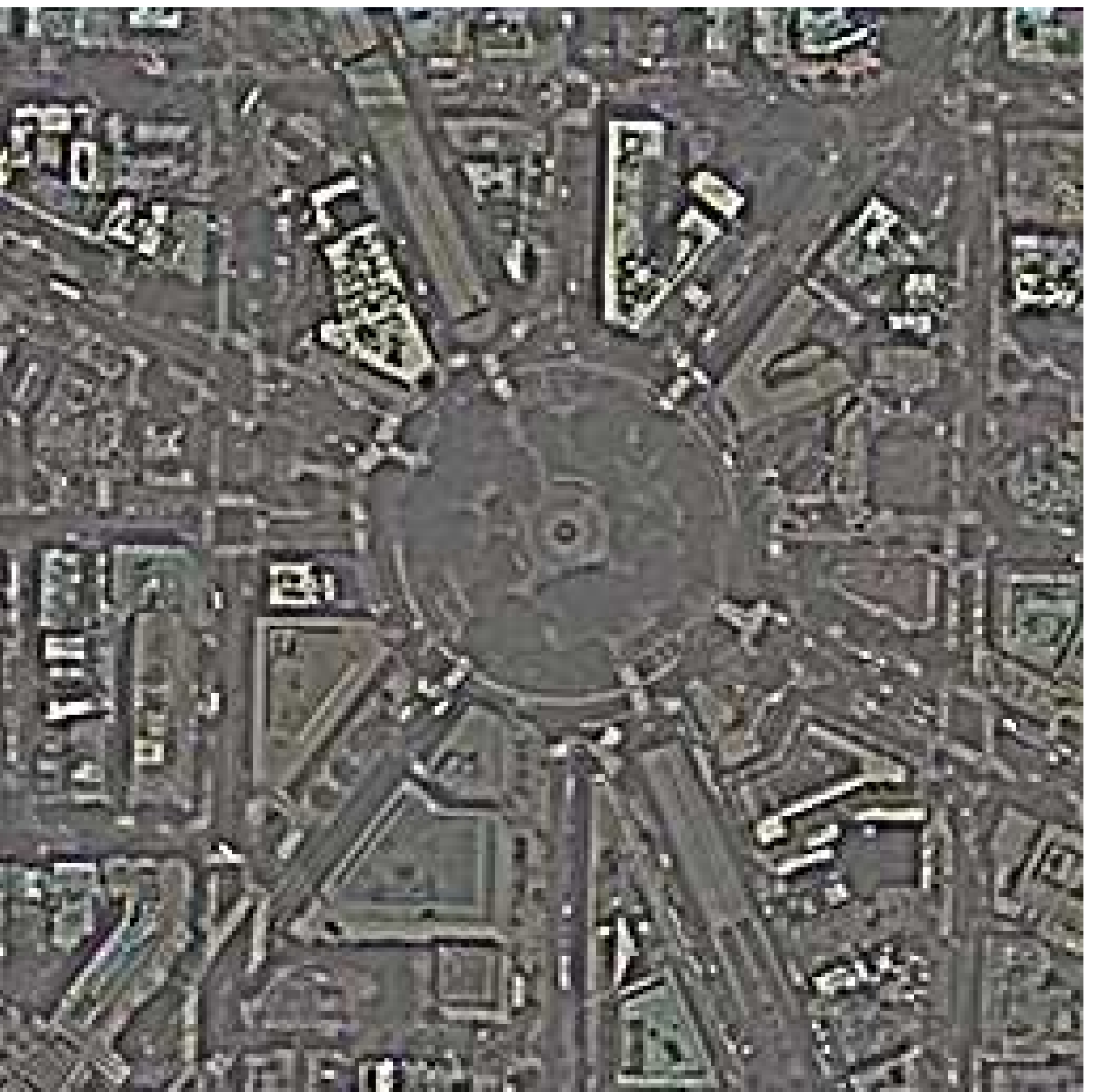} &
\includegraphics[width=0.14\paperwidth]{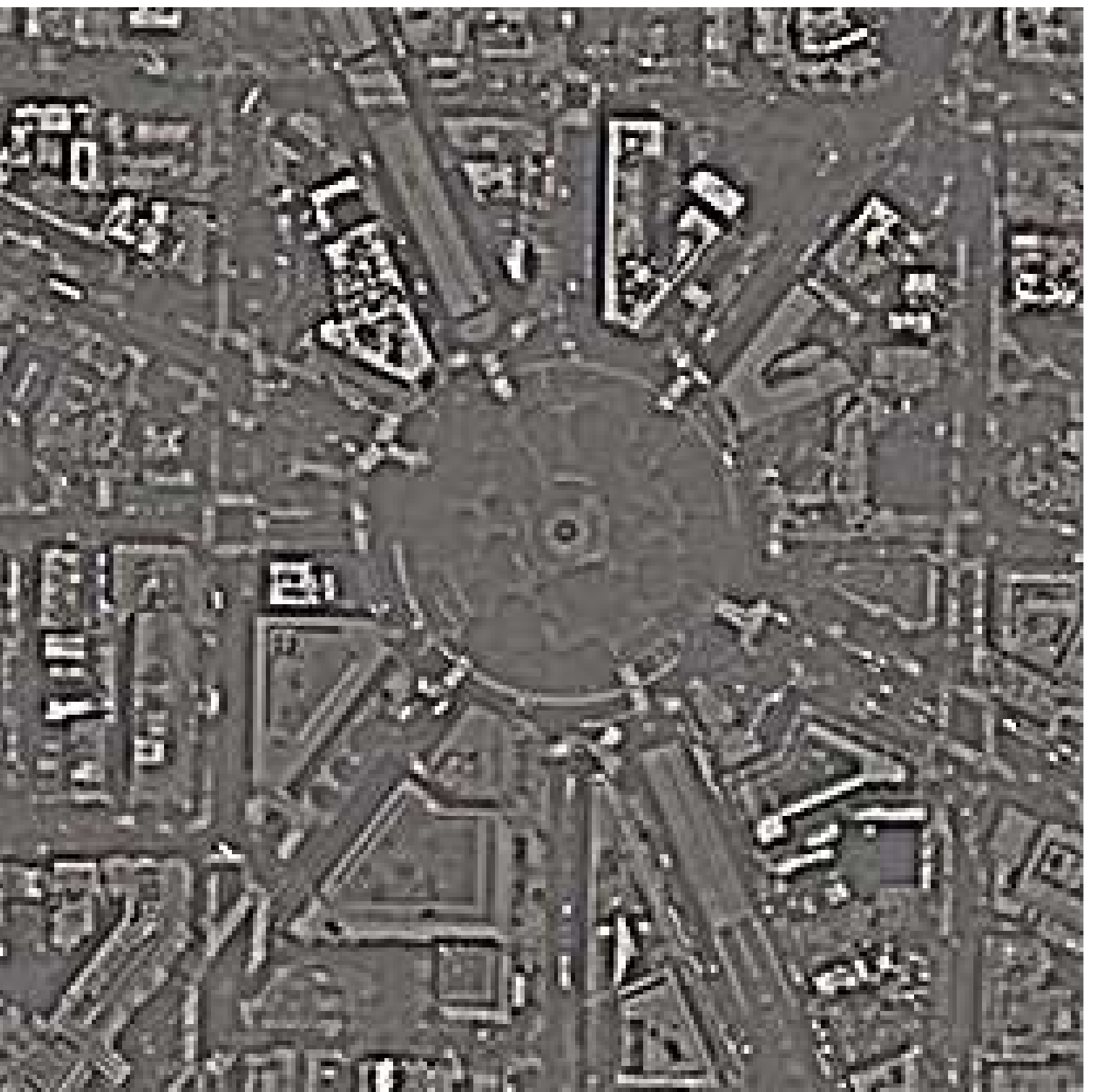} \\
(f) & (g) &(h) &(i) &(j) \\
\\
\end{tabular}
\caption{Detail images of Worldview-2 dataset (a) Ground-truth; (b) GSA; (c) PRACS; (d) ATWT; (e) BDSD; (f) GLP-CBD; (g) PNN; (h) DRPNN; (i)DiCNN1; (j) DiCNN2.}
\label{figure:detailimage:wv}
\end{figure*}

\begin{figure*}[t]\scriptsize
\centering
\begin{tabular}{ccccc}
\includegraphics[width=0.14\paperwidth]{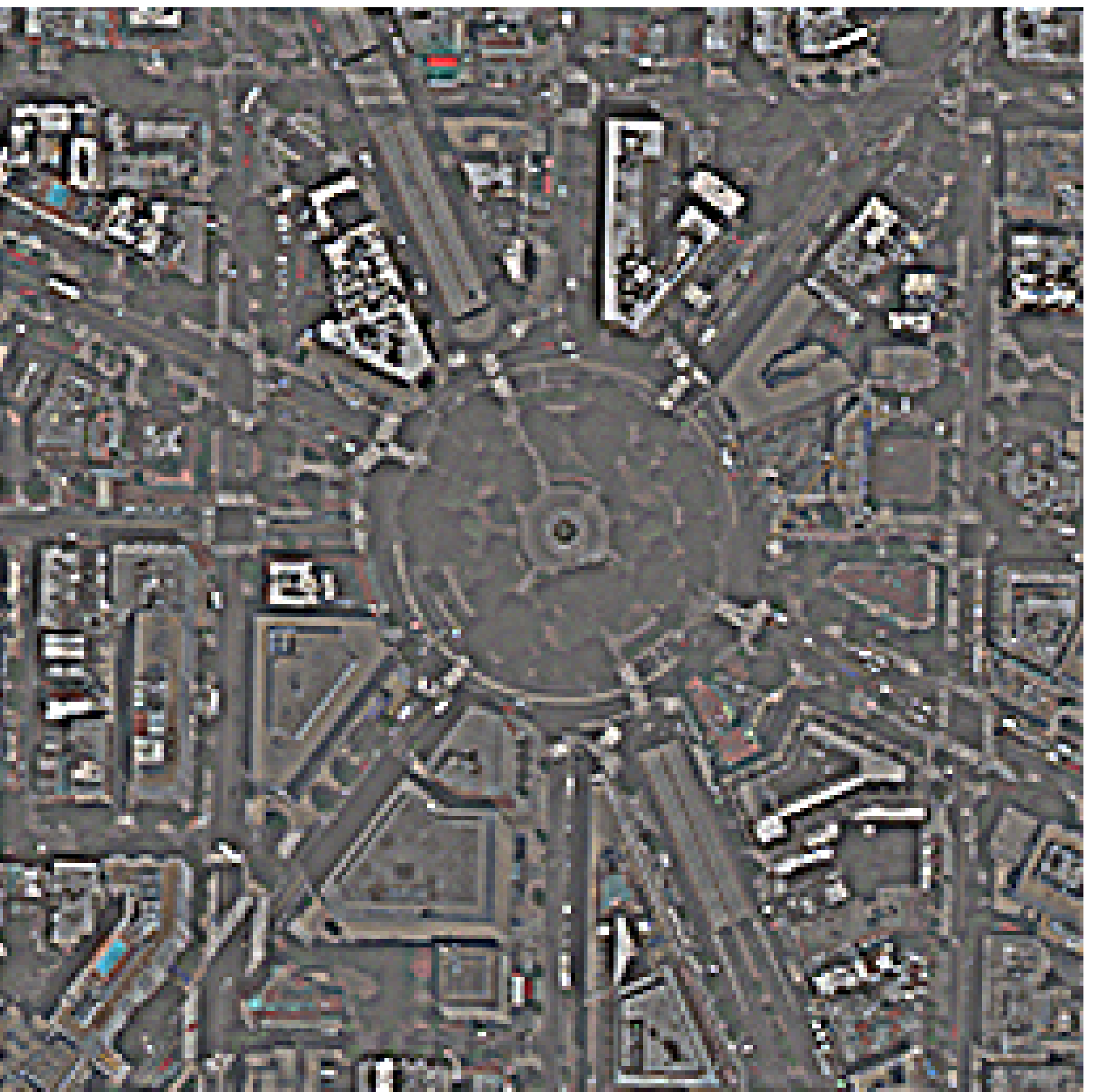} &
\includegraphics[width=0.14\paperwidth]{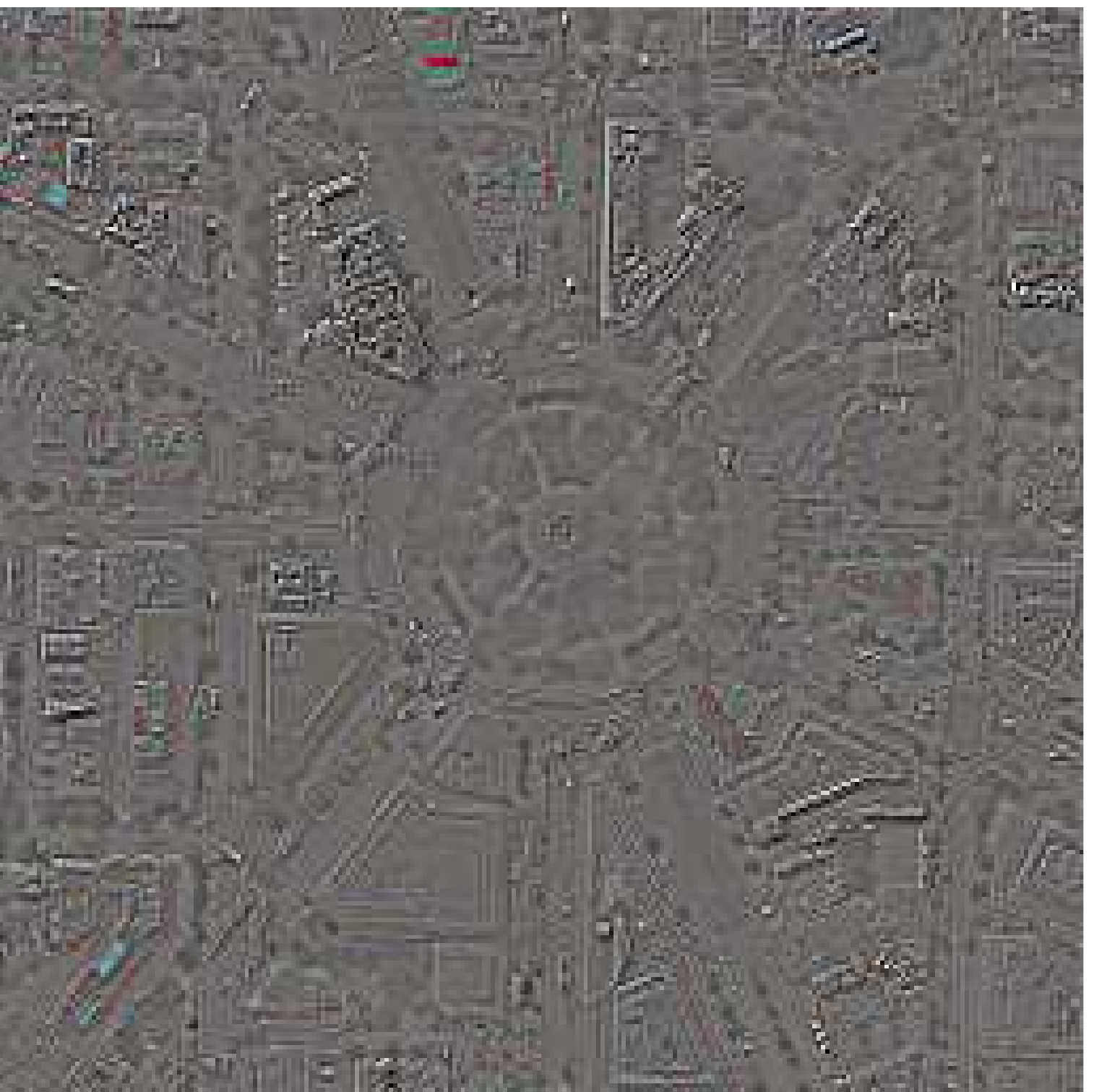} &
\includegraphics[width=0.14\paperwidth]{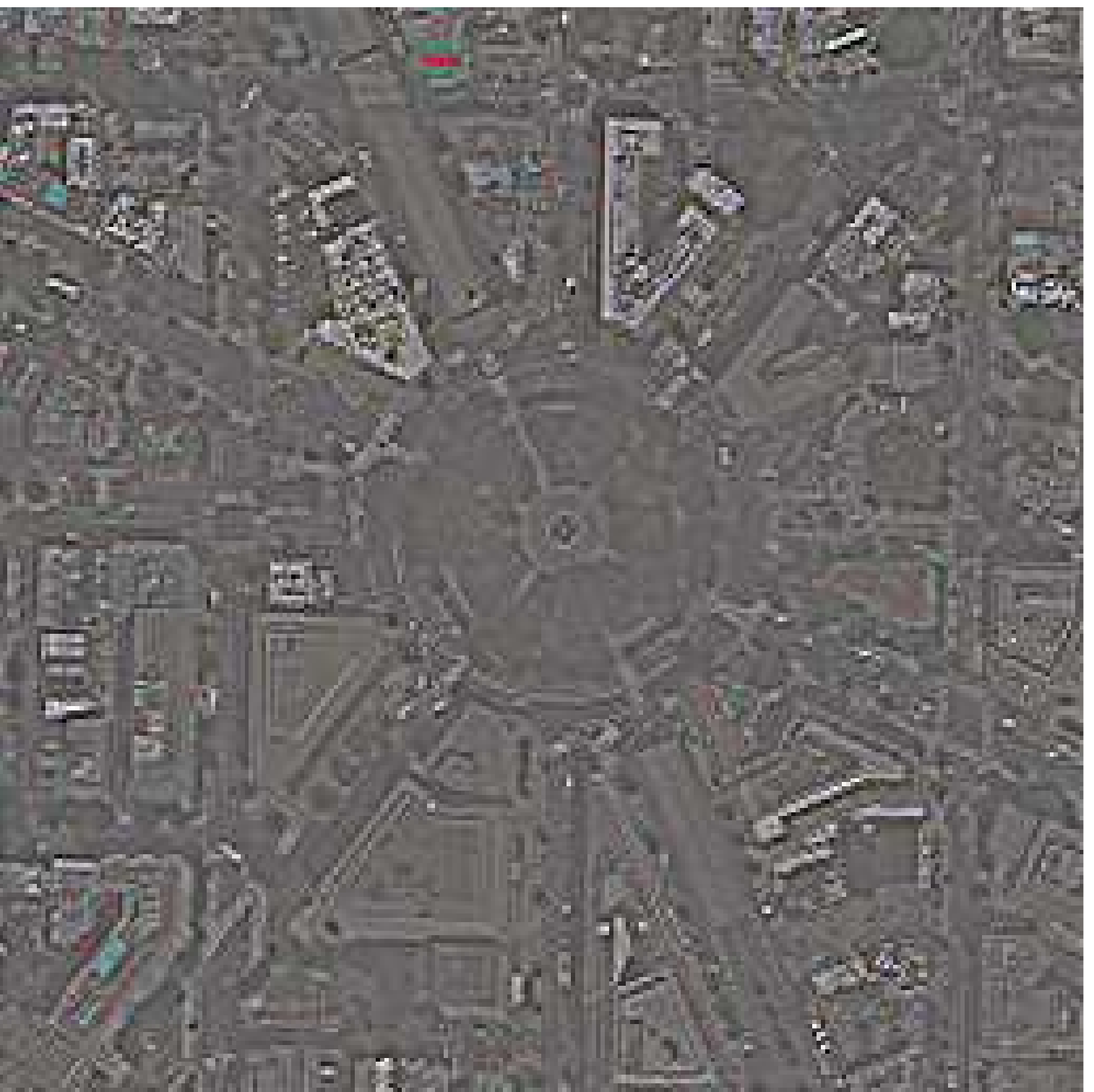} &
\includegraphics[width=0.14\paperwidth]{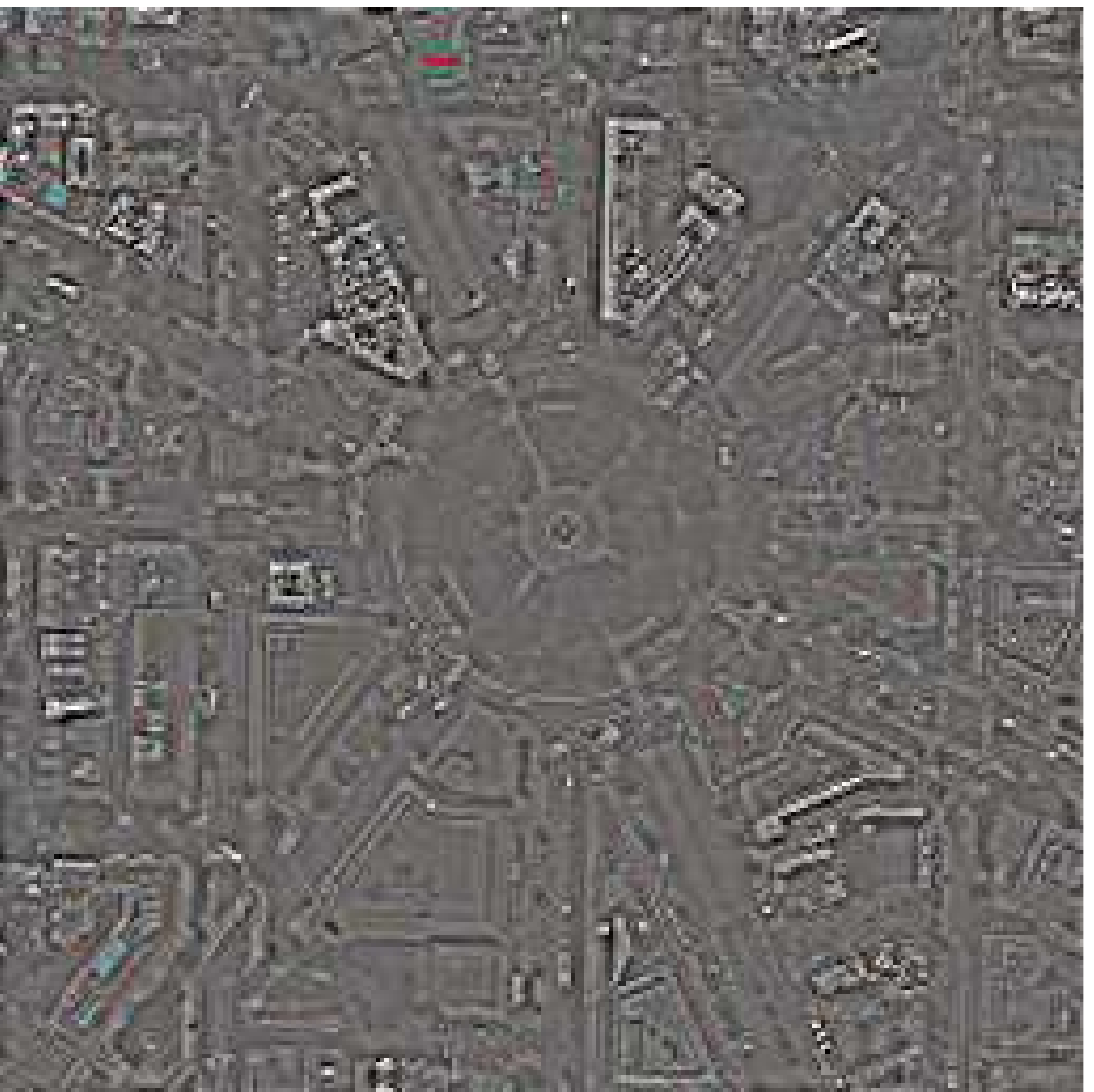} &
\includegraphics[width=0.14\paperwidth]{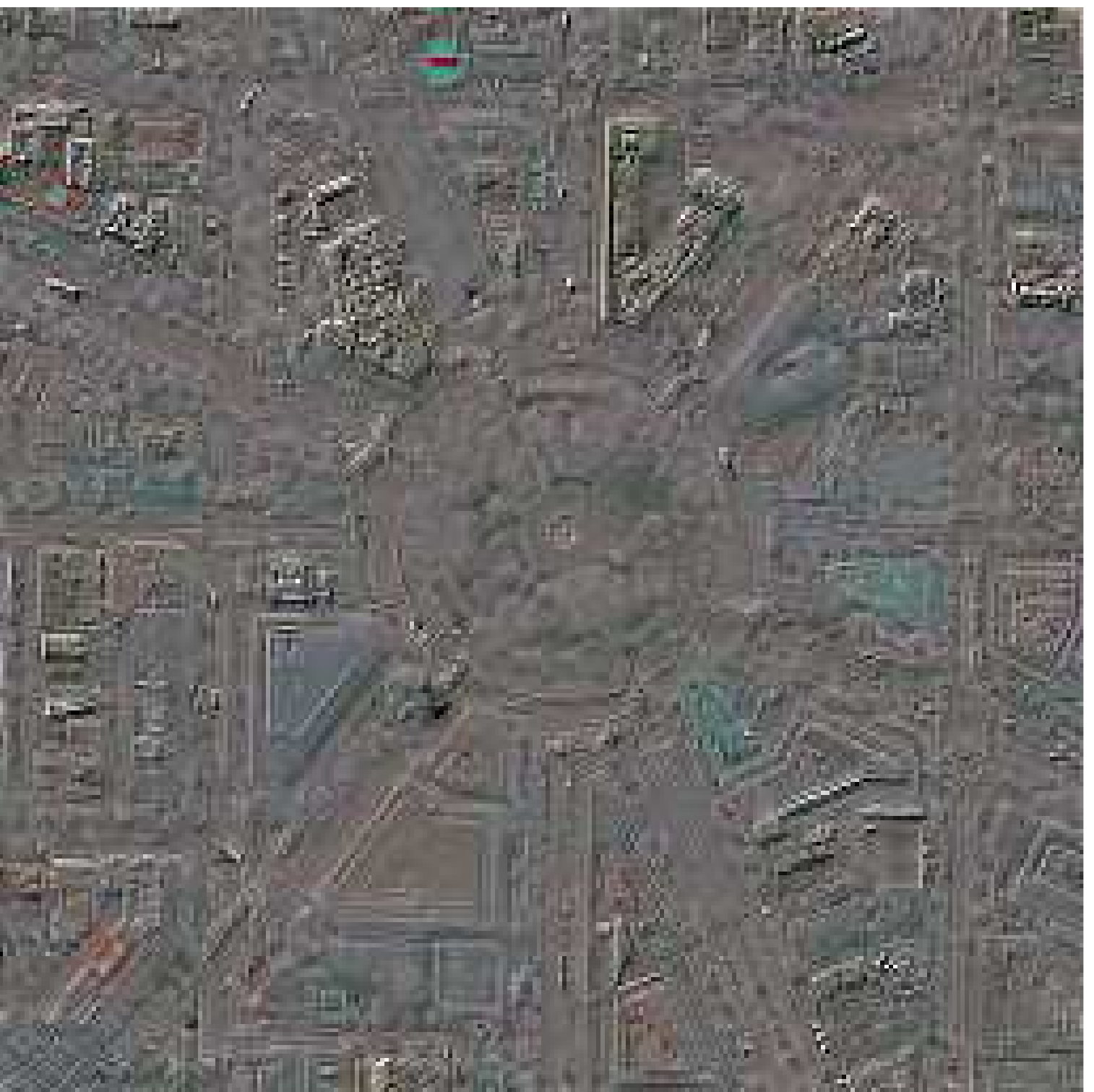} \\
(a) & (b) & (c) &(d) &(e)  \\
\includegraphics[width=0.14\paperwidth]{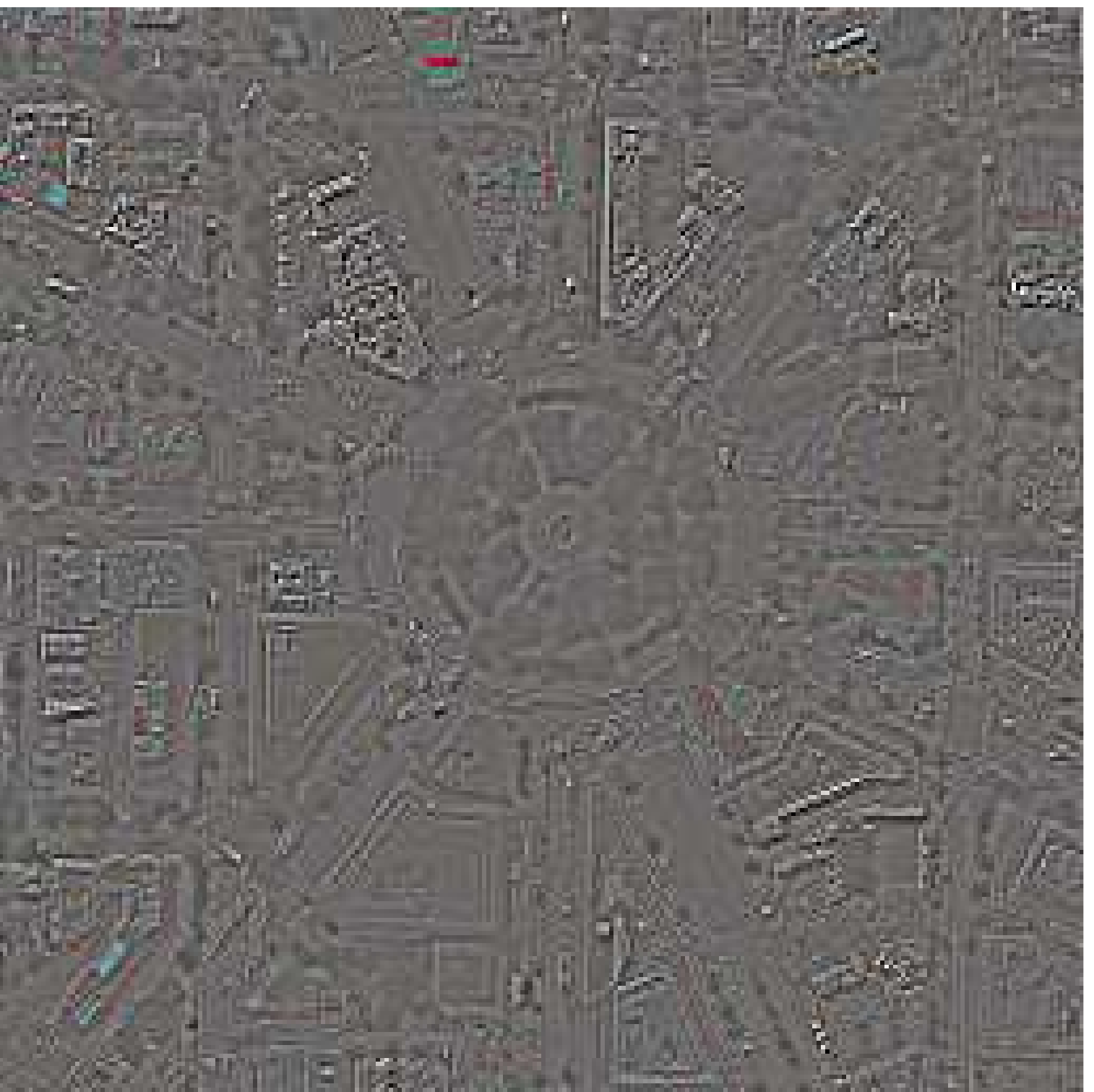} &
\includegraphics[width=0.14\paperwidth]{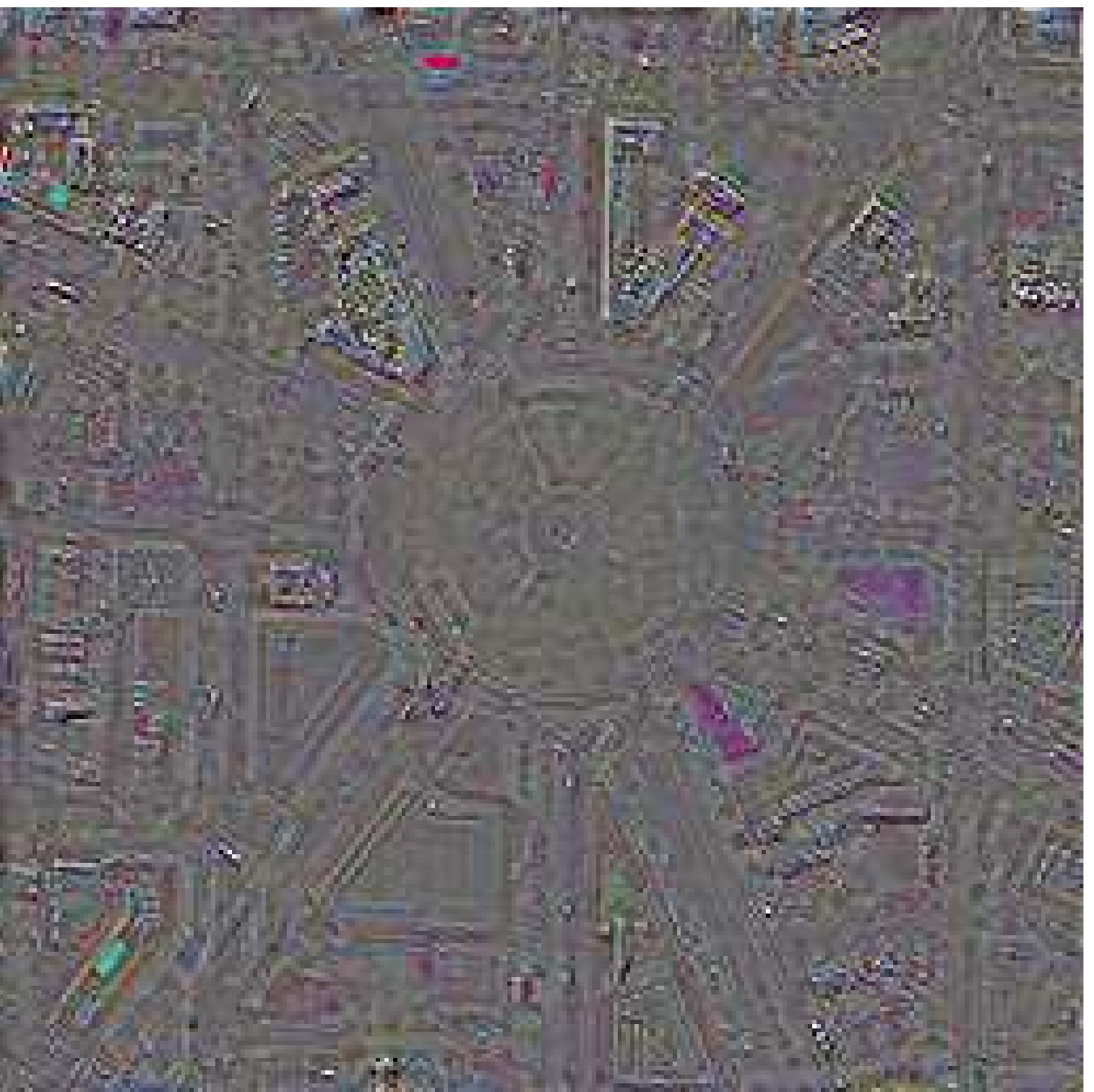} &
\includegraphics[width=0.14\paperwidth]{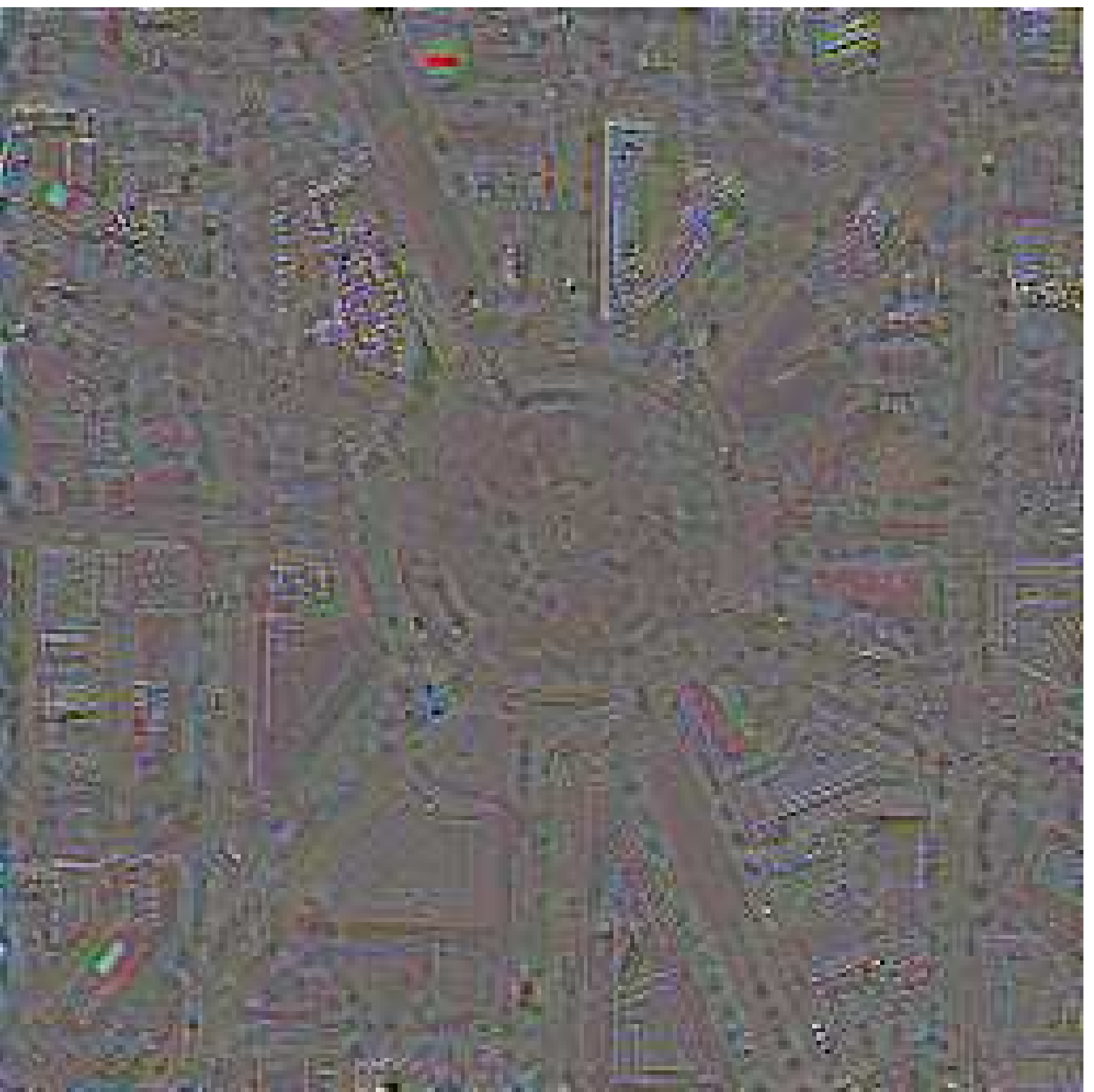} &
\includegraphics[width=0.14\paperwidth]{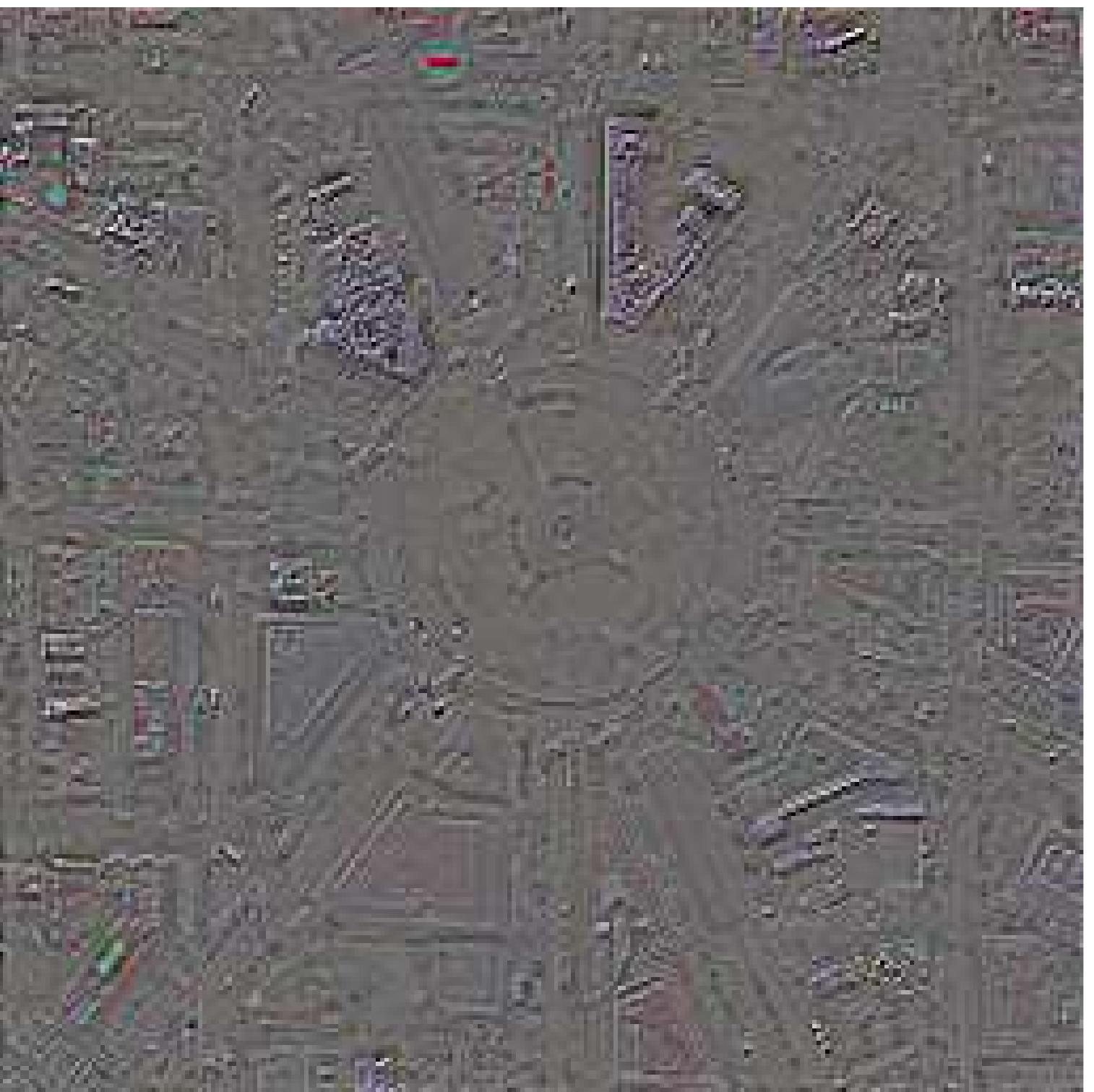} &
\includegraphics[width=0.14\paperwidth]{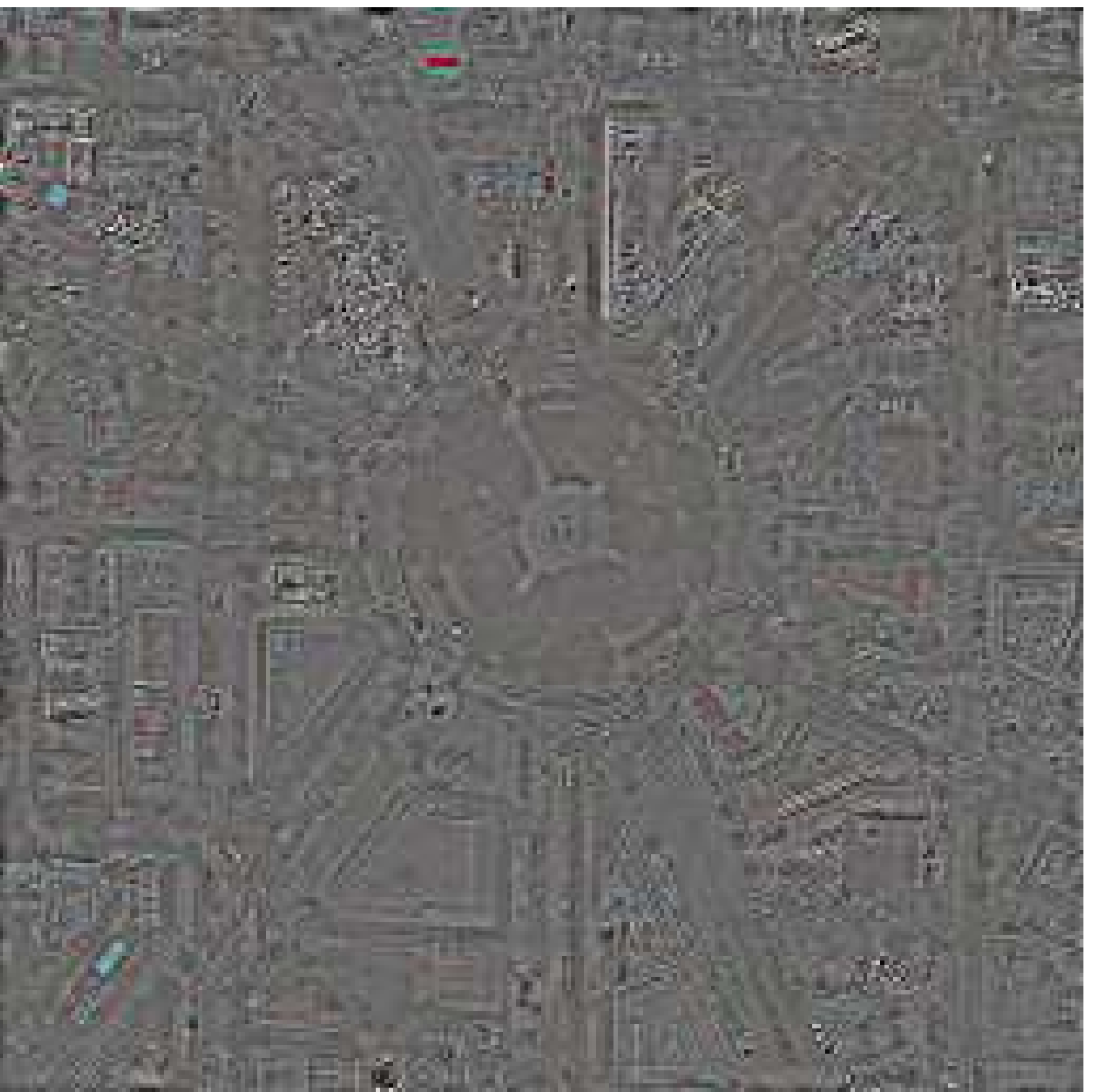} \\
(f) & (g) &(h) &(i) &(j) \\
\\
\end{tabular}
\caption{The differences between pansharpened images and ground-truth of Worldview-2 dataset (a) EXP; (b) GSA; (c) PRACS; (d) ATWT; (e) BDSD; (f) GLP-CBD; (g) PNN; (h) DRPNN; (i)DiCNN1; (j) DiCNN2.}
\label{figure:resimage:wv}
\end{figure*}

The dataset\footnote{Available online: https://www.digitalglobe.com/resources/product-samples} was acquired with the WorldView-2 sensor over an urban area in Washington D.C., which provides a PAN image formed from wavelength $450nm$ to $800nm$, and a MS image with eight bands, including four standard bands (blue, green, red and near infrared 1) and four new bands (coastal, yellow, red edge and near infrared 2). The resolution ratio $R$ is 4 and the radiometric resolution is 11 bits, with the spatial resolution of the PAN image and that of the MS image being $0.46m$ and $1.84m$, respectively. We choose two scenes with $256\times256$ pixels for test in the reduced-resolution and full-resolution experiment separately.

Table \ref{table:qualitywv} shows the results of the reduced-resolution quality assessment. As we can observe, CNN-based methods yield much better pansharpening quality than the CS-based and MRA-based methods. DiCNN1 and DiCNN2 achieve the highest Q8, SAM, ERGAS and SCC scores among all compared methods inclucing CNN-based methods, and meanwhile DiCNN2 is the fastest among all CNN-based methods.

Fig. \ref{figure:reducewv2} displays the images of reduced-resolution experimental results. It shows that the pansharpened images yielded by CNN-based methods look much more similar to the ground-truth, without noticeable artifacts or spectral distortions. Fig. \ref{figure:detailimage:wv} shows the detail images which are produced with the difference between the pansharpened HRMS image and the pre-interpolated LRMS image. The ground-truth details are achieved by the subtraction between the full-resolution MS image and the pre-interpolated one. The detail images are also in favor of the aforementioned observations, as it can be seen in the central circle area. For the CNN-based methods, the performances are hard to distinguish, but by investigating the spectral preservation of ground objects with small sizes, it is clear that DiCNN1 helps to impede spectral distortion more efficiently, as it can be seen in the bottom left part of Fig. \ref{figure:reducewv2}(h)-(k). Fig.\ref{figure:resimage:wv} shows the residual images which are generated by the difference between the pansharpened HRMS image and the ground-truth image.

\begin{figure*}[t]\scriptsize
\centering
  \begin{tabular}{cccc}
\includegraphics[width=0.14\paperwidth]{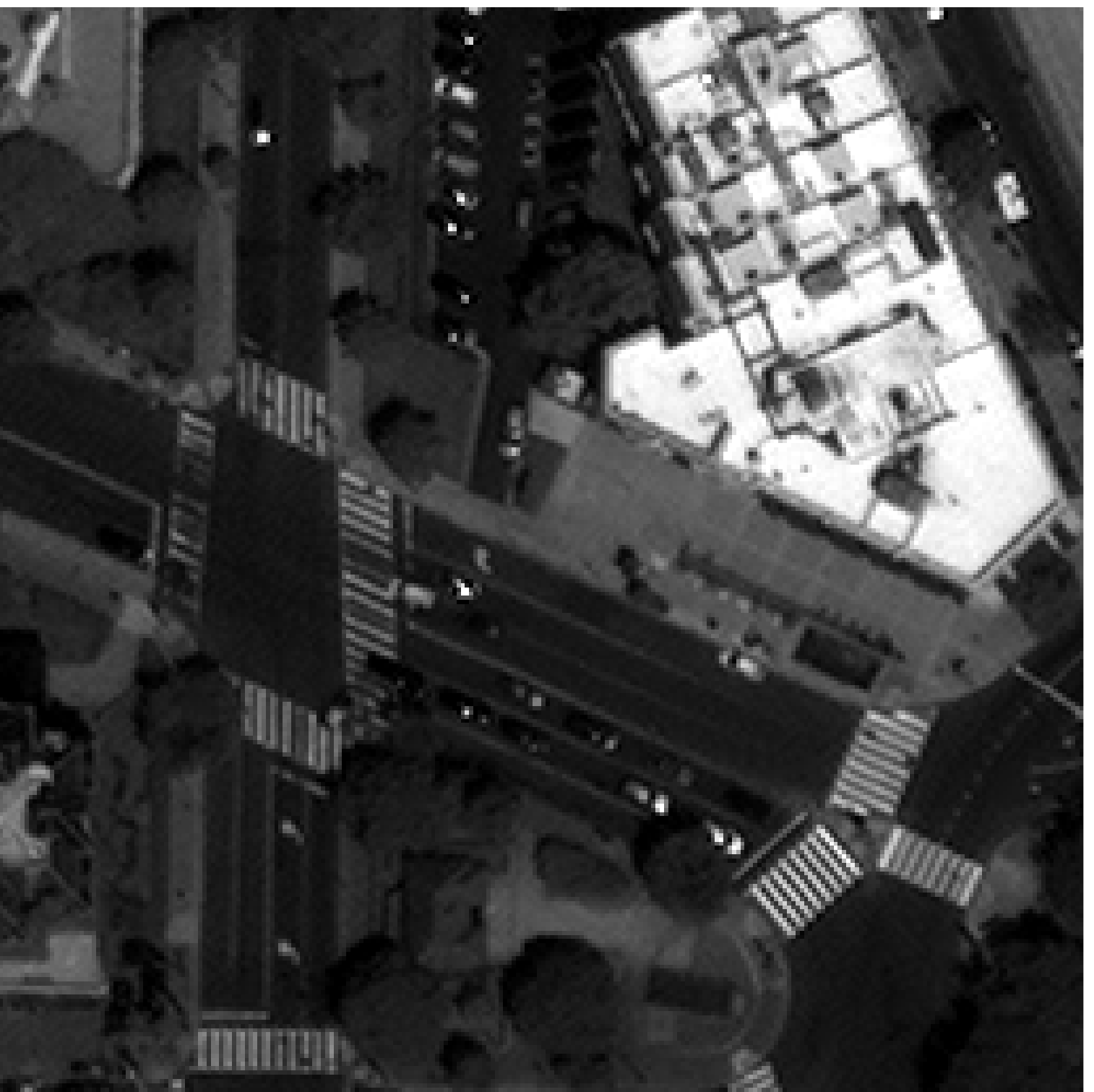} &
\includegraphics[width=0.14\paperwidth]{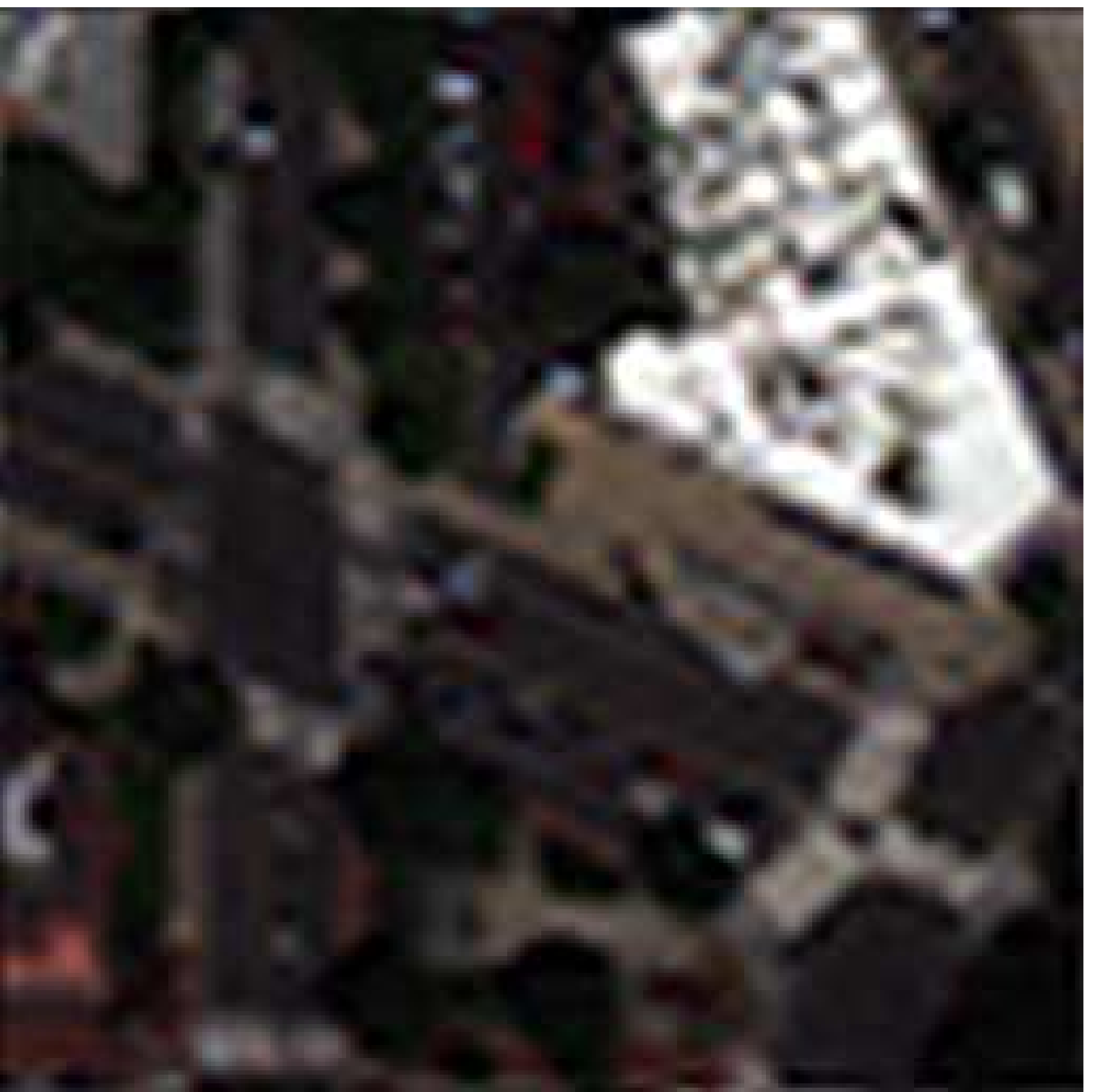} &
\includegraphics[width=0.14\paperwidth]{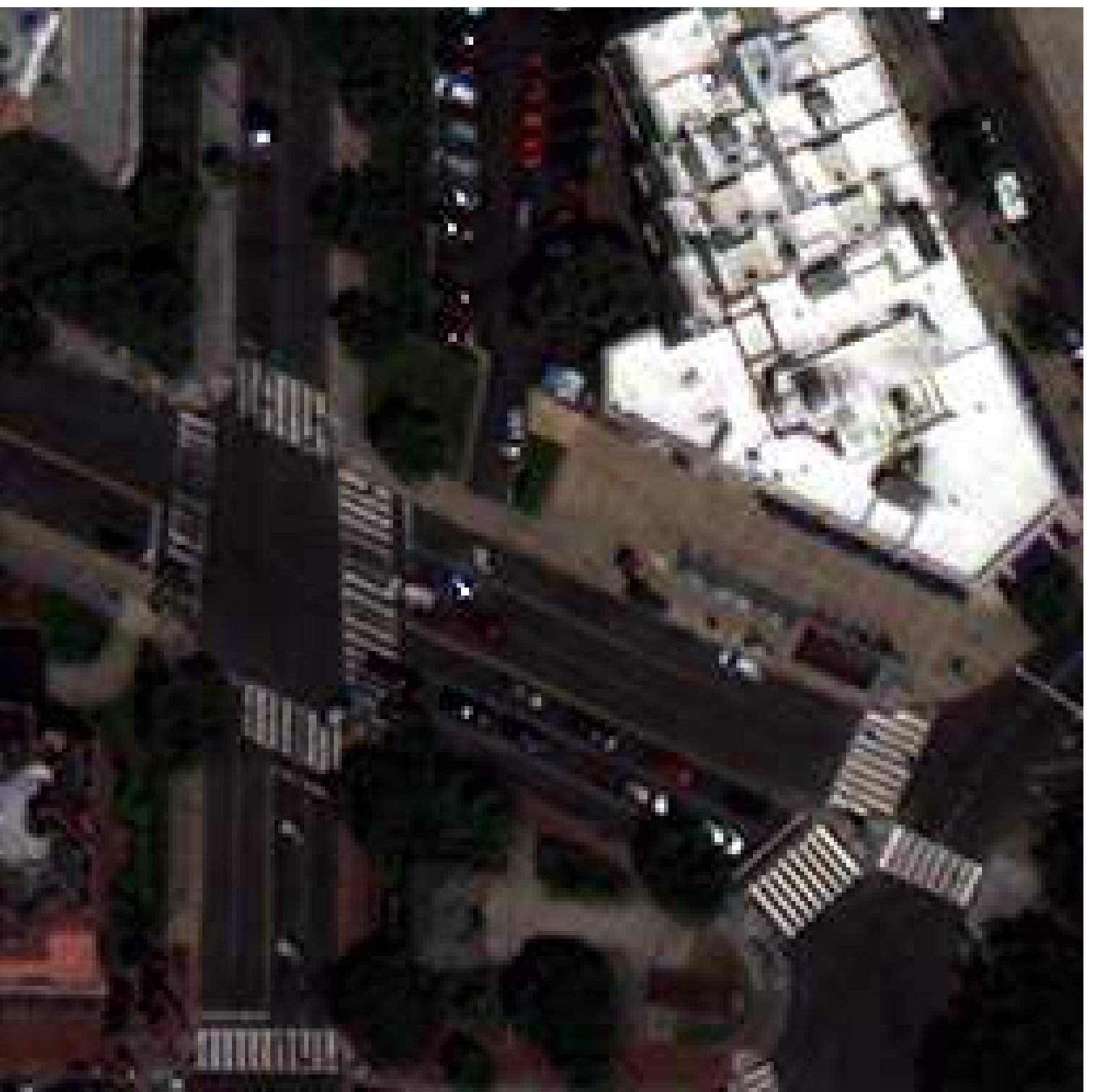} & \includegraphics[width=0.14\paperwidth]{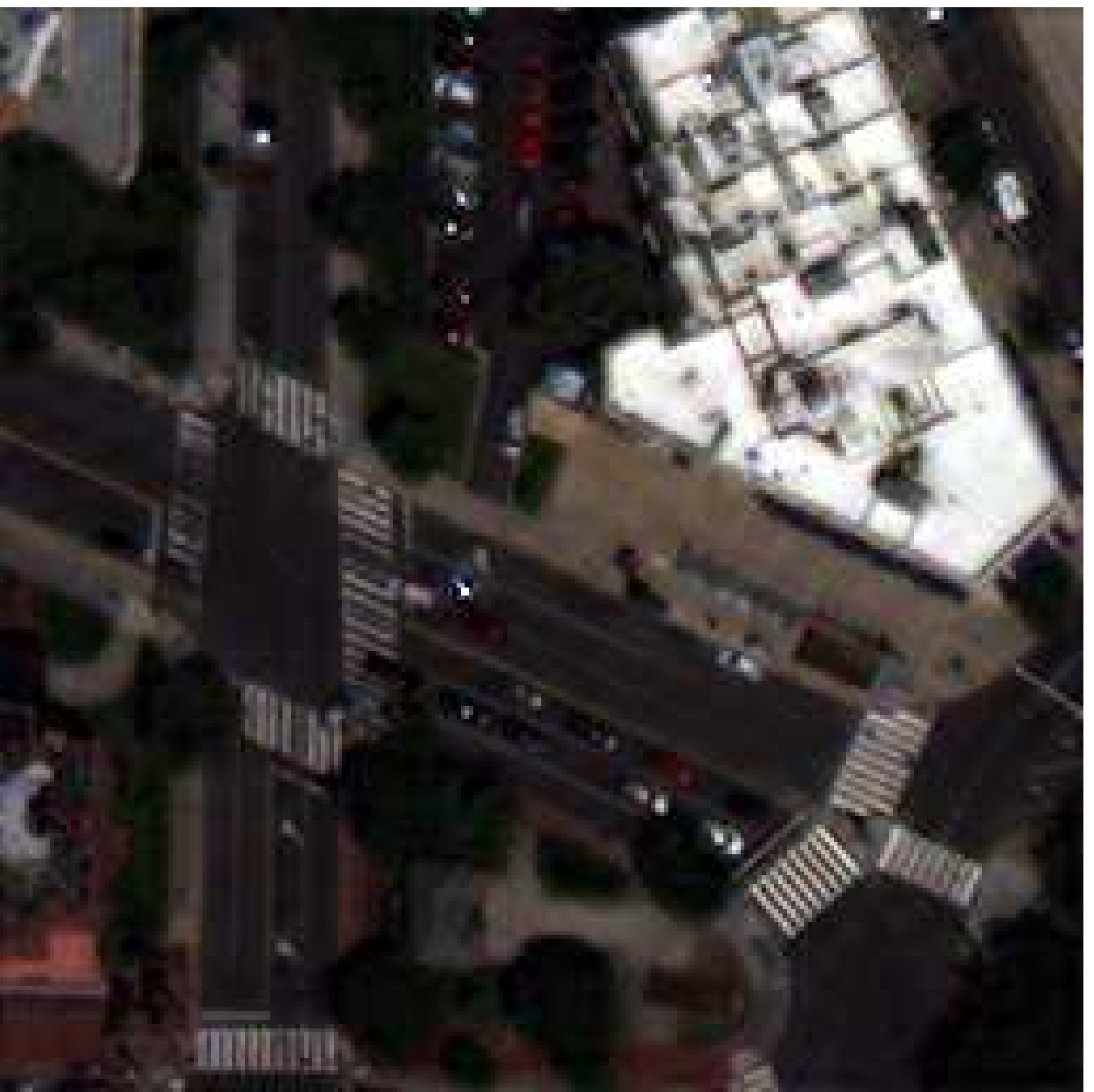} \\
(a) & (b) & (c) & (d) \\
\includegraphics[width=0.14\paperwidth]{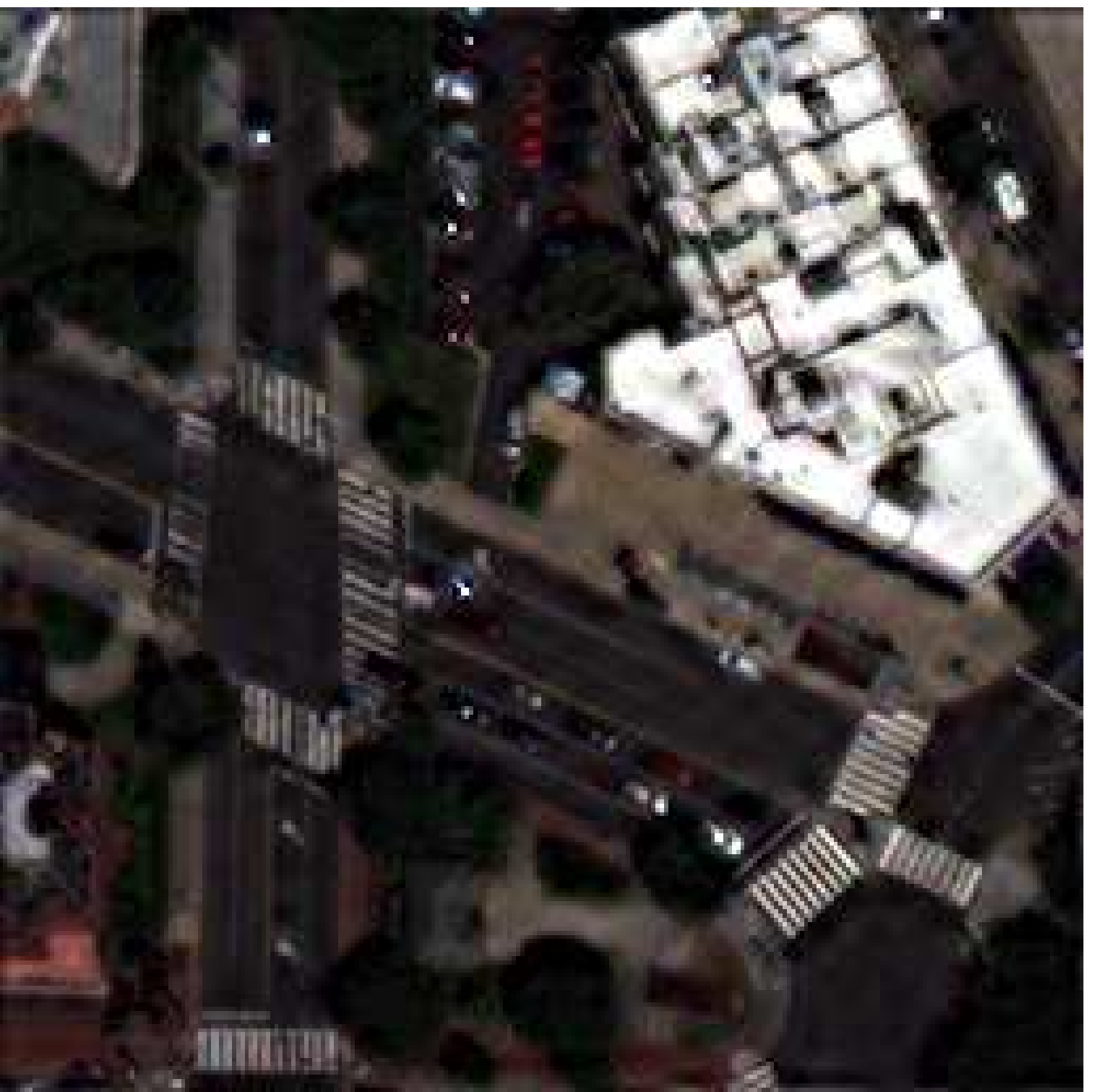} &
\includegraphics[width=0.14\paperwidth]{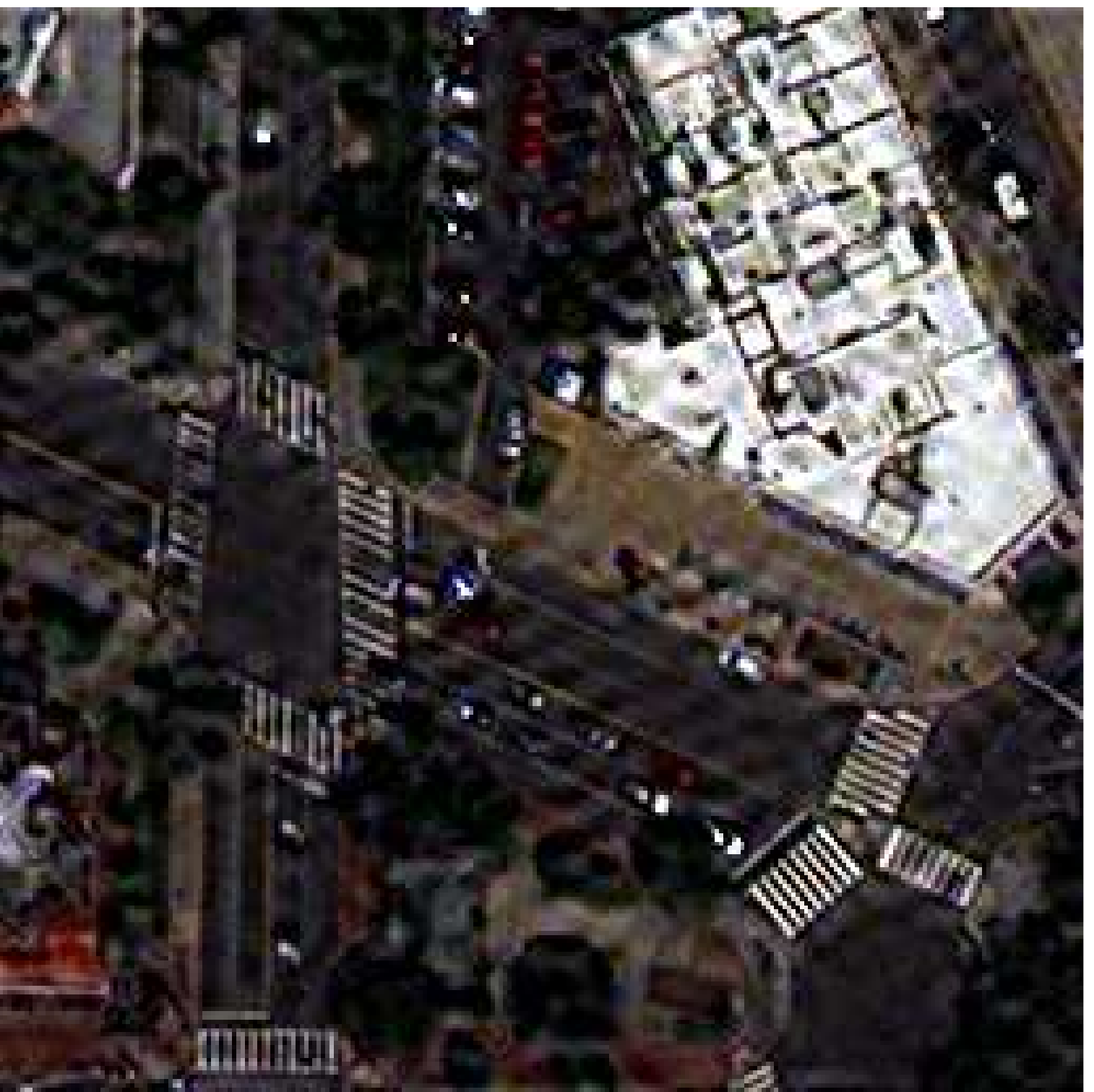} &
\includegraphics[width=0.14\paperwidth]{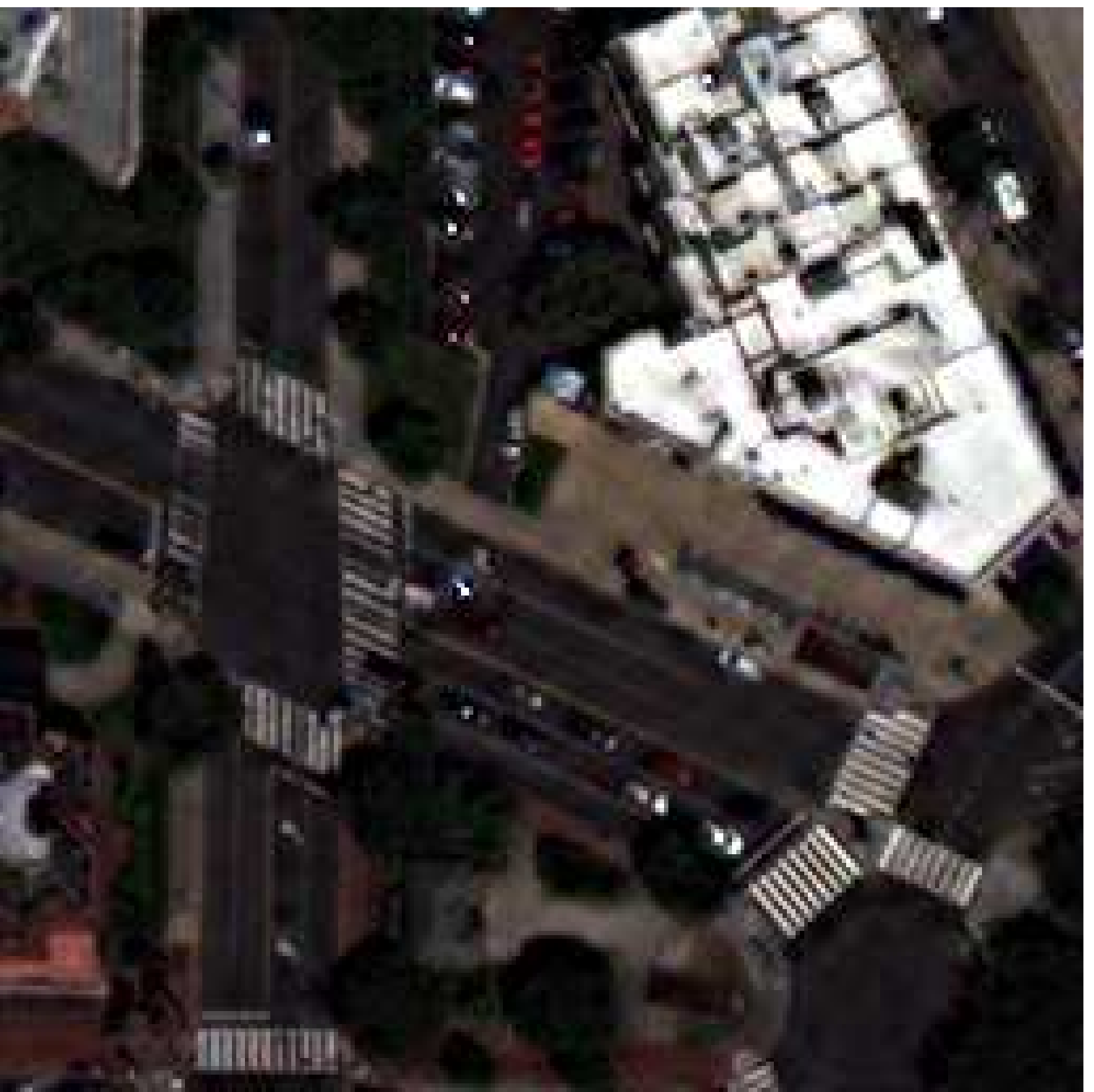} &
\includegraphics[width=0.14\paperwidth]{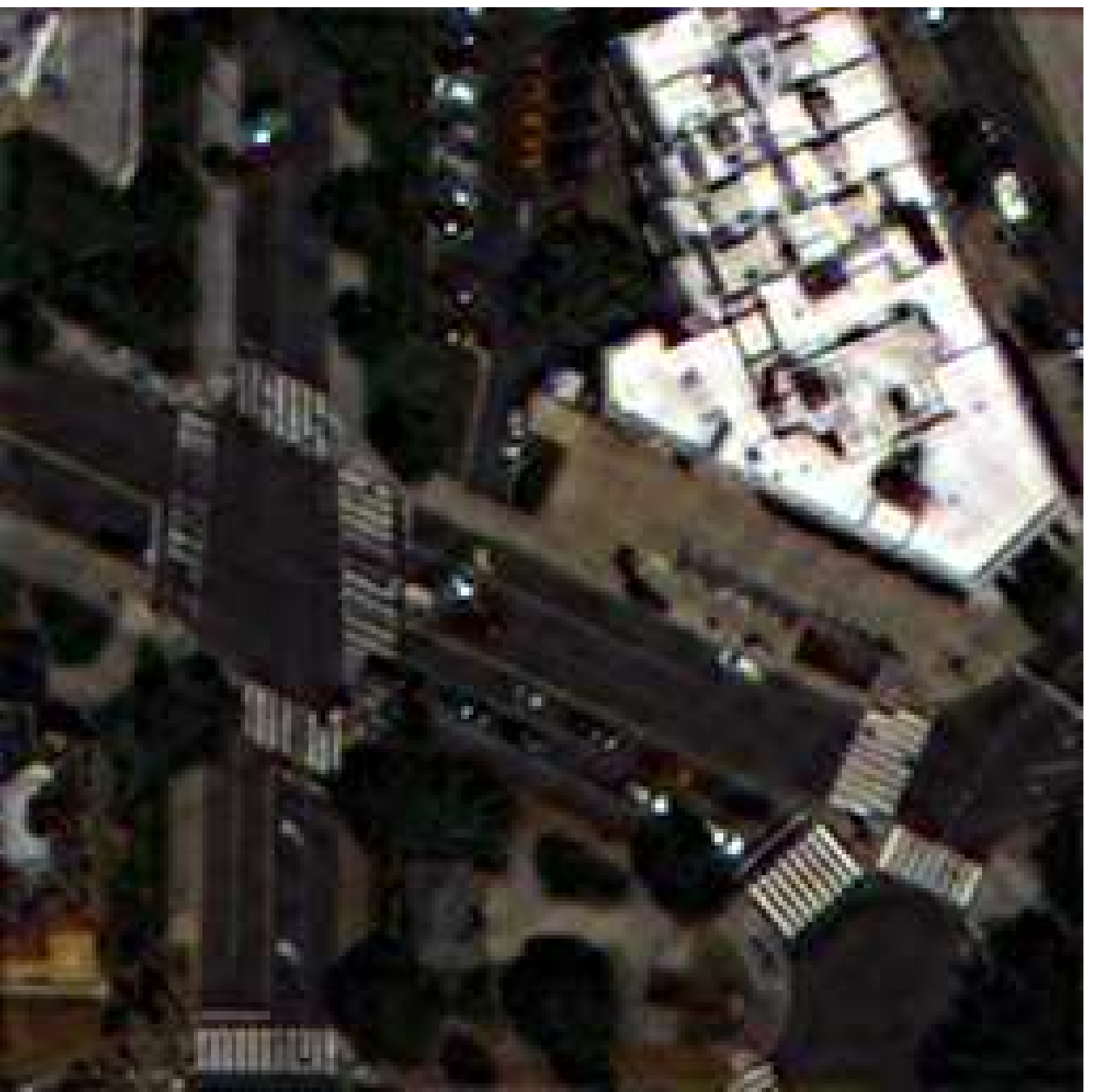} \\
(e) &(f) &(g) &(h) \\
\includegraphics[width=0.14\paperwidth]{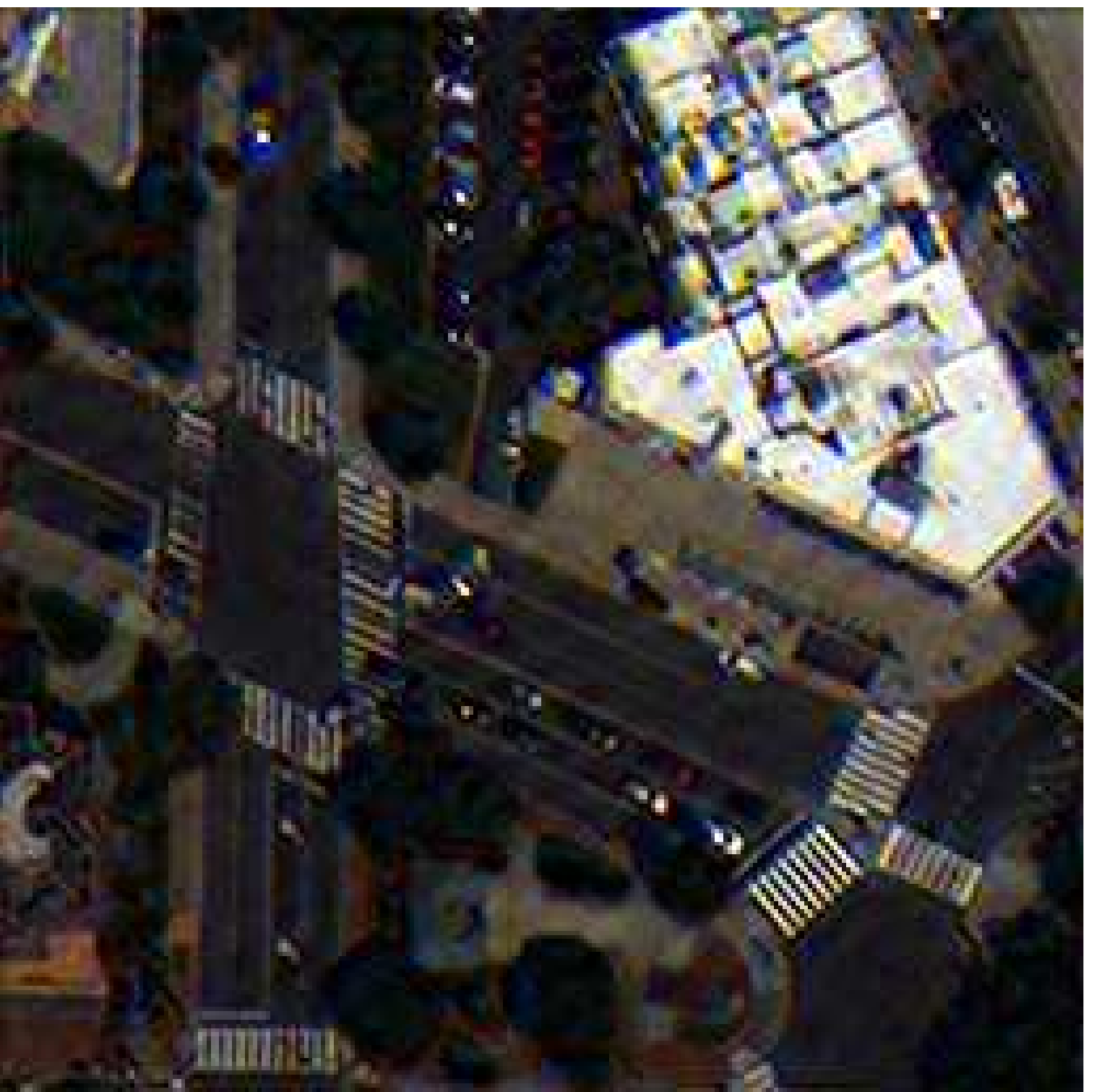} &
\includegraphics[width=0.14\paperwidth]{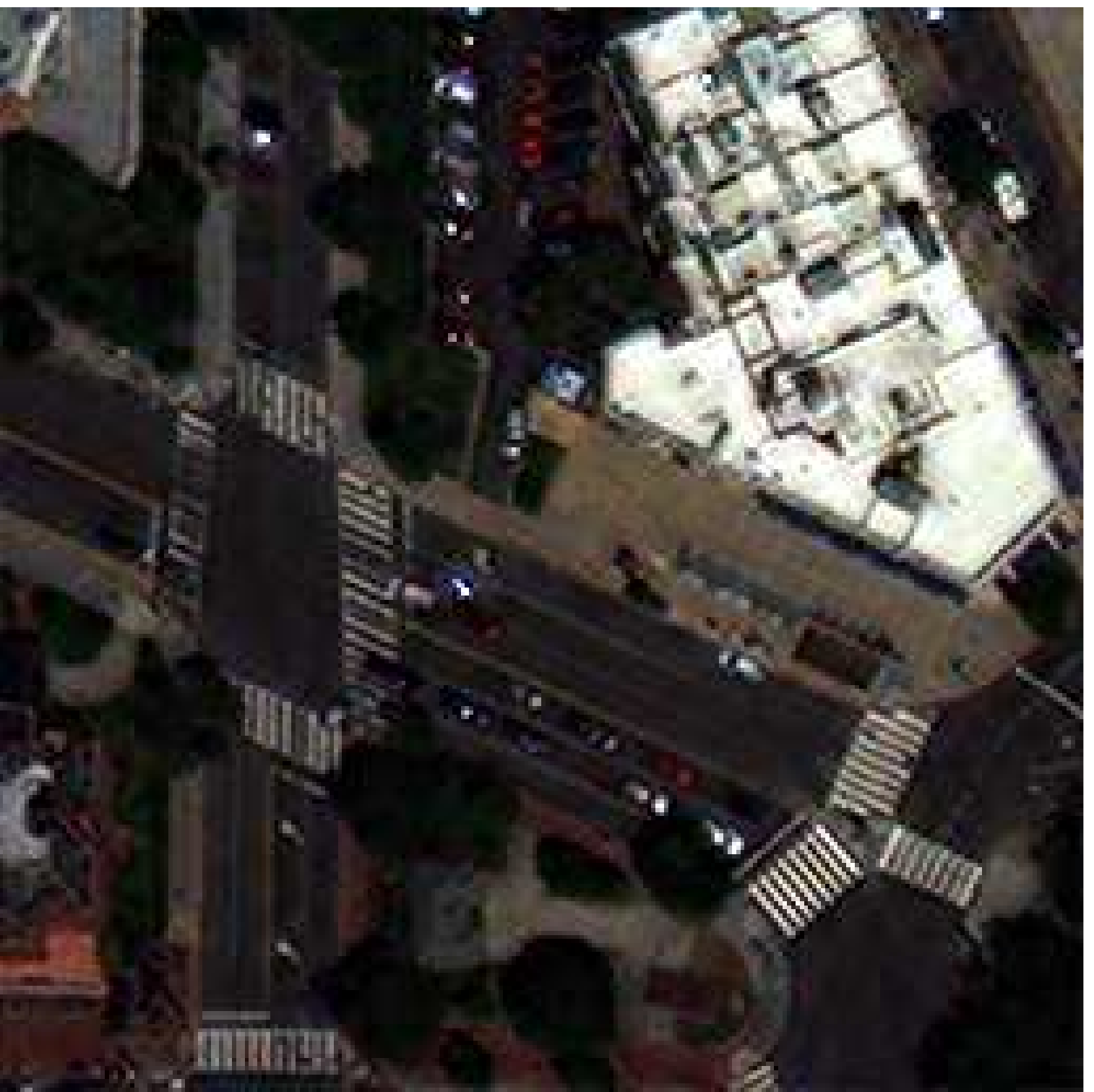} &
\includegraphics[width=0.14\paperwidth]{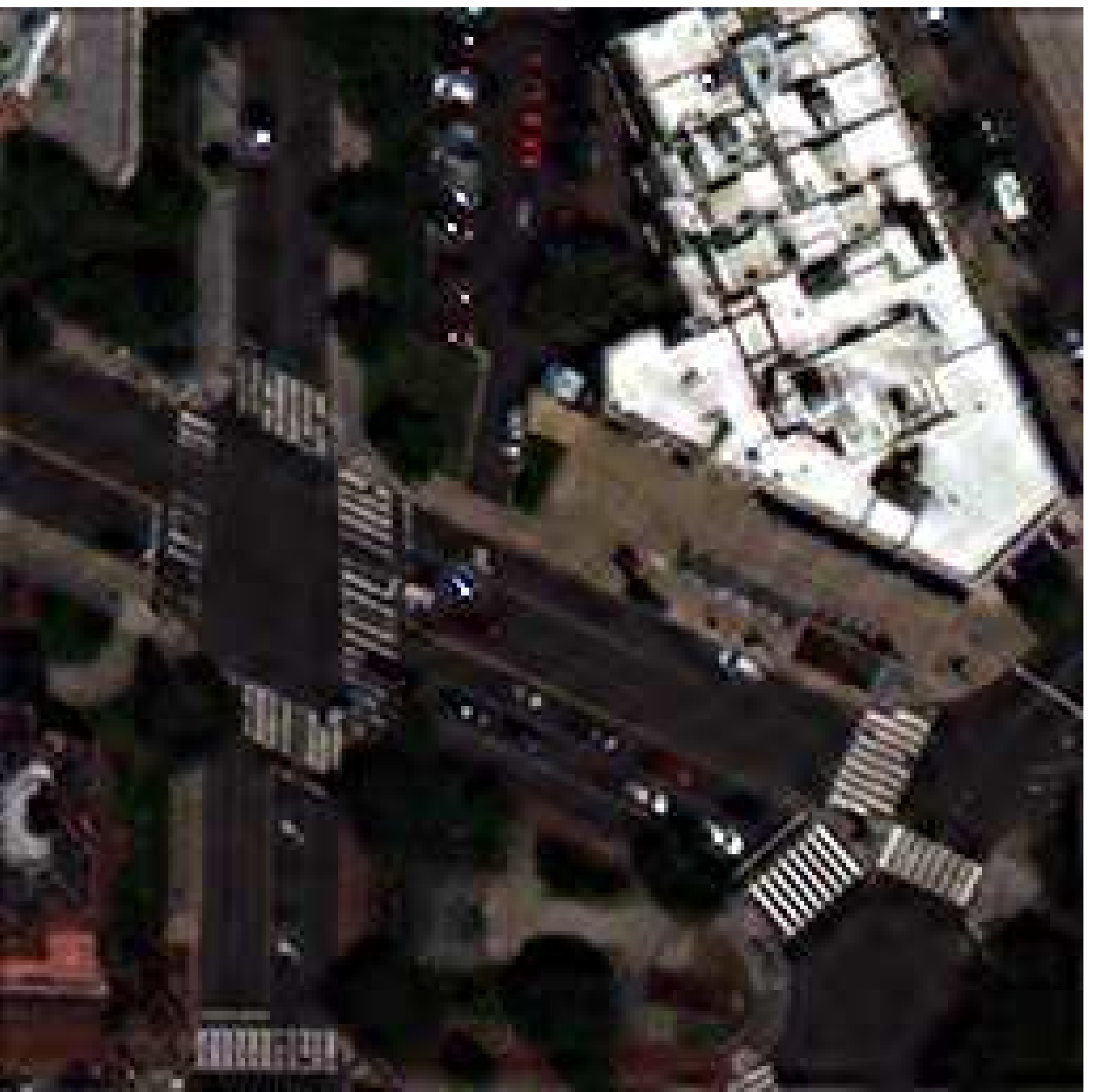} \\
(i) & (j) & (k)  \\
\\
\end{tabular}
\caption{Full-resolution pansharpening results for WorldView-2 dataset: (a) PAN image; (b) EXP; (c) GSA; (d) PRACS; (e) ATWT; (f) BDSD;  (g) GLP-CBD; (h) PNN; (i) DRPNN; (j) DiCNN1; (k) DiCNN2.}
\label{figure:map:full-wv2}
\end{figure*}

Fig. \ref{figure:map:full-wv2} displays the full-resolution experimental results. The CNN-based methods exhibit sharper results than the other tested methods, especially in the vegetation areas. DiCNN1, PNN and DiCNN2 slightly overpass DRPNN in terms of reducing artifacts.

\subsection{Experiment 2: IKONOS Hobart Dataset}

\begin{figure*}[t]\scriptsize
\centering
  \begin{tabular}{cccc}
\includegraphics[width=0.14\paperwidth]{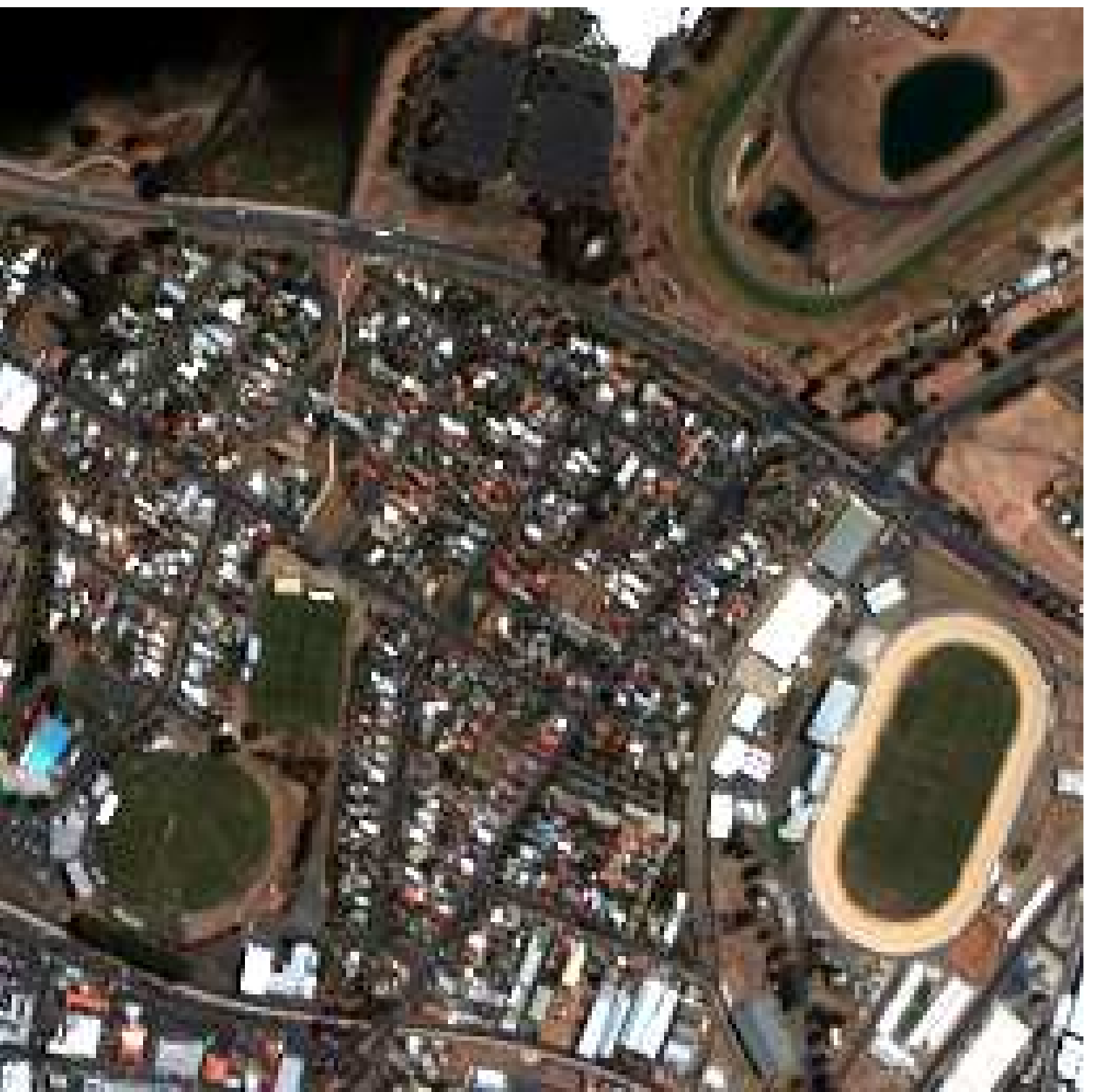} &
\includegraphics[width=0.14\paperwidth]{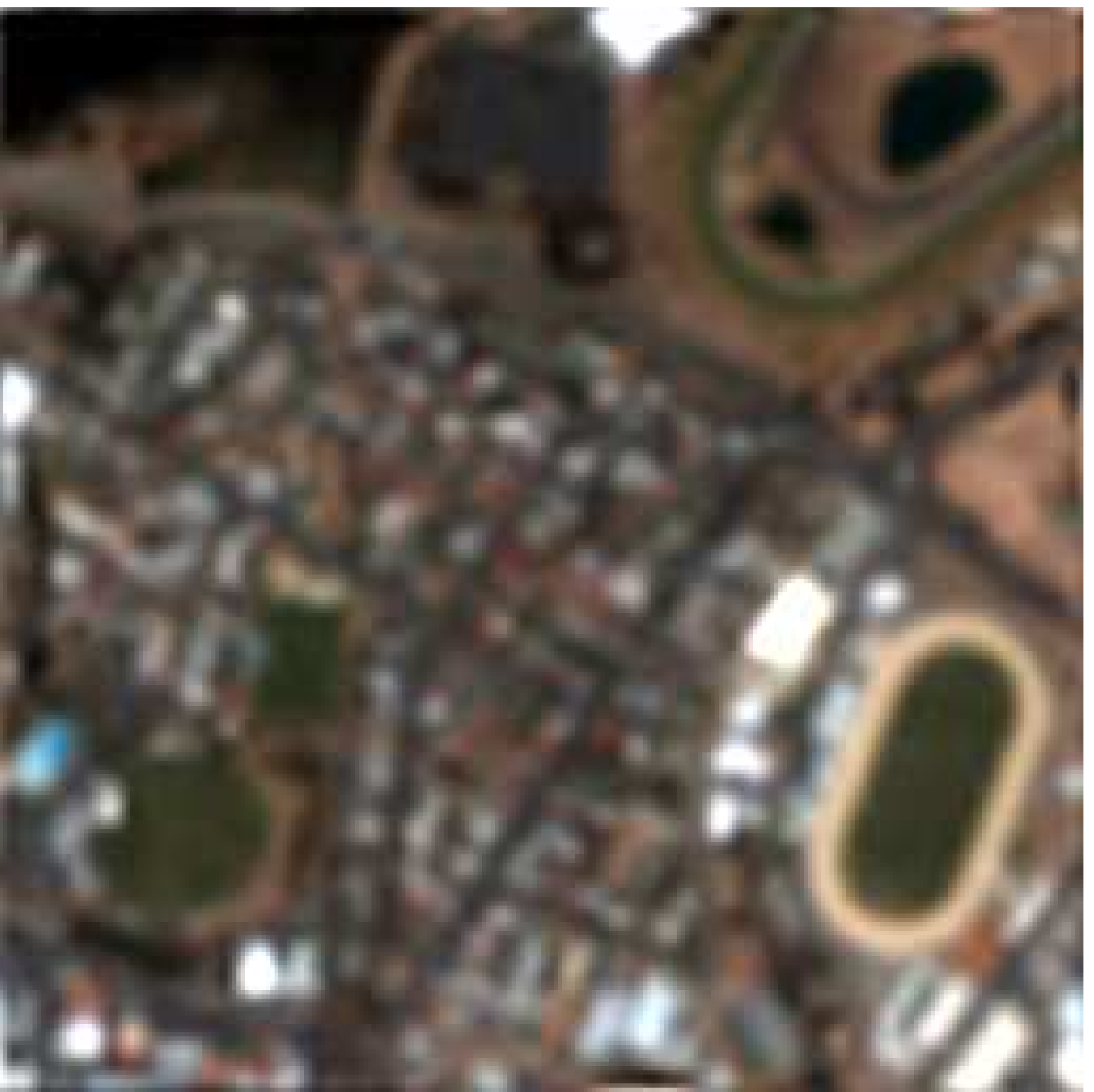} &
\includegraphics[width=0.14\paperwidth]{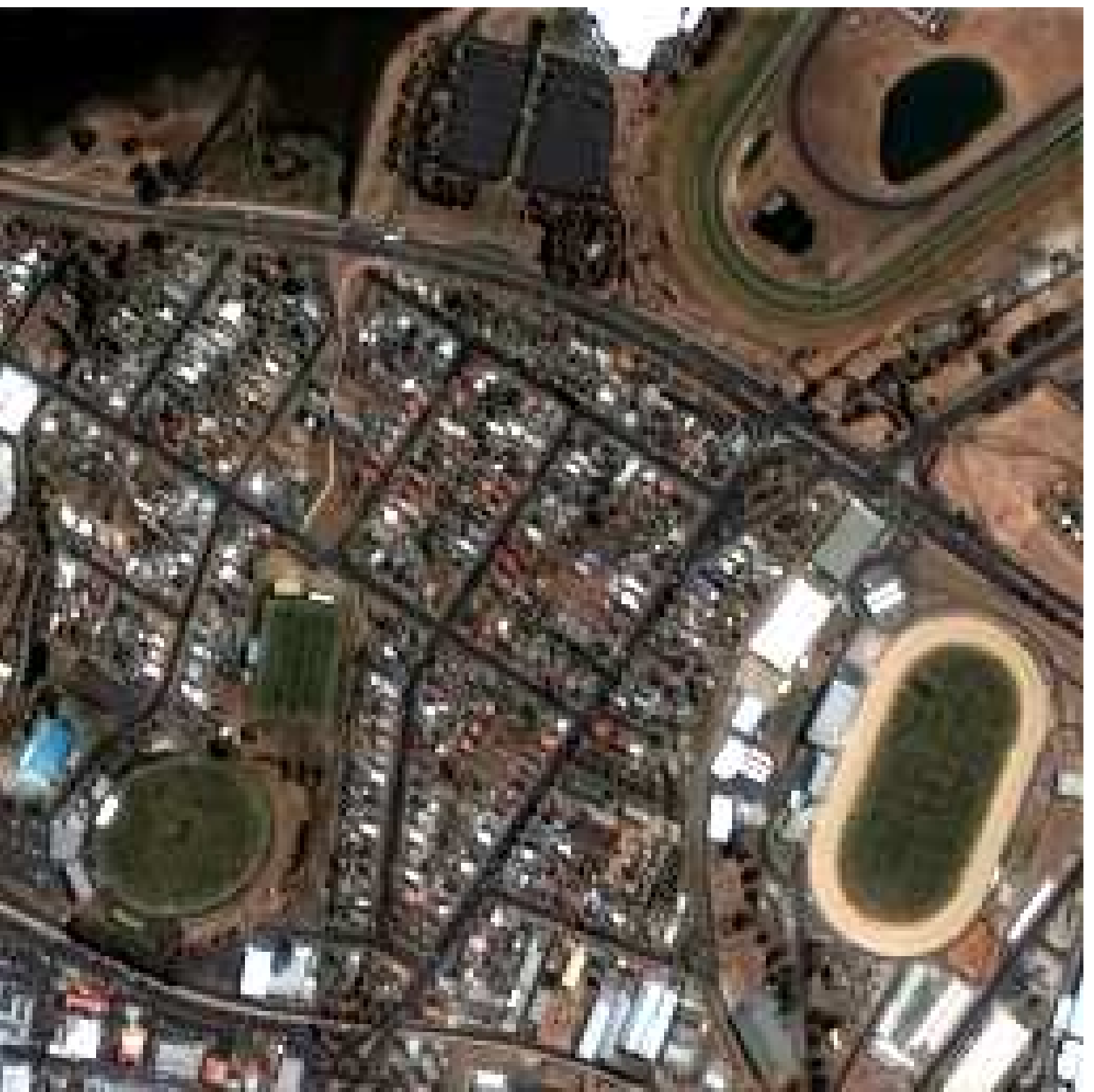} &
\includegraphics[width=0.14\paperwidth]{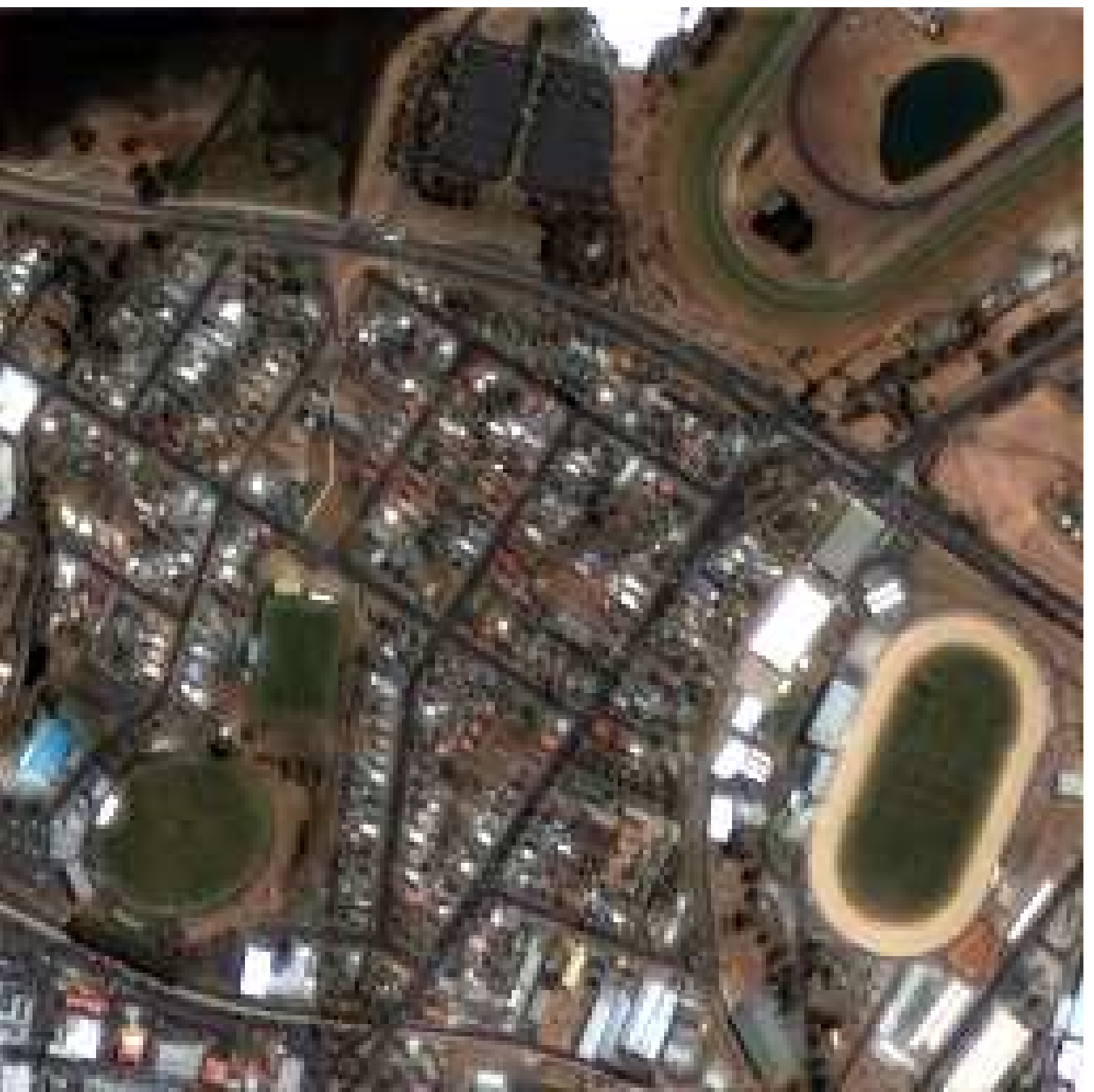} \\
(a) & (b) & (c)  & (d)  \\

\includegraphics[width=0.14\paperwidth]{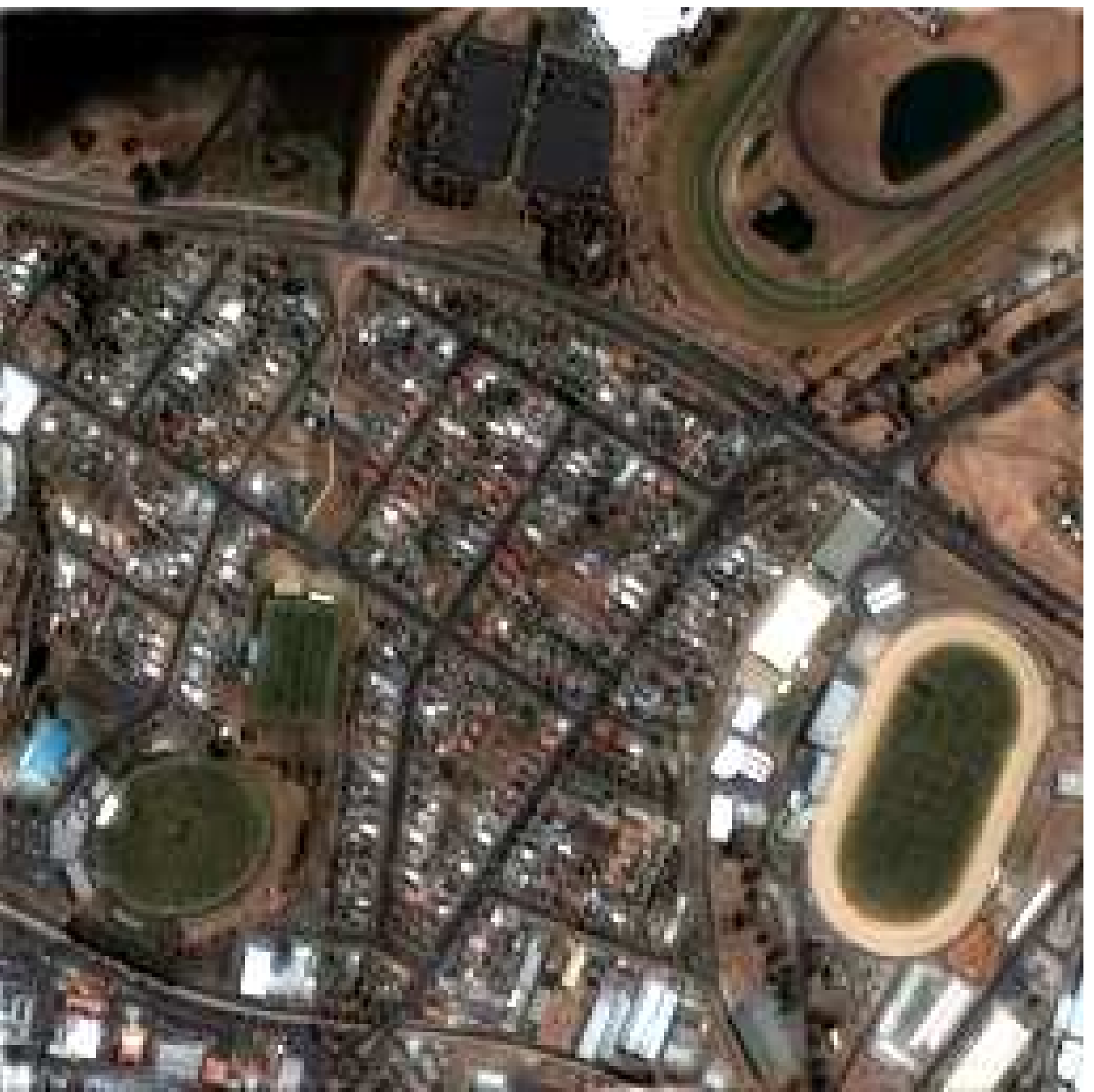} &
\includegraphics[width=0.14\paperwidth]{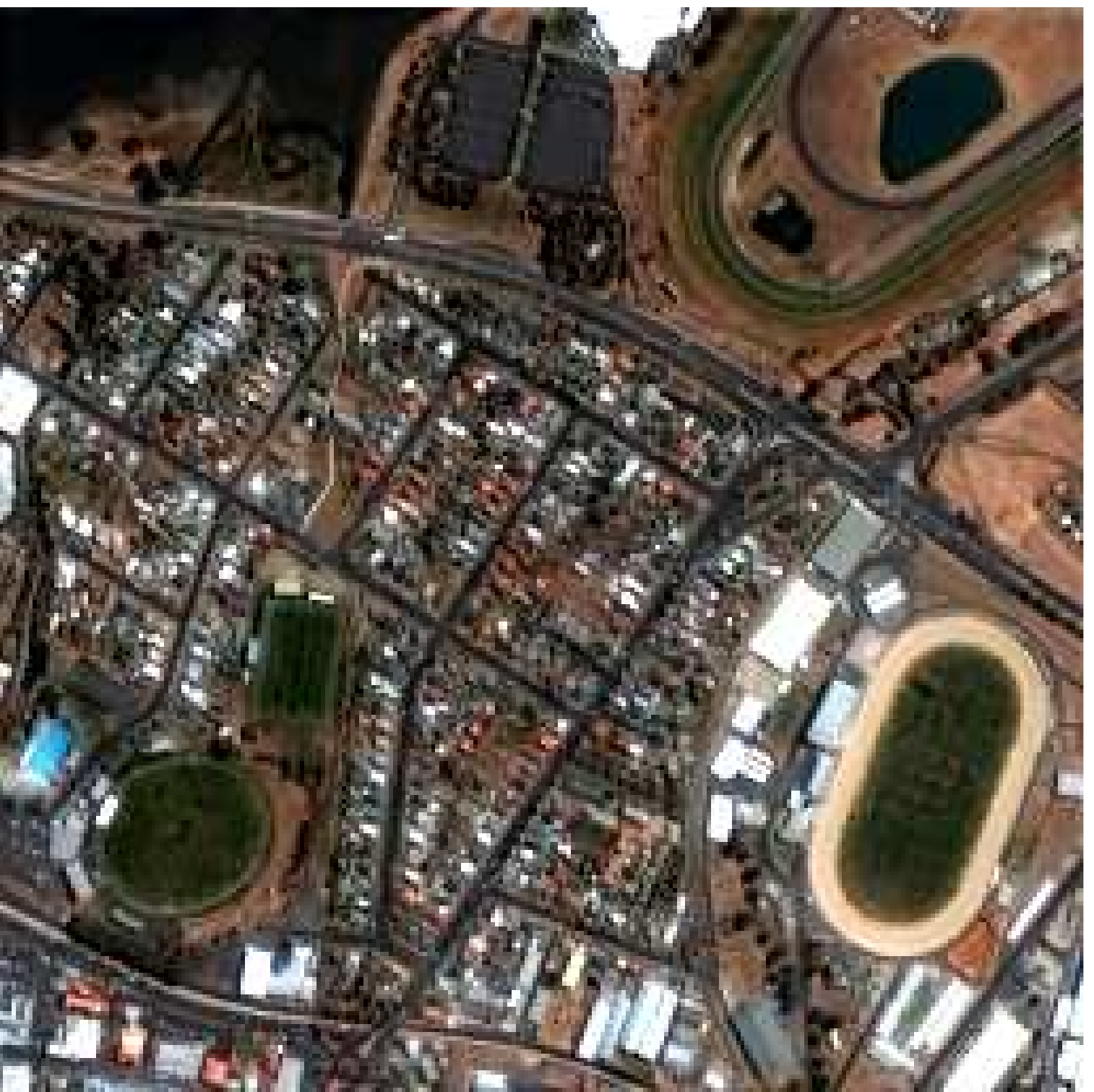} &
\includegraphics[width=0.14\paperwidth]{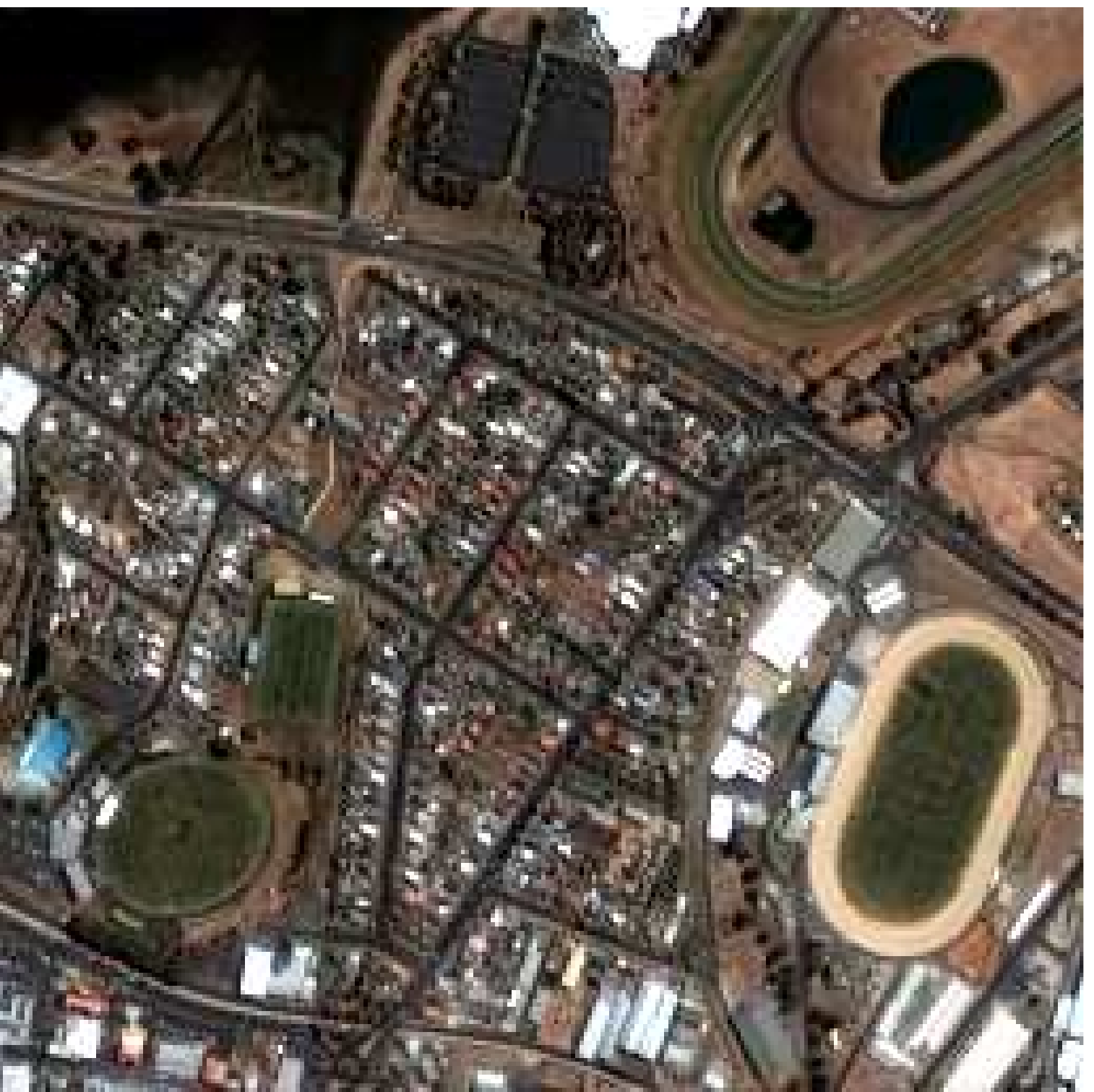} &
\includegraphics[width=0.14\paperwidth]{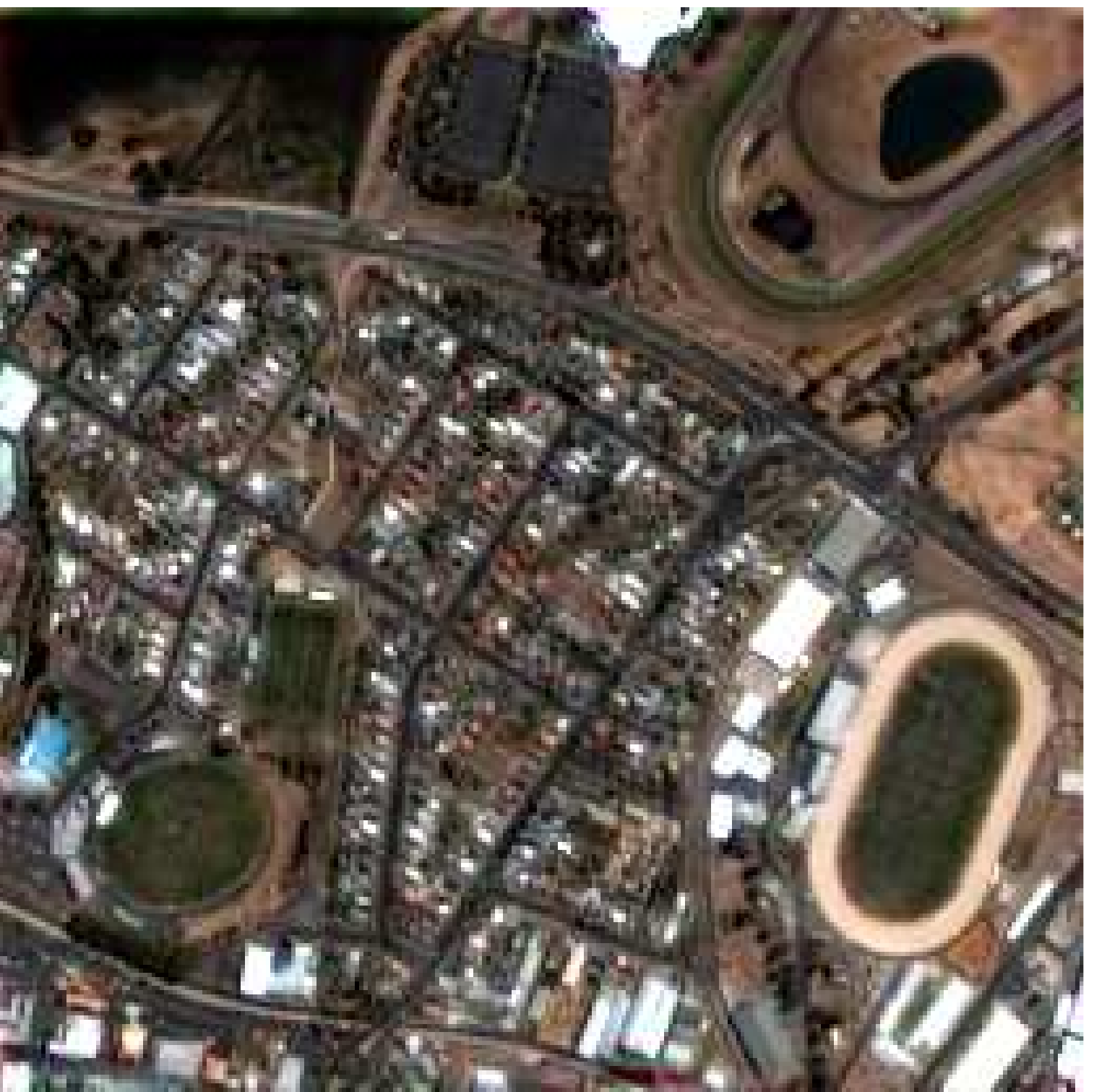} \\
(e) &(f) &(g) & (h) \\

\includegraphics[width=0.14\paperwidth]{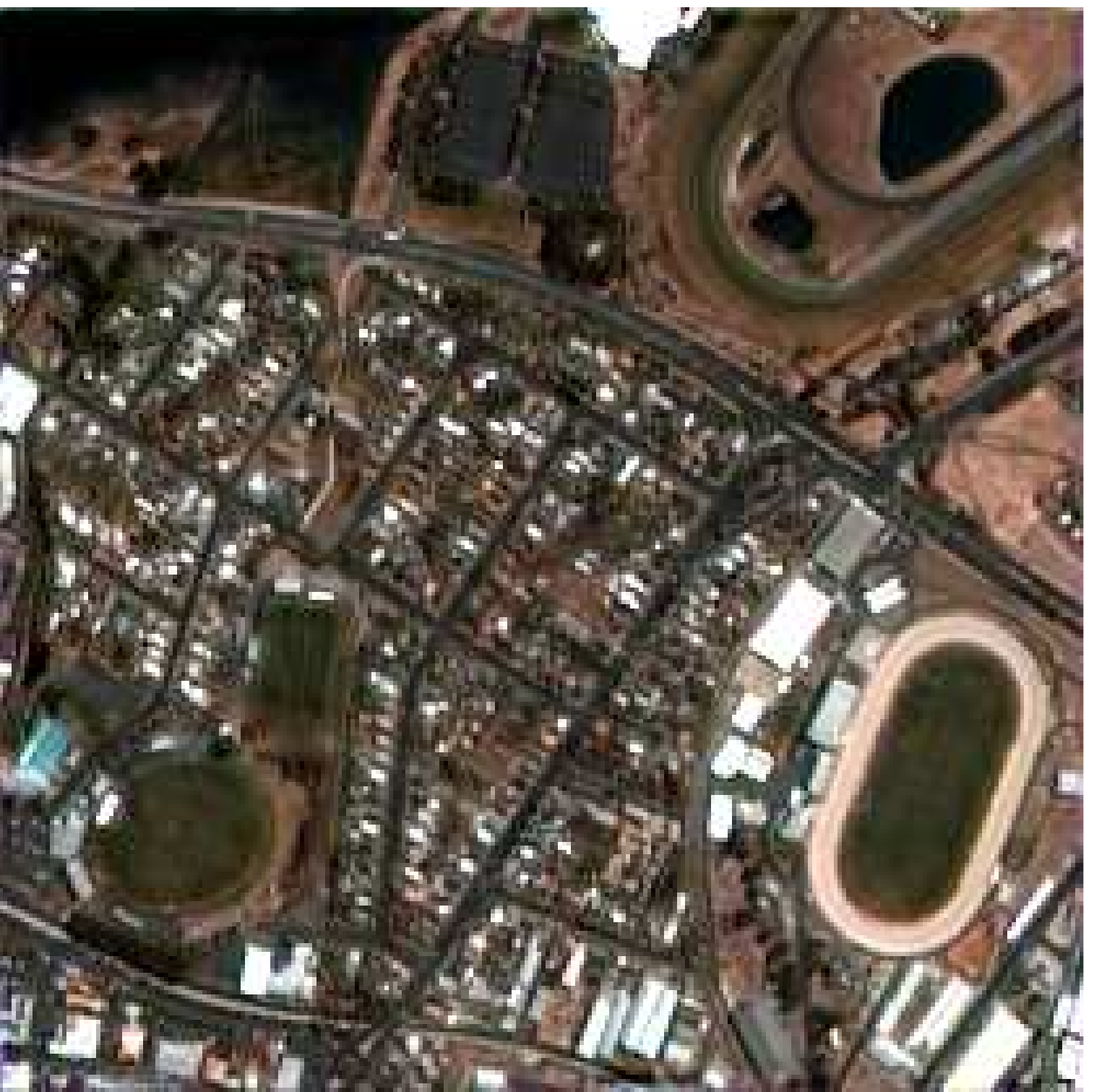} &
\includegraphics[width=0.14\paperwidth]{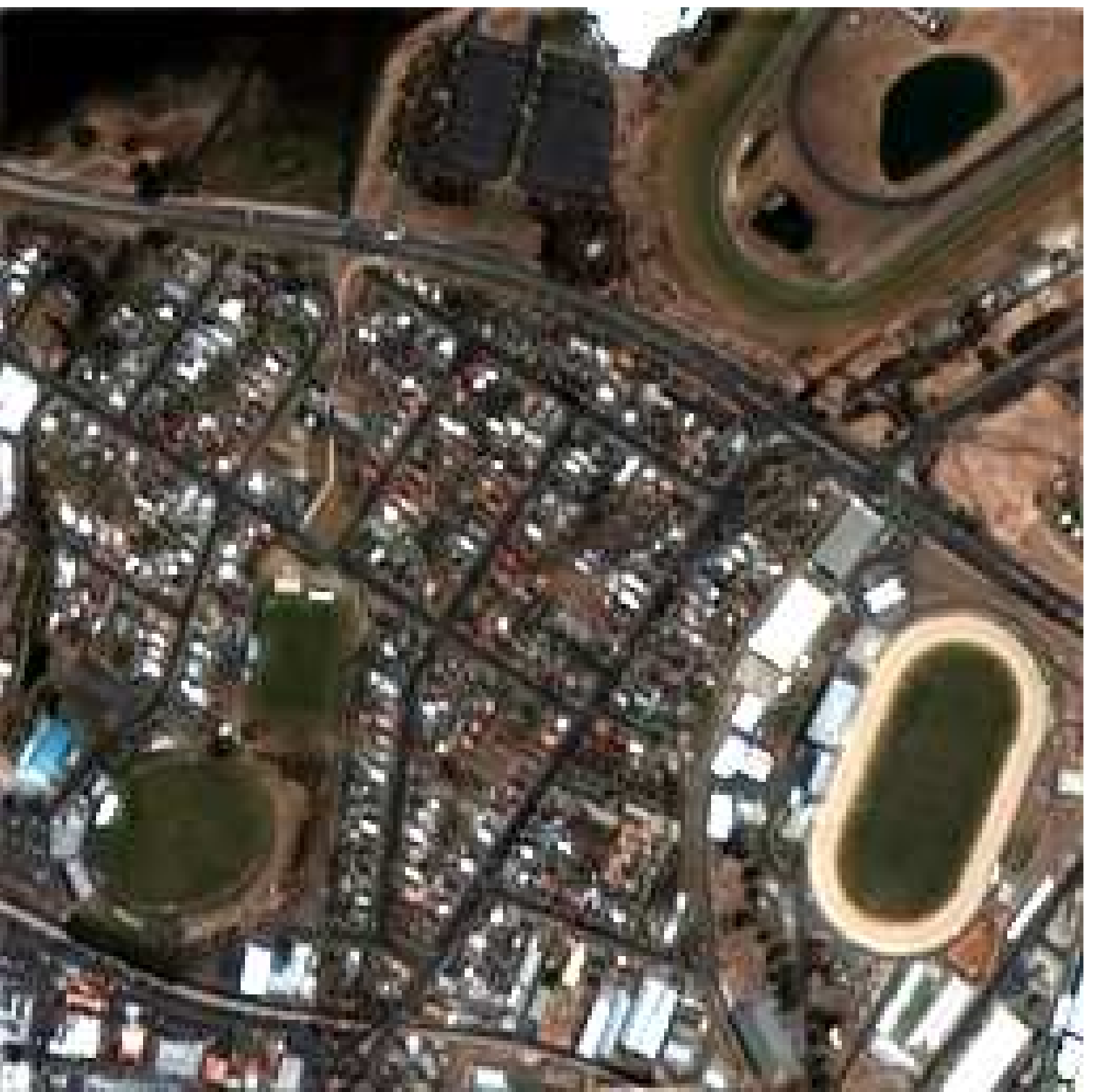} &
\includegraphics[width=0.14\paperwidth]{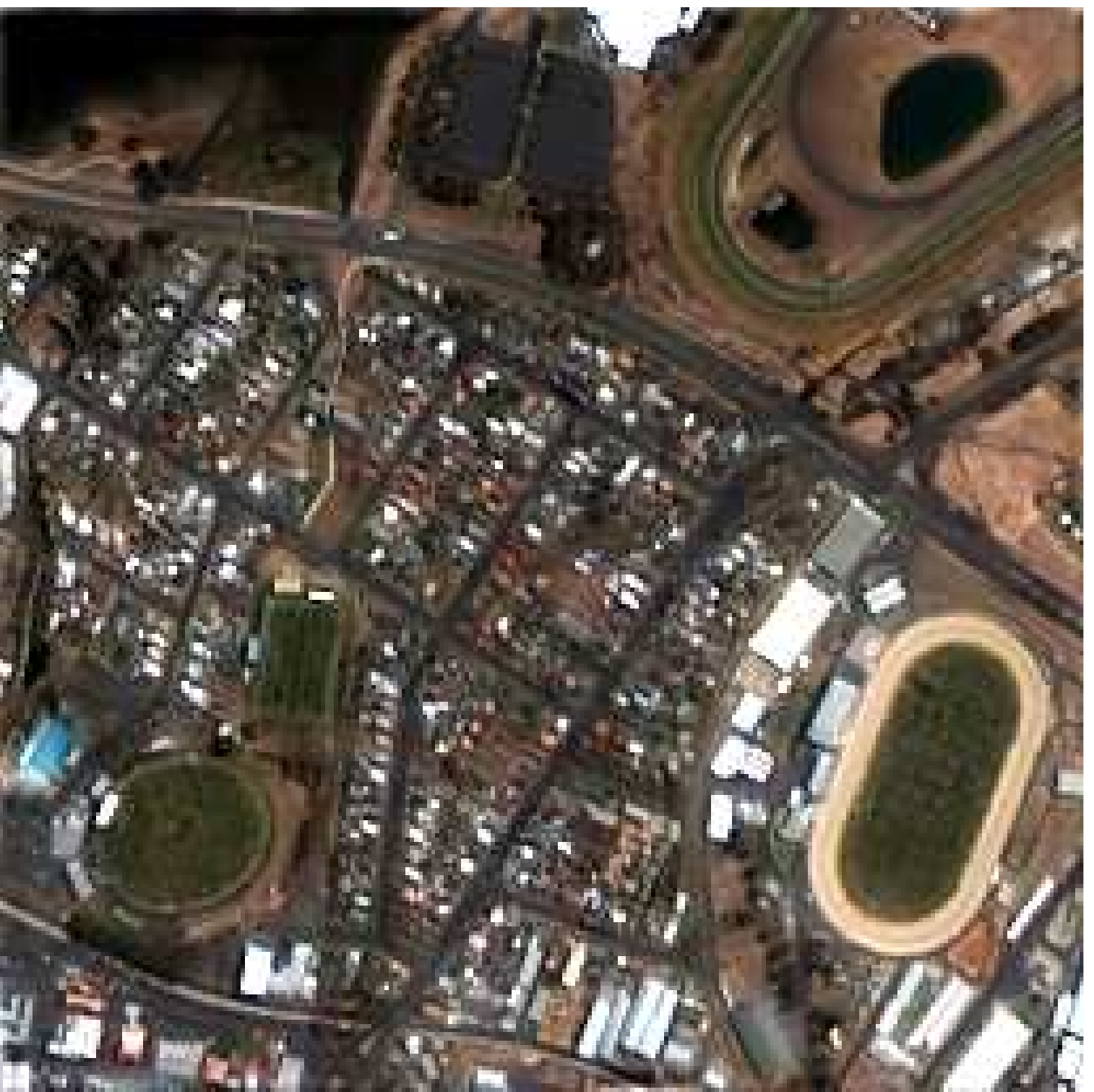} \\
(i) &(j) &(k)  \\
\\
\end{tabular}
\caption{Pansharpening results for IKONOS dataset. (a) Ground-truth; (b)EXP; (c)GSA; (d)PRACS; (e)ATWT; (f)BDSD;  (g)GLP-CBD; (h)PNN; (i)DRPNN; (j)DiCNN1; (k)DiCNN2. }
\label{figure:map:ik}
\end{figure*}

\begin{figure*}[t]\scriptsize
\centering
\begin{tabular}{ccccc}
\includegraphics[width=0.14\paperwidth]{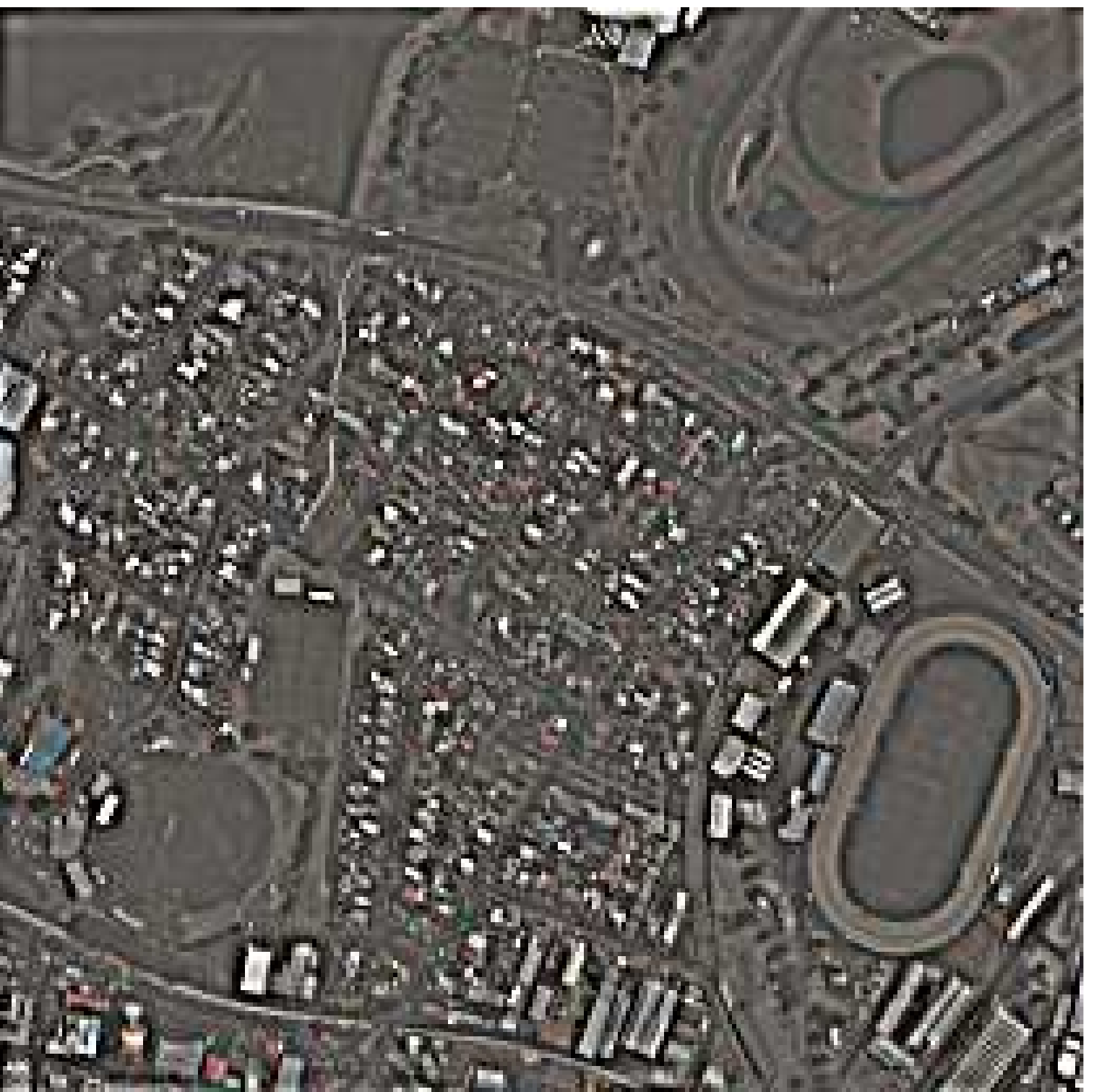} &
\includegraphics[width=0.14\paperwidth]{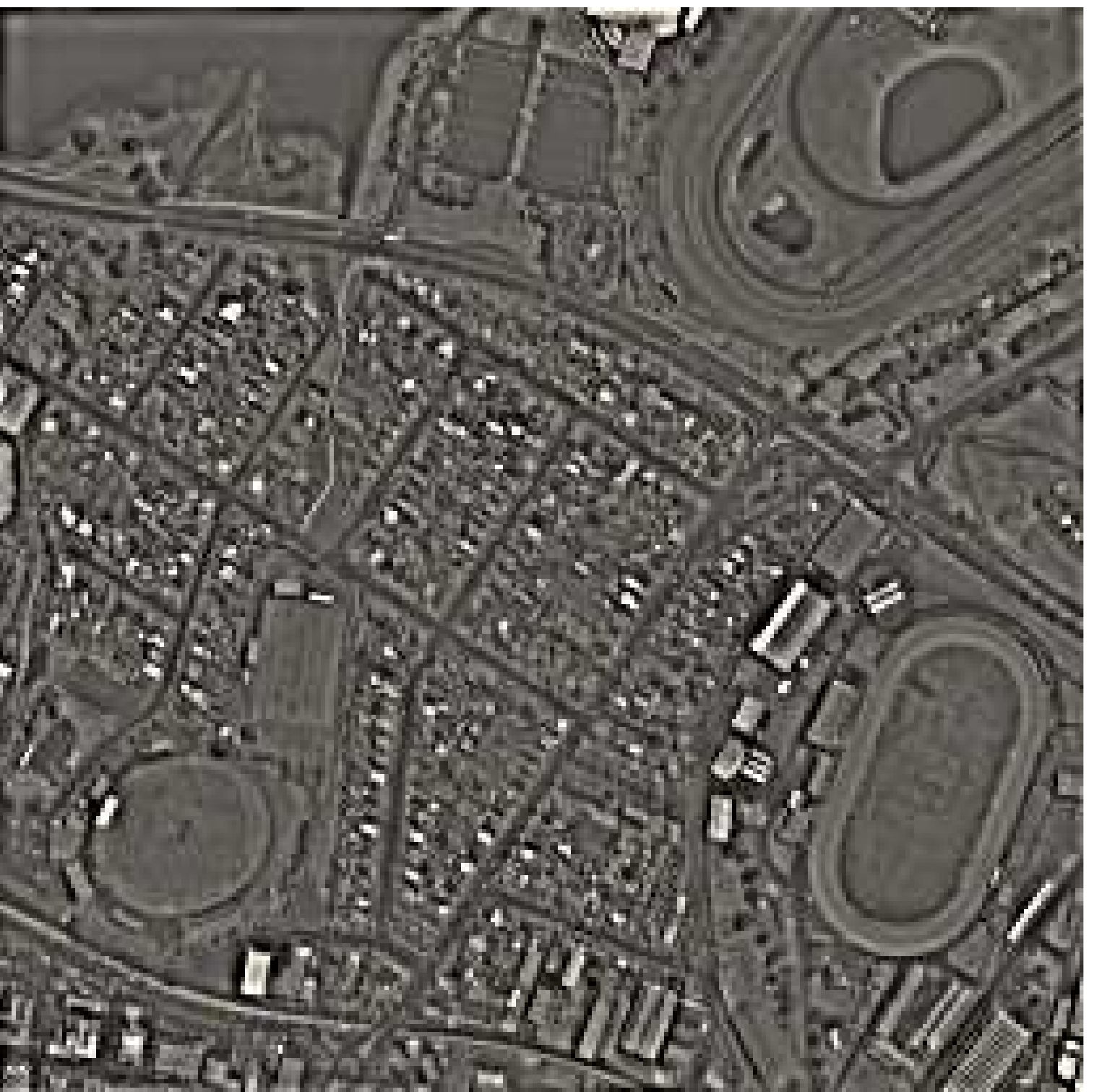} &
\includegraphics[width=0.14\paperwidth]{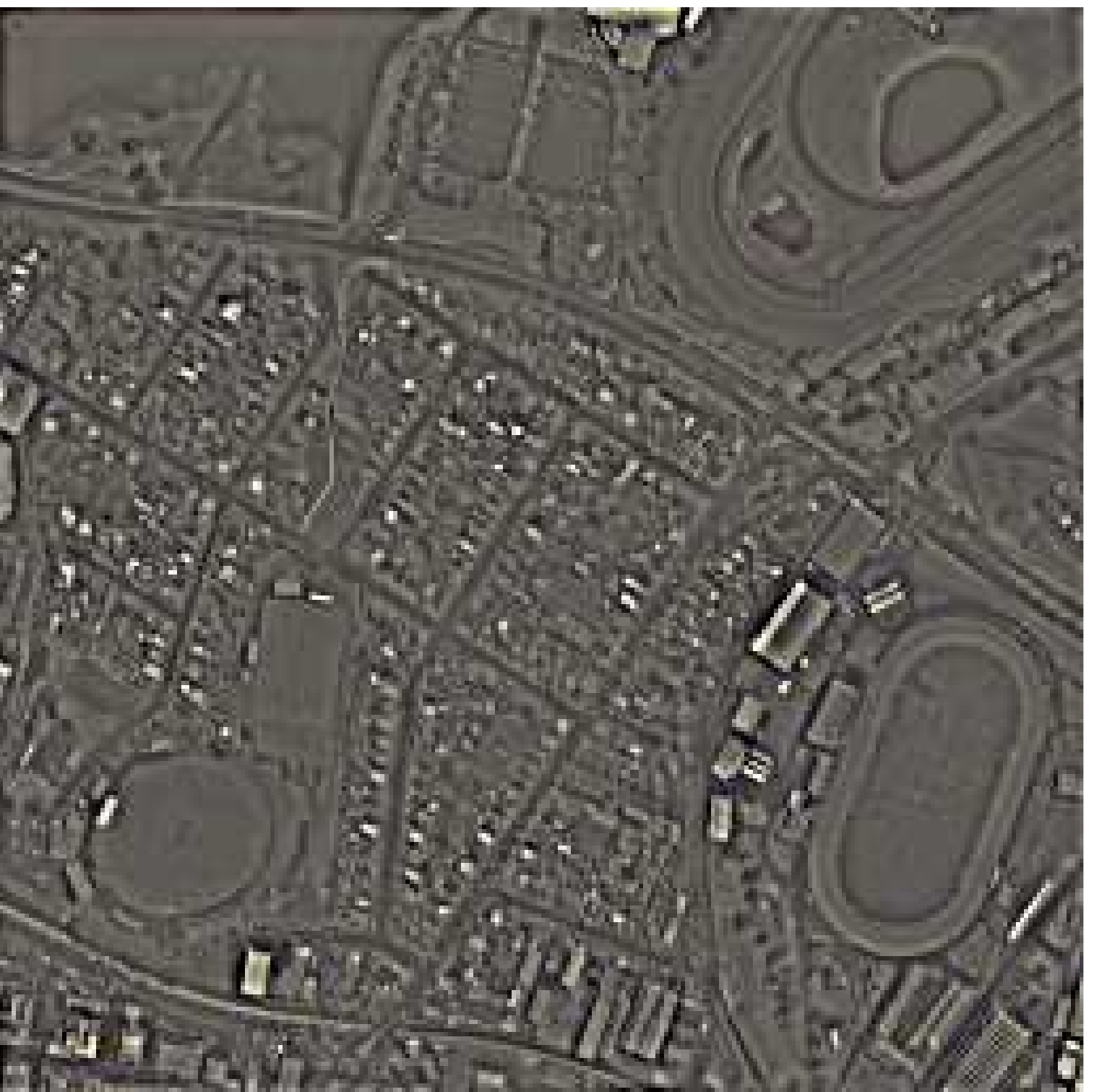} &
\includegraphics[width=0.14\paperwidth]{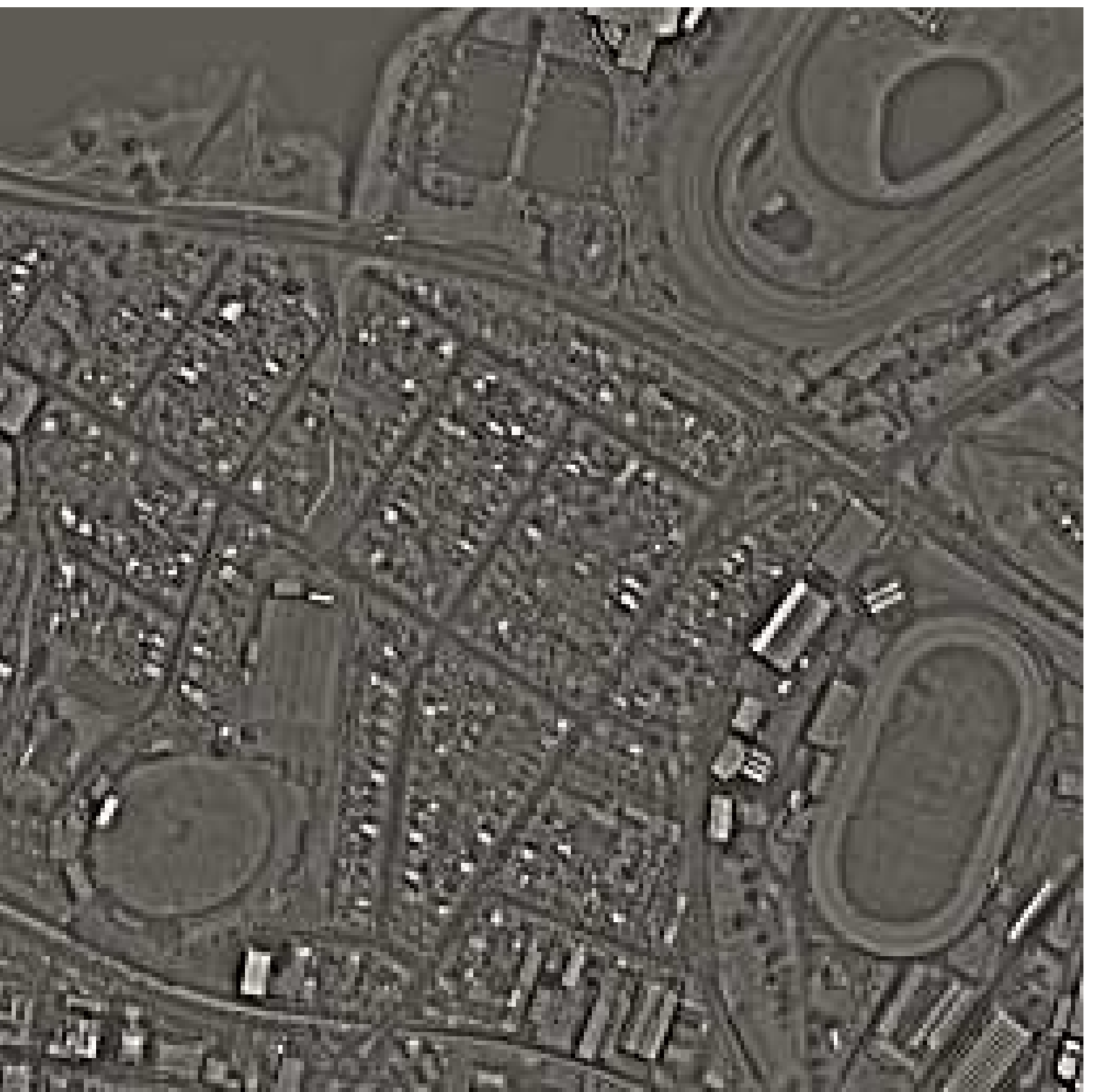} &
\includegraphics[width=0.14\paperwidth]{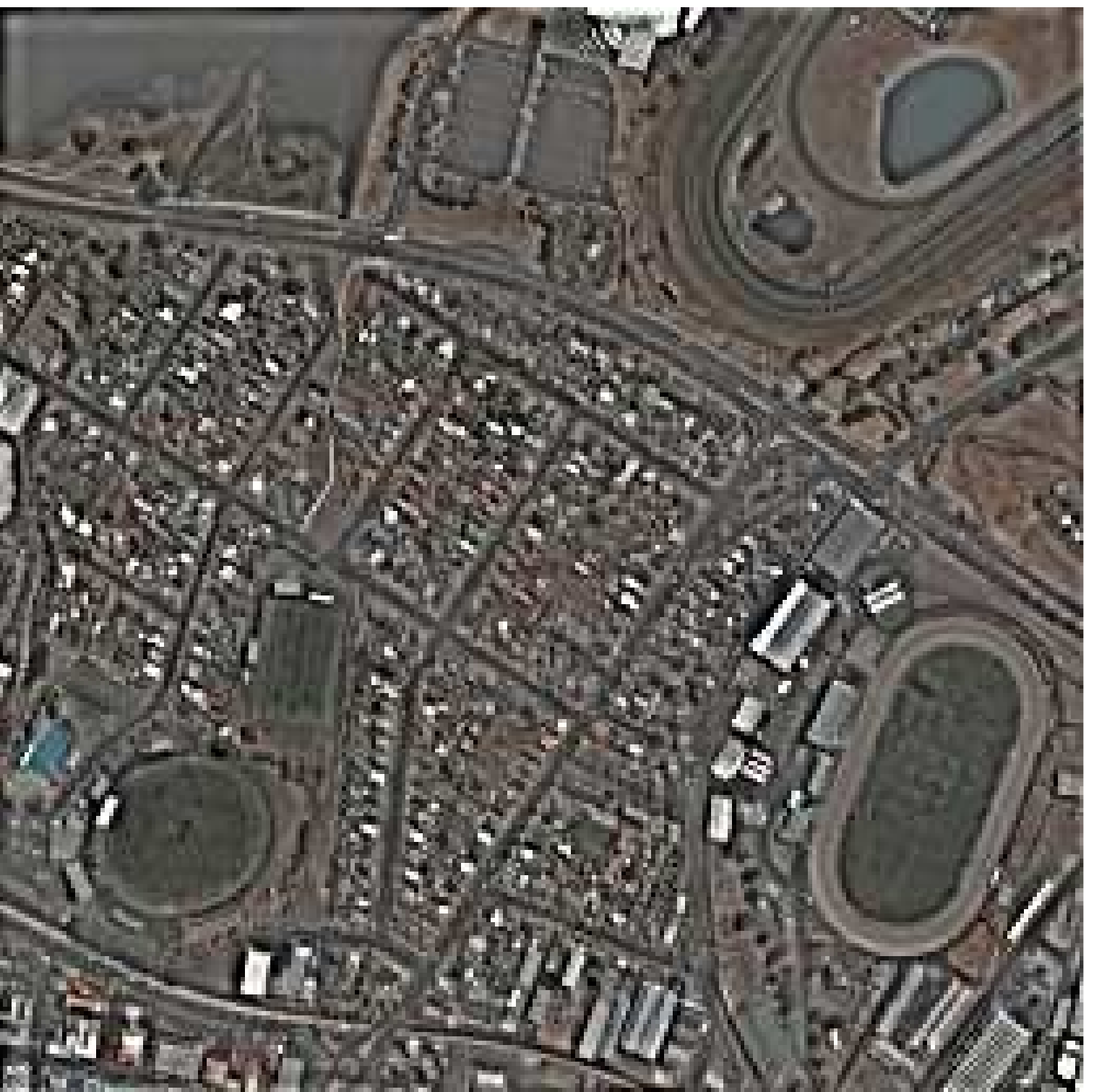} \\
(a) & (b) & (c) &(d) &(e) \\
\includegraphics[width=0.14\paperwidth]{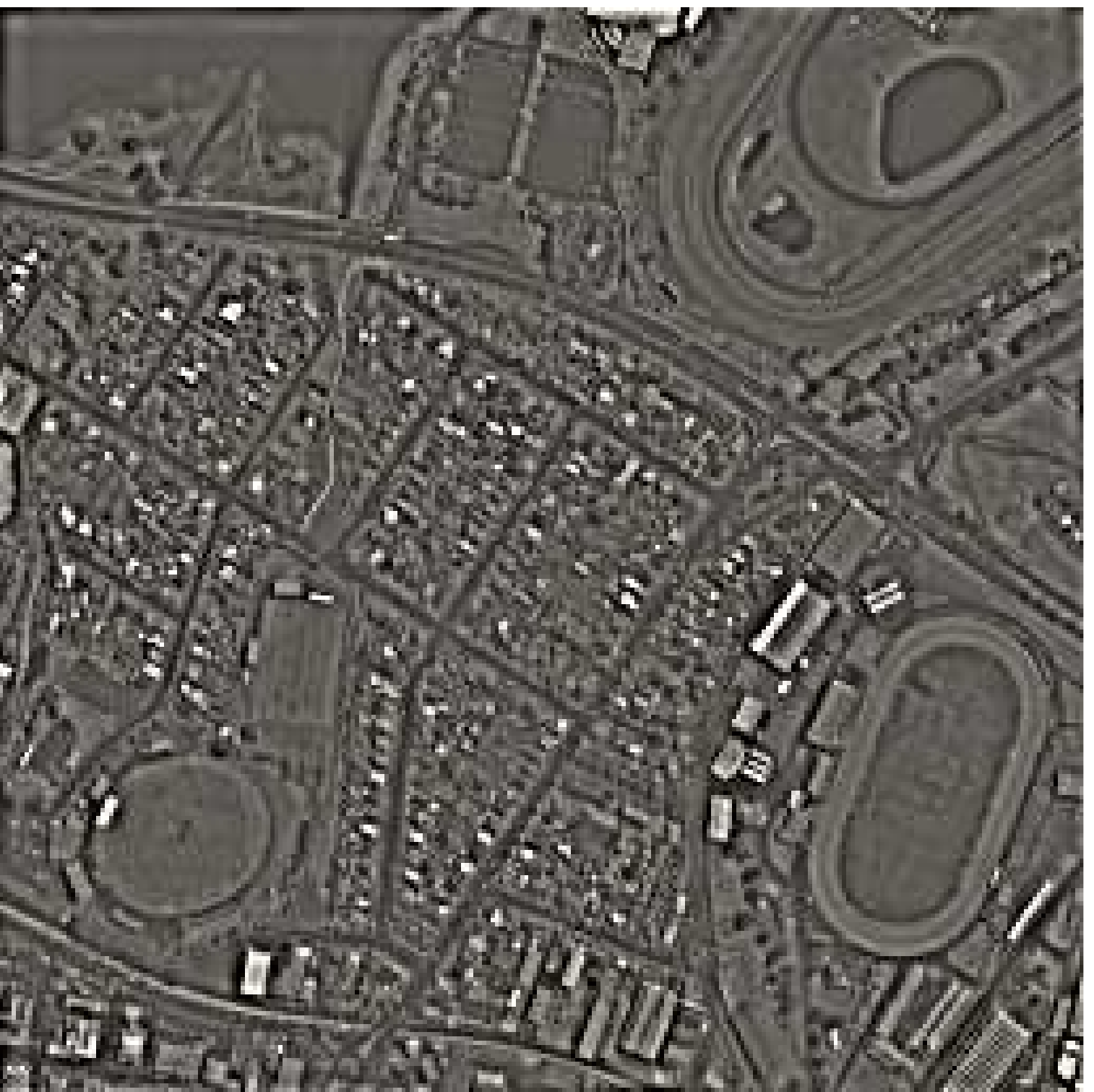} &
\includegraphics[width=0.14\paperwidth]{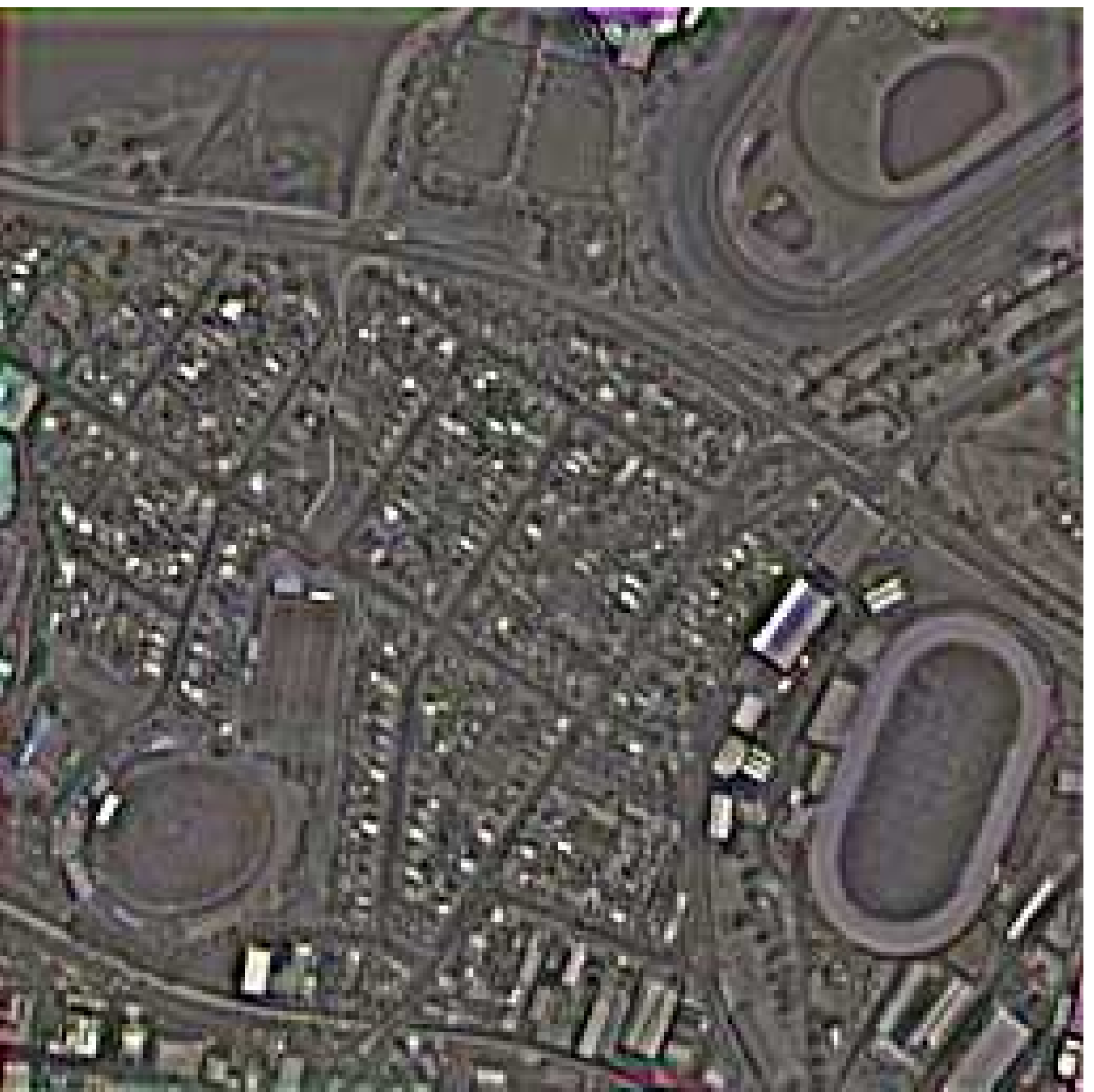} &
\includegraphics[width=0.14\paperwidth]{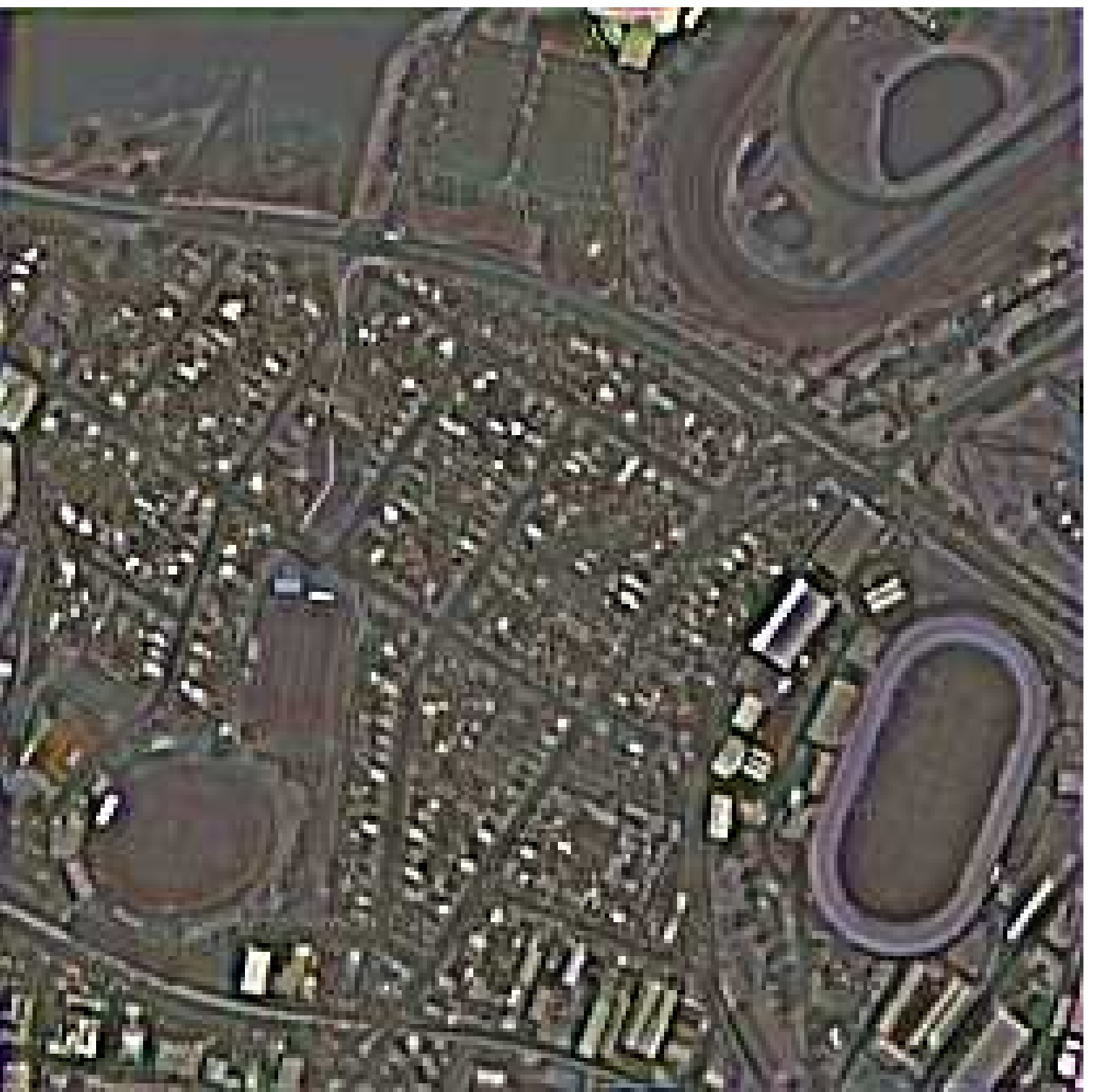} &
\includegraphics[width=0.14\paperwidth]{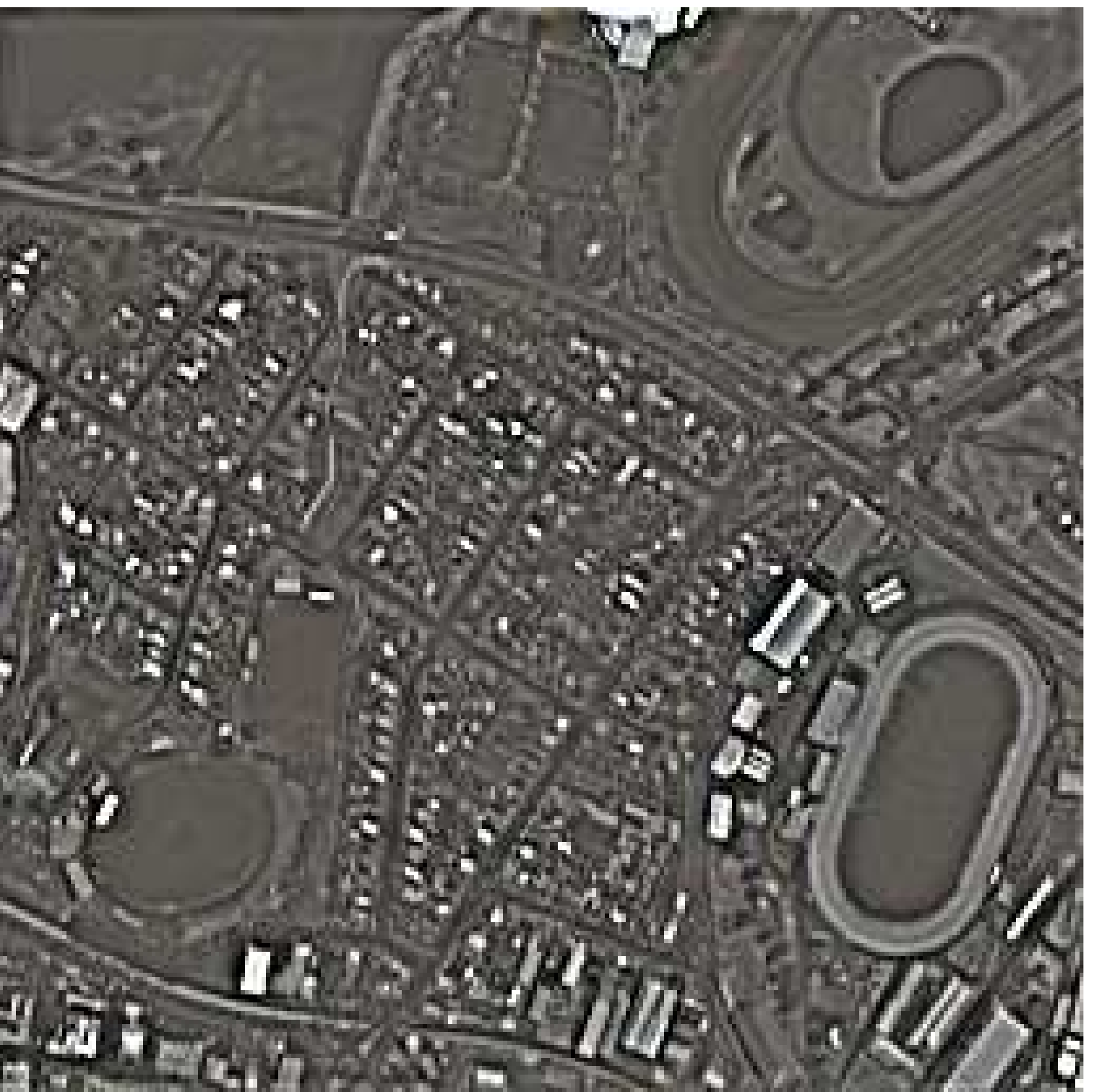} &
\includegraphics[width=0.14\paperwidth]{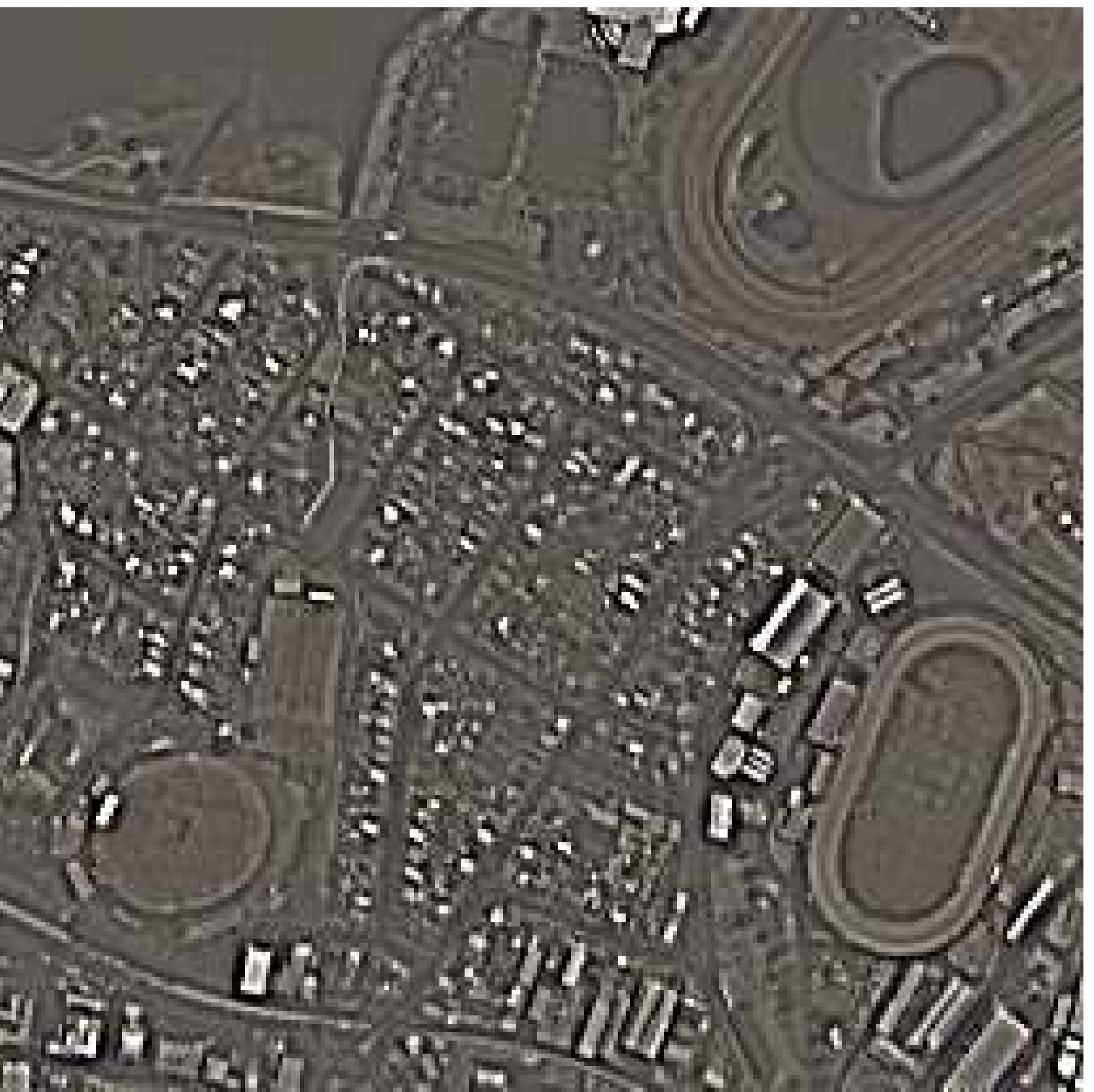} \\
(f) & (g) & (h) &(i) &(j)  \\
\\
\end{tabular}
\caption{Detail images of IKONOS dataset (a) Ground-truth;  (b)GSA; (c)PRACS; (d)ATWT; (e)BDSD; (f)GLP-CBD; (g)PNN; (h)DRPNN; (i)DiCNN1; (j)DiCNN2.}
\label{figure:detailimage:ik}
\end{figure*}

\begin{table}[htp]
\small
\caption{Quality indexes of different pansharpening methods under reduced-resolution quality assessment on a $256\times256$ subscene of IKONOS dataset.}
\centering
\begin{tabular}{c|ccccc}
\hline
{}&Q4&SAM& ERGAS &SCC&Time(s)\\
\hline
Refrence&1 &0 &0 &1&{}\\
\hline
\hline
EXP &0.5791 &5.4338 &5.7489 &0.5453 &{}\\
\hline
GSA&0.8083 &5.1063 &4.1467 &0.7583 &0.73\\
\hline
PRACS&0.7843 &5.1175 &4.2096 &0.7646 &0.61\\
\hline
ATWT&0.8036 &5.1198 &4.0957 &0.7622  &\textbf{0.49}\\
\hline
BDSD&0.8141 &5.4020 &4.2070 &0.7583 &1.15\\
\hline
GLP-CBD&0.8121 &5.0884 &4.0857 &0.7591 &0.52\\
\hline
\hline
PNN &0.8846 &4.8722 &3.1783 &0.8836&2.44\\
\hline
DRPNN &0.8995 &4.5546 &2.9513 &0.9018&1.19\\
\hline
DiCNN1 &\textbf{0.9120} &\textbf{4.3359} &\textbf{2.8532} &\textbf{0.9091}&1.38\\
\hline
DiCNN2 &0.9003 &4.4575 &2.9104 &0.9027&1.06\\
\hline
\end{tabular}
\label{table:reduceik}
\end{table}

The dataset\footnote{http://www.isprs.org/data/default.aspx} represents an urban and harbor area of Hobart in Australia. It was acquired by the IKONOS sensor, which works in visible and near-infrared spectrum ranges. The MS sensor is characterized by four bands (blue, green, red, and near infrared) and also a PAN channel with band range from $450nm$ to $900nm$. The resolution of MS is $4m$ and PAN is $1m$. The radiometric resolution is 11 bits. Different areas with size of $256\times256$ pixels are used for reduced-resolution and full-resolution experiments, respectively.

Table \ref{table:reduceik} tabulates the results of our reduced-resolution quality assessment on the IKONOS Hobart dataset. Similar phenomena to the ones observed with the previous WorldView-2 dataset can be appreciated. Specifically, CNN-based methods achieve better pansharpening quality than the CS-based and MRA-based methods. DiCNN1 achieves the highest Q4, SAM, ERGAS and SCC scores, while PNN is the most time-consuming. DiCNN2 achieves the least computational time among CNN-based methods.

\begin{figure*}[t]\scriptsize
\centering
  \begin{tabular}{ccccc}
\includegraphics[width=0.14\paperwidth]{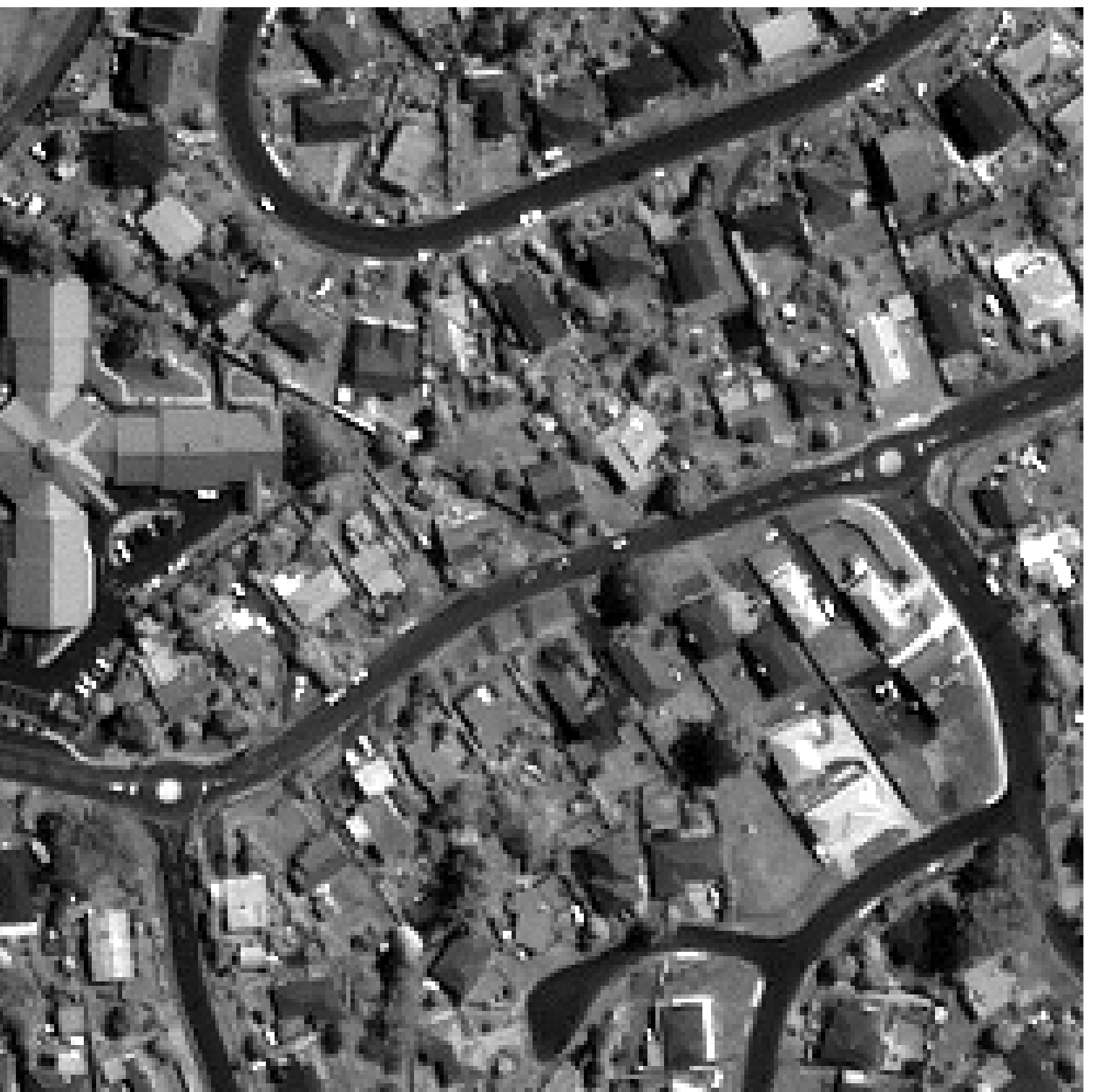} &
\includegraphics[width=0.14\paperwidth]{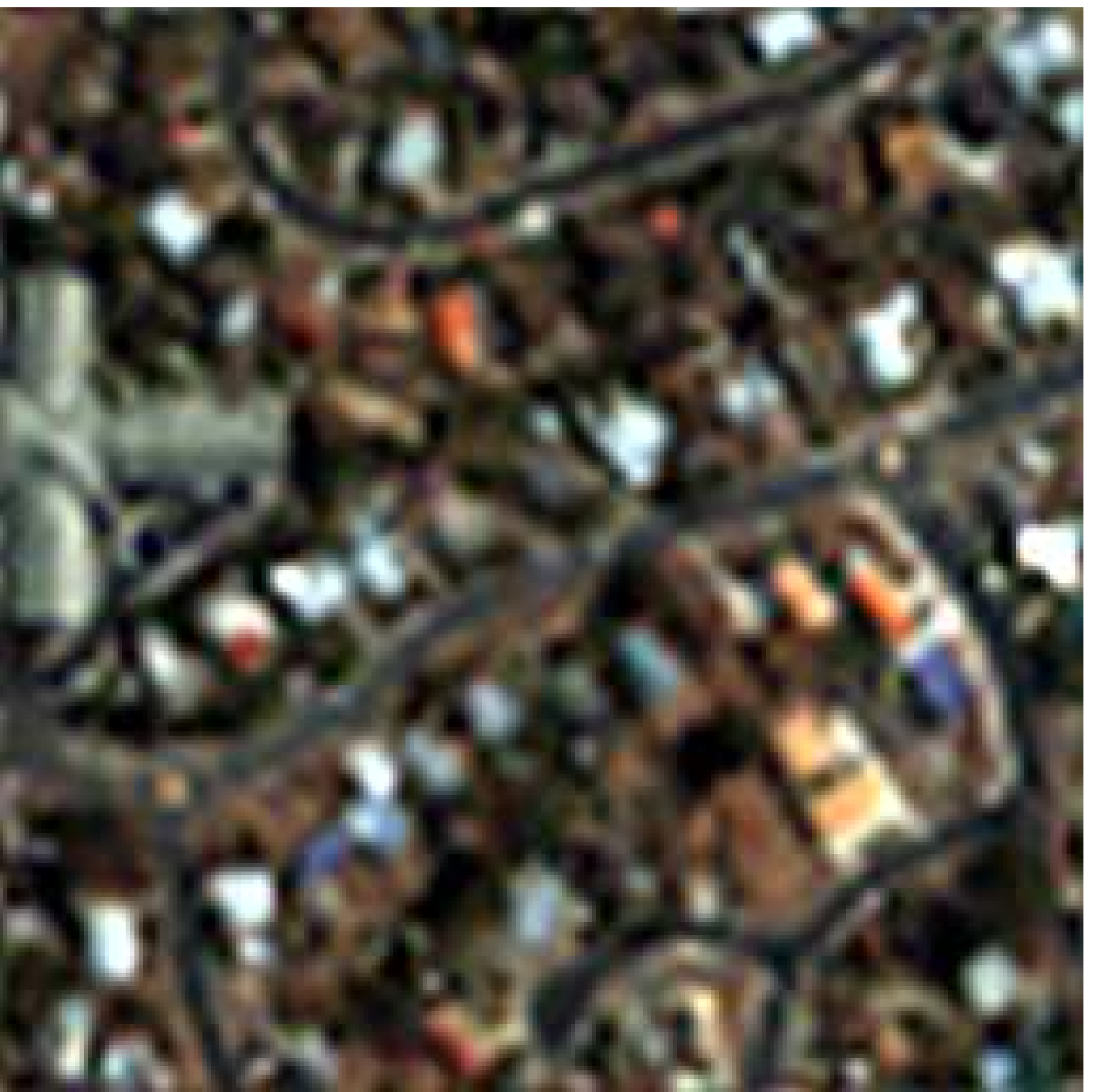} &
\includegraphics[width=0.14\paperwidth]{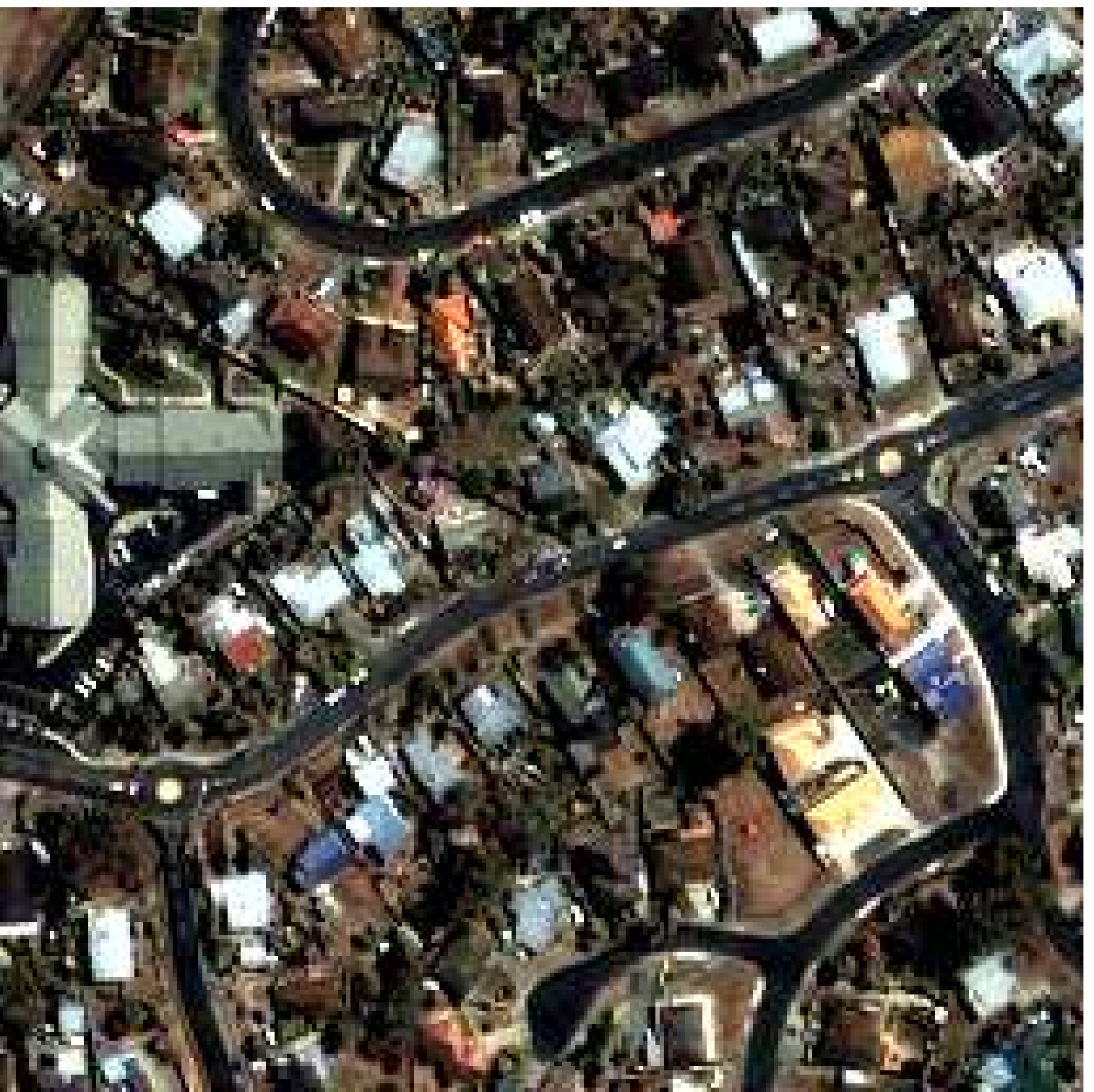} &
\includegraphics[width=0.14\paperwidth]{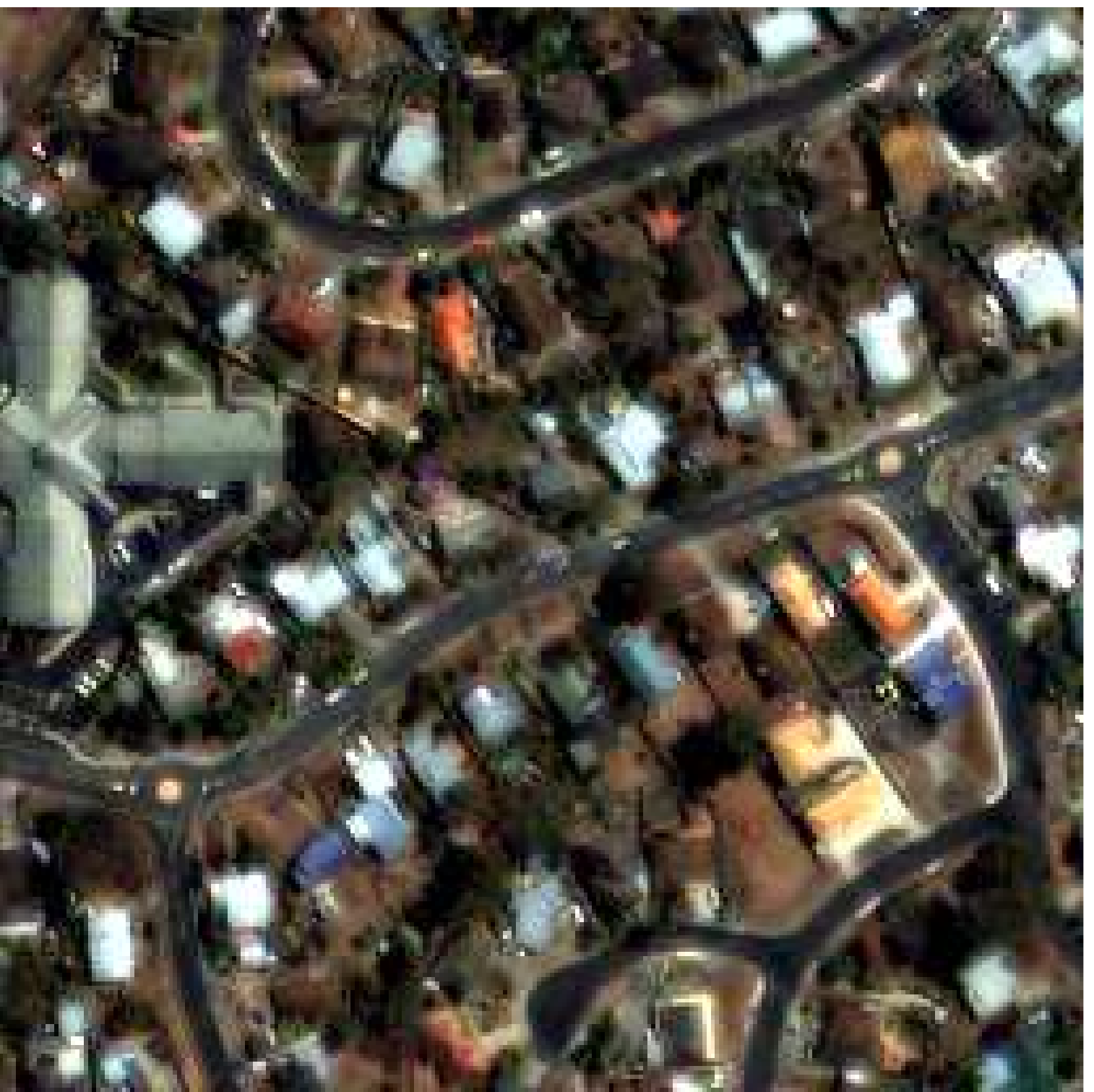} \\
(a) & (b) & (c) & (d) \\
\includegraphics[width=0.14\paperwidth]{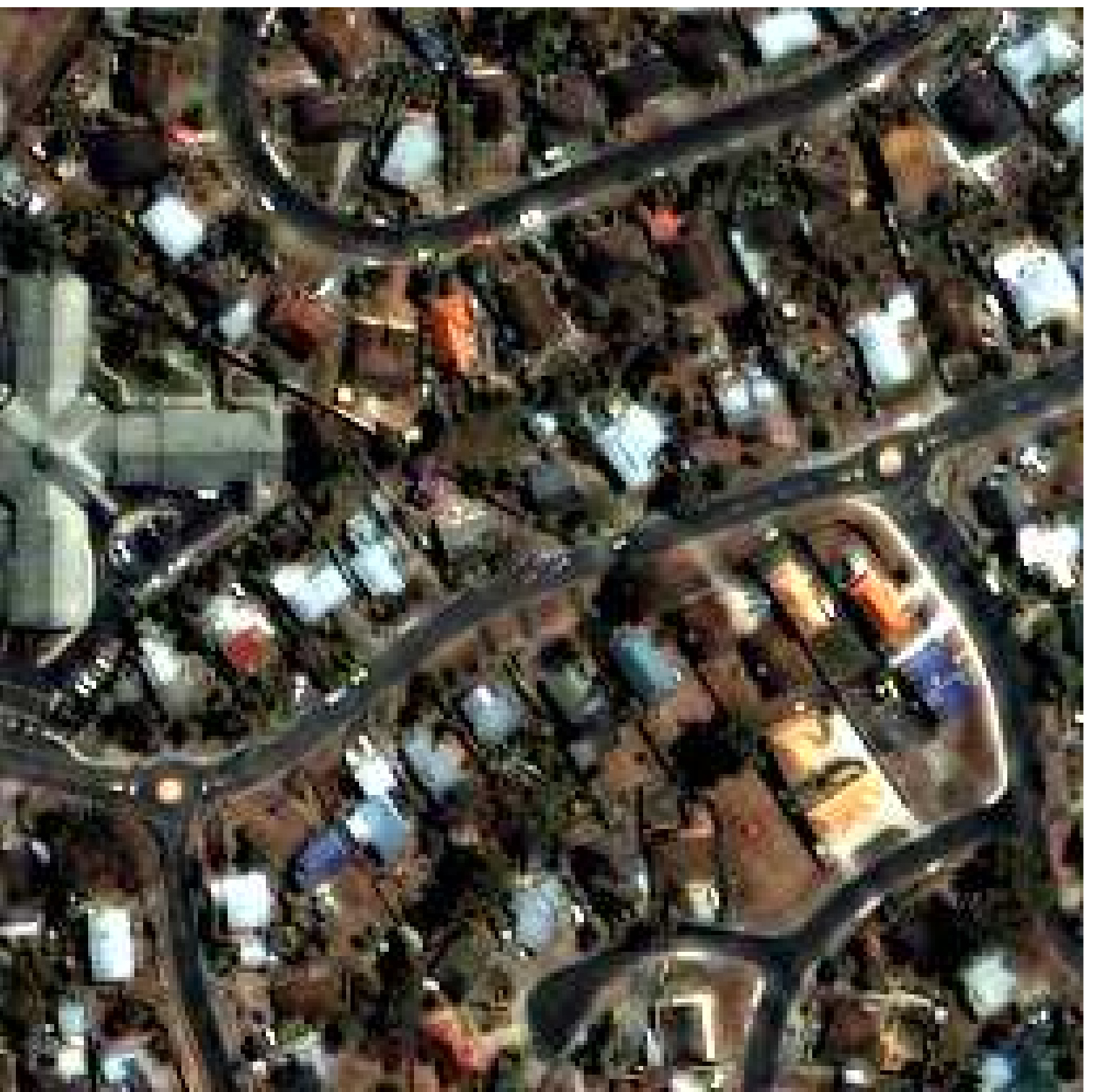}&
\includegraphics[width=0.14\paperwidth]{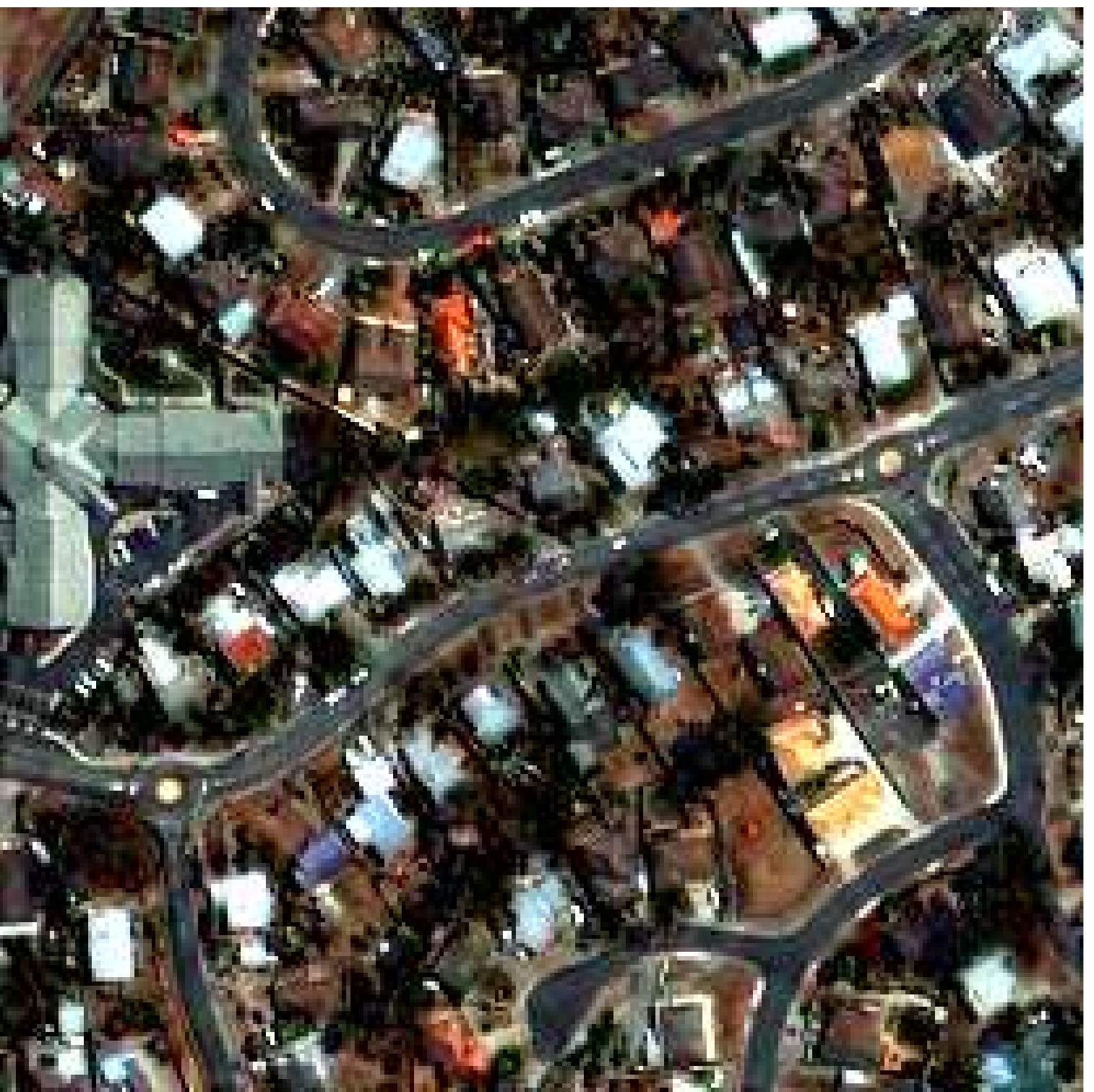}&
\includegraphics[width=0.14\paperwidth]{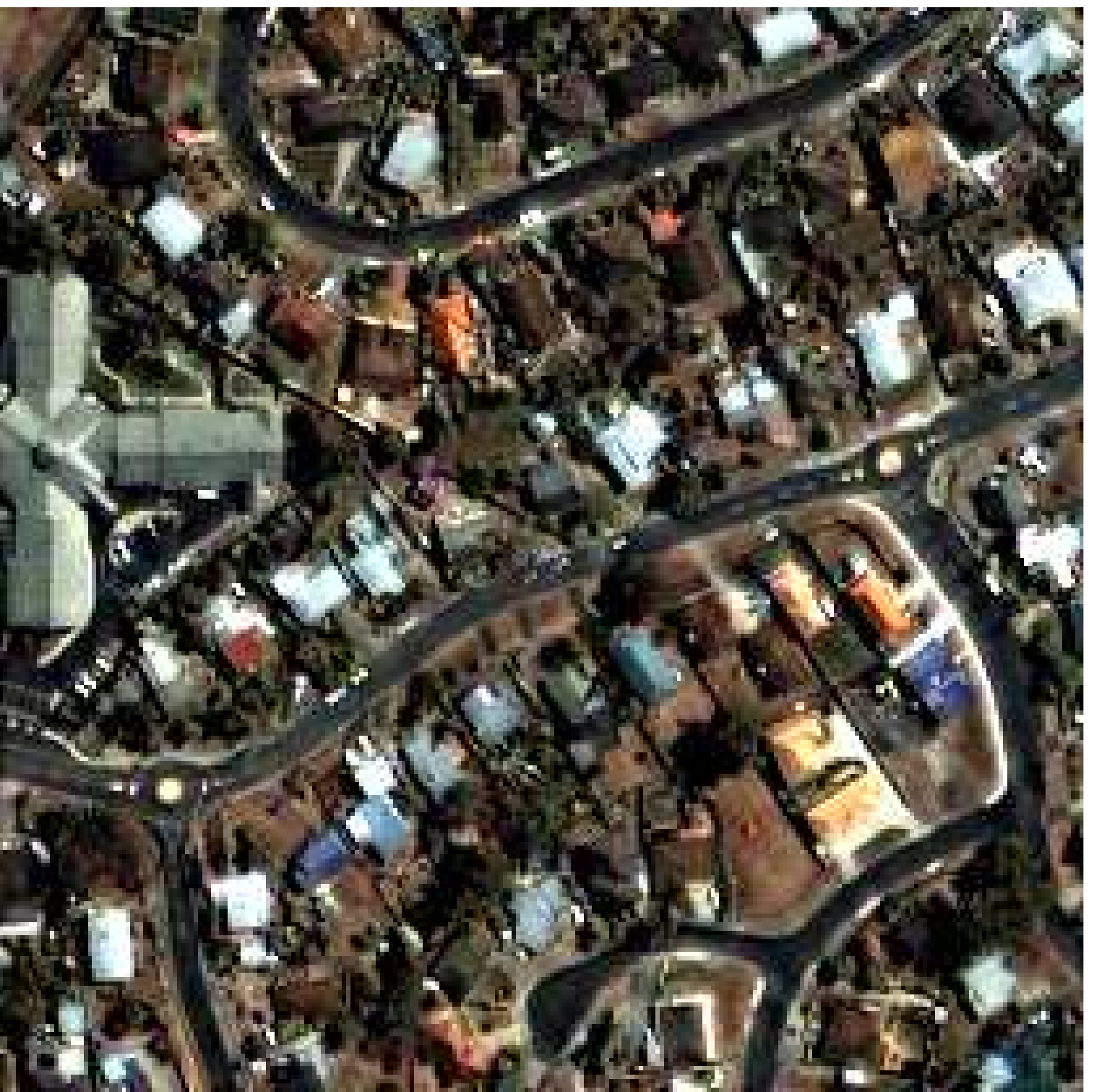}&
\includegraphics[width=0.14\paperwidth]{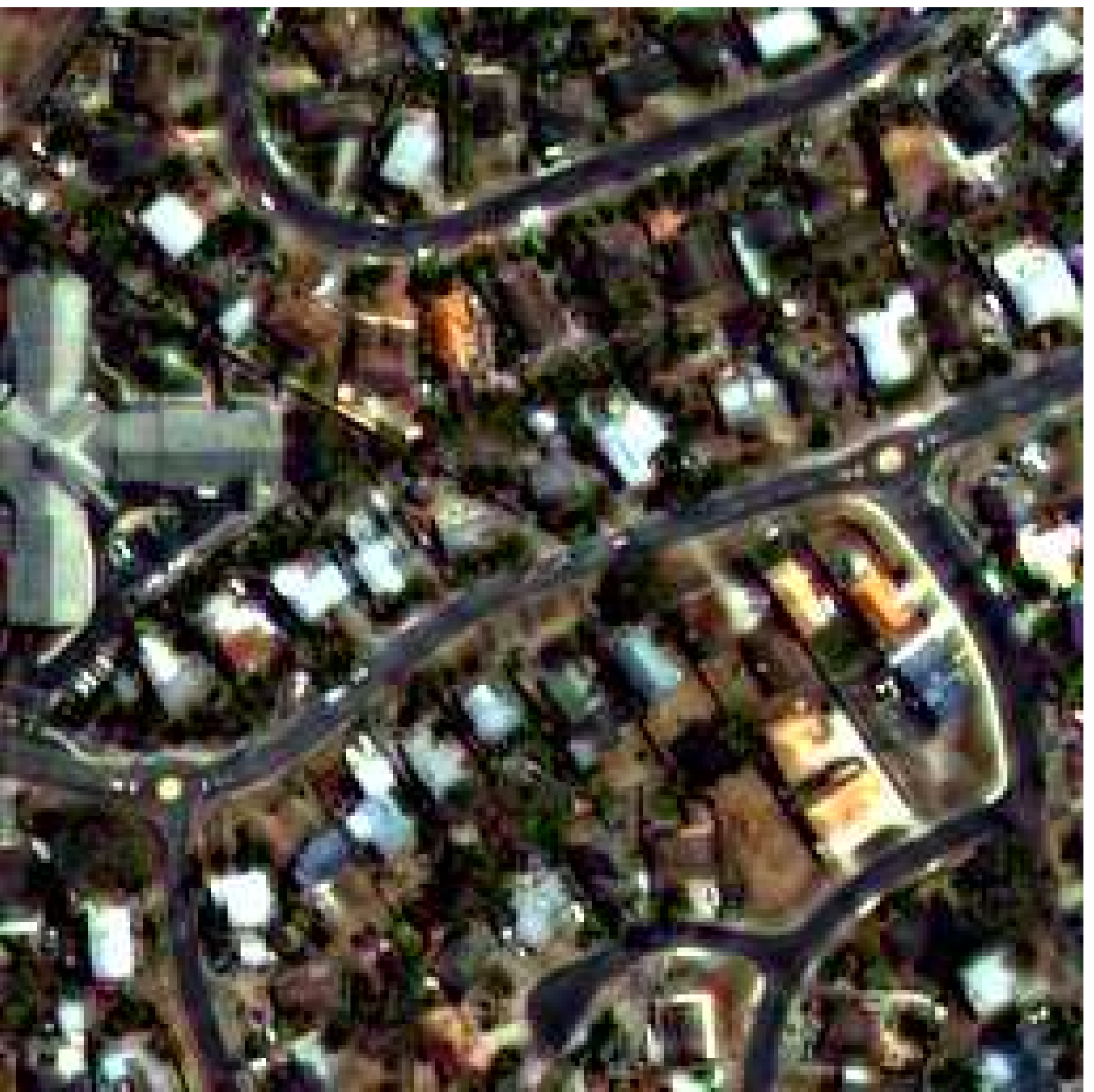} \\
(e) &(f) &(g) &(h) \\
\includegraphics[width=0.14\paperwidth]{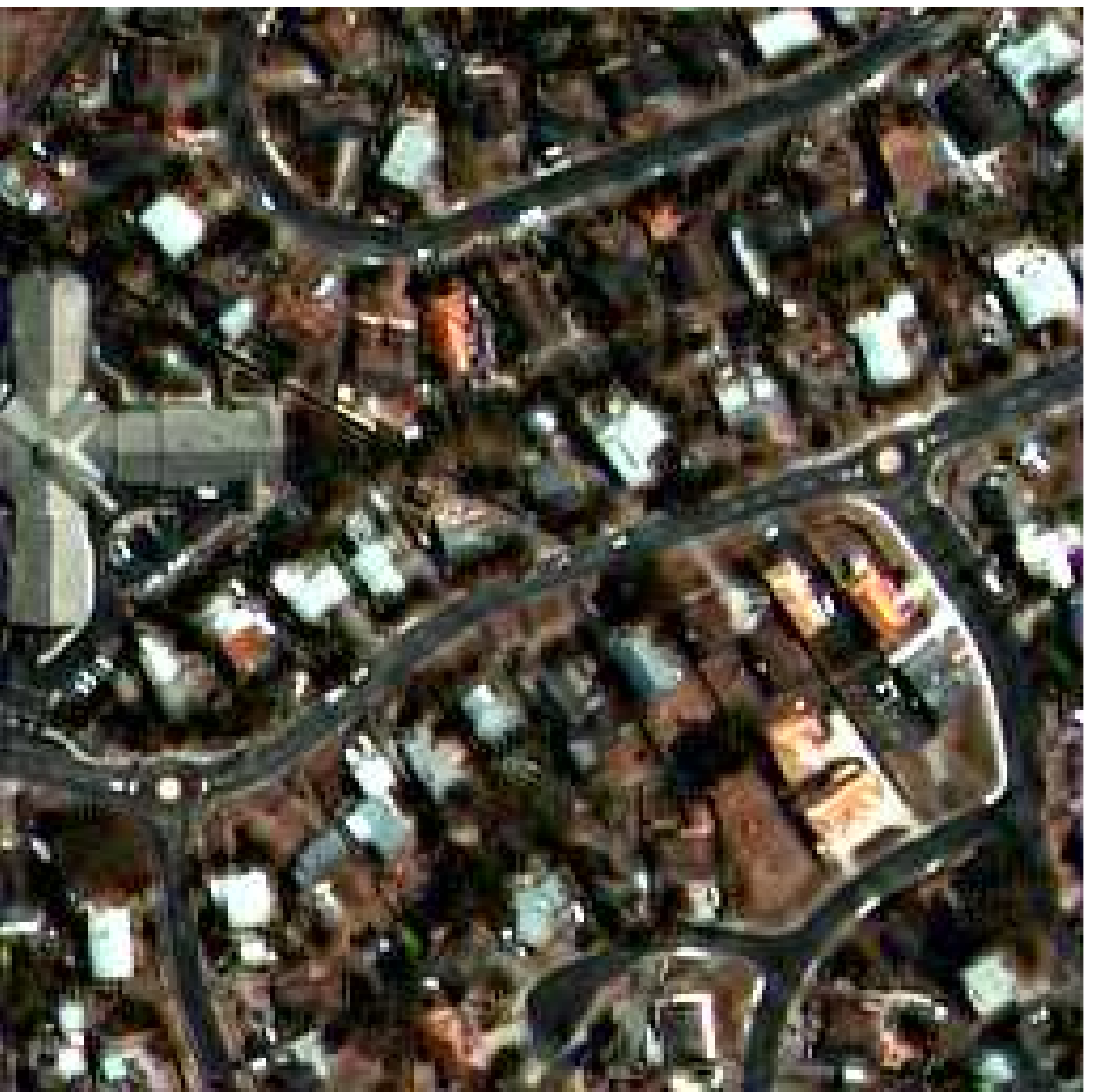} &
\includegraphics[width=0.14\paperwidth]{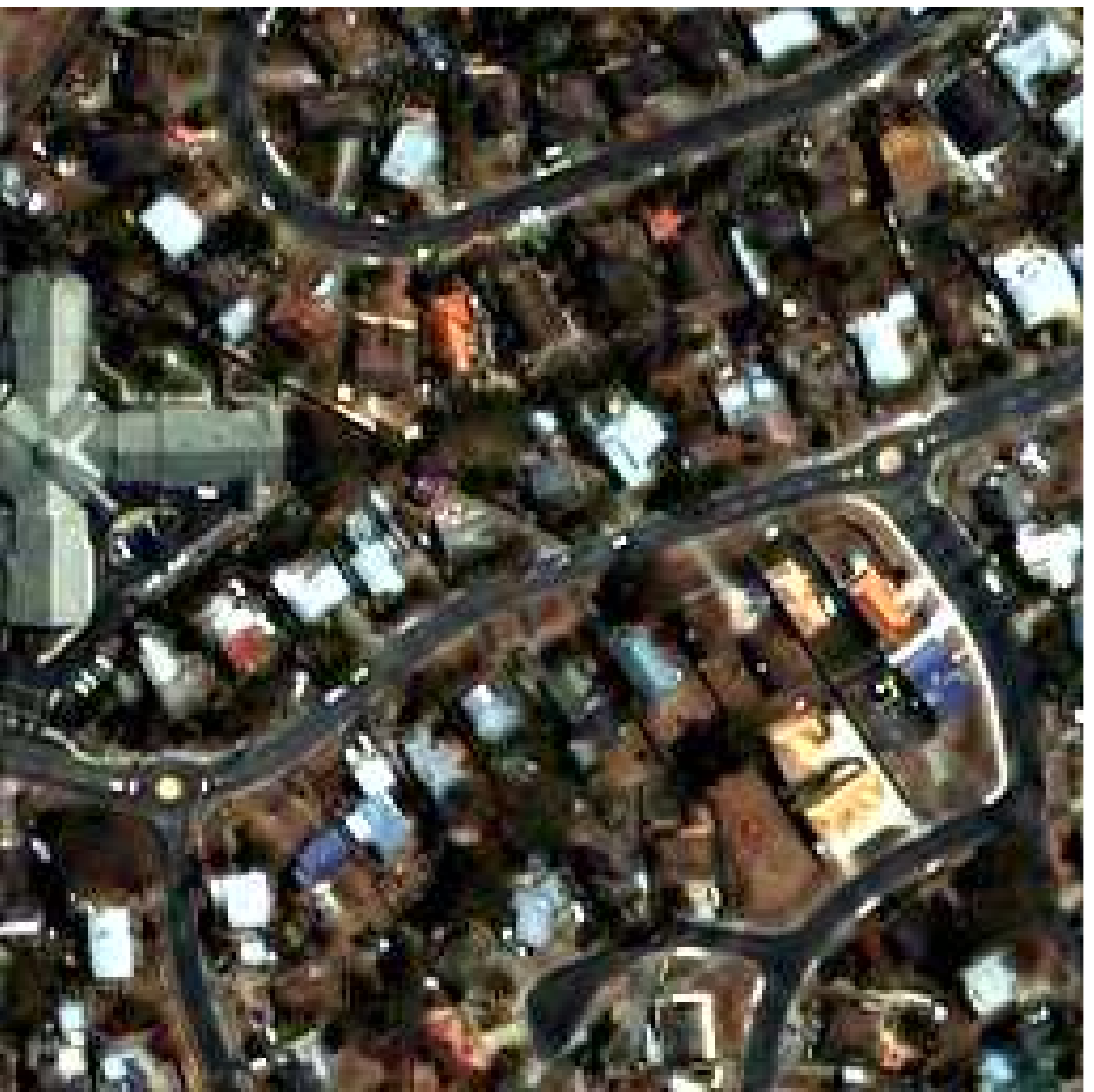} &
\includegraphics[width=0.14\paperwidth]{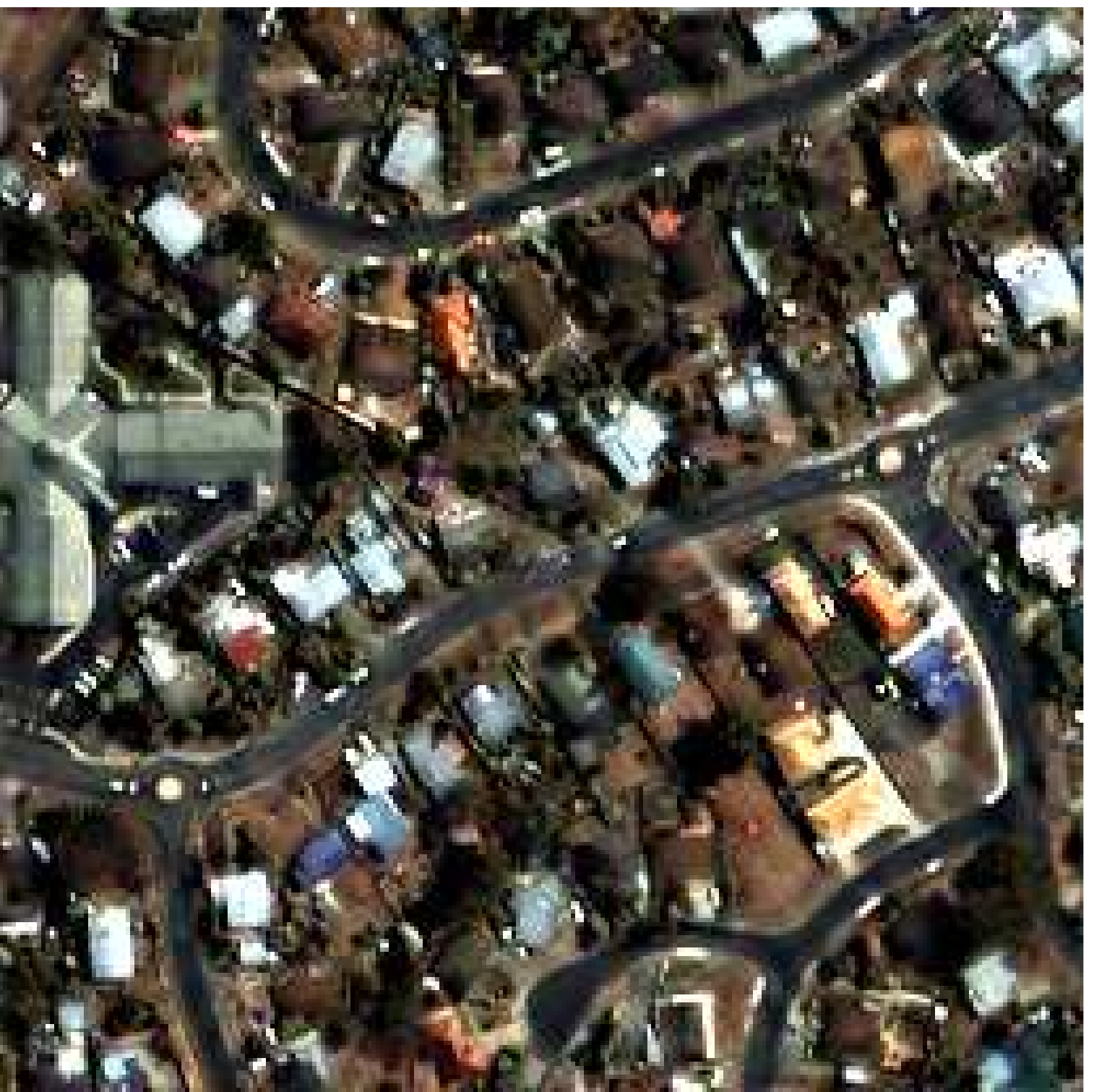} \\
(i) &(j) & (k)  \\
\\
\end{tabular}
\caption{Full-resolution pansharpening results for IKONOS dataset: (a)PAN image; (b)EXP; (c)GSA; (d)PRACS; (e)ATWT; (f)BDSD; (g)GLP-CBD; (h)PNN; (i)DRPNN; (j)DiCNN1; (k)DiCNN2.}
\label{figure:map:full-ik}
\end{figure*}

Fig. \ref{figure:map:ik} displays the reduced-resolution experimental results. As it can be observed, CS/MRA-based methods have poorer pansharpening results than CNN-based methods, as it can be seen in the edges of roofs shown in Fig. \ref{figure:map:ik}(c)-(g). Furthermore, DiCNN1 and DiCNN2 look most similar to the ground-truth in terms of spectral fidelity, as it can be seen in the vegetation area in the top left part of Fig. \ref{figure:map:ik}(j) and (k). Fig. \ref{figure:detailimage:ik} shows the detail images learned from various methods. They also support the previous observations and, additionally, confirm that DiCNN1 performs slightly better than DiCNN2 in terms of edge restoration, as it can be seen in the circle vegetation area in the bottom left part of Fig. \ref{figure:detailimage:ik}(i) and (j). Fig. \ref{figure:map:full-ik} displays the full-resolution experimental results on IKONOS Hobart dataset. Similar observations can be made with regards to the experimental results from the WorldView-2 Washington dataset.

\subsection{Experiment 3: Quickbird Sundarbans Dataset}

\begin{figure*}[t]\scriptsize
\centering
  \begin{tabular}{cccc}
\includegraphics[width=0.14\paperwidth]{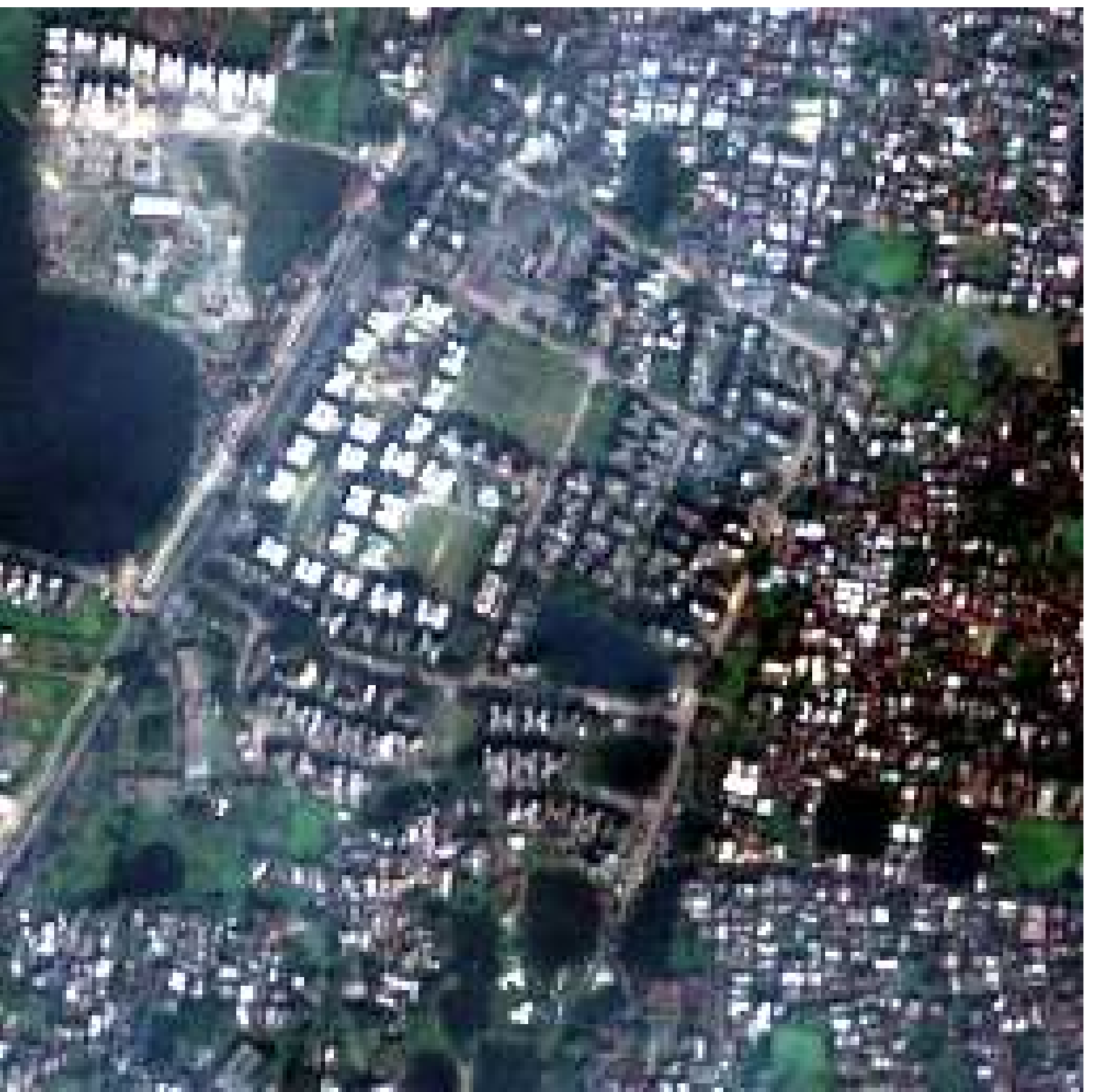} &
\includegraphics[width=0.14\paperwidth]{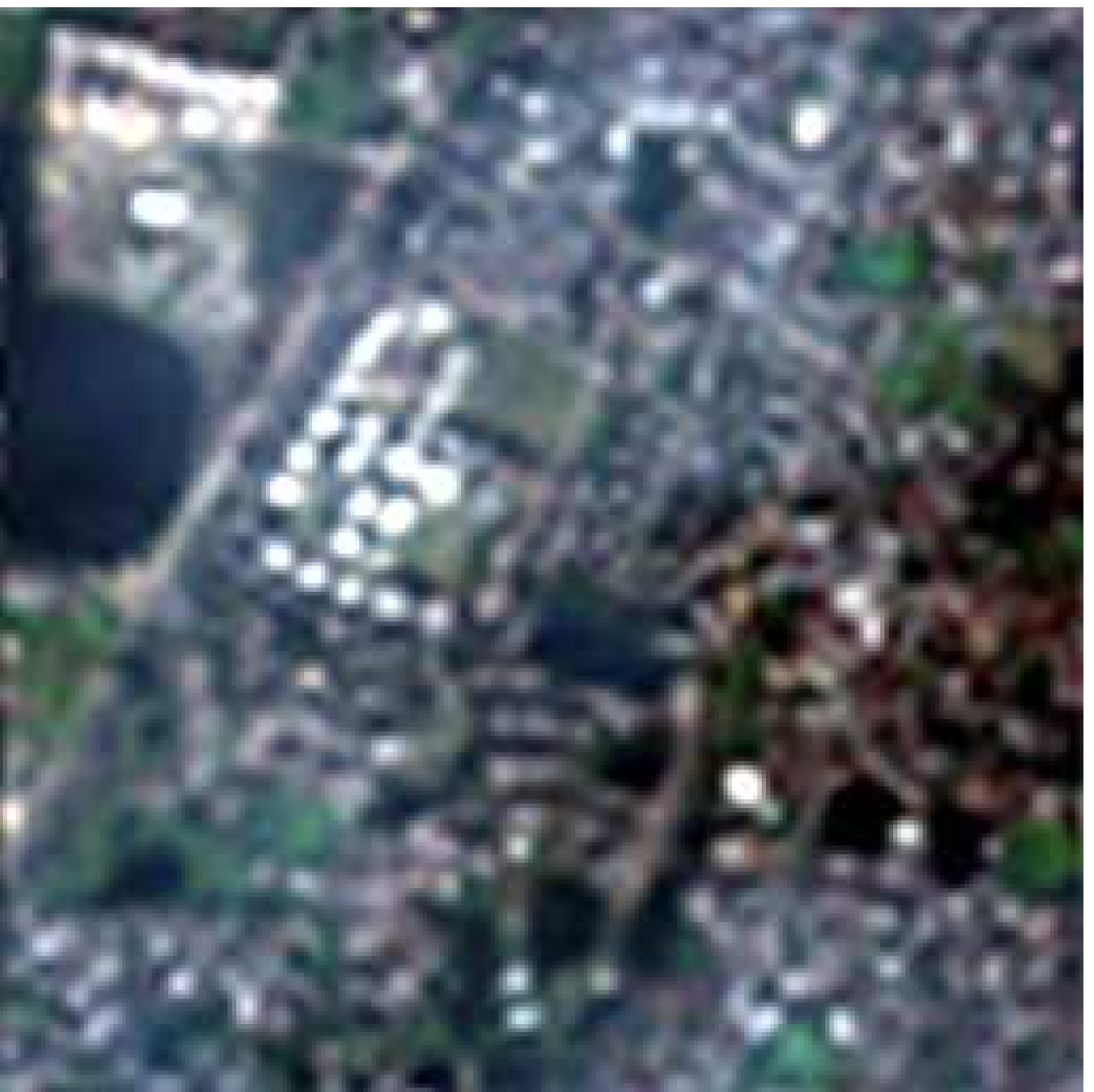} &
\includegraphics[width=0.14\paperwidth]{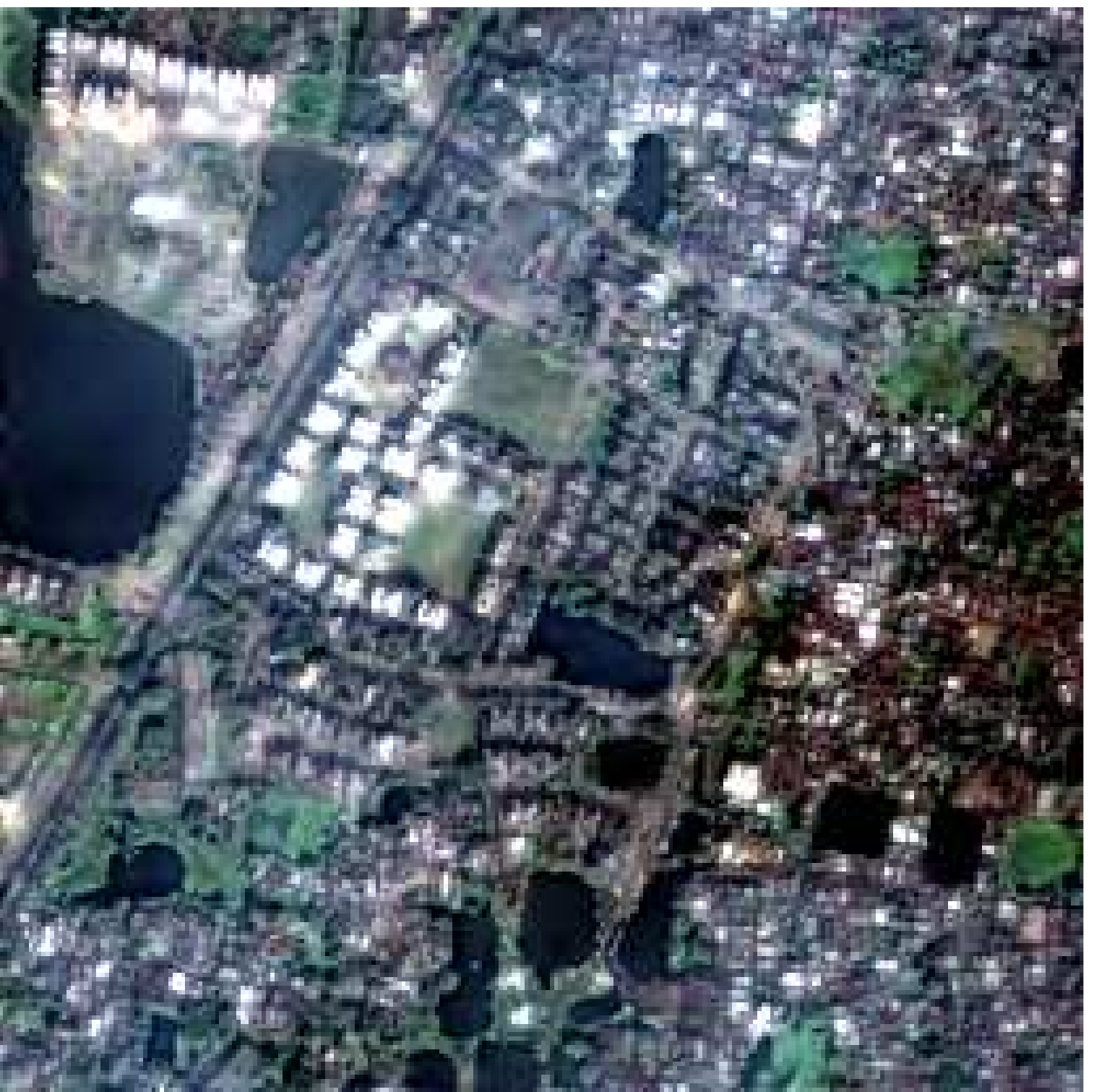} &
\includegraphics[width=0.14\paperwidth]{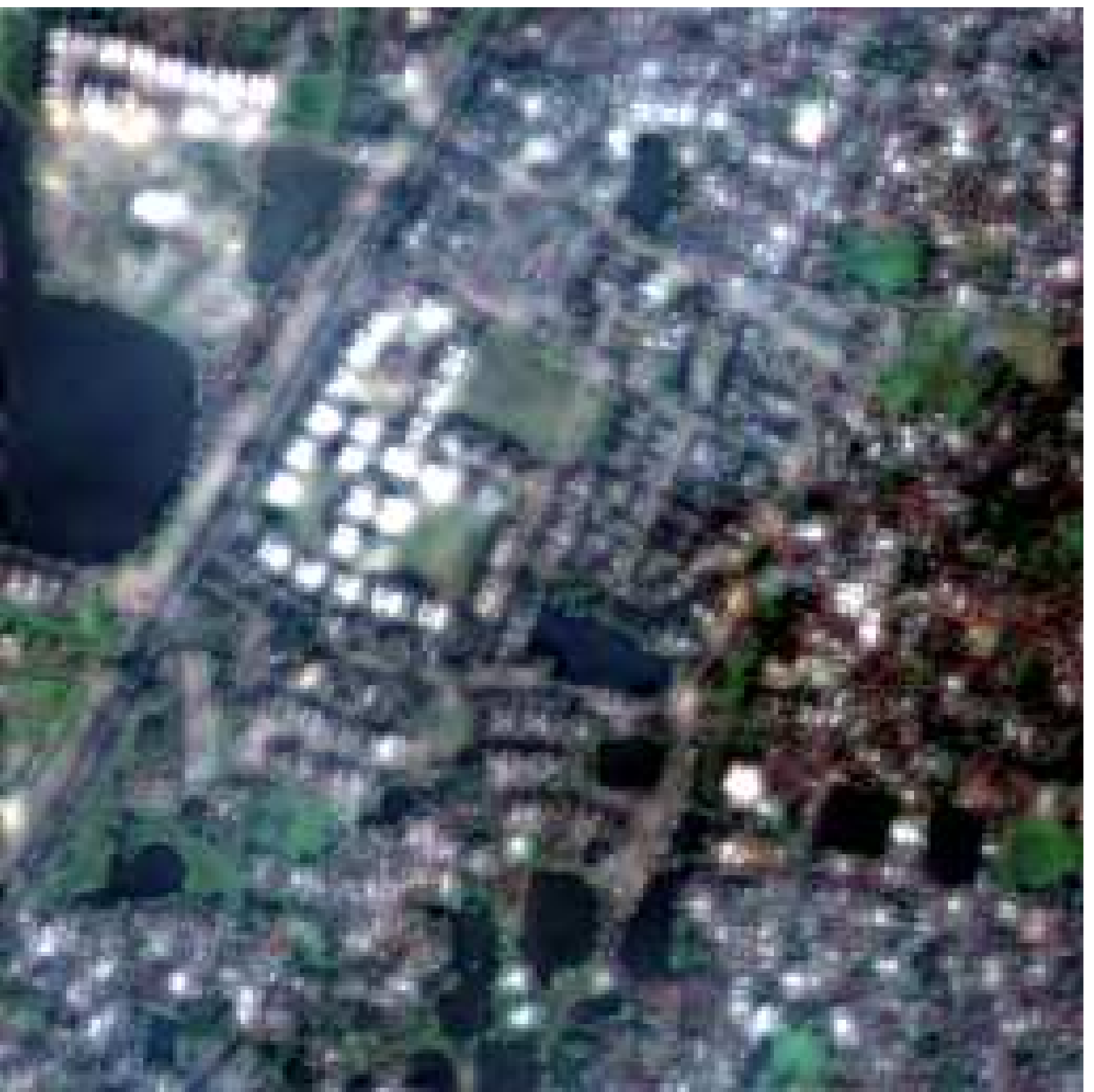} \\
(a) & (b) & (c)  & (d) \\
\includegraphics[width=0.14\paperwidth]{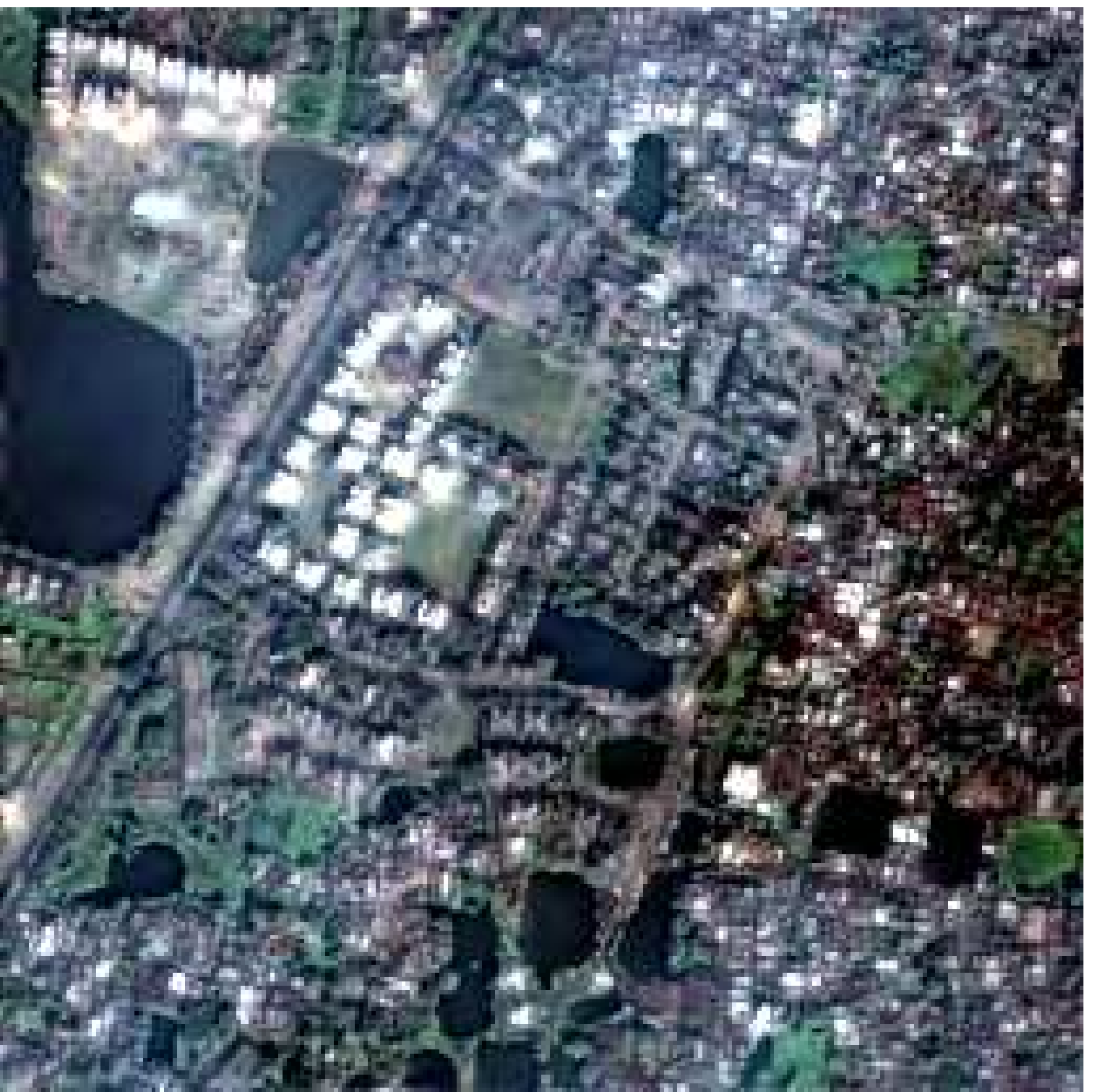}&
\includegraphics[width=0.14\paperwidth]{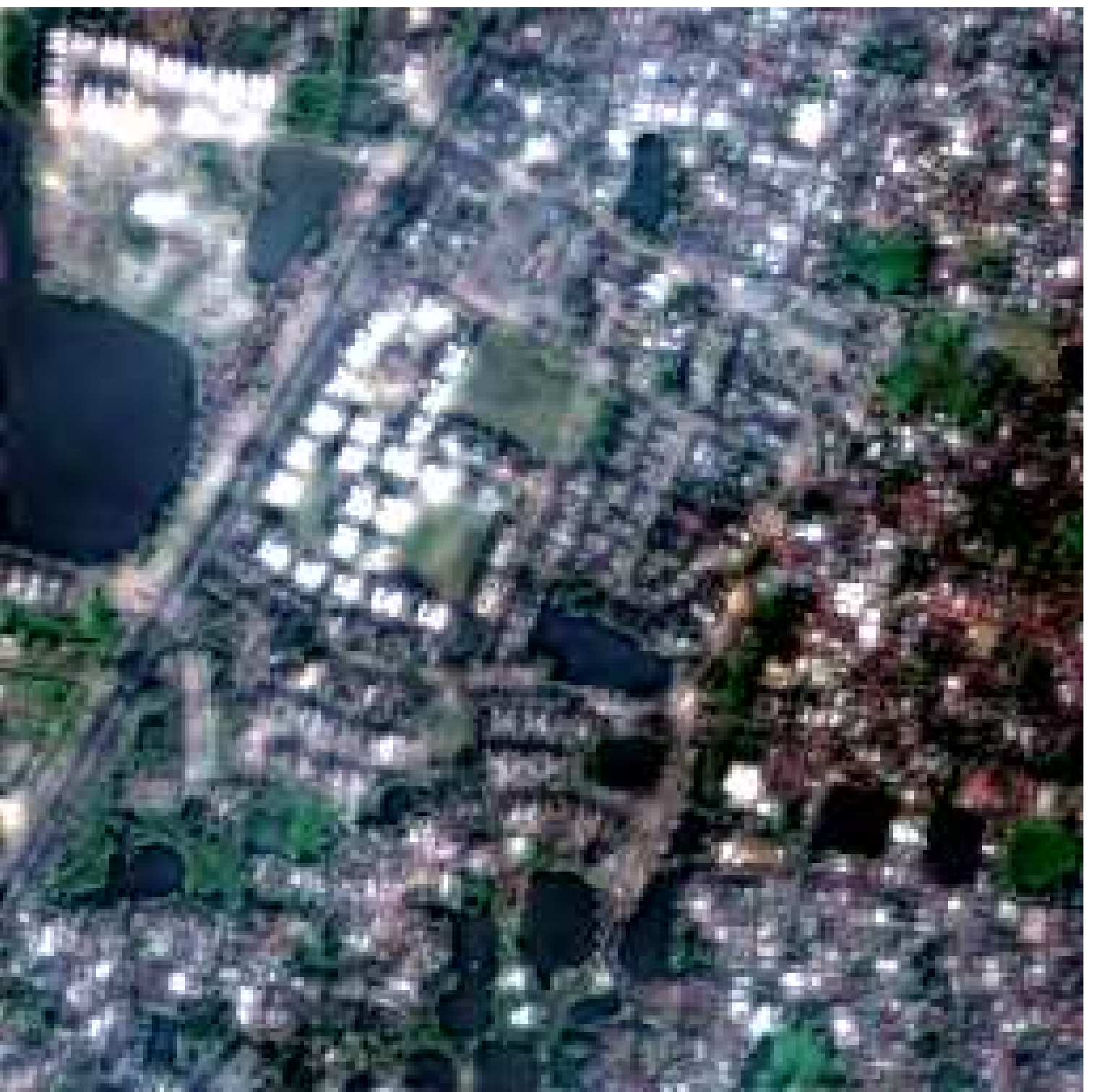} &
\includegraphics[width=0.14\paperwidth]{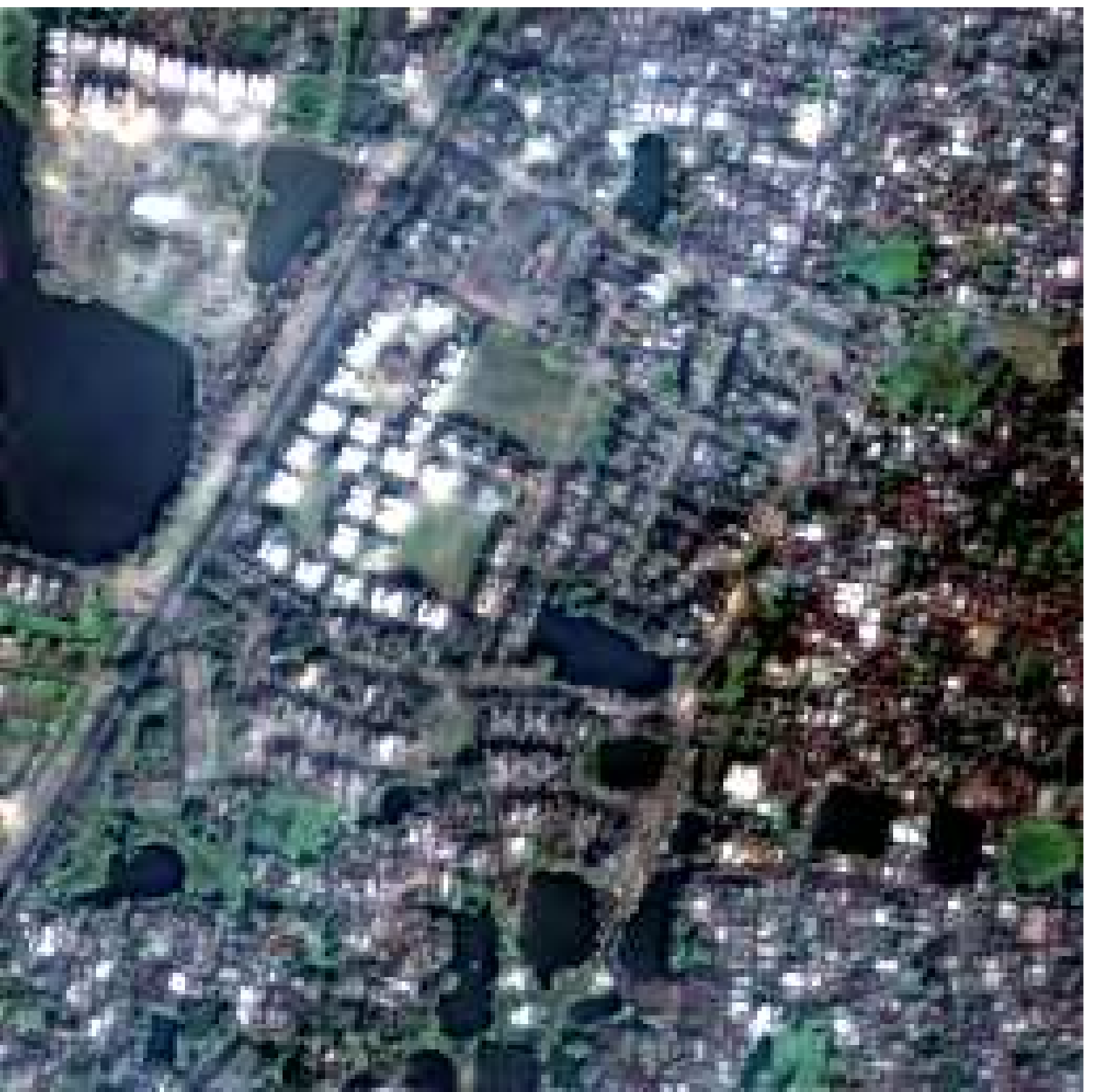} &
\includegraphics[width=0.14\paperwidth]{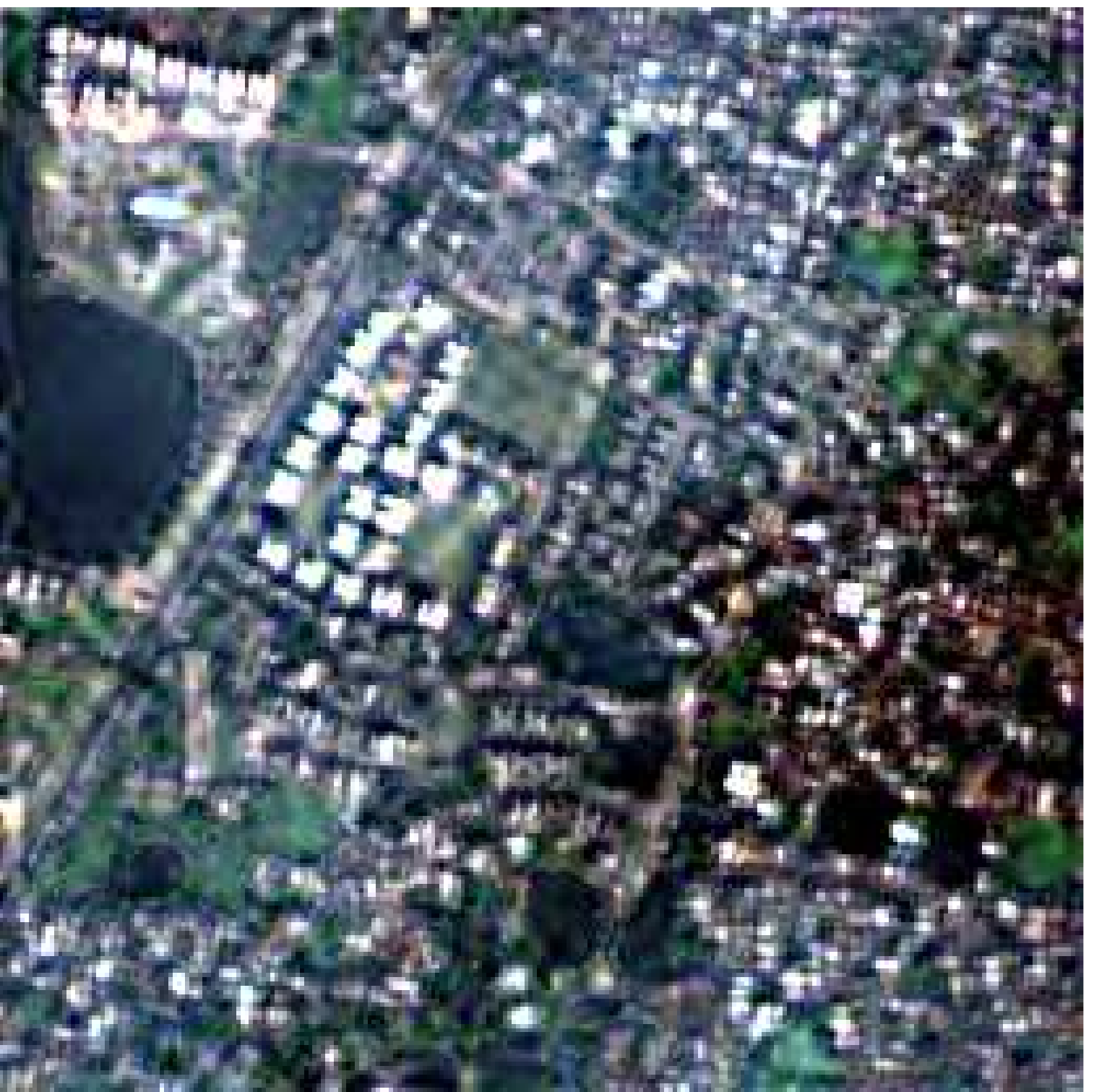}  \\
 (e) & (f) & (g) & (h) \\
\includegraphics[width=0.14\paperwidth]{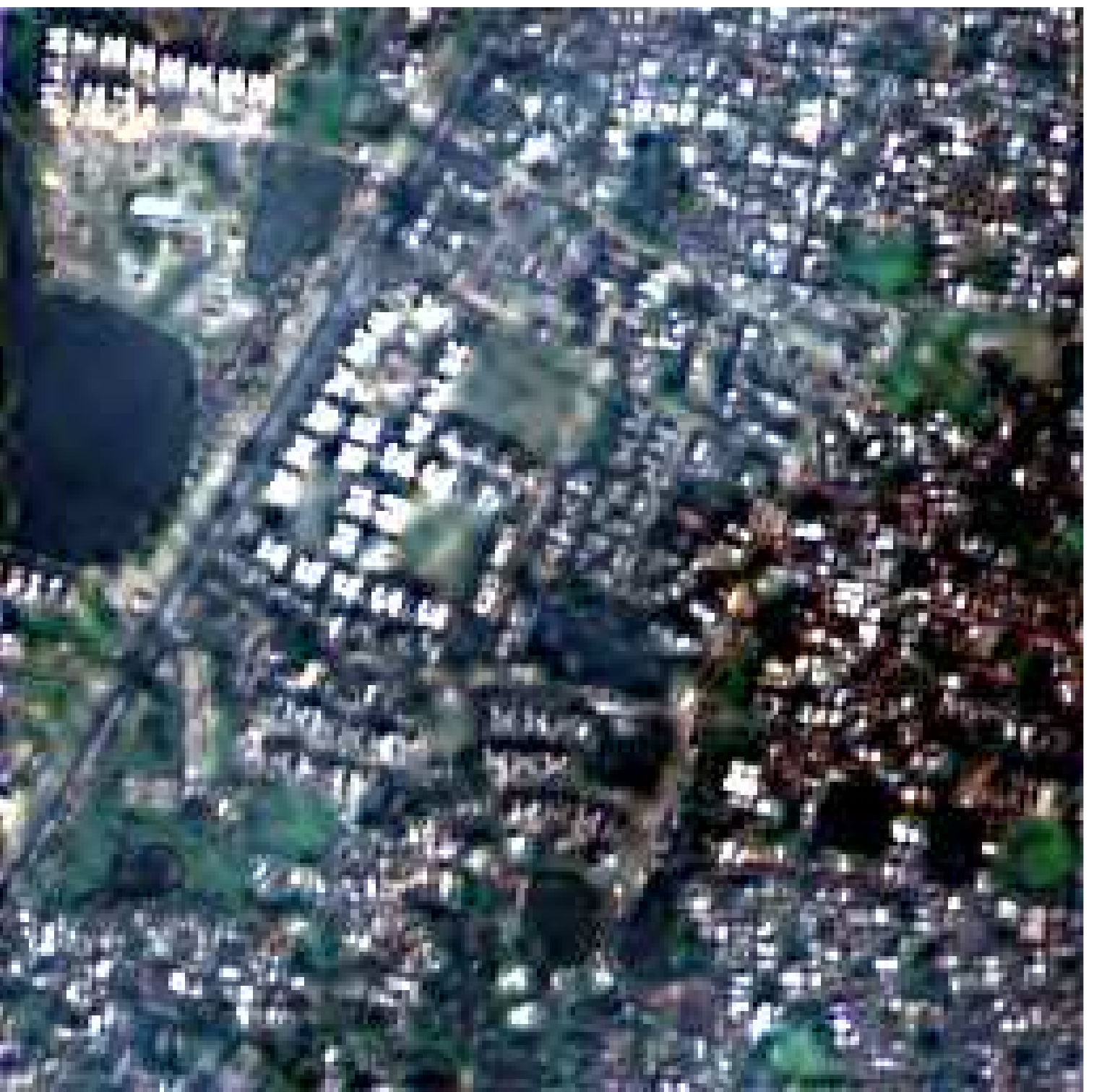} &
\includegraphics[width=0.14\paperwidth]{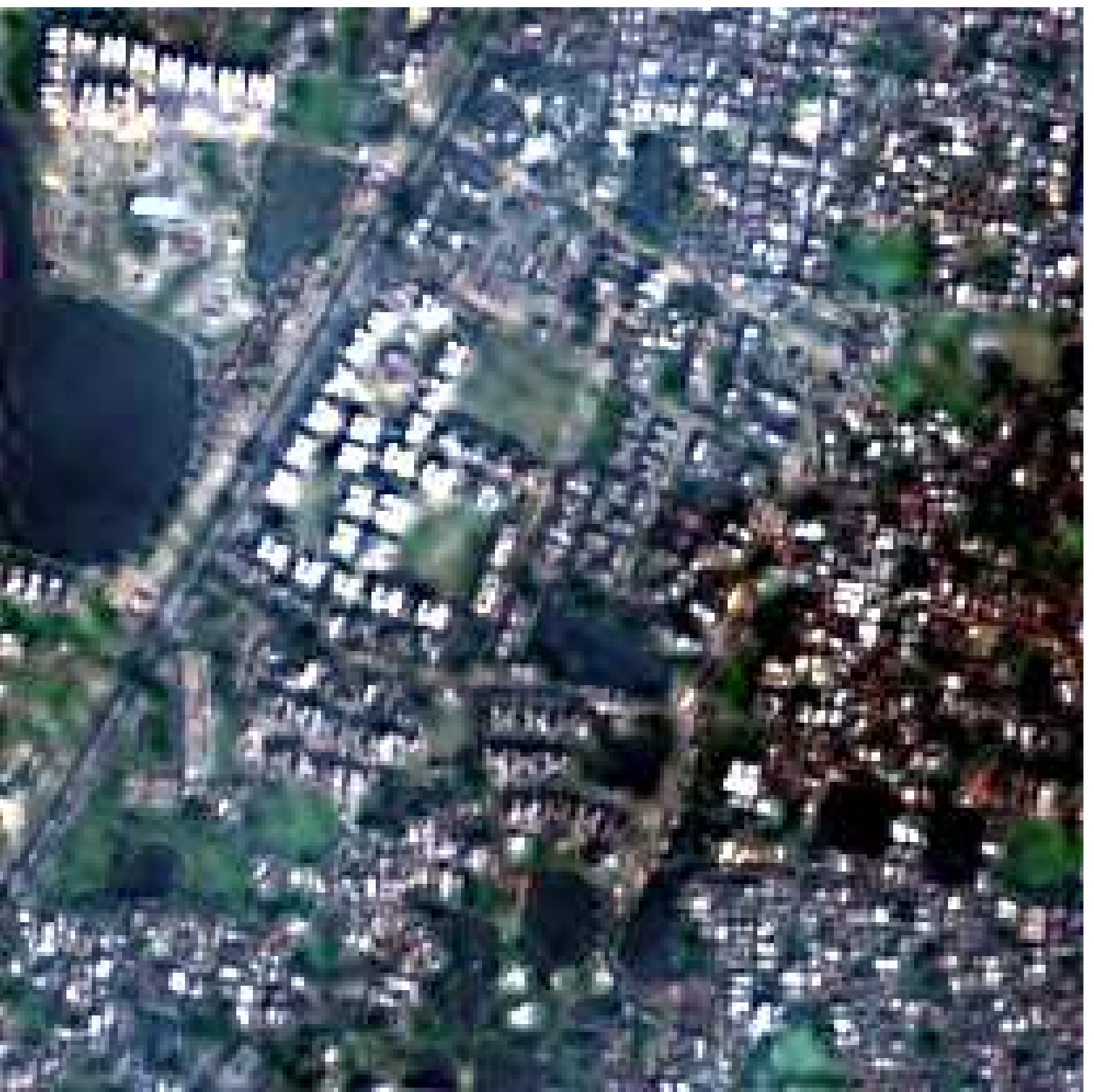} &
\includegraphics[width=0.14\paperwidth]{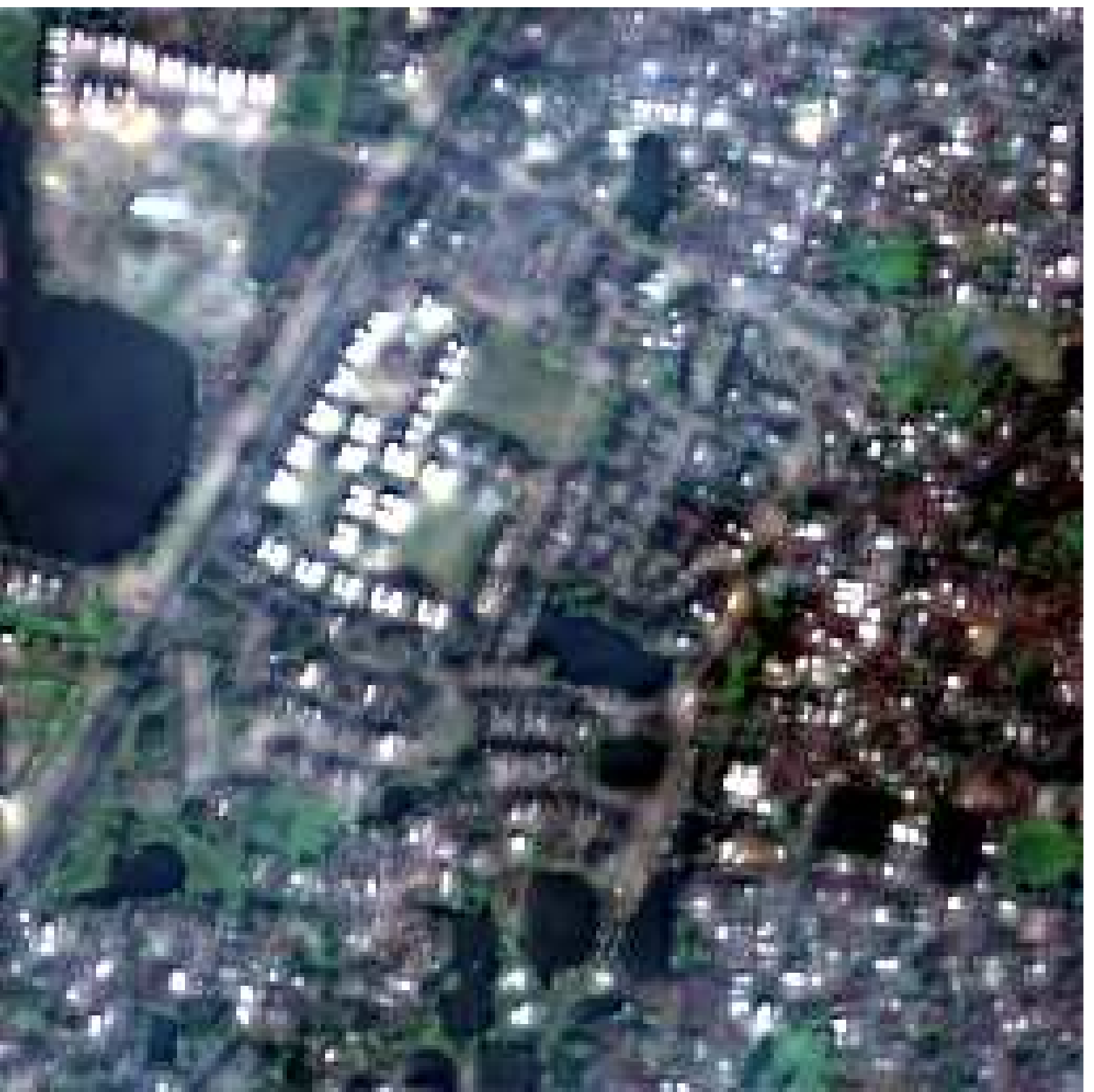} \\
 (i) & (j) & (k)  \\
\\
\end{tabular}
\caption{Pansharpening results for Quickbird dataset. (a) Ground-truth; (b)EXP; (c)GSA; (d)PRACS; (e)ATWT; (f)BDSD;  (g)GLP-CBD; (h)PNN; (i)DRPNN; (j)DiCNN1; (k)DiCNN2.}
\label{figure:map:qb}
\end{figure*}

\begin{figure*}[t]\scriptsize
\centering
  \begin{tabular}{ccccc}
\includegraphics[width=0.14\paperwidth]{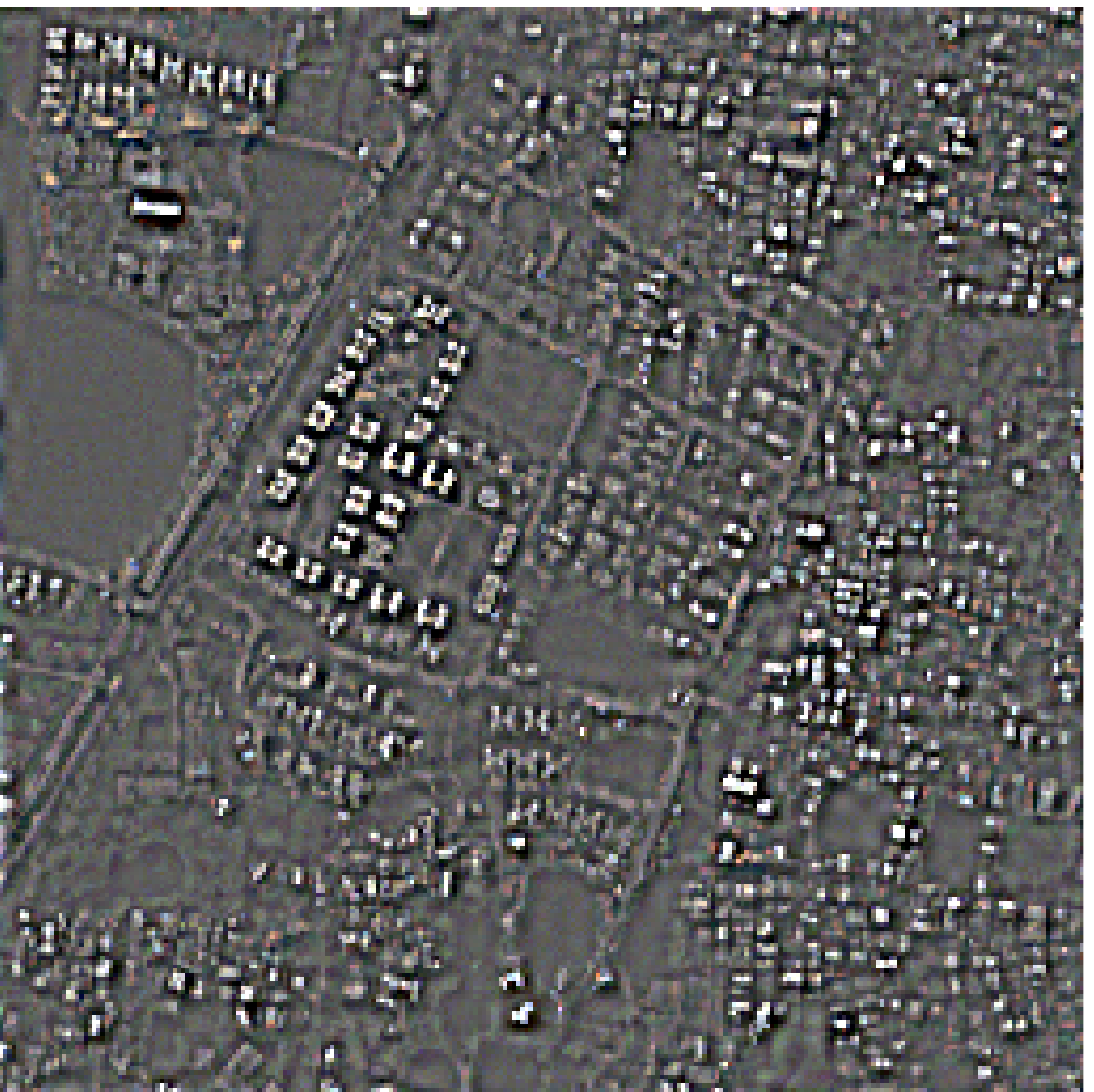} &
\includegraphics[width=0.14\paperwidth]{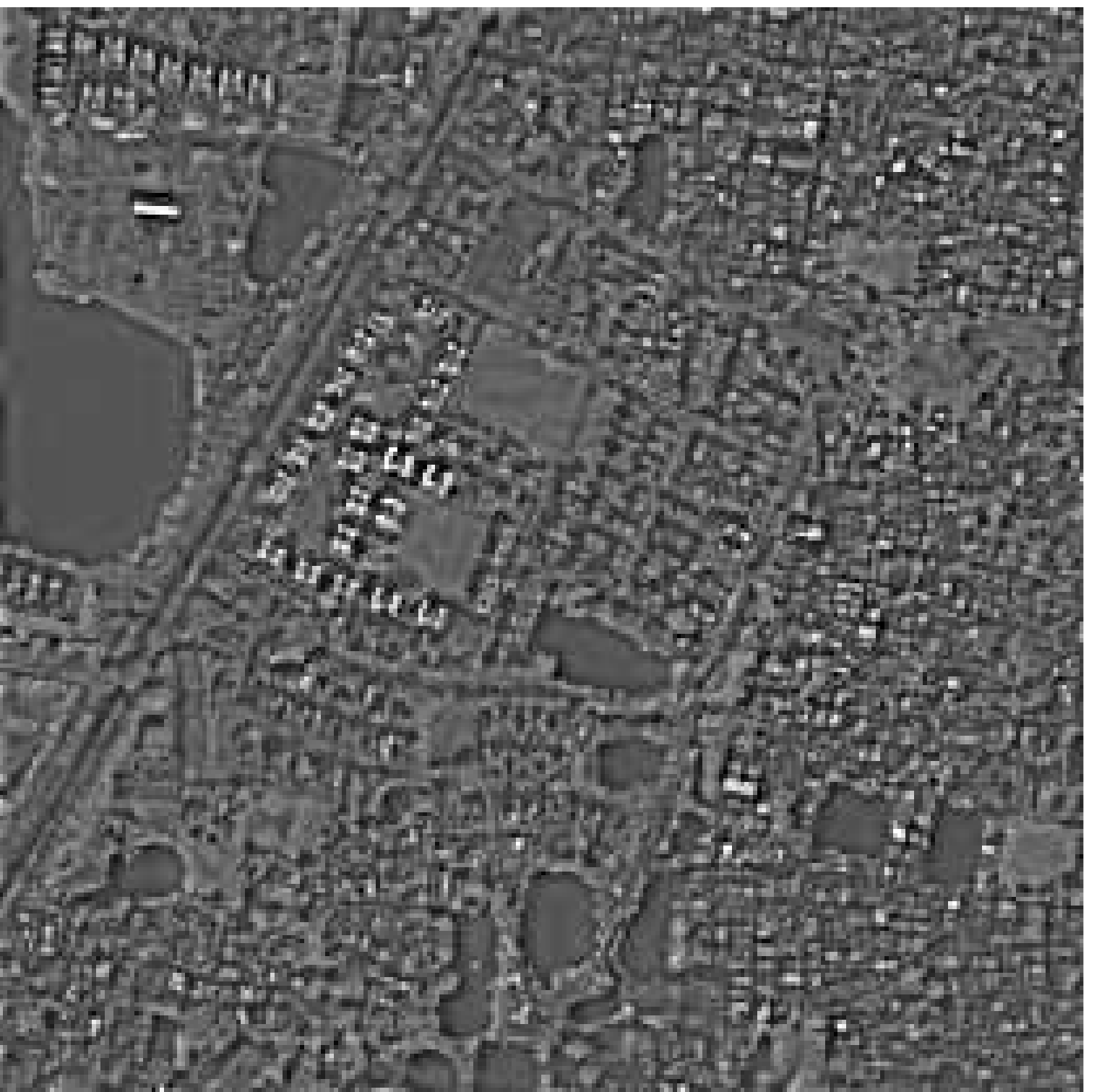} &
\includegraphics[width=0.14\paperwidth]{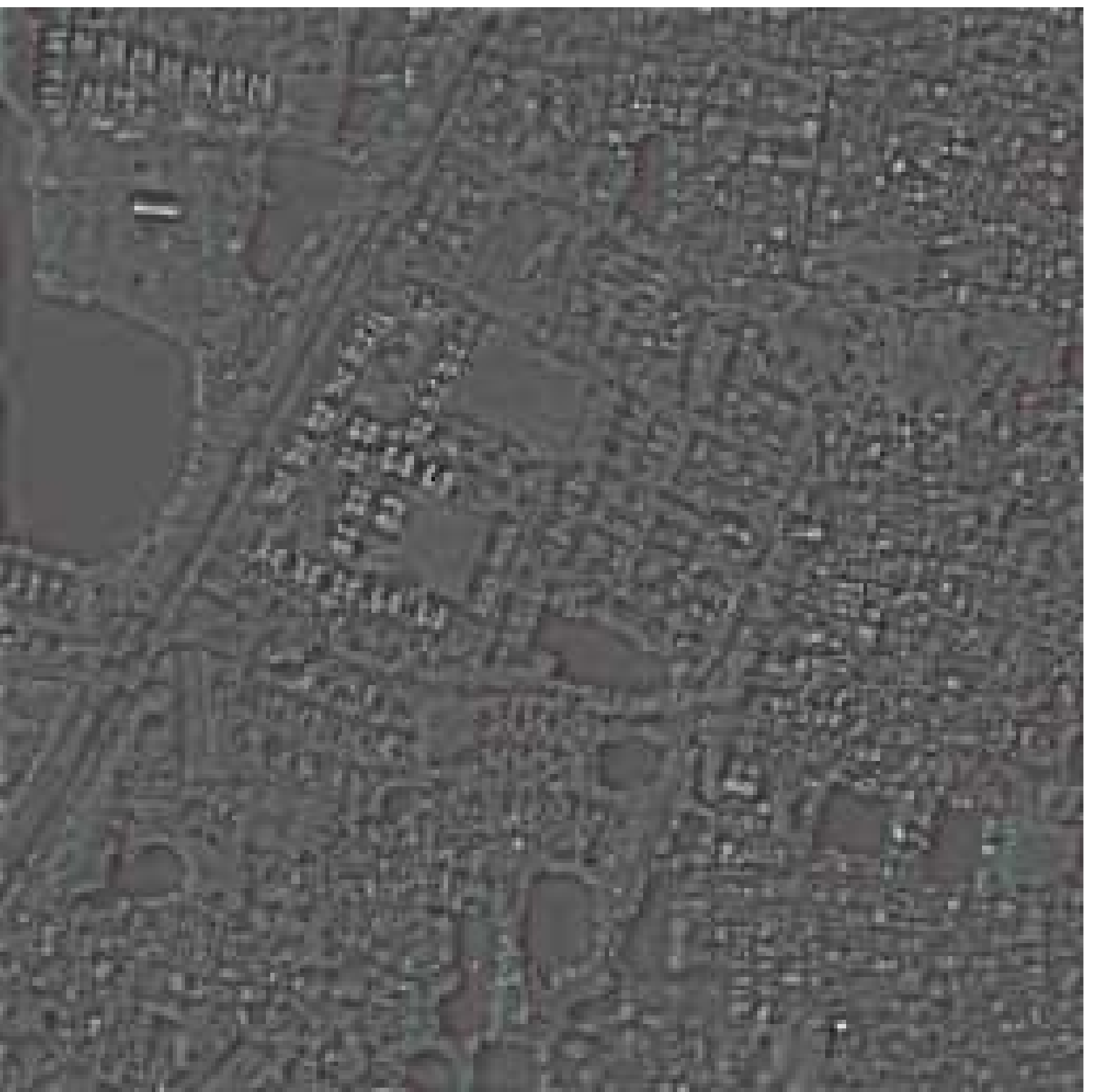} &
\includegraphics[width=0.14\paperwidth]{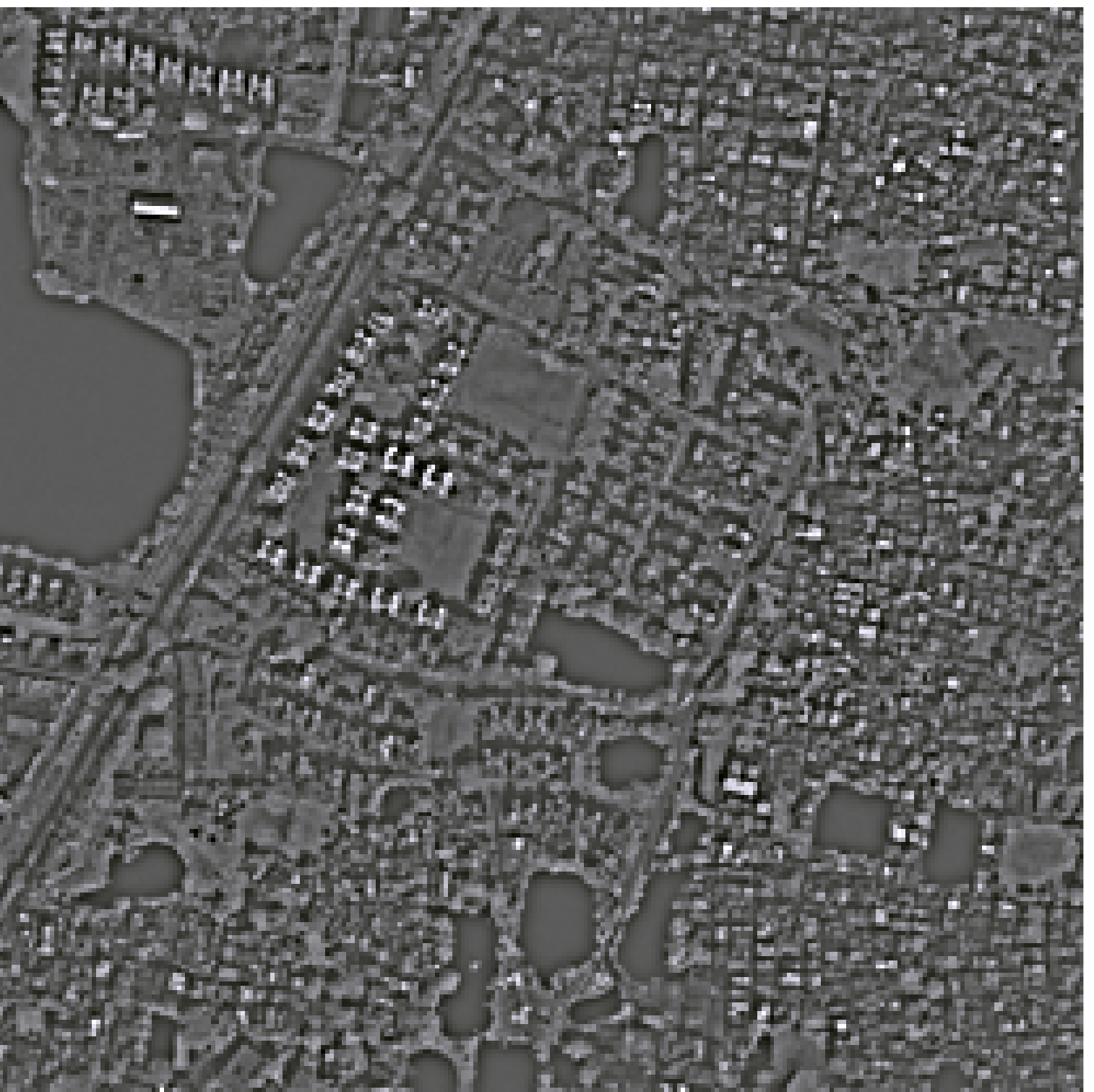}&
\includegraphics[width=0.14\paperwidth]{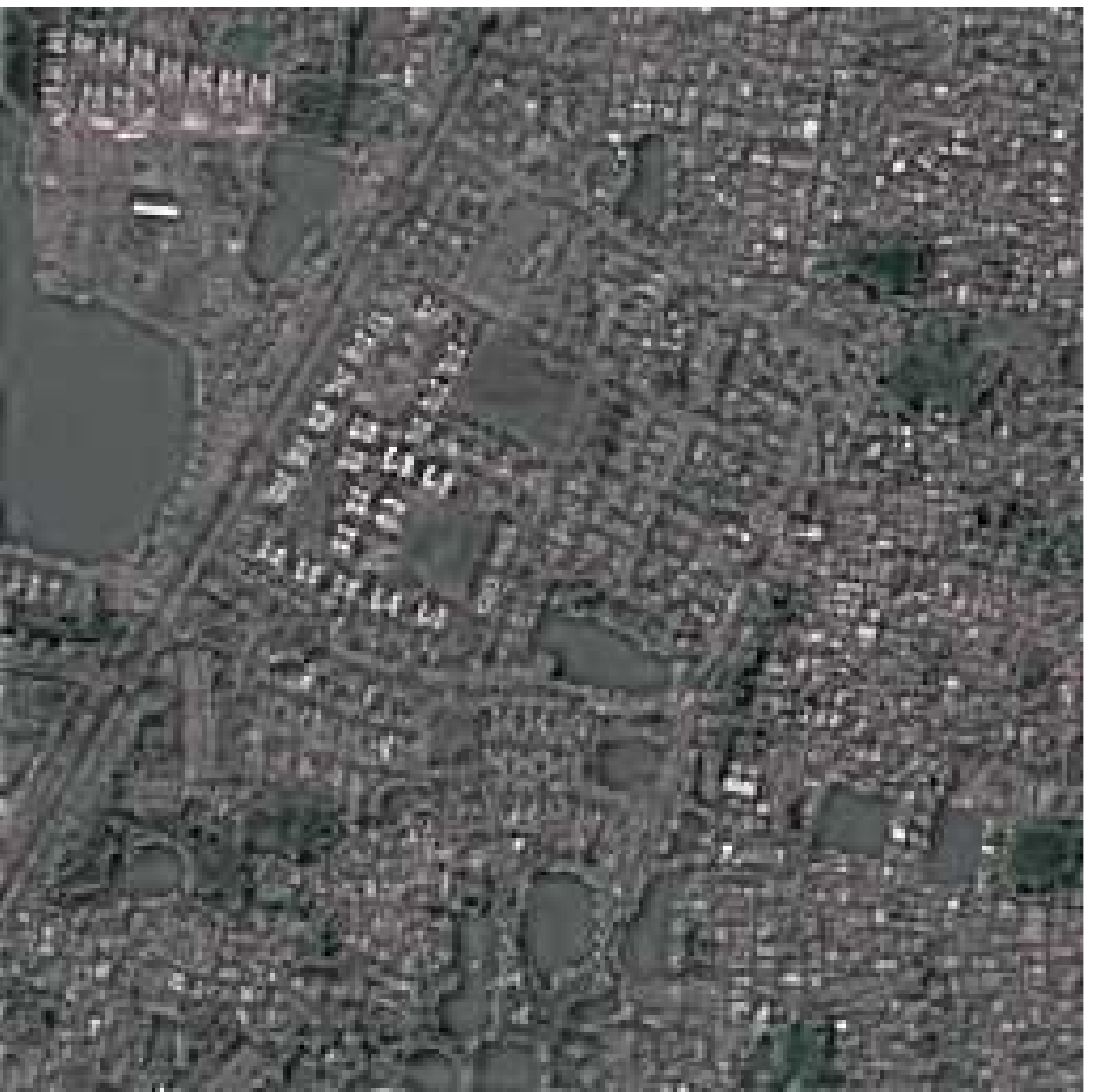} \\
(a) & (b) & (c)  & (d) &  (e)\\
\includegraphics[width=0.14\paperwidth]{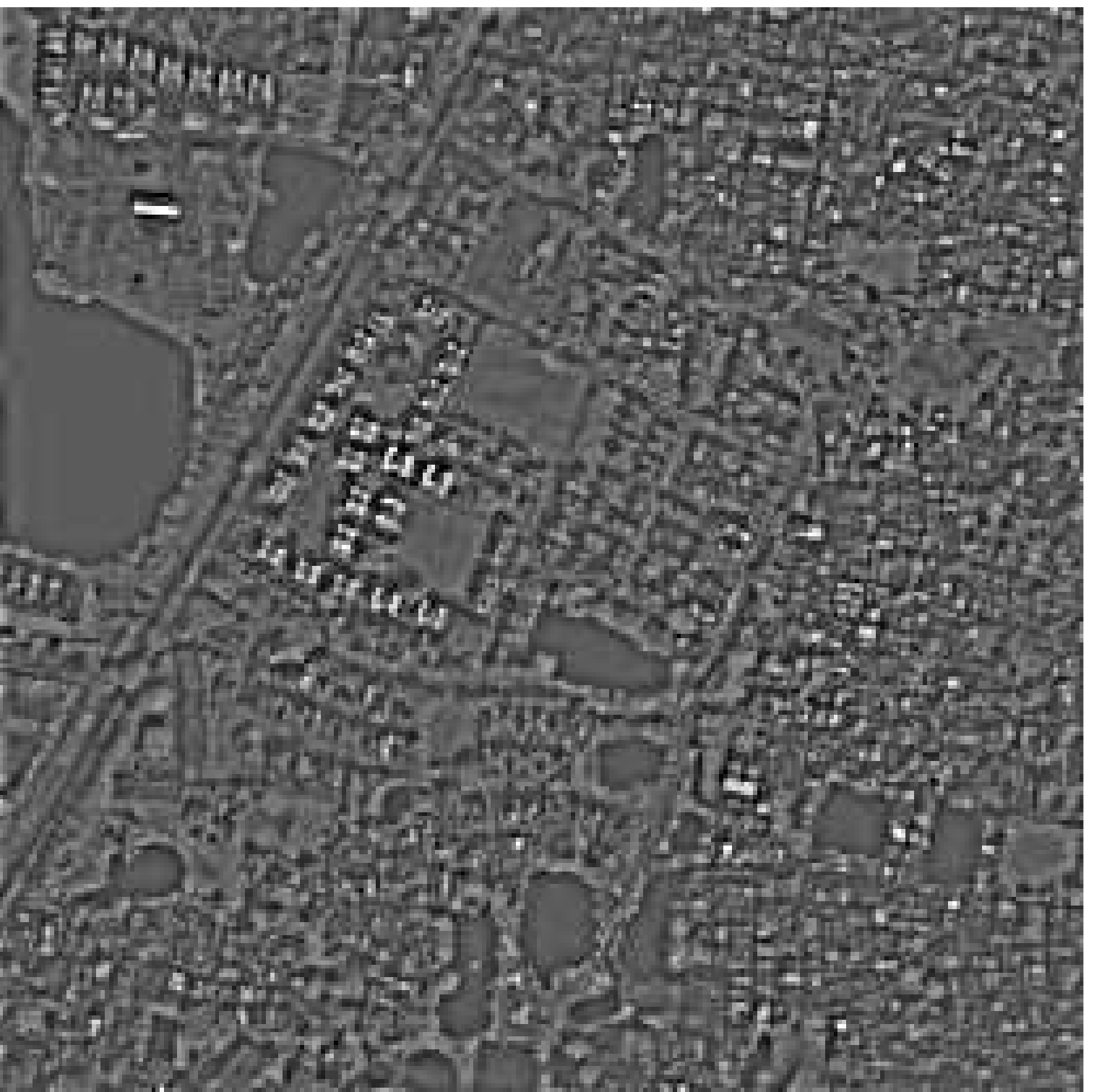} &
\includegraphics[width=0.14\paperwidth]{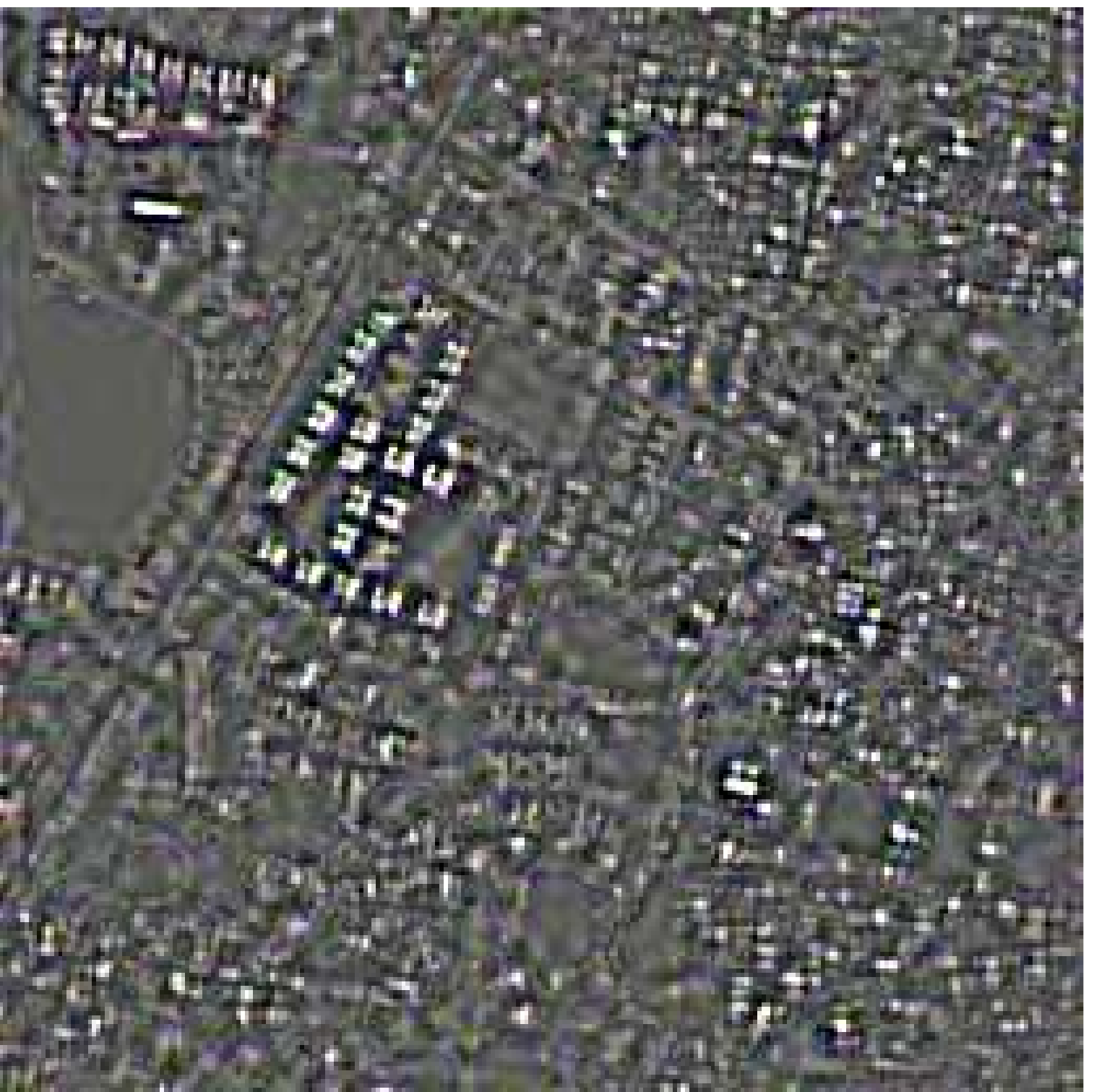} &
\includegraphics[width=0.14\paperwidth]{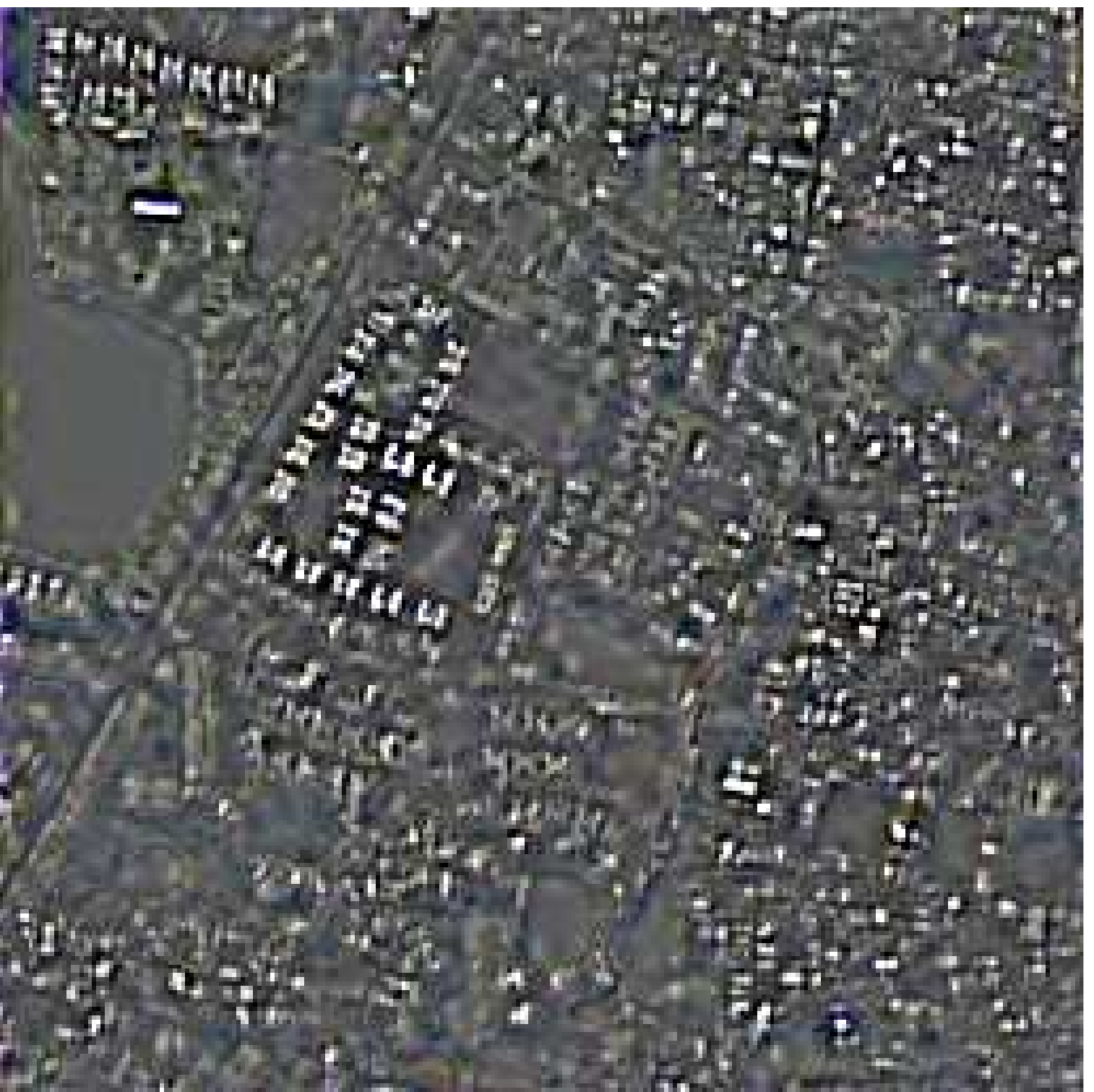} &
\includegraphics[width=0.14\paperwidth]{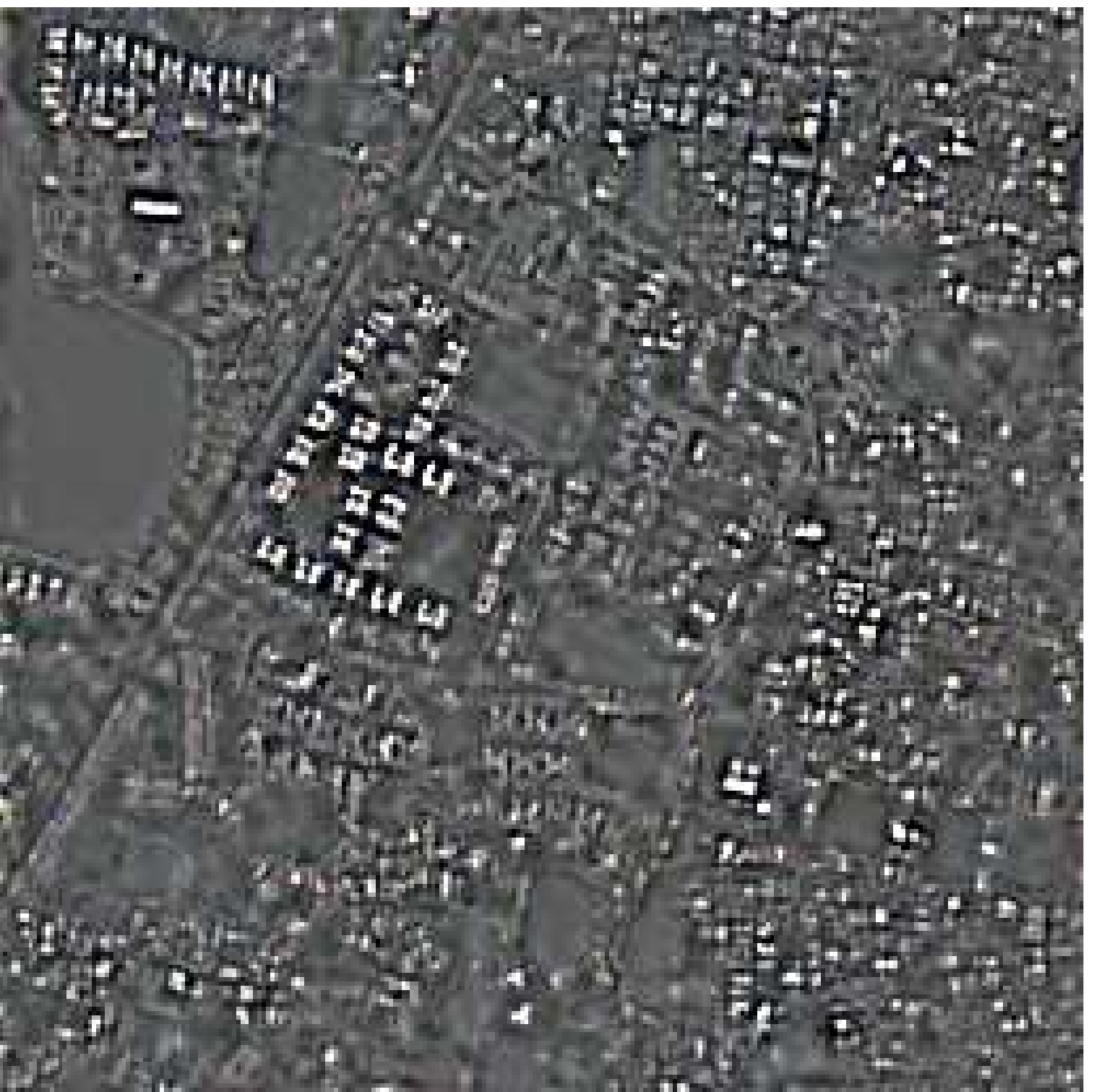} &
\includegraphics[width=0.14\paperwidth]{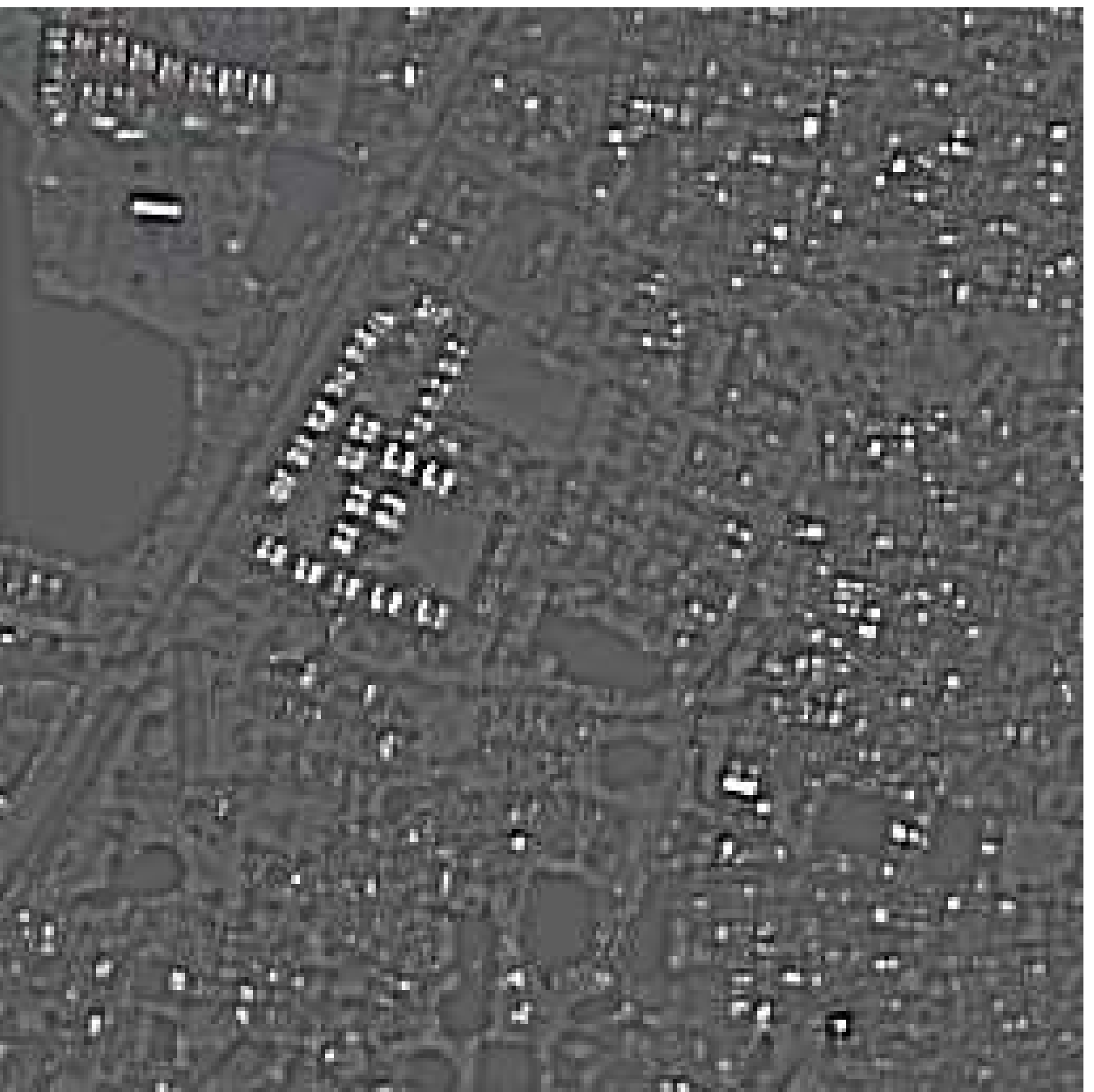} \\
  (f) & (g) & (h) &  (i) & (j)  \\
\\
\end{tabular}
\caption{Detail images of Quickbird dataset:(a) Ground-truth;  (b)GSA; (c)PRACS; (d)ATWT; (e)BDSD;  (f)GLP-CBD; (g)PNN; (h)DRPNN; (i)DiCNN1; (j)DiCNN2. }
\label{figure:detail qb}
\end{figure*}

\begin{table}[htp]
\small
\caption{Quality indexes of different pansharpening methods under reduced-resolution quality assessment on an $256\times256$ subscene of Quickbird data sets}
\centering
\begin{tabular}{c|ccccc}
\hline
{}&Q4&SAM& ERGAS &SCC&Time(s)\\
\hline
Refrence&1 &0 &0 &1&{}\\
\hline
\hline
EXP &0.6521 &3.6555 &3.0620 &0.6615 &{}\\
\hline
GSA&0.8321 &3.4710 &2.4565 &0.8485 &\textbf{0.13}\\
\hline
PRACS&0.7941 &3.0063 &2.2323 &0.8501 &0.20\\
\hline
ATWT&0.8361 &2.9223 &2.1011 &0.8699 &0.29\\
\hline
BDSD&0.8273 &3.8008 &2.6260 &0.8378 &0.15\\
\hline
GLP-CBD&0.8273 &3.5584 &2.5339 &0.8488 &0.41\\
\hline
\hline
PNN &0.8513 &3.2265 &2.0905 &0.9153 &0.31\\
\hline
DRPNN &0.8979 &2.5153 &1.6278 &0.9458 &0.37\\
\hline
DiCNN1 &\textbf{0.9023} &\textbf{2.4674} &\textbf{1.6062} &\textbf{0.9464} &0.32\\
\hline
DiCNN2 &0.8763 &2.7850 &1.7955 &0.9317&0.22\\
\hline
\end{tabular}
\label{table:reduceqb}
\end{table}

The dataset\footnote{http://glcf.umd.edu/data/quickbird/datamaps.shtml} represents a forest area of Sundarbans in India. It is obtained by the QuickBird sensor which provides a high-resolution PAN image with resolution of $0.6m$ and a four-band (blue, green, red and near infrared) MS image with resolution of $2.4m$. The radiometric resolution is also 11 bits. We selected different areas with the size of $256\times256$ pixels for reduced-resolution and full-resolution experiment respectively.

Table \ref{table:reduceqb} shows the reduced-resolution quality assessment on the Chilika Lake dataset. We can easily conclude that similar phenomena also arise in this dataset. CNN-based methods achieve better pansharpening quality than CS-based and MRA-based methods. DiCNN1 overpasses others in terms of Q4, SAM, ERGAS and SCC scores. DiCNN2 still wastes the least time among CNN-based methods, but lags behind DRPNN.

\begin{figure*}[t]\scriptsize
\centering
  \begin{tabular}{cccc}
\includegraphics[width=0.14\paperwidth]{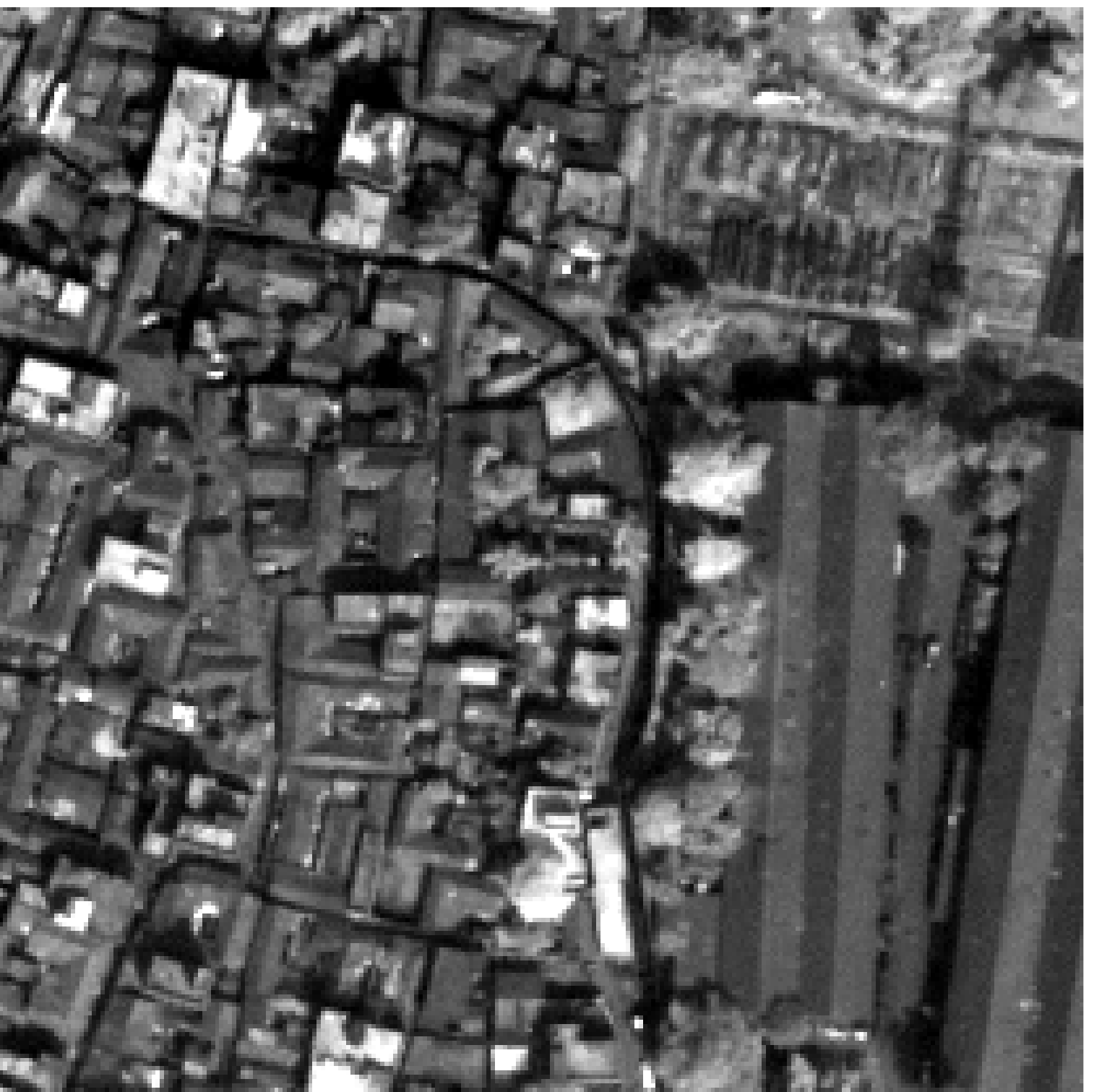} &
\includegraphics[width=0.14\paperwidth]{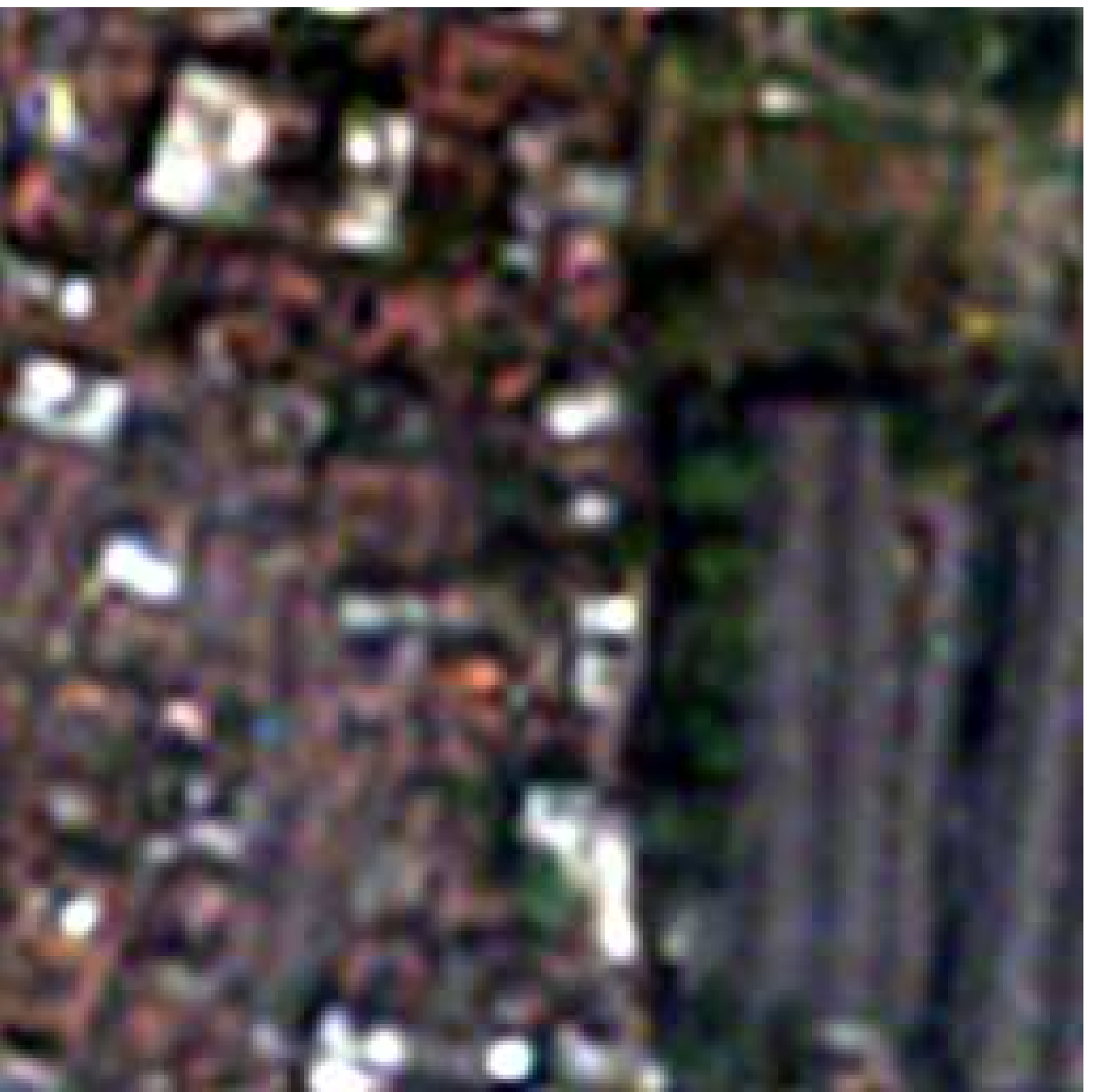} &
\includegraphics[width=0.14\paperwidth]{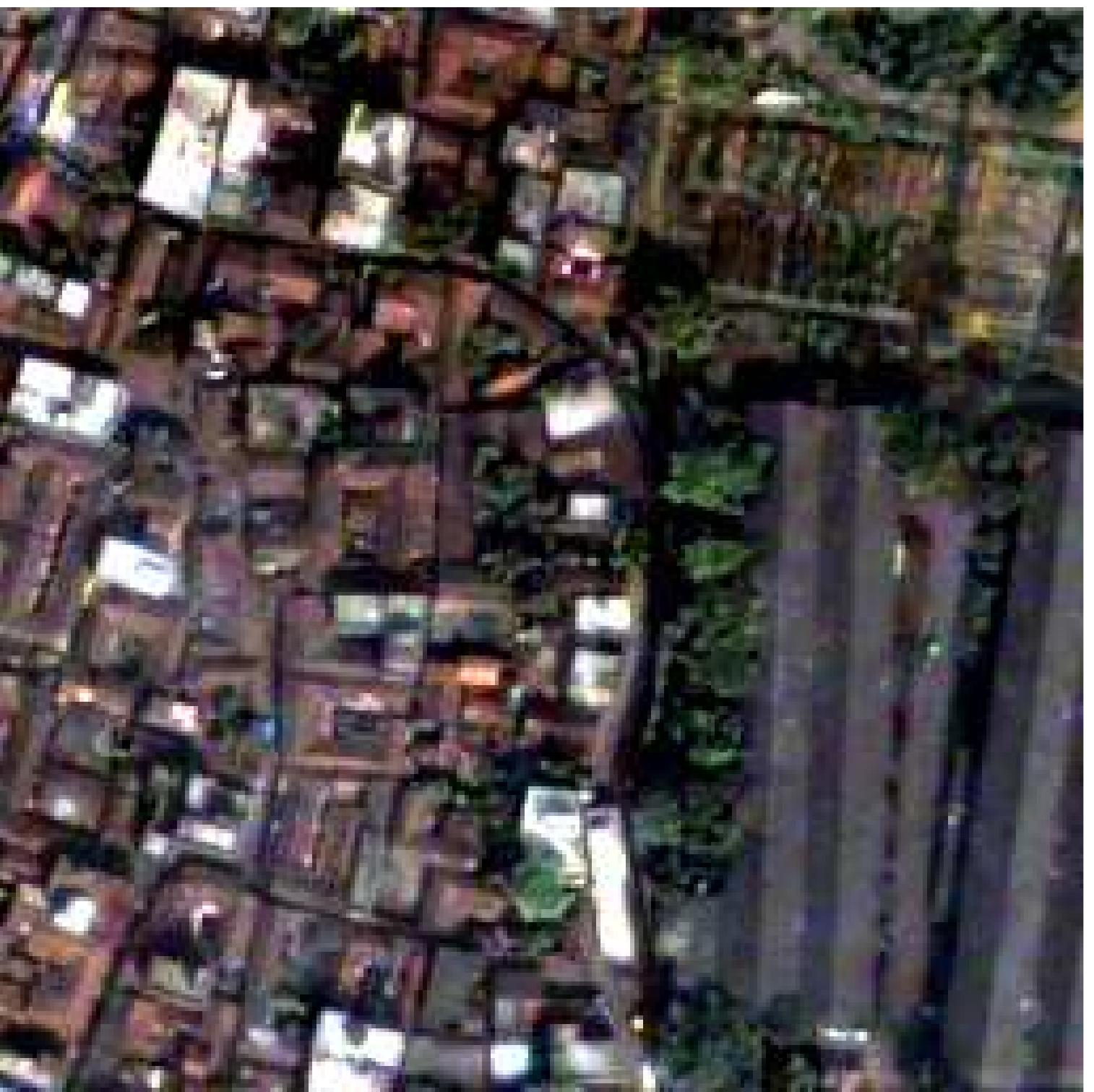} &
\includegraphics[width=0.14\paperwidth]{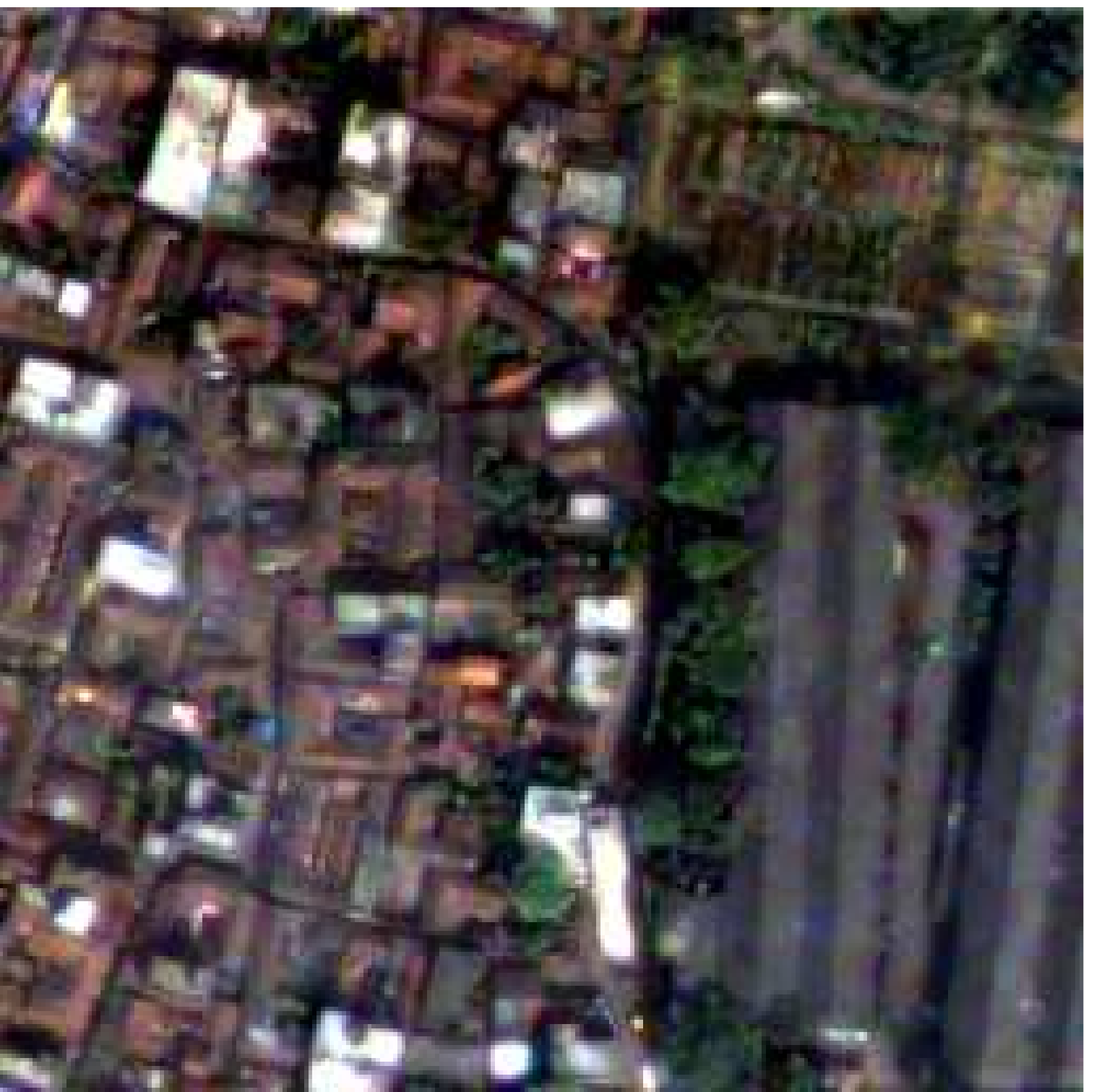} \\
(a) & (b) & (c) & (d) \\

\includegraphics[width=0.14\paperwidth]{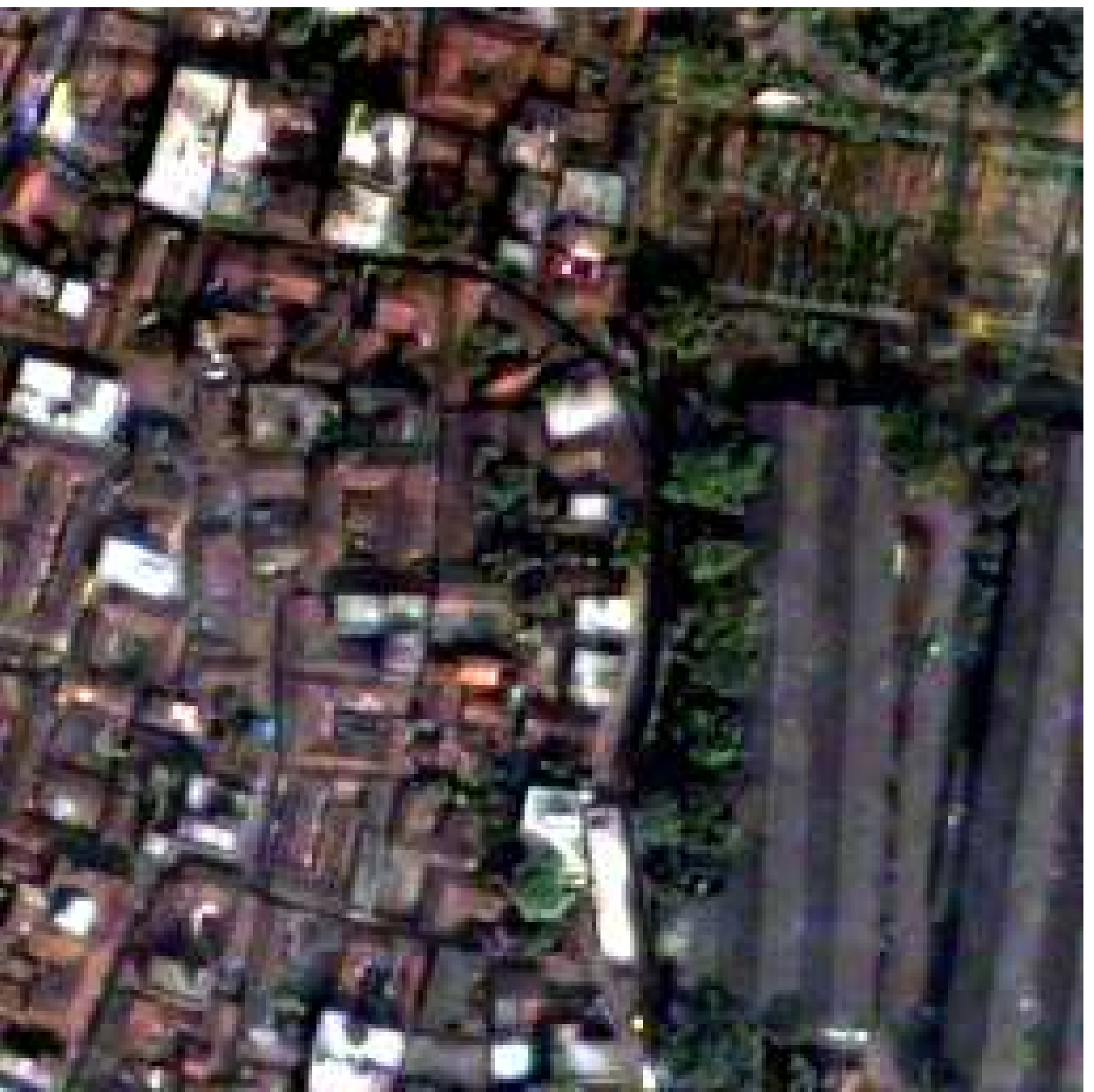} &
\includegraphics[width=0.14\paperwidth]{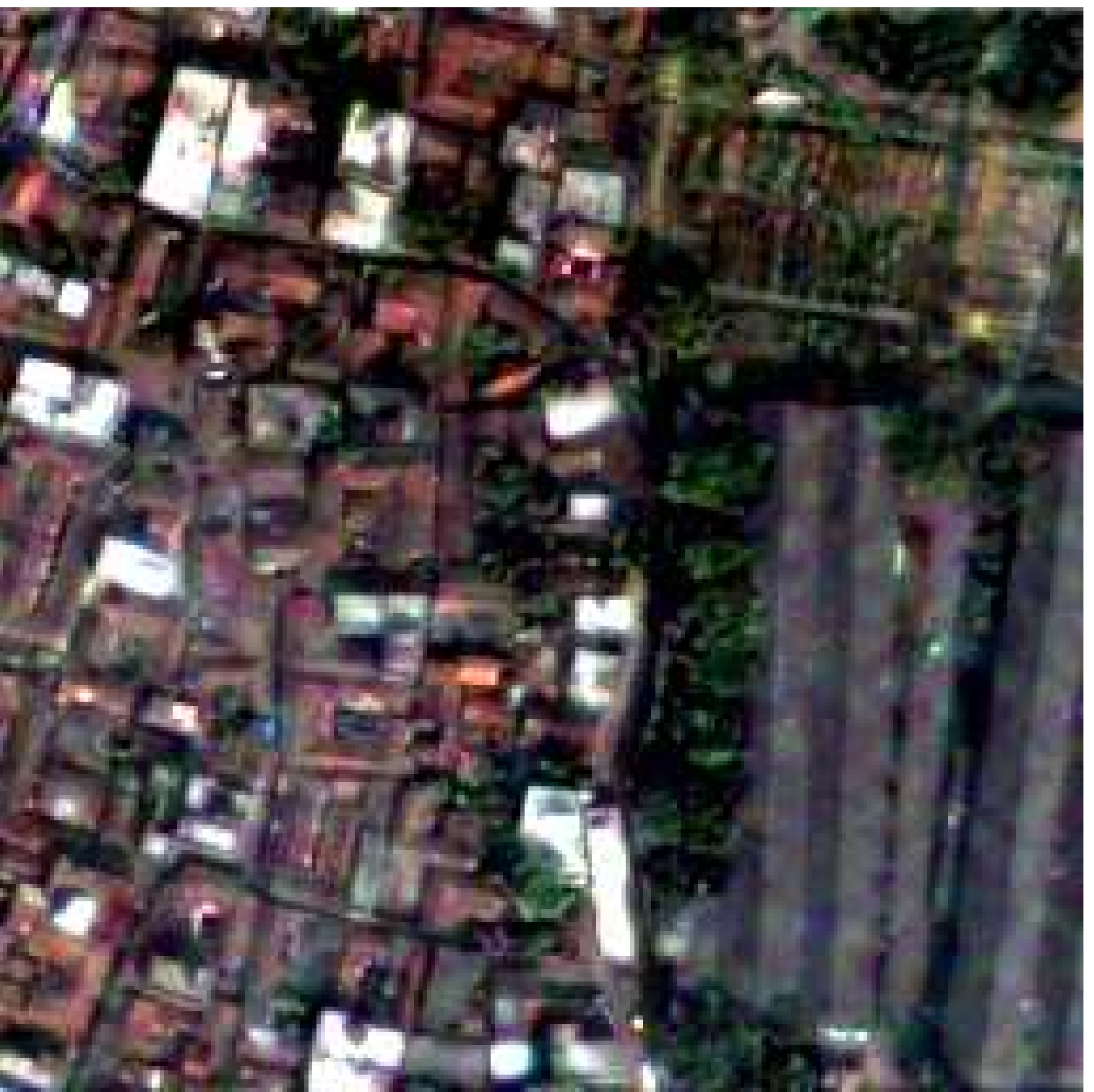} &
\includegraphics[width=0.14\paperwidth]{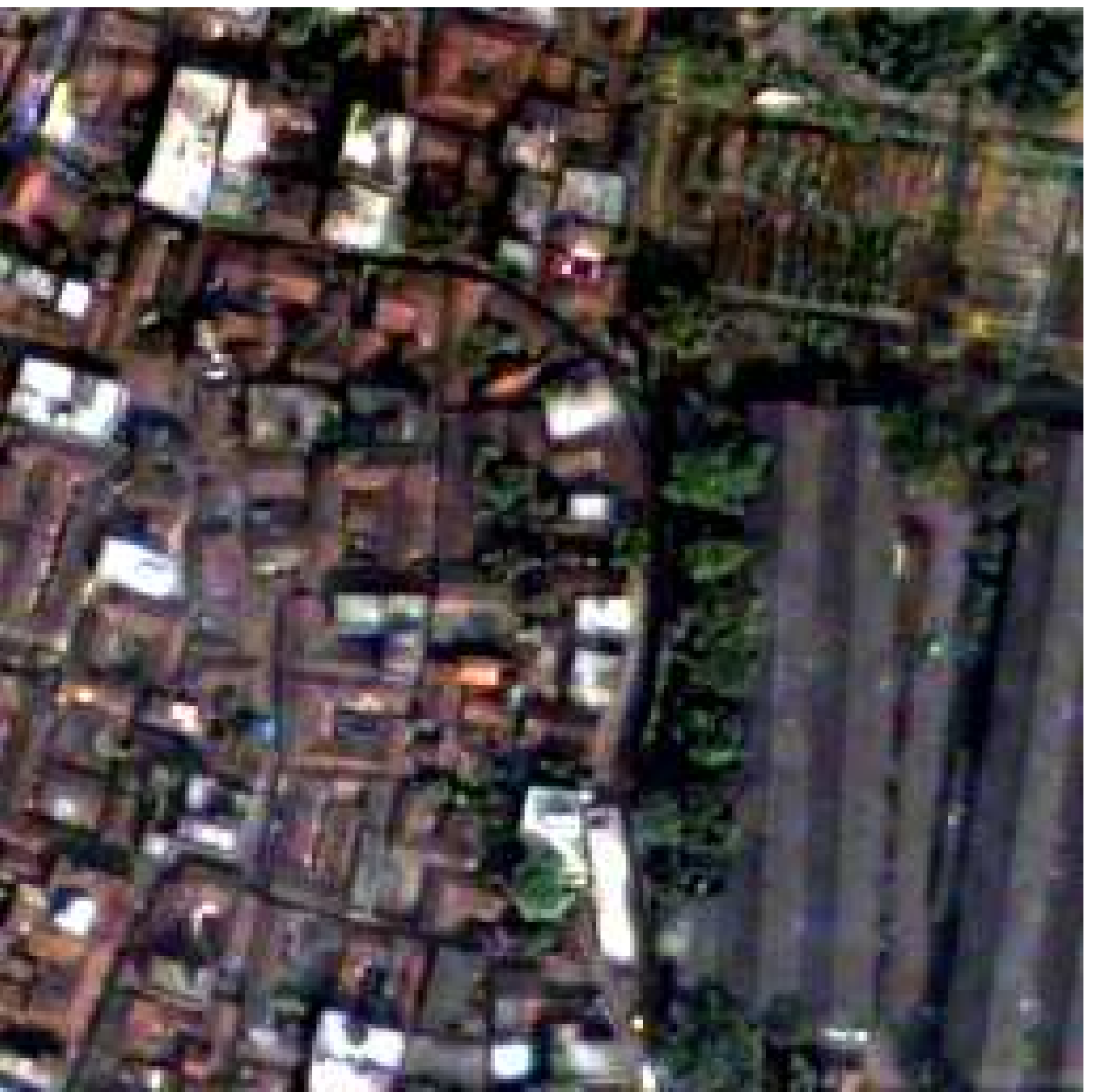} &
\includegraphics[width=0.14\paperwidth]{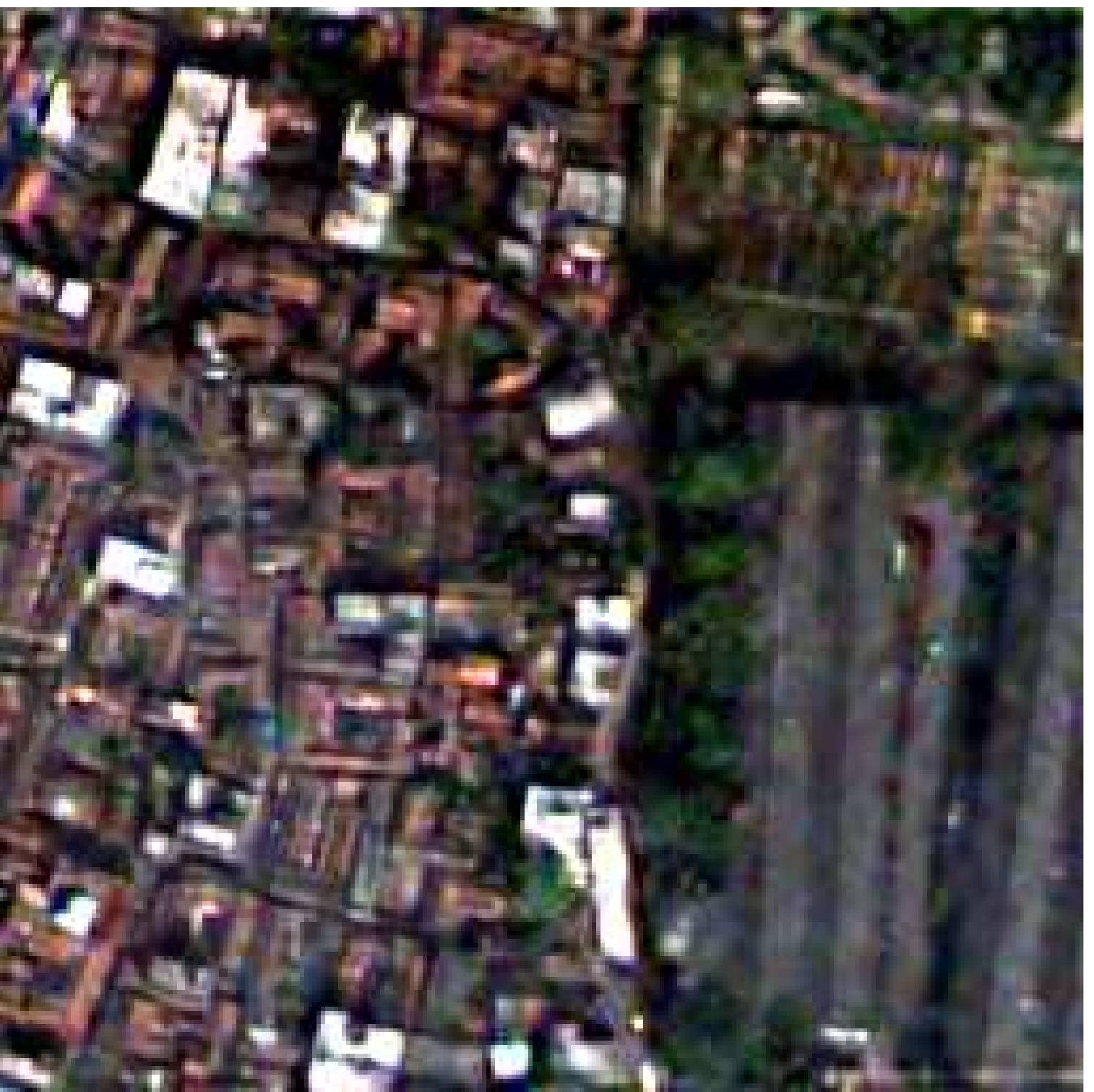} \\
 (e) & (f) &(g) & (h)\\

\includegraphics[width=0.14\paperwidth]{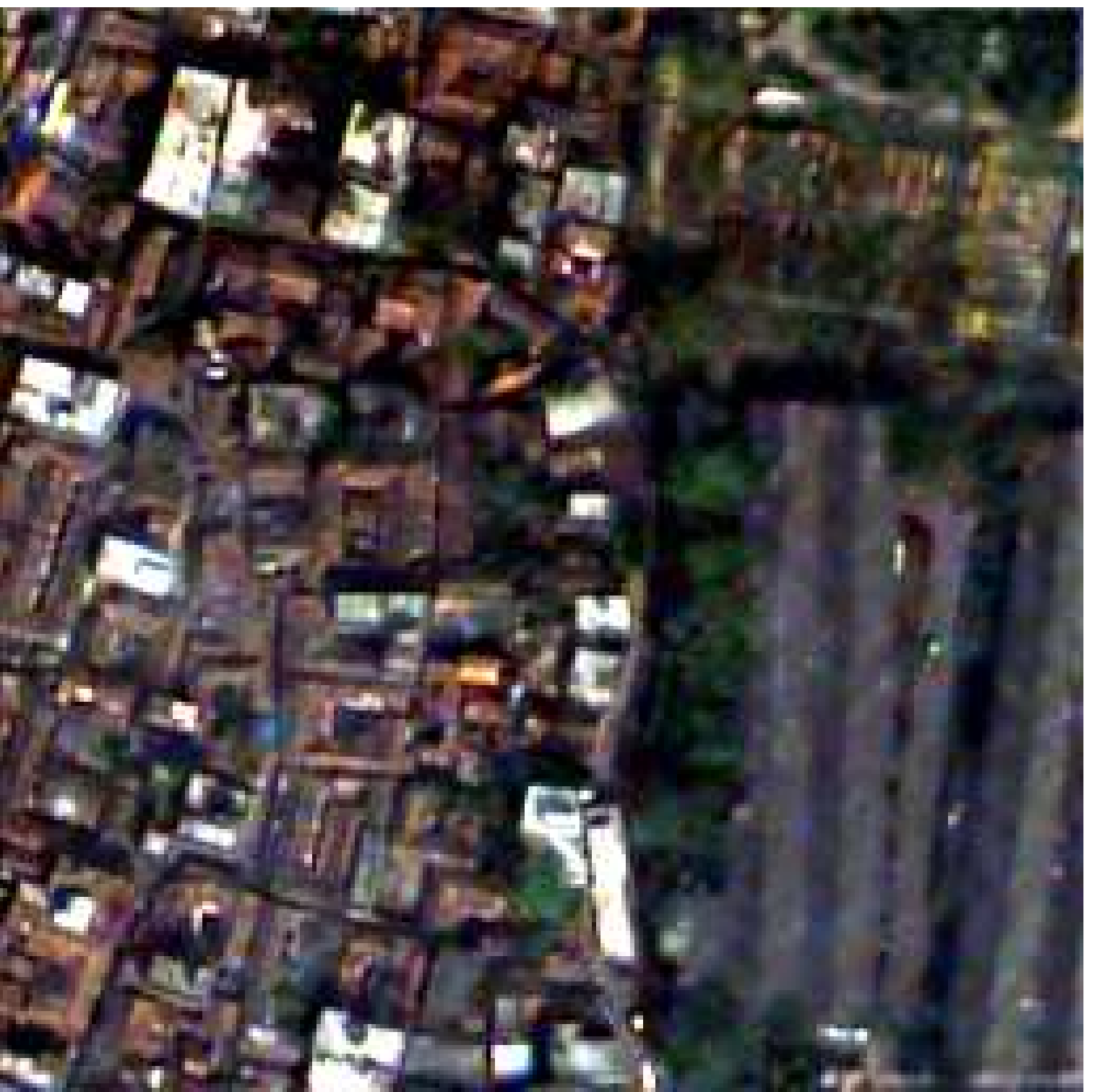} &
\includegraphics[width=0.14\paperwidth]{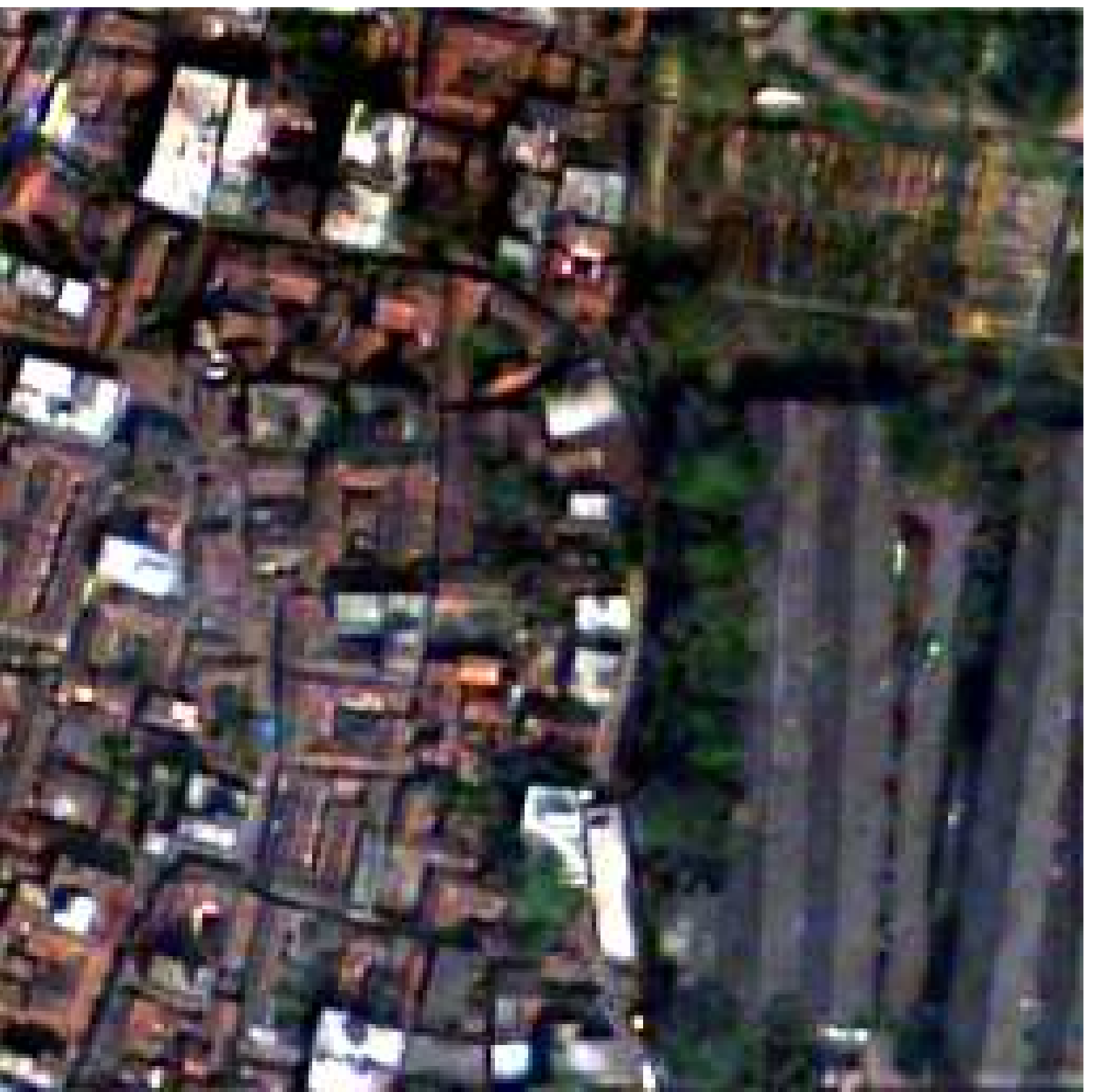} &
\includegraphics[width=0.14\paperwidth]{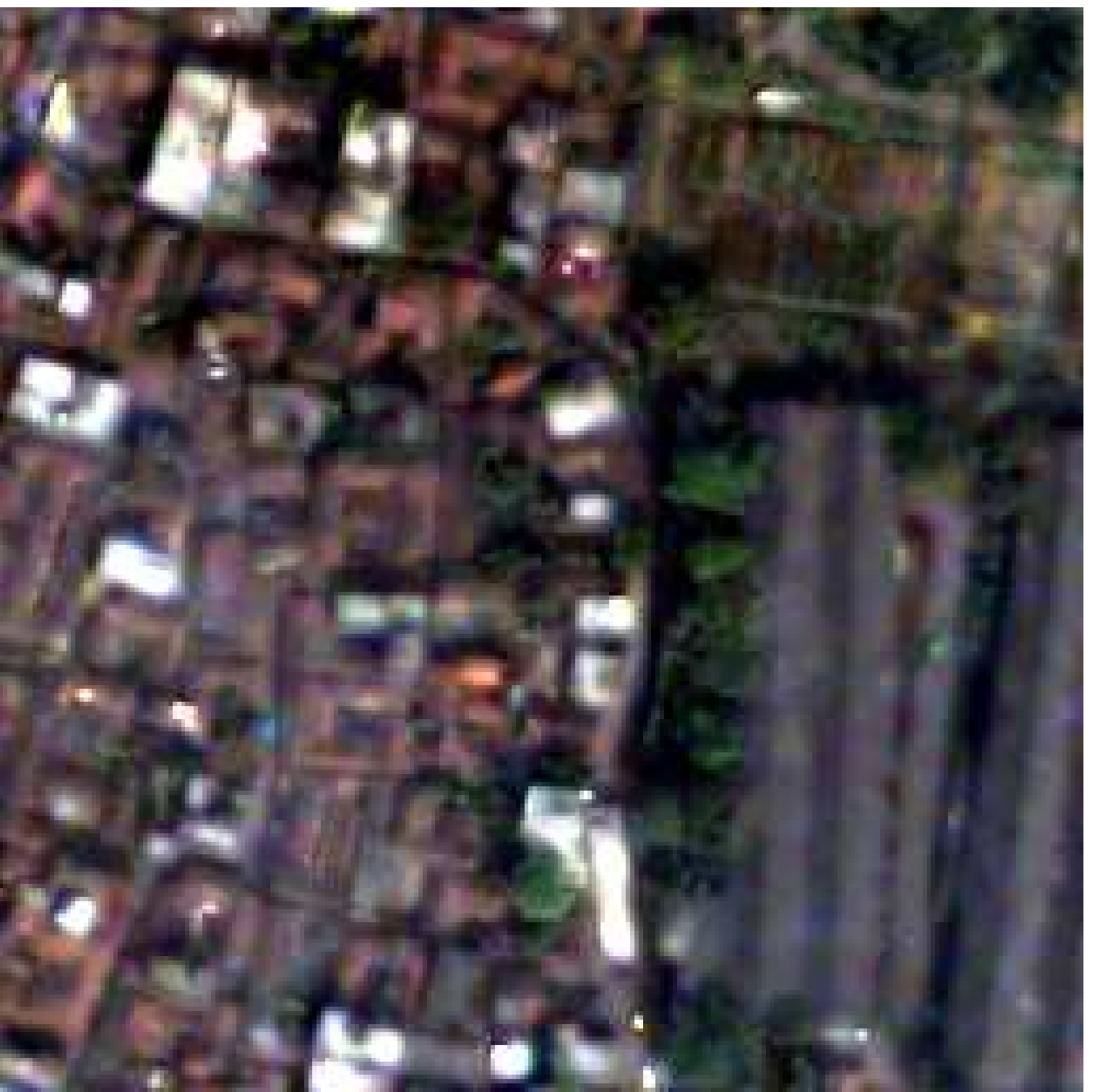} \\
 (i) & (j) & (k)  \\
\\
\end{tabular}
\caption{Full-resolution pansharpening results for Quickbird dataset: (a) Ground-truth; (b)EXP; (c)GSA; (d)PRACS; (e)ATWT; (f)BDSD;  (g)GLP-CBD; (h)PNN; (i)DRPNN; (j)DiCNN1; (k)DiCNN2.}
\label{figure:map:full-qb}
\end{figure*}

Fig. \ref{figure:map:qb} displays the reduced-resolution experimental results. DiCNN1, DiCNN2 and DRPNN look much more similar to the original MS image, but DiCNN2 exhibits less ringing artifacts, such as the edges of the lakes in the leftmost part of Fig. \ref{figure:map:qb}(i)-(k). This phenomenon occurs more frequently in PNN. Fig. \ref{figure:detail qb} shows the detail images learned from various methods, which also support the observations above. Fig. \ref{figure:map:full-qb} displays the full-resolution experimental results. PRACS and DiCNN2 introduce more spatial blurring than PNN, while DiCNN1 exhibits less artifacts than DRPNN and PNN.

\subsection{Experiment 4: Transfer Learning}

To demonstrate the robustness of DiCNN2 under the situation that the number of bands of the test MS image has varied, we use the WorldView-2 Washington dataset and IKONOS Hobart Dataset in this experiment. Here, DiCNN2 is first trained on the original dataset. Then some of the MS bands are removed and the final convolution layers are fine-tuned to accommodate the current number of bands with $1.0 \times 10^4$ training iterations, much less than that in the previous training step. For WorldView-2 Washington dataset with 8 MS bands, 4 bands are removed. For IKONOS Hobart Dataset with 4 MS bands, 1 band is removed.

\begin{table}[htp]
\small
\caption{Quality indexes of CNN-based methods on an $256\times256$ subscene of four-band WorldView-2 dataset}
\centering
\begin{tabular}{c|ccccc}
\hline
{}&Q4&SAM& ERGAS &SCC&Time(s)\\
\hline
Refrence&1 &0 &0 &1&{}\\
\hline
\hline
PNN &0.9308 &3.4808 &2.5678 &0.9343&222\\
\hline
DRPNN &0.9462 &3.0384 &2.4160 &0.9383&360\\
\hline
DiCNN1 &0.9497 &2.8630 &2.3080 &0.9407 &327\\
\hline
DiCNN2 &\textbf{0.9499} &\textbf{2.7148} &\textbf{2.2853}  &\textbf{0.9420} &\textbf{173}\\
\hline
\end{tabular}
\label{table:transwv}
\end{table}

Table \ref{table:transwv} shows a quantitative assessment result on WorldView-2 Washington dataset. As shown in the table, DiCNN2 yields the best scores in all evaluation metrics. It is remarkable that the time DiCNN2 needs for the training phase is less than half of the longest one, which results from the fact that DiCNN2 only needs to fine-tune the final convolutional layer.

We also apply a similar experiment using the IKONOS data. Since the IKONOS dataset consists of 4 bands, we randomly choose three of them for testing. The four-band dataset is used to train DiCNN2, while the three-band one is applied to fine-tune the last layer of DiCNN2 and train other CNN-based methods.

\begin{table}[htp]
\small
\caption{Quality indexes of CNN-based methods on an $256\times256$ subscene of three-band IKONOS dataset}
\centering
\begin{tabular}{c|ccccc}
\hline
{}&Q4&SAM& ERGAS &SCC&Time(s)\\
\hline
Refrence&1 &0 &0 &1&{}\\
\hline
\hline
PNN &0.8748 &2.4828 &3.0206 &0.8988&176\\
\hline
DRPNN &0.8928 &2.8445 &3.0683 &0.9093&355\\
\hline
DiCNN1 &\textbf{0.8989} &\textbf{1.9336} &2.7109 &0.9144 &335\\
\hline
DiCNN2 &0.8986&2.0317 &\textbf{2.6908}  &\textbf{0.9181} &\textbf{160}\\
\hline
\end{tabular}
\label{table:transik}
\end{table}

Table \ref{table:transik} depicts the pansharpening results obtained by different CNN-based methods. As can be observed, DiCNN2 outperforms others in most quality indexes. In addition, although DiCNN1 attains comparative results with regards to DiCNN2, the training time of the latter is far less than the former.



\section{Conclusions and Future Lines}
\label{sec:Conclusions}

In this paper, we have developed two CNN-based pansharpening methods, i.e., DiCNN1 and DiCNN2, based on the detail injection framework (DiPAN) which classical CS/MRA-base pansharpening methods can be ascribed into. In DiCNN1 and DiCNN2, the MS details are learned in an end-to-end manner, which has explicit physical meaning and avoids separately dealing with injection gains and PAN details as it is the case in traditional CS and MRA methods. Our DiCNN1 and DiCNN2 methods can gain low initial loss, which tends to  yield faster convergence, and exhibit excellent pansharpening performance. Particularly, DiCNN2 can additionally realize transfer learning when the type of the MS image or the PAN image changes, which is a highly desirable property. In the future, we will explore the possibility of designing pansharpening CNNs with more hidden layers and more complex inter-connections among multiple convolution layers.


\ifCLASSOPTIONcaptionsoff
  \newpage
\fi

\bibliographystyle{IEEEtran}
\bibliography{PanRef}

\end{document}